\documentclass[12pt]{article}
\usepackage{a4wide,latexsym,amsfonts,amsmath,amssymb,epsf,bbm}
\usepackage[dvips]{graphics}
\usepackage[mathscr]{eucal}

\newcounter{defthm}
\newtheorem{defthm}{\whattheorem}[section]
\newcommand\dt[1]  {\noindent\def\whattheorem{#1}\pagebreak[0]\begin{defthm}{}%
                   \samepage{$\!\!${\rm:}\nopagebreak\\[-1.91em]{}}\end{defthm}}
\newcommand\dtl[2] {\noindent\def\whattheorem{#1}\pagebreak[0]\begin{defthm}{}%
                   \samepage{$\!\!${\rm:}\label{#2}\nopagebreak\\[-1.91em]{}}%
                   \end{defthm}}

\newcommand\includeourbeautifulpicture[1] {{\begin{picture}(0,0)(0,0)
                   \scalebox{.38}{\includegraphics{#1.eps}} \end{picture}}}
\newcommand\includeourbeautifulsmallpicture[1] {{\begin{picture}(0,0)(0,0)
                   \scalebox{.29}{\includegraphics{#1.eps}} \end{picture}}}

\global\def\draftcontrol{0}

\def\internal {\begin{quote}\small INTERNAL REMARK:\\ \bf }
\def\eternal {\end{quote} \mbox{$\ $}\normalsize\rm }

\def\calc  {{\mathcal C}}
\def\cald  {{\mathcal D}}

\def\calh  {{\mathcal H}}
\def\cali  {{\mathcal I}}

\def\calk  {{\mathcal K}}

\def\caln  {{\mathcal N}}

\def\cals  {{\mathcal S}}

\def\calu  {{\mathcal U}}

\def\calz  {{\mathcal Z}}
\def\complex       {{\mathbbm C}}
\def\complexx      {{\mathbbm C}^{\times}_{}}

\def\rationals     {{\mathbb Q}}

\def\zet           {{\mathbb Z}}

\def\AB            {{\rm A}\!{\rm B}}
\def\alg           {algebra}
\def\alph          {\alpha}
\def\Atop          {\mbox{$A_{\rm top}$}}

\newcommand\bb[2]  {{{}_{#2}b}}
\newcommand\bbb[2] {{{}_{#2\!}b}}
\def\bc            {boundary condition}
\def\be            {\begin{equation}}
\def\bea           {\begin{equation}\begin{array}l}
\def\bearl         {\begin{array}{l}}
\def\bearll        {\begin{array}{ll}}
\def\berta         {Schellekens algebra}
\def\Beta          {{\lambda}}
\def\Betabar       {{\bar\lambda}}
\def\bfe           {{\bf1}}
\def\bicals        {{\mathcal S}}
\def\bicalu        {{\mathcal U}}
\newcommand\biphi[2]{{{}_{#1\!}\phi_{#2}}}
\newcommand\bivarphi[2]{{{}_{#1\!}\varphi_{#2}}}
\newcommand\bL[3]  {{{}_{#2}b_{#3}}}
\newcommand\brdiYabYc[5]{\begin{picture}(0,0)
                   \put(-2,-8){\scriptsize$#1$}
                   \put(28,-8){\scriptsize$#2$}
                   \put(50,-8){\scriptsize$#3$}
                   \put(25.1,71.1){\scriptsize$#4$}
                   \put(15.4,40.8){\scriptsize$#5$} \end{picture}}  
\newcommand\brdiYaYbc[5]{\begin{picture}(0,0)
                   \put(-2,-8){\scriptsize$#1$}
                   \put(21,-8){\scriptsize$#2$}
                   \put(50,-8){\scriptsize$#3$}
                   \put(25.1,71.1){\scriptsize$#4$}
                   \put(34.5,40){\scriptsize$#5$} \end{picture}}  
\newcommand\brdiYacYb[5]{\begin{picture}(0,0)
                   \put(-2,-8){\scriptsize$#1$}
                   \put(21,-8){\scriptsize$#2$}
                   \put(50,-8){\scriptsize$#3$}
                   \put(25.1,71.1){\scriptsize$#4$}
                   \put(16.1,41.2){\scriptsize$#5$} \end{picture}}  
\newcommand\brdiYbYac[5]{\begin{picture}(0,0)
                   \put(-2,-8){\scriptsize$#1$}
                   \put(21.3,-8){\scriptsize$#2$}
                   \put(50,-8){\scriptsize$#3$}
                   \put(25.1,71.1){\scriptsize$#4$}
                   \put(37.7,42.3){\scriptsize$#5$} \end{picture}}  
\def\cala          {{\mathfrak A}}
\def\calcA         {{\mathcal C}_{\!A}}
\def\calcAA        {{\mathcal C}_{\!A|A}}
\def\calpic        {{\cal P}\!\mbox{\sl ic}}
\def\calsA         {\cals_{\!A}}
\def\cat           {category}
\def\cats          {categories}
\def\cft           {conformal field theory}
\def\cfts          {conformal field theories}
\def\central       {central} 
\def\Chi           {{\mathcal X}}

\def\citei         {[\,I\,]}  

\def\citeii        {[\,II\,]} 

\def\citeiv        {\cite{fuRsXX}}
\def\citeIaII      {[I,\,II]} 
\def\citeOaI       {[1,\,I]} 
\def\citeOaIaII    {[1,\,I,\,II]} 
\def\cir           {\,{\circ}\,}

\def\Dim           {{\rm Dim}}
\def\dimAA         {{\rm dim}_{\!A|A}}
\def\dimc          {{\rm dim}_\complex}

\def\dyd           {Dynkin diagram}
\def\ee            {\end{equation}}
\def\eE            {{\rm e}}
\def\eear          {\end{array}}
\def\EL            {{\rm E}\!{\rm M}}

\def\End           {{\rm End}}
\def\EndA          {{\rm End}_{\!A}}

\def\EndAAv        {{\rm End}_{\!{A_1}|{A_2}}}
\newcommand\epicture[2] {\end{picture}\\{}\\[#1.#2em]\end{array}}
\def\eps           {\varepsilon}
\def\eq            {\,{=}\,}
\newcommand\erf[1] {(\ref{#1})}

\def\FF            {{\sf F}}
\def\findim        {fi\-ni\-te-di\-men\-si\-o\-nal}
\newcommand\Frac[2]{\mbox{\large$\frac{#1}{#2}$}}
\newcommand\Fs[6]  {{\sf F}_{\,{#5}\,{#6}}^{\,({#1}\,{#2}\,{#3})\,{#4}}}
\def\fts           {field theories}
\newcommand\fusiYaYbc[5]{\begin{picture}(0,0)
                   \put(-2,-8){\scriptsize$#1$}
                   \put(25,-8){\scriptsize$#2$}
                   \put(50,-8){\scriptsize$#3$}
                   \put(24.8,71.1){\scriptsize$#4$}
                   \put(35.2,38.7){\scriptsize$#5$} \end{picture}}  
\newcommand\fusiYabYc[5]{\begin{picture}(0,0)
                   \put(-2,-8){\scriptsize$#1$}
                   \put(25,-8){\scriptsize$#2$}
                   \put(50,-8){\scriptsize$#3$}
                   \put(24.8,71.1){\scriptsize$#4$}
                   \put(15.4,38.9){\scriptsize$#5$} \end{picture}}  
\newcommand\fusiYaYbYcd[7]{\begin{picture}(0,0)
                   \put(-2,-8){\scriptsize$#1$}
                   \put(24,-8){\scriptsize$#2$}
                   \put(50,-8){\scriptsize$#3$}
                   \put(76,-8){\scriptsize$#4$}
                   \put(37.8,65.5){\scriptsize$#5$}
                   \put(48.5,34){\scriptsize$#6$}
                   \put(60.5,21){\scriptsize$#7$} \end{picture}}  
\newcommand\fusiYaYbcd[7]{\begin{picture}(0,0)
                   \put(-2,-8){\scriptsize$#1$}
                   \put(24,-8){\scriptsize$#2$}
                   \put(50,-8){\scriptsize$#3$}
                   \put(76,-8){\scriptsize$#4$}
                   \put(37.8,65.5){\scriptsize$#5$}
                   \put(48.5,33){\scriptsize$#6$}
                   \put(40,20.3){\scriptsize$#7$} \end{picture}}
\newcommand\fusiYabcYd[7]{\begin{picture}(0,0)
                   \put(-2,-8){\scriptsize$#1$}
                   \put(24,-8){\scriptsize$#2$}
                   \put(50,-8){\scriptsize$#3$}
                   \put(76,-8){\scriptsize$#4$}
                   \put(37.8,65.5){\scriptsize$#5$}
                   \put(25.8,33){\scriptsize$#6$}
                   \put(36,20.3){\scriptsize$#7$} \end{picture}} 
\newcommand\fusiYabYcYd[7]{\begin{picture}(0,0)
                   \put(-2,-8){\scriptsize$#1$}
                   \put(24,-8){\scriptsize$#2$}
                   \put(50,-8){\scriptsize$#3$}
                   \put(76,-8){\scriptsize$#4$}
                   \put(37.8,65.5){\scriptsize$#5$}
                   \put(14.8,20){\scriptsize$#6$}
                   \put(27.2,34){\scriptsize$#7$} \end{picture}} 
\newcommand\fusiYabYcd[7]{\begin{picture}(0,0)
                   \put(-2,-8){\scriptsize$#1$}
                   \put(27,-8){\scriptsize$#2$}
                   \put(47,-8){\scriptsize$#3$}
                   \put(76,-8){\scriptsize$#4$}
                   \put(37.8,65.5){\scriptsize$#5$}
                   \put(22.8,28){\scriptsize$#6$}
                   \put(55.2,28){\scriptsize$#7$} \end{picture}}

\def\Hom           {{\rm Hom}}
\def\HomA          {{\rm Hom}_{\!A}}
\def\HomAA         {{\rm Hom}_{\!A|A}}
\def\HomAAv        {{\rm Hom}_{\!{A_1}|{A_2}}}
\def\hsfa          {haploid special Frobenius algebra}
\newcommand\hsp[1] {\mbox{\hspace{#1 em}}}
\def\hy            {$\mbox{-\hspace{-.66 mm}-}$}
\def\ibar          {{\bar\imath}}
\def\iBar          {{\overline\imath}}
\def\id            {\mbox{\sl id}}
\def\ii            {{\rm i}}
\def\II            {{\mathcal I}}

\def\iN            {\,{\in}\,}

\def\IndA          {{\rm Ind}_{\!A}}
\def\IndAA         {{\rm Ind}_{\!A|A}}
\def\IndAAv        {{\rm Ind}_{\!{A_1}|{A_2}}}

\def\iralpha       {\beta}
\def\irbeta        {\alpha}
\def\ircalc        {{\mathcal D}}

\def\Io            {I_\circ}
\def\JJ            {\mathcal J}
\def\kapa          {\varkappa} 
\def\kappb         {{\kappa'}}

\def\ksb           {Kreu\-zer\hy Schel\-le\-kens bihomomorphism} 
\long\def\labl#1   {\label{#1}\ee }

\def\llb           {\big(}
\def\lrb           {\big)}

\def\Mho           {\Omega}
\def\modinv        {modular invarian}
\def\Modinv        {Modular invarian}
\newcommand\nxt[1] {\\\raisebox{.12em}{\rule{.35em}{.35em}}\hsp{.6}#1}
\newcommand\Nxt[1] {\raisebox{.12em}{\rule{.35em}{.35em}}\hsp{.6}#1}
\def\NXT           {\raisebox{.12em}{\rule{.35em}{.35em}}}
\def\obj           {{\mathcal O}bj}
\def\one           {{\bf 1}}
\def\onedim        {one-dimen\-sional}
\def\oti           {\,{\otimes}\,}
\def\Oti           {{\otimes}}
\def\Pic           {{\rm Pic}}
\def\Pico          {{\rm Pic}^{\circ\!}}
\def\Pics          {\mbox{\sl Pic}}
\def\PF            {Per\-ron\hy Fro\-be\-ni\-us} 
\def\qed           {\hfill\checkmark\medskip}
\def\QF            {{\rm Q}\!{\rm F}}

\long\def\query#1{\hskip 0pt{\vadjust{\everypar={}\small\vtop to 0pt{\hbox{}%
     \vskip -13pt\rlap{\hbox to 47.0pc{\hfil{\vtop{\hsize=8pc\tolerance=6000%
     \hfuzz=.5pc\rightskip=0pt plus 3em\noindent#1}}}}\vss}}}}%
\def\r             {\lambda_{\rm PF}} 
 
\def\rep           {representation} 

\def\rhs           {right hand side}
\def\rmA           {{\rm A}}
\def\rmd           {{\rm d}}
\def\RR            {{\sf R}}
\newcommand\Rs[4]  {{\sf R}^{#1\,(#2\,#3)#4}}
\newcommand\Rsm[3] {{\sf R}^{{}^{\!-}(#1\,#2)\,#3}}
\newcommand\Rss[3] {{\sf R}^{(#1\,#2)\,#3}} 
\newcommand\sect[1]{\section{#1}\setcounter{equation}0\setcounter{defthm}0}
\def\sfa           {special Frobenius algebra}
\def\Sg            {{\mathscr S}^g} 
\def\ssfa          {symmetric special Frobenius algebra}

\def\th            {\theta}
\def\Ti            {{\boxtimes}}
\newcommand\tildeb[2]{{b_{#2}}}
\def\Times         {\,{\times}\,}
\def\To            {\mapsto}

\def\tr            {{\rm tr} \, }
\def\tSg           {\widetilde{\mathscr S}^g} 
\def\tP            {\widetilde{P}} 
\def\tM            {\widetilde{M}} 

\def\U             {L}

\def\wzwm          {WZW\hy model}

\def\XI            {\Xi}

\catcode`\@=11

\newif\if@fewtab\@fewtabtrue
{\count255=\time\divide\count255 by 60
\xdef\hourmin{\number\count255}
\multiply\count255 by-60\advance\count255 by\time
\xdef\hourmin{\hourmin:\ifnum\count255<10 0\fi\the\count255}}
\def\ps@draft{\let\@mkboth\@gobbletwo
    \def\@oddhead{}
    \def\@oddfoot{\hbox to 7 cm{\tiny \versionno
       \hfil}\hskip -7cm\hfil\rm\thepage \hfil {\tiny\draftdate}}
    \def\@evenhead{}\let\@evenfoot\@oddfoot}
\def\draftdate{\number\month/\number\day/\number\year\ \ \ \hourmin }

\global\def\draftcontrol{0}
\def\citen#1{\if@filesw \immediate\write \@auxout {\string\citation{#1}}\fi%
\@tempcntb\m@ne \let\@h@ld\relax \def\@citea{}%
\@for \@citeb:=#1\do {\@ifundefined {b@\@citeb}%
    {\@h@ld\@citea\@tempcntb\m@ne{\bf ?}%
    \@warning {Citation `\@citeb ' on page \thepage \space undefined}}%
    {\@tempcnta\@tempcntb \advance\@tempcnta\@ne
    \setbox\z@\hbox\bgroup\ifcat0\csname b@\@citeb \endcsname \relax
    \egroup \@tempcntb\number\csname b@\@citeb \endcsname \relax
    \else \egroup \@tempcntb\m@ne \fi \ifnum\@tempcnta=\@tempcntb
    \ifx\@h@ld\relax \edef \@h@ld{\@citea\csname b@\@citeb\endcsname}%
    \else \edef\@h@ld{\hbox{--}\penalty\@highpenalty
    \csname b@\@citeb\endcsname}\fi
    \else \@h@ld\@citea\csname b@\@citeb \endcsname \let\@h@ld\relax \fi}%
\def\@citea{,\penalty\@highpenalty\hskip.13em plus.13em minus.13em}}\@h@ld}
\def\@citex[#1]#2{\@cite{\citen{#2}}{#1}}%
\def\@cite#1#2{\leavevmode\unskip\ifnum\lastpenalty=\z@\penalty\@highpenalty\fi%
  \ [{\multiply\@highpenalty 3 #1%
  \if@tempswa,\penalty\@highpenalty\ #2\fi}]}   %
\makeatother 

\def\draftcite#1{\ifnum\draftcontrol=1#1\else{}\fi}
\def\@lbibitem[#1]#2{\item{}\hskip -3\hbox to 2cm
{\hfil$\scriptstyle\draftcite{#2}$}\hskip
1cm[\@biblabel{#1}]\if@filesw
     {\def\protect##1{\string ##1\space}\immediate
      \write\@auxout{\string\bibcite{#2}{#1}}}\fi\ignorespaces}

\def\@bibitem#1{\item\hskip -3cm \hbox to 2cm
{\hfil {\footnotesize\draftcite{#1}}}\hskip 1cm
\if@filesw \immediate\write\@auxout
       {\string\bibcite{#1}{\the\value{\@listctr}}}\fi\ignorespaces}

\catcode`\@=12

\begin{document}


\begin{flushright}  {~} \\[-12mm]
{\sf hep-th/0403158}\\[1mm]{\sf HU-EP-04-12}\\[1mm]
{\sf Hamburger$\;$Beitr\"age$\;$zur$\;$Mathematik$\;$Nr.$\;$191}\\[1mm]
{\sf LPTHE-04-05} / {\sf STR-02-042}
\\[2mm]$\simeq$ 4 March 2004 
  \end{flushright} 

\begin{center} \vskip 14mm
{\Large\bf TFT CONSTRUCTION OF RCFT CORRELATORS}\\[4mm]
{\Large\bf III: SIMPLE CURRENTS}\\[20mm] 
{\large 
J\"urgen Fuchs$\;^1$ \ \ \ Ingo Runkel$\;^2$ \ \ \ Christoph 
Schweigert$\;^3$}
\\[8mm]
$^1\;$ Institutionen f\"or fysik, Karlstads Universitet\\
Universitetsgatan 5, \,S\,--\,651\,88\, Karlstad\\[5mm]
$^2\;$ Institut f\"ur Physik, HU Berlin \\
Newtonstra\ss{}e 15, \ D\,--\,12\,489\, Berlin  \\[5mm]
$^3\;$ Fachbereich Mathematik, Universit\"at Hamburg \\
Bundesstra\ss{}e 55, \,D\,--\,20\,146\, Hamburg
\end{center}
\vskip 20mm

\begin{quote}{\bf Abstract}\\[1mm]
We use simple currents to construct symmetric special
Frobenius algebras in modular tensor categories. We classify such 
simple current type algebras with the help of abelian group cohomology.
We show that they lead to the modular invariant torus partition functions 
that have been studied by Kreuzer and Schellekens. We also classify 
boundary conditions in the associated conformal field theories and show 
that the boundary states are given by the formula proposed in hep-th/0007174.
Finally, we investigate conformal defects in these theories.
\end{quote}
\vfill
\newpage 

\tableofcontents\newpage


\sect{Introduction}

In a series of papers \citeOaIaII, we are developing a description of 
correlation functions
of rational conformal field theory that is based on the combination of
algebra and representation theory in modular tensor categories
with topological field theory in three dimensions. A central
ingredient in our construction are {\em symmetric special Frobenius
algebras\/} in these tensor categories.

The main purpose of the present paper is to discuss a specific class of 
examples, namely those based on algebras built from invertible objects or, 
in CFT terminology, {\em simple currents\/} \cite{scya,scya6,intr}. 
The general theory is developed further only to the extent that is necessary
to understand the special features of such simple current algebras.
While one of the important virtues
of Frobenius algebras is that they allow for a unified treatment of exceptional
modular invariants and simple current modular invariants, simple
currents are predominant in applications. One reason is the fact that 
mutually local simple currents with trivial twist (i.e., integral conformal 
weight) can be used in particular to implement various projections. 

For a discussion of the interplay between simple currents and projections we refer 
to \cite{fusw,Scfw} in the context of string theory, and to \cite{fpsw} in the 
context of universality classes of quantum Hall fluids. Simple currents
with non-trivial twist, on the other hand, play an important role in the
description of those symmetries that cannot be incorporated in a
(bosonic) chiral algebra, such as supersymmetries or parafermionic
symmetries. In string theory, they can also be used to implement mirror symmetry 
for Gepner models \cite{fkllsw}. The methods of \citeOaIaII\ allow in particular 
for the calculation of the structure constants of operator product expansions.
For certain classes of \cfts\ with torus partition function of simple current
type structure constants for bulk fields had been considered before
e.g.\ in \cite{jf12,fukl,petk2,pezu,rest}.
Structure constants for boundary fields in such theories were first
studied, for the case of Virasoro minimal models, in \cite{runk2}.
Our results provide a rigorous basis for the tools used in these applications.

Another important role of simple currents is their use in the construction
of different Klein bottle amplitudes. Actually we expect them to provide 
particularly strong relations in the case of Azumaya algebras (corresponding 
to pure automorphism modular invariants). An analysis of this aspect of simple 
currents requires, however, further concepts from the general theory
and hence will be presented elsewhere.

There is also a more theoretical motivation: (isomorphism classes of) 
simple currents span a subring of the fusion ring, and this subring is 
isomorphic to the group ring of a finite abelian group. The full
subcategory whose objects are (direct sums of) simple currents
can  be seen as a {\em categorification\/} of this group ring. 
Quite generally, finding a categorification of an algebraic object should be
a problem of cohomological nature. In the case of simple currents, this
can be made precise: group cohomology controls categorifications
while abelian group cohomology \cite{eima1x} controls braided categorifications. 
Any general theory of categorifications should be a generalization of these
cohomology theories; the fact that the categorification of simple currents 
leads to good cohomology theories provides a partial mathematical explanation 
of the computational power of simple currents which makes them so useful in many
applications.  

This paper is organized as follows. In Section 2 we discuss braided categories
whose objects are all invertible. This is the category-theoretic analogue
of the study of line bundles over, say, a manifold; 
accordingly we refer to the resulting theory as
{\em braided Picard theory\/}.
Section 3 is devoted to the study of haploid symmetric special Frobenius
algebras all of whose simple subobjects are invertible. We show that they
give rise precisely to the class of modular invariant partition functions
studied by Kreuzer and Schellekens \cite{krSc}. Section 4 deals with
the representation theory of such algebras. Modules correspond to
boundary conditions; therefore our formalism allows us in particular to give a
rigorous proof of the formulae of \cite{fhssw} for boundary states in
conformal field theories with simple current modular invariant.
In section 5 we develop the theory of bimodules of these algebras;
bimodules describe conformal defects in the corresponding 
conformal field theories.

\bigskip

In our notation we follow \citei\ and \citeii\ whenever possible; in
particular we use the notation for the duality, braiding and twist morphisms
introduced there, see e.g.\ the list (I:2.8).
We refer to formulas, theorems etc.\ from \citei\ and \citeii\ by 
the numbering used there, preceded by the symbol `I:' and `II:', respectively.


\sect{Braided Picard theory}

Let us start with some basic definitions.
\\[-2.3em]

\dtl{Definition}{invertible}
(i)~\,\,An object $V$ of a tensor category $\calc$ is called {\em invertible\/},
    or a {\em simple current\/},\,%
 \footnote{~This term was introduced in the physics literature \cite{scya},
    while the qualification `invertible' is standard in the mathematics literature.}
    iff there exists an object $V'$ such that $V\oti V'$ is isomorphic to the
    tensor unit $\bfe$.  \\[.2em]
(ii)~\,A tensor category is called {\em pointed\/} \cite{etno}
    iff every simple object is invertible. 
\\[.2em]
(iii)~A {\em theta-category\/} \cite{FRke} is a braided pointed tensor category.
\\[.2em]
(iv)~The {\em Picard category\/}\,%
 \footnote{~The term `Picard category' is also used for other, sometimes
 related, mathematical objects in algebra, stable homotopy theory and 
 category theory. Since there is no danger of confusing those with the present
 use of the term, we refrain from commenting on the relation to its various 
 other uses.}
$\calpic(\calc)$ of tensor category $\calc$ is the full tensor subcategory 
of $\calc$ whose objects are direct sums of invertible objects of $\calc$.  
\\[.2em]
(v)~\,\,A simple object $U$ of $\calpic(\calc)$ is said to be of finite order
iff some tensor power of $U$ is isomorphic to the tensor unit. The
{\em order\/} of such an object $U$ is the smallest positive integer $N_U$
such that $U^{\otimes N_U} \,{\cong}\,\one$.

\medskip

Note that in these definitions it is not assumed that the category
also has a duality and a twist or, in case it does, whether they (together
with the braiding) make the category into a sovereign category. However, in 
the present context of rational CFT, all pointed and theta-categories arise 
as full subcategories of sovereign (and in fact modular) categories and are
therefore sovereign themselves. Indeed, for the purposes of this paper
we consider only categories with the following additional properties:
\\[-2.3em]

\dtl{Convention}{categories}
All categories in this paper are assumed to be small abelian
$\complex$-linear semisimple sovereign tensor categories, and to possess 
the following additional properties:
\nxt The morphism spaces are \findim\ complex vector spaces.
\nxt The tensor unit $\one$ is simple. 
\nxt The dimension of any object is a non-negative real number.
\nxt There are only finitely many isomorphism classes of simple objects.
\\[.1em]
    Moreover, with the exception of the concrete categories $\calc(G,\psi)$
    and $\calc(G,\psi,\Omega)$ (to be introduced in lemma \ref{xl12} below),
    these categories are assumed to be {\em strict\/} tensor categories.

\medskip

Also recall that braided sovereign tensor categories are also known as 
{\em ribbon categories\/}. For the applications to conformal field theory
we have in mind, we are interested in (subcategories of) modular tensor 
categories.  

When taking into account the convention \ref{categories},
the following observations are straightforward:
\\[-2.3em]

\dtl{Remark}{invertible:remark}
(i)~\,\,Invertible objects are simple.
\\[.2em]
(ii)~\,For any invertible object $V$ the relations
  \be  b_V \circ \tilde d_V = \id_{V \otimes V^\vee}
  \qquad{\rm and}\qquad  \tilde b_V \circ d_V = \id_{V^\vee\otimes V}
  \labl{sc-rel}
for the left and right duality morphisms (as defined in (I:2.8) and (I:2.12))
are valid.
\\[.2em]
(iii)~Existence of an object $V'$ such that $V\oti V'\,{\cong}\,\one$ is 
equivalent to the existence of an object $V''$ such that
$V''\oti V\,{\cong}\,\one$. (In short, `right-invertibility' is equivalent
to `left-invertibility'.) Indeed, both the object $V'$ in definition
\ref{invertible}(i) and the object $V''$ in question can be taken to be
the dual object $V^\vee$, since one has 
$V\oti V^\vee\,{\cong}\,\one\,{\cong}\,V^\vee\oti V$.
\\[.2em]
(iv)~\,It follows in particular that
an object $V$ is invertible iff the dual object $V^\vee$ is invertible.  
Because of $V\oti V^\vee\,{\cong}\,\one$, the dimension of an invertible
object is thus an invertible number. And because of $\dim(V)\eq \dim(V^\vee)$, 
it must be $\pm1$ and hence, by the positivity assumption on dimensions, 
$\dim(V)\eq 1$. Indeed, simple currents can alternatively be characterized as 
simple objects of dimension 1. Such objects have been classified \cite{jf15}
for certain classes of modular tensor categories, like the ones based on an 
untwisted affine Lie algebra at integral level (compare also \cite{fuge,happy}). 
\\[.2em]
(v)~\,\,\,If $\calc$ is sovereign, then $\calpic(\calc)$ is sovereign, too, and
it is a pointed category.
\\[.2em]
(vi)~\,If $\calc$ is braided, then $\calpic(\calc)$ is braided as well, 
and hence it is a theta-category.
\\[.2em]
(vii)~Even when $\calc$ is modular, the Picard category $\calpic(\calc)$ is,
in general, {\em not\/} modular.

\medskip

The following result is a direct consequence of the definitions: \\[-2.3em]

\dtl{Proposition}{prop:fingroup}
(i)~\,\,The Grothendieck ring of a pointed category $\ircalc$ is isomorphic to
the group ring of a finite group $G$, $K_0(\calc)\,{\cong}\,\zet G$.  
\\[.3em]
(ii)~\,The fusion ring of a theta-category is isomorphic to the group ring
of a finite abelian group.

\dtl{Definition}{def-center}
Let $\calc$ be a tensor category.
The {\em Picard group\/} of $\calc$, denoted by $\Pic(\calc)$, is the group
of all isomorphism classes of invertible objects in $\calc$, with the product
being the one of the Grothendieck ring $K_0(\calc)\,{\supset}\,\Pic(\calc)$.

\dt{Remark}
(i)~\,\,$\Pic(\calc)$ is a finite group, and we have
$K_0(\calpic(\calc))\,{\cong}\,\zet\,\Pic(\calc)$ as a ring over $\zet$.
When the category $\calc$ is braided, then the Picard group $\Pic(\calc)$
is a finite abelian group.
\\[.2em]
(ii)~\,In the physics literature, the Picard group $\Pic(\calc)$ of a 
{\em braided\/} tensor category is also called the
{\em simple current group\/} or {\em center\/} of $\calc$. This notion of
center should not be confused with the Drinfeld center 
(see e.g.\ chapter XIII.4 of \cite{KAss}) of the \cat\ $\calc$. Also note that
the latter is a category of global dimension $\Dim(\calc)^2$ \cite{ostr},
while the global dimension of the Picard category $\calpic(\calc)$
is bounded by $\Dim(\calc)$.  
\\
(The global dimension of $\calc$ is defined as the number
$\Dim(\calc)\,{:=}\,\llb\sum_{i\in \II} \dim (U_i)^2{\lrb}^{1/2}_{}$, with $U_i$, 
$i\iN \II$ a set of representatives for the isomorphism classes of 
simple objects of $\calc$.)
\\[.2em]
(iii)~The same notion of Picard group is used for instance in stable homotopy 
theory, see e.g.\ \cite{hoSa}.  In the literature there is also a different 
notion of Picard group, namely the tensor subcategory consisting of all 
invertible objects, which is in particular a {\em categorical group\/}, or 
{\em cat-group\/} (see e.g.\ \cite{cace,may2,vita4}).

\bigskip

To obtain examples of pointed categories and theta-categories, we select a 
finite group $G$, which for theta-categories will be required to be abelian.
Consider the category of \findim\ $G$-graded complex vector spaces with the
grade-respecting linear maps as morphisms. This is an abelian semisimple
category whose isomorphism classes of simple objects $\U_g$ 
are in bijection 
to the elements $g$ of $G$. A general object $V$ can be written as a direct sum
$\bigoplus_{g\in G}\! V_g$ of \findim\ vector spaces. Every three-co\-cycle 
$\psi\iN Z^3(G,\complexx)$ defines an associativity constraint by
  \be  \begin{array}{lrcl}
  \alpha_{V,V',V''}:\; & \left(V^{}_{g_1}\oti V'_{g_2}\right) \oti V''_{g_3}
  &\!\to\!& V^{}_{g_1} \oti \left( V'_{g_2}\oti V''_{g_3}\right)
  \\{}\\[-.7em]
  & (v\oti v')\oti v'' &\!\mapsto\!& \psi(g_1,g_2,g_3)^{-1}\, v\oti (v'\oti v'')
  \,. \end{array} \labl{xl1}
To obtain a theta-category, we endow this category with a braiding;
this is possible only if $G$ is abelian.
It turns out ($\!$\cite{joSt,joSt6}, see also \cite{FRke,rOse2})
that a representative $(\psi,\Omega)$ of the third abelian group cohomology 
(as defined in \cite{eima1x} and summarized in appendix \ref{Abeliangc})
contains exactly the relevant data; we can define the braiding by
  \be  \begin{array}{lrcl}
  c_{V,V'}:\;  & V^{}_{g_1}\oti V'_{g_2} &\!\!\to\!\!&  V'_{g_2} \oti V^{}_{g_1}
  \\{}\\[-.7em]
  & v\oti v' &\!\!\mapsto\!\!&  \Omega(g_2,g_1)^{-1}\, v'\oti v \,.
  \end{array} \labl{xl2}

\medskip

For a discussion of equivalences between such categories we introduce
the following notion.  
\\[-2.3em]

\dtl{Definition}{def:marked}
Let $R$ be a ring. A pair consisting of a sovereign tensor category $\calc$ 
and an isomorphism $f{:}\; R \,{\to}\, K_0(\calc)$ is called
an $R$-{\em marked\/} tensor category. If the ring is evident, we simply call
the category {\em marked\/}.

\medskip

In a marked category one has a distinguished correspondence between elements 
of the ring $R$ and isomorphism classes of simple objects in $\calc$. 
In the example of $G$-graded vector spaces treated above, we naturally get
a marked category by taking $R$ to be the group ring $\zet G$ and choosing
$f(g) \eq [V_g]$. 

\dtl{Lemma}{xl12}
(i)~\,\,Let $G$ be a finite group.
The associativity constraint $\alpha$ given in \erf{xl1}
defines the structure of a pointed tensor category $\calc(G,\psi)$. 
\\[.2em]
(ii)~\,The categories $\calc(G,\psi)$ and $\calc(G,\psi')$ are equivalent as 
       $\zet G$-marked tensor categories if and only if the 
       three-co\-cycles $\psi$ and $\psi'$ are cohomologous.
\\[.2em]
(iii)~ Let $G$ be a finite abelian group. The morphisms given in \erf{xl1} and
\erf{xl2} define the structure of a theta-category $\calc(G,\psi,\Omega)$.
\\[.2em]
(iv)~The categories $\calc(G,\psi,\Omega)$ and $\calc(G,\psi',\Omega')$
are equivalent as $\zet G$-marked
ribbon categories if and only if the abelian
three-co\-cycles $(\psi,\Omega)$ and $(\psi',\Omega')$ are cohomologous.

\medskip\noindent
Proof:\\
(i)~\,\,The associativity constraint $\alpha$ obeys the pentagon condition, 
because $\psi$ is closed. 
\\[.2em]
(iii)~The morphisms in \erf{xl2} indeed furnish a braiding; in particular
they satisfy the relevant compatibility conditions with the associator 
\erf{xl1}, i.e.\ the two hexagon diagrams.
\\[.2em]
(ii)~\,and~\,(iv)~\,
Since we are only concerned with functorial isomorphism between $\calc$
and $\calc'$ as marked tensor (respectively, ribbon) categories, we can
assume that the relevant functor acts as the identity on objects.
The statements then follow directly by comparison with the relevant notions 
from group cohomology and abelian group cohomology, respectively. (These 
notions are summarized in appendix \ref{appcoho}.) 
\qed

\medskip

We now show that these examples already exhaust the class of pointed 
categories and theta-categories, respectively. To this end we need, 
besides the result given in proposition \ref{prop:fingroup},
the notion (see e.g.\ theorem 2.3 in \cite{crye4}) of categorification.
To introduce this, we first  recall that a {\em based\/} ring 
$R$ over the integers $\zet$ is a unital ring over $\zet$ together with a basis 
of $R$ in which the structure constants are non-negative; we can then state
\\[-2.3em]

\dtl{Definition}{categorification}
Let $R$ be a based ring.
\\[.1em]
(i)~\,\,A {\em categorification\/} of $R$ is a sovereign tensor category $\calc$ 
   such that $K_0(\calc)\,{\cong}\, R$.
\\[.2em]
(ii)~\,A {\em braided categorification\/} of $R$ is a ribbon category $\calc$ 
   such that $K_0(\calc)\,{\cong}\, R$.

\dtl{Remark}{remark:categorification}
(i)~\,\,In the definition, the category $\calc$ is required to be sovereign.
        This property guarantees that the tensor product functor $\otimes$ is
        exact and thus induces a product on the Grothendieck group $K_0(\calc)$.
\\[.2em]
(ii)~An additional necessary condition for a ring $R$ to have a braided 
     categorification is that $R$ is abelian.
\\[.2em]
(iii)~The tools used in much of the physics literature on conformal field 
theory 
are essentially (fusion) rings and the twist (appearing as the exponentiated
conformal weight). Other aspects of categorification are usually not
taken into account.

\dtl{Proposition}{prop:structure}
(i)~\,\,Let $G$ be a finite group. 
     For every categorification $\calc$ of the group ring $\zet G$
     there exists some $\psi\iN Z^3(G,\complexx)$ such that $\calc$ is
     equivalent to $\calc(G,\psi)$ as a sovereign tensor category. 
\\[.2em]
(ii)~\,Let $G$ be a finite abelian group. 
     For every braided categorification $\calc$
     of $\zet G$ there is some abelian three-co\-cycle $(\psi,\Omega)$ 
     such that $\calc$ is equivalent to $\calc(G,\psi,\Omega)$ as a 
     theta-category.
\\[.2em]
(iii)~The equivalence classes of marked
     categorifications of $\zet G$ are in bijection with the group
     $H^3(G,\complexx)$, and the equivalence classes of braided
     categorifications of $\zet G$ are in bijection
     with $H^3_{{\rm ab}}(G,\complexx)$. 
\\[.2em]
(iv)~These categorifications exhaust the class of pointed categories (for an 
    arbitrary finite group $G$) and of theta-categories (for $G$ finite 
    abelian), respectively.

\medskip\noindent
Proof: \\
(i) The proof can essentially be found in appendix E of \cite{mose3},
compare also \cite{quin3,yama7}.
The isomorphism classes of simple objects are labeled by group elements
$g\iN G$. We fix representatives $\U_g$; this will always be done in such a 
way that the representative $\U_e$ of the unit element is the tensor unit
$\one$. The morphism spaces $\Hom(\U_{g_1}\Oti\, \U_{g_2}, \U_{g_1g_2})$ 
are one-di\-men\-si\-onal. We select a basis 
  \be  \bL_{g_1\!}{g_2}  \,\in\, \Hom(\U_{g_1}\Oti\,\U_{g_2},\U_{g_1g_2})
  \,.  \labl{bL-def}
(There is no distinguished basis; group cohomology is
the appropriate tool for formulating basis independent statements.
Our conventions for basis choices are collected in appendix \ref{appC}.)
\\
The associator is described by the family of isomorphisms
  \be \begin{array}{lrcl}
  \phi_{g_1,g_2,g_3}:\,&
  \Hom(\U_{g_1}\Oti(\U_{g_2}\Oti\U_{g_3}),\U_{g_1g_2g_3})
  &\!\!\to\!\!& \Hom((\U_{g_1}\Oti\U_{g_2})\Oti\U_{g_3}),\U_{g_1g_2g_3})
  \\{}\\[-.7em]
  & f &\!\!\mapsto\!\!& f \cir \alpha_{\U_{g_1},\U_{g_2},\U_{g_3}}
  \end{array} \ee
Since the two morphism spaces are \onedim, the isomorphism $\phi_{g_1,g_2,g_3}$
is described, in the chosen basis, by a non-zero number $\psi(g_1,g_2,g_3)$:
  \be  \phi\big( \bL_{g_1\!}{g_2 g_3} \cir (\id_{\U_{g_1}} \Oti\, \bL_{g_2}{g_3})
  \big) = \psi(g_1,g_2,g_3)^{-1} \bL_{g_1 g_2}{g_3} \cir (\bL_{g_1\!}{g_2} \oti
  \id_{\U_{g_3}}) \,. \labl{eq:phi-psi}
The pentagon axiom is easily seen to be equivalent to the assertion that
$\psi$ is a three-co\-cycle. Different choices of basis lead to cohomologous
three-co\-cycles. We conclude that the pointed category is equivalent to
$\calc(G,\psi)$.
\\[.3em]
(ii)~\,These statements can be found in \cite{rOse2}.
For the case of theta-categories, we use the fact that the braiding
is described by the isomorphism that it induces on morphism spaces,
  \be \begin{array}{rcl}
  \Hom(\U_{g_1}\Oti\, \U_{g_2},\U_{g_1g_2}) & \!\!\to\!\! &
  \Hom(\U_{g_2}\Oti\, \U_{g_1},\U_{g_1g_2})  \\
  \varphi &\!\!\mapsto\!\!& \varphi\circ c_{\U_{g_2},\U_{g_1}}  \, . 
  \end{array} \ee
The action of this isomorphism on a basis element is described by the 
two-cochain $\Omega$ on $G$ with values in $\complexx$:
  \be  \bL_{g_1\!}{g_2} \circ c_{\U_{g_2}, \U_{g_1}} = 
  \Omega (g_1,g_2)^{-1}\, \bL_{g_2}{g_1} \,. \labl{d1}
It is straightforward to check that the pentagon axiom and the two hexagon 
axioms imply the two constraints \erf{app1neu}, so that $(\psi,\Omega)$ is an
abelian three-co\-cycle. Cohomologous three-co\-cycles give rise to equivalent
theta-categories.
\\[.3em]
(iii)~The statements follow from (i) and (ii), respectively, together with
      lemma \ref{xl12}.
\\[.3em]
(iv)~The statements follow from combining (i)\,--\,(iii) with  
proposition \ref{prop:fingroup}.  
\qed

\dtl{Remark}{rem:FR}
(i)~\,\,The matrix elements of the associator $\phi$ for general simple objects 
are, in a chosen basis, the elements of the fusing matrices \FF, or 
$6j$-symbols, while those of $\Omega$ are the elements of the braiding matrices 
\RR. To make contact with the corresponding notation for fusing and braiding
used in \citei, let us compare formulas (I:2.36) and (I:2.41)
to the above relations \erf{eq:phi-psi} and \erf{d1}; we have
  \be 
  \Fs rs{\,t}{\,r{\cdot}s{\cdot}t}{s{\cdot}t}{\,\,r{\cdot}s} = \psi(r,s,t)^{-1}
  \qquad {\rm and} \qquad \Rs{}rs{\,r{\cdot}s} = \Omega(s,r)^{-1} \,,  \ee 
where for better readability we have indicated the product in the group $G$
by the symbol `$\cdot$'. It is straightforward to check that a pair $(\FF,\RR)$ 
obeys the pentagon and the two hexagon relations if and only if $(\psi,\Omega)$ 
is an abelian three-co\-cycle (for details, see appendix \ref{appC}).
\\[.2em]
(ii)~\,In proposition \ref{prop:structure}(iii) we considered equivalence 
classes of {\em marked\/} categorifications of $\zet G$. The equivalence classes
of categorifications of $\zet G$, on the other hand, are in one-to-one 
correspondence with elements of $H^3(G,\complexx) / {\rm Out}(G)$, respectively 
$H^3_{{\rm ab}}(G,\complex) / {\rm Out}(G)$ in the braided case, 
see e.g.\ proposition 1.21 of \cite{caet}.
\\[.2em]
(iii)~Later on we need some understanding of the possible associators and 
braidings on the theta-category $\calpic(\calc)$ that arises as a subcategory 
of a modular category $\calc$. In comparing two associators or braidings, we 
will always leave the relevant objects fixed. This explains our interest in 
{\em marked\/} categorifications.  

\medskip

The group of abelian 3-cocycles is isomorphic to the group of quadratic forms 
on $G$ (see appendix \ref{Abeliangc}). What makes theta-categories 
particularly accessible is that this quadratic form can easily be computed in 
concrete CFT models from the conformal weights. To this end, we need the 
notion of balancing isomorphism or twist, which is introduced in
\\[-2.3em]

\dtl{Definition}{balancing}
(i)~\,\,For every object $U$ of a ribbon category $\calc$, the {\em twist\/}
$\theta_U$ is the endomorphism
  \be \theta_U := (d_U \otimes \id_U)  \circ (\id_{U^\vee}\oti c_{U,U})
  \circ (\tilde b_U \oti \id_U) \ \in \End(U)\, . \labl{def:theta}
(ii)~\,For simple objects $U\eq U_i$ we set
  \be  \theta_{U_i} =: \th_i\, \id_{U_i} \labl{thi}
with $\th_i\iN\complexx$. The number $\th_i$ is also called the twist, or 
the {\em balancing phase\/},\,%
  \footnote{~The numbers $\th_i$ are indeed phases, and even roots of unity,
  see e.g.\ \cite{etin3}. \label{phasephootnote}}
of $U_i$.
\\
When the simple object is invertible, $U\eq\U_g$, we write 
$\theta_{\U_g} \;{=:}\, \th_g\,\id_{\U_g}$. 

\medskip

For any object $V$ the twist $\theta_V$ is an {\em iso\/}morphism (so that in
particular the definition \erf{thi} of the balancing phase makes sense); 
it is also called the {\em balancing isomorphism\/}.
The balancing phases of isomorphic simple objects, as well as of their
dual objects, coincide; in particular, 
   \be  \th_{g^{-1}} \eq\th_g \qquad{\rm for~all~~} g\iN G \,.  \ee

\medskip

The balancing phases are related to the braiding as follows.
\\[-2.2em]

\dtl{Proposition}{prop:form}
(i)~\,\,The function
  \be \begin{array}{lrcl}
  \delta:\,  & G &\!\to\!& \complexx \\{}\\[-1.1em] & g &\!\mapsto\!& \th_g
  \end{array} \labl{delta}
is a quadratic form on the group $G$. It is the inverse of the quadratic 
form $q{:}\; G\,{\to}\,\complexx$ that (as described in formula \erf{EL}) 
characterizes the isomorphism class of a marked theta-category:
    \be  q(g)^{-1} \equiv \Omega(g,g)^{-1} = \th_g = \delta(g)\,.  \labl{pf1}
In particular, 
  \be  c_{\U_g,\U_g} = \th_g\, \id_{\U_g\otimes \U_g} \, . \labl{pf2}
(ii)~\,The bihomomorphism associated to the quadratic form $\delta$ is given by 
  \be \begin{array}{lrcl}
  \beta:\,  & G \times G &\!\to\!& \complexx
  \\{}\\[-.7em]
  & (g_1,g_2) &\!\mapsto\!&
  \delta(g_1g_2)\,\delta(g_1)^{-1}_{}\delta(g_2)^{-1}_{} \,,
  \end{array} \labl{beta} 
and one has the identity
  \be  c_{\U_{g_2},\U_{g_1}} \circ c_{\U_{g_1},\U_{g_2}} = \beta(g_1,g_2)
  \, \id_{\U_{g_1}\otimes \U_{g_2}} \, . \labl{pf3}

\medskip\noindent
Proof: \\
(i)~\,\,Note that for $g_1\eq g_2\,{=:}\,g$, the relation \erf{d1} for the 
braiding of basis morphisms implies that 
$c_{\U_g,\U_g} \eq \Omega(g,g)^{-1}\, \id_{\U_g\oti \U_g}$.
For $U\,{\cong}\,\U_g$ an invertible object of a ribbon category, the
definition \erf{def:theta} of the twist then gives
  \be  \theta_{\U_g} = \Omega(g,g)^{-1}\, \dim (\U_g)\, \id_{\U_g}
  = \Omega(g,g)^{-1}\, \id_{\U_g} \,.  \ee
This establishes both \erf{pf1} and \erf{pf2}.
\\[.3em]
(ii)~\,The first statement is just the definition of an associated 
bihomomorphism (see formula \erf{A11}). The relation \erf{pf3}
follows directly from the compatibility condition
  \be \theta_{V\otimes W} = c_{W,V}\circ c_{V,W} \circ (\theta_V\oti\theta_W)
  \labl{compat}
between braiding and twist (it is also implied by formula \erf{app2}).
\qed


\dtl{Remark}{DeltaQ}
When $\calc$ is the modular tensor \cat\ associated to some rational
CFT, then the balancing phases are related by
  \be  \th_i = \exp(-2\pi\ii \Delta_i)  \labl{vDelta}
with the conformal weights $\Delta_i\iN\rationals$ of the primary fields of
the CFT. Note that for the \cat\ $\calc$ only the fractional part of the
conformal weight matters. In particular, for $g\iN G$ we have
$\delta(g)\eq\exp(-2\pi\ii \Delta_g)$, and thus the isomorphism class of the 
theta-category is given by the exponentiated conformal weights. Moreover, 
the associated bihomomorphism $\beta$ obeys
  \be 
  \beta(g_1,g_2) = \exp( 2\pi\ii\, Q_{g_1}(g_2))  \labl{mono}
for $g_1,g_2\iN G$, where
  \be
  Q_{g_1}(g_2) := \Delta_{g_1} + \Delta_{g_2} - \Delta_{g_1 g_2}
  \bmod \zet \,. \labl{Qdef}
In CFT, $Q_{g_1}(g_2)$ is known as the {\em monodromy charge\/}
of the simple current $\U_{g_1}$ with respect to $\U_{g_2}$ (or vice versa)
\cite{scya6}. Also note that $\delta$ contains more information than $\beta$ 
(see section \ref{Abeliangc}).
In the CFT setting, this means that the fractional parts of the conformal
weights contain more information than the monodromy charges.

\medskip

In a braided setting, a certain subgroup of the Picard group plays a
particularly important role. We need the 
\\[-2.2em]

\dtl{Definition}{def:eff}
Let $G$ be an abelian group and $\psi$ a three-co\-cycle
on $G$. We call a subgroup $H$ of $G$ $\psi$-{\em trivializable\/} 
iff there exists a two-cochain $\omega$ on $H$ such that 
   \be  \rmd \omega = \psi_{|H} \,.  \labl{domega}
The two-cochain $\omega$ is then called a {\em trivialization\/} of $H$.

\medskip

For braided categories, one should rather start with an
{\em abelian\/} three-co\-cycle $(\psi,\Omega)$. Then $\psi$-trivializable 
subgroups of $G$ possess a simple characterization, as a corollary
to the following result.
\\[-2.2em]

\dt{Lemma} 
Let $G$ be a finite abelian group and $(\psi,\Omega)$ an abelian three-co\-cycle
on $G$. Then the cohomology class $[\psi]$ of $\psi$ in the ordinary group 
cohomology $H^3(G,\complexx)$ is equal to the trivial class $[1]$ if and only 
if for each element $g\iN G$ one has $(\Omega(g,g))^{N_g}\eq 1$ with $N_g$ 
the order of $g$.

\medskip\noindent
Proof: \\
The map $\EL{:}\; H^3_{{\rm ab}}(G,\complex^*) \,{\to}\, \QF(G,\complexx)$
given by
  \be 
  [(\psi,\Omega)] \mapsto q \qquad{\rm with}\quad q(g) := \Omega(g,g) 
  \labl{eq:EL2}
is well defined and 
an isomorphism of abelian groups, see appendix \ref{Abeliangc}.
\\[.2em]
Suppose that $[\psi] \eq [1]$. For the abelian cohomology class
this implies $[(\psi,\Omega)] \eq [(1,\tilde\Omega)]$ for some
appropriate two-cochain $\tilde\Omega$. According to formula \erf{app1neu},
$(1,\tilde\Omega)$ is an abelian three-co\-cycle iff $\tilde\Omega$
is a bihomomorphism. Now
\be
 \Omega(g,g) = \EL( [(\psi,\Omega)] )(g) = \EL( [(1,\tilde\Omega)] )(g)
 = \tilde\Omega(g,g) \,, \ee
and $\tilde\Omega(g,g)^{N_g} \eq \tilde\Omega(g,g^{N_g}) \eq 1$,
because $\tilde\Omega$ is a bihomomorphism. Thus indeed
$\Omega(g,g)^{N_g} \eq 1$ for all $g\iN G$.
\\[.2em]
Suppose now that, conversely, $\Omega(g,g)^{N_g}\eq 1$ for all $g\iN G$.
To establish that this implies $[\psi] \eq [1]$, it is sufficient
to construct a bihomomorphism $\tilde\Omega(g,h)$ such
that $\tilde\Omega(g,g) \eq \Omega(g,g)$ for all $g\iN G$. 
Once we have such an $\tilde\Omega$, we know that
$(1,\tilde\Omega)$ is an abelian three-co\-cycle obeying
  \be
  \EL( [(1,\tilde\Omega)] )(g) = \tilde\Omega(g,g) =
  \Omega(g,g) = \EL( [(\psi,\Omega)] )(g) \,.  \ee
Since the map \erf{eq:EL2} is an isomorphism, this implies
$[(1,\tilde\Omega)] \eq [(\psi,\Omega)]$ in abelian group cohomology,
and thus in particular $[1]\eq[\psi]$ in ordinary cohomology.
\\[.2em] 
Let us proceed to construct an $\tilde\Omega$ with the desired
properties. Define $q(g) \,{:=}\, \Omega(g,g)$, which by
\erf{eq:EL2} is a quadratic form. 
Select a set $g_1,g_2,...\,,g_r$ of generators of the group $G$. 
For $a,b\,{\in} \{1,2,...\,,r\}$ choose numbers $X_{ab}$ such that
  \be   \left\{\begin{array}{ll}
  \exp( 2 \pi \ii X_{aa} ) = q(g_a) & {\rm for~all~} a \,,  \\[.3em]
  \exp( 2 \pi \ii X_{ab} ) = q(g_a g_b) \,[q(g_a)\, q(g_b)]^{-1}_{}
  & {\rm for~} a\,{<}\,b \,,  \\[.3em]
  X_{ab}=0 & {\rm for~} a\,{>}\,b \,.
  \eear\right. \labl{3case}
For two group elements 
$g\eq\prod_a (g_a)^{m_a}$ and $h\eq\prod_a (g_a)^{n_a}$, 
we would then like to set
  \be
  \tilde\Omega(g,h) := \exp(2\pi\ii \sum_{a,b=1}^k m_a X_{ab} n_b)\,.
  \labl{eq:tOm}
For this to furnish a well-defined map $G\,{\times}\, G \,{\to}\,\complexx$, 
shifting $m_a \,{\mapsto}\,{\pm} N_a$ or $n_a \,{\mapsto}\,{\pm} N_a$ must not 
affect the value of the right hand side of \erf{eq:tOm}. We consider only the 
former shift, the latter being analogous; the right hand side of \erf{eq:tOm} 
changes by the factor
  \be 
  \exp \big({2\pi\ii \sum_{b=1}^k N_a X_{ab} n_b} \big)
  = \prod_{b=1}^r \big( \eE^{2\pi\ii N_a X_{ab}} \big)^{\!n_b} \,.  \ee
Now in fact  each of the factors $\exp(2\pi\ii N_a X_{ab})$ is equal to one. 
To see this consider the three cases in \erf{3case} separately. For $a\,{>}\,b$ 
the statement is trivial. For $a\eq b$, the statement follows because by 
assumption $q(g_a)^{N_a}\eq 1$. Finally, for $a\,{<}\,b$, the statement is 
implied by the identity 
  \be
  1 = \beta_q(g_a^{N_a},g_b) = \beta_q(g_a,g_b)^{N_a}
  = \Big( \frac{ q(g_a g_b) }{q(g_a)\, q(g_b)} \Big)^{\! N_a} 
  \labl{eq:tOm-aux1}
for the bihomomorphism $\beta_q$ \erf{A11} that is associated to $q$.
Here in the first two steps we use $g_a^{N_a}\eq e$ and the bihomomorphism
property of $\beta_q$, while in the last step the definition
of $\beta_q$ in terms of $q$ is inserted.
\\[.2em]
Thus \erf{eq:tOm} gives a well-defined
map $\tilde\Omega{:}\; G \,{\times}\, G \,{\to}\,\complexx$. From the
definition it is obvious that $\tilde\Omega$ is a bihomomorphism.
This implies that $\tilde q(g) \eq \tilde\Omega(g,g)$ is a quadratic
form on $G$. It remains to show that
$q(g) \eq \tilde q(g)$ for all $g \iN G$. By lemma \ref{lem:QF}, 
it is sufficient to verify that $q(g_a) \eq \tilde q(g_a)$ for all $a$ and 
$q(g_a g_b) \eq \tilde q(g_a g_b)$ for all $a\,{\neq}\,b$. The first equality 
holds by definition. To show the second, we may assume that $a\,{<}\,b$; then
  \be
  \tilde q(g_a g_b) = 
  \tilde \Omega(g_a g_b, g_a g_b)  =
  \tilde \Omega(g_a, g_a)\, \tilde\Omega(g_b, g_b)\,
  \tilde \Omega(g_a, g_b)\, \tilde\Omega(g_b, g_a) = q(g_a g_b) \,.  \ee
Here in the first step the definition of $\tilde q$ is inserted, in the 
second step it is used that $\tilde\Omega$ is a bihomomorphism, and in the 
last step the definition of $\tilde\Omega$ is substituted.
\qed

\dtl{Corollary}{cor:eff}
Let $G$ be an abelian group and $(\psi,\Omega)$ an abelian three-co\-cycle
on $G$. A subgroup $H\,{\le}\,G$ is $\psi$-trivializable iff for each element 
$h\iN H$ one has $(\Omega(h,h))^{N_h}\eq 1$ with $N_h$ the order of $h$.

\dtl{Remark}{square0}
The condition that $(\delta(h))^{N_h}\eq (\Omega(h,h))^{-N_h}$ is equal to 1 
is fulfilled for every $h\iN G$ that can be written as a square. 
This is implied by the following argument from \cite{scya6}, which for
the convenience of the reader is reformulated using the present terminology.
First note that $\beta(g,g^{n-1})\eq\delta(g^n)[\delta(g^{n-1})
\delta(g)]^{-1}$ for $g\iN G$ and any integer $n$, implying that
  \be  \delta(g^n) = \delta(g^{n-1})\,\delta(g)\,(\beta(g,g))^{n-1}_{} \,.
  \labl{indkstep}
For $n\eq N_g$ this gives, after choosing a square root
$\sqrt{\beta(g,g)}$ of $\beta(g,g)$, 
  \be  \delta(g) = \epsilon_g\,\big(\sqrt{\beta(g,g)}\,{\big)}^{-N_g+1}_{}\ee
with $\epsilon_g\iN\{\pm1\}$. With the help of \erf{indkstep}, one can now
prove by induction that
  \be  \delta(g^n) = \epsilon_g^n\,
  \big(\sqrt{\beta(g,g)}\,{\big)}^{-n(N_g-n)}_{} . \labl{dgn}
Let now $h\iN G$ be a square, $h\eq g^2$, so that $N_g \eq 2 N_h \iN 2\zet$.
Since $\beta(g,g)$ is an $N_g$th root
of unity, by applying \erf{dgn} with $n\eq2$ it follows that
  \be  (\delta(h))^{N_h}_{} = (\delta(g^2))^{N_g/2}_{} 
  = \llb \epsilon_g^2\,\big(\sqrt{\beta(g,g)}\,\big)^{-2(N_g-2)}
  {\lrb}^{N_g/2}_{} = 1 \,.  \ee

\dtl{Definition}{Picardeff}
Suppose the Picard category of a tensor category $\calc$ is equivalent
to $\calc(G,\psi,\Omega)$. Then the {\em effective center\/} $\Pico(\calc)$ of
the category $\calc$ is the union (as sets) of all $\psi$-trivializable 
subgroups of $G$.

\smallskip

\dtl{Remark}{square}
(i)~\,\,Note that if $g$ has order $N_g$, then $g^n$ has order
$N_{g^n} \eq m N_g /n$, with $m$ the smallest positive integer such
that $m N_g /n \iN \zet$. If $\theta_g^{\,N_g}\eq 1$, then by lemma
\ref{lem:QF}(ii) we also have
  \be
  (\theta_{g^n})^{N_{g^n}}_{} = (\theta_g)^{n^2 m N_g/ n}_{}
  =(\theta_g^{\,N_g})^{n m}_{} = 1 \,.  \ee
Thus, if $\theta_g^{\,N_g}\eq 1$ for some $g\iN\Pic(\calc)$, then
also $\theta_h^{\,N_h}\eq 1$ for all $h$ in the cyclic subgroup generated
by $g$. Together with corollary \ref{cor:eff}, it follows that the effective
center $\Pico(\calc)$ consists precisely of those $g\iN\Pic(\calc)$ for which
$\theta_g^{\,N_g}\eq 1$. In CFT, where the twist is related to the conformal
weights by \erf{vDelta}, this means that
  \be  \Pico(\calc) = \{ g\iN G \,|\, N_g\Delta_g\iN\zet \} \,.  \ee
This is the original definition \cite{gasc} of the effective center.
\\[.2em]
(ii)~\,The effective center is, in general, {\rm not\/} a group.\,%
 \footnote{~We thank Bert Schellekens for pointing this out to us.}
For example, one checks that the function $q{:}\; G \,{\to}\, \complexx$
given by $q(g_1^{m_1} g_2^{m_2})\,{:=}\, \exp(\pi\ii\, m_1^{\,2} / 2)$
is a quadratic form on the group $G\eq \zet_2 \Times \zet_4$, with
the two generators denoted by $g_1$ and $g_2$.
The associated bihomomorphism $\beta_q$ acts as
$\beta_q(g_1^{m_1} g_2^{m_2},g_1^{n_1} g_2^{n_2})\eq (-1)^{m_1 n_1}_{}$.
Let $[(\psi,\Omega)] \,{:=}\, \EL^{-1}(q) \iN H^3_{{\rm ab}}(G,\complex)$
be the abelian cohomology class determined by $q$ via the isomorphism
\erf{EL}, with representative $(\psi,\Omega)\iN Z^3_{{\rm ab}}(G,\complex)$,
and  set $\cald \,{:=}\,\calc(G,\psi,\Omega)$.
\\
The two elements $g\,{:=}\,g_2$ and $h\,{:=}\,g_1 g_2$ of $G$ both have
order four, and $q(g)^4 \eq 1 \eq q(h)^4$, while $gh$ has order two, and
$q(gh)^2 \eq {-}1$. Together with part (i), this implies that $g$ and
$h$ are in $\Pico(\cald)$, while $gh$ is not. Thus $\Pico(\cald)$ is not
a subgroup of $\Pic(\cald) \eq G$. In fact, in this example one has 
$\Pico(\cald) \eq \{ e, g_2^{}, g_2^{\,2}, g_2^{\,3}, g_1^{}g_2^{}, 
g_1^{}g_2^{\,3} \}$.  
\\[.2em]
(iii)~According to remark \ref{square0},
every square in $\Pic(\calc)$ is even in $\Pico(\calc)$.
Also note that in particular every group element $h$ of odd order can be 
written as a square, namely as $ h \eq \llb h^{(N_h+1)/2}_{}{\lrb}^2$.
Thus simple currents of odd order are always in the effective center.
\\[.2em] 
(iv)~The quotient $\omega/\omega'$ of two trivializing two-cochains $\omega$ 
and $\omega'$ is closed, but not necessarily exact.  
\\[.2em]
(v)~\,The Picard group of the product $\calc_1\Ti\calc_2$ of two
categories is the product of the Picard groups, 
  \be  \Pic(\calc_1\Ti\calc_2) \cong \Pic(\calc_1) \times \Pic(\calc_2) \,.
  \ee
In contrast, a similar identity does {\em not\/} hold for the effective 
center. Here is a simple counter example: 
take for $\calc_1\eq\calc_2\eq\calc$ the modular tensor category
for the WZW theory based on $A_1^{(1)}$ at level 1. The effective 
center of $\calc$ is trivial, whereas the product has effective 
center $\Pico(\calc\Ti\calc)\,{\cong}\, \zet_2$.


\sect{Algebras}

We now proceed to discuss symmetric special Frobenius algebras. Such algebras
in modular tensor categories play an important role in conformal field theory
\citeOaI. An important subclass is the one of 
simple symmetric \sfa s, which contains in particular the class of \hsfa s:\,%
  \footnote{~A haploid special Frobenius algebra is automatically
  symmetric, see corollary I:3.10.} 
\\[-2.2em]

\dtl{Definition}{haploid}
An algebra $A$ in a tensor category $\calc$ is called {\em haploid\/}
\cite{fuSc16} iff it contains the tensor unit with multiplicity one, i.e.\ iff
  \be  \Hom(\one,A) \cong \complex \, . \labl{hapl}
$A$ is called {\em simple\/} iff \cite{ostr,ffrs}
  \be  \HomAA(A,A) \cong \complex \, . \labl{simpl}

Haploid algebras can be characterized as the algebras that are simple
as left modules over themselves, and simple algebras as those that are simple as
bimodules over themselves. A haploid algebra is in particular simple. Also, 
every simple algebra in $\calc$ is Morita equivalent to a haploid algebra 
\cite{ostr}. 
Since in the construction of a full CFT on oriented world sheets for a given 
modular tensor category only the Morita class of an algebra matters, this
justifies to restrict, as we will do later on, our attention to haploid algebras. 

The goal of this section is to provide a systematic construction of \hsfa s from
invertible elements and to show that these algebras give rise to most of 
the known module categories -- the ones that are of ``D-type'' in A-D-E type 
classifications. In subsection 3.1 we prove general estimates on haploid
algebras in modular tensor categories. In subsection 3.2 we show that
the restriction to invertible simple subobjects gives rise to subalgebras. These
subalgebras can be classified by the theory of algebras in theta-categories
that we develop in subsection 3.3. In subsection 3.4 we show that isomorphism
classes of such algebras are classified by certain bihomomorphisms which
we call {\em \ksb s\/}. Finally, in subsection 3.5,
we compute the modular invariant torus partition function that results from
such an algebra and show that it is of the form studied by Kreuzer
and Schellekens.

\subsection{Simple algebras in modular tensor categories}

In this subsection we establish some estimates, involving the dimensions of
simple objects in $\calc$, for quantities related to simple (in particular,
to haploid) symmetric special Frobenius algebras in modular tensor categories.
Some of these are similar to considerations in \cite{gann17},
where they are formulated at the level of ``modular data'' rather than for
modular tensor categories. They use some basic results from \PF\ theory (for a 
summary of \PF\ theory see e.g.\ chapter XIII of \cite{GAnt}), in particular
\\[-2.3em] 

\dtl{Lemma}{lemma:gann0}
Let ${\rm R}\eq ({\rm R}_{ij})$ be an $n\,{\times}\, n$ matrix with non-negative 
real entries. Then the diagonal entries of $\rm R$ are bounded by the \PF\ 
eigenvalue $\r$ of $\rm R$: $\r\,{\ge}\, \max_{1\leq i\leq n}\{{\rm R}_{ii}\}$.

\medskip\noindent
Proof: \\
Since the matrix $\rm R$ has non-negative entries, the diagonal entries 
$({\rm R}^m)_{ii}$ of its $m$th power are bounded below by $({\rm R}_{ii})^m$: 
$({\rm R}^m)_{ii}\,{\ge}\, ({\rm R}_{ii})^m$. The eigenvalues of 
${\rm R}^m$, on the other hand, are powers $(\lambda_j)^m$ of the eigenvalues
$\lambda_j$ of $\rm R$. So we have the inequality
  \be  ({\rm R}_{ii})^m \le \sum_k({\rm R}^m)_{kk} = \sum_j (\lambda_j)^m
  \le n\, \r^m  \labl{l1}
for any $i$. If the entry ${\rm R}_{ii}$ were larger than the \PF\
eigenvalue $\r$, then for sufficiently large $m$ 
the left hand side of \erf{l1} would surpass the right hand side.
\qed

A symmetric special Frobenius algebra $A$ has only finitely many equivalence
classes of simple left modules (see theorem I:5.18); we select a set 
$\{M_\kappa\}$ of representatives for these, labeled by $\kappa\iN \JJ$.  
We now assume that $A$ is haploid; then $A$ is a simple left module over
itself, and we choose it as one of the representatives, denoting the
corresponding label in $\JJ$ by $0$. We also introduce, for every $i\iN \II$, 
the $|\JJ|\,{\times}\,|\JJ|$\,-matrix $\rmA_i$ with entries the non-negative 
integers 
  \be  (\rmA_i)_\kappa^\kappb \equiv \rmA_{i\kappa}^{\;\kappb}
  := \dimc\HomA( M_\kappa\oti U_i, M_\kappb) \,.  \labl{ma}
These matrices, which in \cft\ yield the annulus partition function,
are known to furnish a NIM-rep of the fusion rules (see theorem I:5.20),
and their eigenvalues are $S_{i,m}/S_{0,m}$; thus the \PF\ eigenvalue of 
$\rmA_i$ is given by $\dim(U_i)$.

\dtl{Proposition}{absch}
Let $A$ be a haploid \sfa\ in a modular tensor category $\calc$. Then
  \be \dimc\Hom(U,A) \,\leq\, \dim(U)  \labl{e2}
for every object $U$ of $\calc$.

\medskip\noindent
Proof: \\
Due to the semisimplicity of $\calc$ it is enough to establish \erf{e2} for
simple objects $U\eq U_i$ only. For these one has the Frobenius reciprocity
relation $\Hom(U_i,A) \,{\cong}\,
\HomA(\IndA(U_i),A)$, where $\IndA(U_i)$ is an induced $A$-module, which
as an object of $\calc$ is $A\oti U_i$. Thus
  \be  \dimc\Hom(U,A) = \dimc\HomA(\IndA(U_i),A) = \rmA_{i\,0}^{\;0} \,,
  \ee
and by lemma \ref{lemma:gann0} this integer is indeed bounded above by 
the \PF\ eigenvalue $\dim(U_i)$ of $\rmA_i$.
\qed 

\dt{Remark}
(i)~\,\,The result \erf{e2} implies in particular that a simple current appears
    in a \hsfa\ with multiplicity at most one. 
\\[.2em]
(ii)~\,The result also implies that in a modular tensor category there 
are only finitely many equivalence classes 
of objects that can carry the structure of a \hsfa.
\\[.2em]
(iii)~It can actually be shown (see formula (3.9b) of \cite{gann17}) that 
{\em all\/} entries of the matrix $\rmA_i$,
not only the diagonal ones, are bounded by $\dim(U_i)$:
  \be  \rmA_{i\kappa}^{\;\kappb} \,\leq\, \dim(U_i)  \labl{A-bound}
for all $i\iN \II$ and all $\kappa,\kappb\iN \JJ$.
This can be used to obtain upper bounds on $\dimc\Hom(U,A)$ when $A$ is
simple, but not haploid. For instance, using that $A$ is isomorphic 
\cite{ostr} to the internal End \underline{\rm End}$(A)$ of $A$ 
(regarded as a left module over itself) it follows that
  \be \dimc\Hom(U,A) \le \llb \dimc(\Atop)\lrb^2 \dim(U) \,, \labl{E2}
where as in \citei\ by $\Atop$ we denote the $\complex$-\alg\ $\Hom(\one,A)$.

\medskip

Next we wish to show that each object in a modular tensor category can be
endowed with the structure of a \hsfa\ in at most finitely many ways.
To this end, we need the following result, which was first obtained in 
\cite{boev} (see also a related estimate in \cite{gann}) with the help of 
\PF\ theory. The proof given here is of independent interest.
We derive an upper bound for the multiplicities $Z_{ij}\eq Z_{ij}(A)$
that appear in the modular invariant torus partition function associated
to $A$. According to theorem I:5.1, $Z_{ij}$ can be expressed
as the invariant of a certain ribbon graph in the three-manifold 
$S^2{\times}S^1$, see formula (I:5.30); $Z_{ij}$ can also
be interpreted as the dimension of a space of $A$-bimodule morphisms
between alpha-induced $A$-bimodules, see section I:5.4.
The interest of the estimate lies in the fact that it is independent of the 
choice of $A$ and involves only data of the tensor category $\calc$.

\dtl{Lemma}{est:pf}
For $A$ a {\em simple\/} symmetric special Frobenius algebra in a modular 
tensor category $\calc$, the integers $Z_{ij}\eq Z_{ij}(A)$ satisfy 
  \be  Z_{ij} \le \dim (U_i) \dim (U_j) \, . \labl2

\noindent
Proof: \\
(i)~\,We start with the observation that 
  \be  \dimc\Hom (U,V) \leq \dim(U) \dim(V)  \labl{l3}
for any two objects $U$ and $V$ of
a semisimple sovereign tensor category $\calc$. To see this, use the 
semisimplicity to write the objects $U,V$ as direct sums of simple objects, 
$U\,{\cong}\,\bigoplus_{i\in \II} n_i^U U_i$ and $V\,{\cong}\,\bigoplus_
{i\in \II}n_i^V U_i$. Since dimensions are bounded below by 1, we have
$\sum_{i\in \II} n_i^U \,{\le}\,\dim(U)$ and
$\sum_{i\in \II} n_i^V \,{\le}\,\dim(V)$. We thus obtain the estimate
  \be  \bearll \dimc\Hom (U,V) = \displaystyle\sum_{i\in \II} n_i^U n_i^V
  \!\! &  \le \displaystyle\sum_{i,j\in \II} n_i^U n_j^V
  \\{}\\[-.7em]
  &= \llb \displaystyle\sum_i n_i^U\lrb \llb\sum_j n_j^V\lrb \leq \dim(U)\dim(V)
  \,. \eear \ee
(ii) We apply this result to the category $\calcAA$ of $A$-bimodules, which
is again a semisimple tensor category.  
Let us first explain that $\calcAA$ is even a sovereign tensor category.
If $X\eq(\dot X,\rho,\tilde\rho)$ is an $A$-bimodule, then the left
action $\rho$ of $A$ on $\dot X$ can be used to define a right action
of $A$ on the dual object $\dot X^\vee$, while the right action
$\tilde\rho$ of $A$ on $\dot X$ gives a left action on $\dot X^\vee$.
Furthermore, using sovereignty of $\calc$ one checks that the two
actions commute, and hence one has an $A$-bimodule structure on $\dot X^\vee$;
in \citeii, this bimodule structure is denoted by $X^v$, compare (II:2.40).
The assignment $X\,{\mapsto}\,X^v$, together with 
morphisms $d^A_X$ and $b^A_X$ that are analogous to those described
e.g.\ in formula (3.51) of \cite{ffrs} for the case of $A$-modules,
furnishes a left duality of $\calcAA$; analogously one has a right duality. 
Finally one checks, by arguments similar to those in section 5.3 of 
\cite{fuSc16}, that this way $\calcAA$ becomes a sovereign tensor category.
\\
Further, alpha-induction provides two tensor functors $\alpha_{\!A}^\pm$ 
from $\calc$ to $\calcAA$ that preserve the dualities and thus the dimensions 
(provided that the tensor unit of $\calcAA$ is simple, which is the case iff 
$A$ is simple as an algebra). Combining this result with \erf{l3}, we obtain
  \be  \begin{array}{ll}
  Z_{ij} \!\!& = \dim \HomAA (\alpha_{\!A}^-(U_j) , \alpha_{\!A}^+(U_i^\vee) )
  \\{}\\[-.7em]
  & \le \dimAA(\alpha_{\!A}^-(U_j)) \, \dimAA(\alpha_{\!A}^+(U_i^\vee))
  = \dim (U_i) \dim (U_j) \,,  \end{array} \ee
which establishes the estimate \erf{2}.
\qed

\medskip

Recall from \citei\ that, by our conventions, the counit and coproduct of a 
symmetric special Frobenius algebra are normalized such that
$\eps \cir \eta \eq \dim(A) \,\id_{\one}$.

\dtl{Proposition}{prop:finite}
(i)~\,An object in a modular tensor category can be endowed with the 
structure of a haploid \sfa\ in at most finitely many inequivalent ways. \\
(ii)~A modular tensor category admits only finitely many inequivalent
     haploid \sfa s. 

\medskip\noindent
Proof: \\
According to theorem I:5.18 
the number $|\JJ|$ of inequivalent simple left modules over a symmetric special 
Frobenius algebra $A$ is given by $\sum_i Z_{i \ibar }(A)$, where $\ibar $
labels the simple object isomorphic to $U_i^\vee$.
When combined with the estimate \erf2, we thus have
  \be  |\JJ| = \sum_{i\in \II} Z_{i \ibar } \,\le \sum_{i\in \II} \dim (U_i)^2 
  = {\Dim(\calc)}^2_{} \,.  \ee
Now according to corollary 2.22 of \cite{etno}, for every (multi-)fusion category 
$\calc$ the number of module categories over $\calc$ with a prescribed number 
of non-isomorphic simple objects is finite. 
Thus let $\{ \mathcal{M}_\ell \}$, with $\{\ell\}$ some finite index set, be a 
list of non-isomorphic module categories, each of which has at most 
$\Dim(\calc)^2$ simple objects. Every simple object in each category 
$\mathcal{M}_\ell$ gives rise to a haploid algebra in $\calc$ \cite{ostr}. 
This provides us with a finite list $\{A_{\ell,p}\}$ (with $p$ labeling the
isomorphism classes of simple objects in $\mathcal{M}_\ell$) of haploid 
algebras in $\calc$. Now take an arbitrary haploid \sfa\ $A$ in $\calc$. 
Then $\calcA \,{\cong}\,\mathcal{M}_\ell$ for some value of $\ell$. Since $A$ 
is haploid, it is simple as a left module over itself. Any simple object in 
$\mathcal{M}_\ell$ isomorphic to $A$, regarded as a left module over itself,
gives an algebra in $\calc$ that is isomorphic (as an algebra)
to $A$. Thus $A$ is isomorphic to $A_{\ell,p}$ for some $p$. Since the list 
$\{A_{\ell,p}\}$ is finite, this implies the finiteness assertions (i) and (ii).
\qed

\subsection{Restriction to invertible subobjects}

We now study invertible objects in \hsfa s. We start with the 
\\[-2.3em]

\dtl{Definition}{def:pic}
(i)~\,\,Writing an object $U$ of $\calc$ as a direct sum of simple objects $U_i$,
  \be  U \cong \bigoplus_{i\in \II} n_i\, U_i 
  \equiv \bigoplus_{i\in \II}U_i^{\oplus n_i} \,,  \ee
we denote by $\Io\,{\subseteq}\,I$ the subset containing those labels $i\iN \II$
such that $U_i$ is an invertible object.
The {\em Picard subobject\/} $\Pics(U)$ of $U$ is defined as
  \be  \Pics(U) := \bigoplus_{i\in \Io} n_i\, U_i \,.  \ee
(ii)~\,An object $U$ of $\calc$ is called {\em of simple current type\/}
    iff $U\,{\cong}\, \Pics(U)$. 
\\[.2em]
(iii)~The modular invariant torus partition function derived from a \hsfa\ of
    simple current type is called a {\em simple current modular invariant\/}.
\\[.2em]  
(iv)~We call a \hsfa\ of simple current type a {\em \berta\/}.

\dtl{Remark}{rem}
(i)~\,Whenever it is convenient, we identify $\Pics(U)$ with an object
         of the Picard category $\calpic(\calc)$. 
\\[.2em]
(ii)~\,Every simple algebra $A$ of simple current type in $\calc$ is Morita 
equivalent to a \berta. Indeed, any such $A$ is also a simple algebra in 
$\calpic(\calc)$ and hence, by general arguments, Morita equivalent (in 
$\calpic(\calc)$) to a haploid algebra in $\calpic(\calc)$. That haploid 
algebra, in turn, is also an algebra in
$\calc$, in fact a \berta, and it is Morita equivalent in $\calc$ to $A$.
(The argument also shows that the interpolating bimodules of the associated
Morita context are of simple current type as well.)
\\[.2em]
(iii)~As we will see, with our notion of simple current type algebras we 
obtain precisely the modular invariants that were discussed in \cite{krSc}. 
Thus the notion of a modular invariant of simple current type used here is 
more restrictive than the one given in \cite{gasc,gasc2}. It is not known if
modular invariants that are simple current invariants in the sense of
\cite{gasc,gasc2}, but not simple current invariants in our sense, 
can appear as torus partition functions of consistent \cfts.
(A class of examples of modular invariants of this type that are
definitely unphysical has been found in section 4 of \cite{fusS}.)

\medskip

A construction similar to the one of definition \ref{def:pic}(i) can be 
performed for any semisimple full tensor subcategory $\calc'$ of $\calc$. 
We refer to the object $U'$ in $\calc'$ associated in this way to 
an object $U$ in $\calc$ as the {\em truncation\/} of $U$ to $\calc'$. 
\\[-2.2em]

\dt{Proposition}
Let $\calc'$ be a semisimple full tensor subcategory of $\calc$. Let 
$A'$ be the truncation of an object $A\iN\obj(\calc)$ to $\calc'$, and 
select embedding and restriction morphisms $e\iN \Hom(A',A)$ and 
$r\iN \Hom(A,A')$ such that $r\cir e\eq\id_{A'}$. 
\\[.2em]
(i)~\,\,If $(A,m,\eta)$ is an algebra in $\calc$, then
     $(A', r\cir m\cir(e\oti e), r \cir\eta)$ is an algebra in $\calc'$
     (and in $\calc$).
\\[.2em]
(ii)~\,If $\calc$ is a ribbon \cat\ and $(A,m,\eta,\Delta,\eps)$ is a 
symmetric Frobenius algebra in $\calc$, then $(A',r\cir m\cir(e\oti e), 
r\cir\eta, \zeta^{-1} (r\oti r)\cir\Delta\cir e, \zeta \eps\cir e)$ 
is a symmetric Frobenius algebra in $\calc'$, for any $\zeta\iN\complexx$. 
\\[.2em]
(iii)~Let $A$ and $A'$ be as in (ii). Suppose that in addition $A$ and $A'$
are haploid. Then if $A$ is special, so is $A'$.

\medskip\noindent
Proof:\\
(i)~\,Introducing bases in the relevant morphism spaces, the associativity
property of the multiplication $m$ takes the form given in (I:3.8). In that 
form, the associativity relations for $A$ contain as a subset the associativity 
relations for $A'$. A similar statement applies to the unit properties.
\\[.2em]
(ii)~\,An analogous reasoning allows one to extend the result to
symmetric special Frobenius algebras.
\\[.2em]
(iii)~From (ii) we already know that $A'$ is a symmetric Frobenius algebra. To 
show that $A'$ is special, by lemma I:3.11 it is sufficient to verify that the 
counit $\eps'$ of $A'$ obeys $\eps'\eq\gamma \eps'_\natural$ for some 
$\gamma\iN\complexx$, where $\eps'_\natural$ is the morphism defined in
(I:3.46). Since $A'$ is haploid, the relation $\eps'\eq\gamma \eps'_\natural$ 
holds for some $\gamma\iN\complex$, so it remains to show that $\gamma\,{\ne}
\,0$. Composing both sides of $\eps'\eq\gamma \eps'_\natural$ with $\eta'$,
on the \rhs\ we obtain $\gamma \dim(A')$. To establish that $\gamma\,{\ne}\,0$
we must check that the left hand side is non-zero. Now note that 
since both $A$ and $A'$ are haploid, we have
$\eps'\eq\lambda \, \eps$ and $\eta'\eq\tilde\lambda \, \eta$ for some
$\lambda, \tilde\lambda \iN \complexx$. It follows that
$\eps' \cir \eta'\eq\lambda \tilde\lambda \, \eps \cir \eta
\eq\lambda \tilde\lambda \, \dim(A) \,{\neq}\, 0$.
\qed

\medskip

As a consequence of this result, the classification of \berta s
in modular tensor categories amounts to classifying \hsfa s in 
theta-categories. At the same time, such a classification fixes the algebra
structure of a general \hsfa\ on its simple current type subobjects.

\subsection{Algebras in theta-categories}

We start with a discussion of (associative, unital) algebras in 
theta-categories. The following construction provides examples of algebras 
in the pointed category $\calc(G,\psi)$. 
We start with a closed three-co\-cycle $\psi\iN Z^3(G,\complexx)$ and
a $\psi$-trivializable subgroup $H$ of $G$, with trivializing two-cocycle
$\omega$. This cocycle is, in general, not closed, so modifying the
convolution product of the group algebra $\complex H$ by $\omega$
according to
  \be  b_g \star_\omega^{} b_{g'} \equiv m(b_g \oti b_{g'})  
  := \omega(g,g')\, b_{g g'}  \labl{mCoH1} 
does {\em not\/}, in general, yield an associative algebra 
$\complex_\omega H$, at least not in the category of vector spaces. However, 
since the violation of associativity is nothing but $\rmd\omega\eq\psi$ and
thus a closed three-co\-cycle, it can be canceled by simply changing the notion
of associativity -- i.e., by regarding $\complex_\omega H$ as an object in 
the tensor category $\calc(G,\psi)$ of $G$-graded vector spaces with 
associativity constraint given by $\psi$. 
The following lemma asserts that in this category the  object
  \be  \complex_\omega H =: A(H,\omega) \,\in \obj(\calc(G,\psi))
  \labl{CoH}
{\em is\/} in fact an associative algebra. By a slight abuse of terminology,
we will still refer to \erf{CoH} as a {\em twisted group algebra\/}.

\dtl{Lemma}{def:algtheta}
Let $G$ be a finite group.\\[.2em]
(i)~\,\,For $\psi\iN Z^3(G,\complexx)$, fix a $\psi$-trivializable subgroup 
    $H$ of $G$ with trivialization $\omega$. Then the twisted group 
    algebra $\complex_\omega H \,{\equiv}\, A(H,\omega)$
    with product given by \erf{mCoH1} is a haploid algebra in $\calc(G,\psi)$.
\\[.2em]
(ii)~Any two trivializations of $H$ differ by a two-cocycle on $H$. If
    the two-cocycle is a coboundary, the two associated twisted group algebras
    are isomorphic. As a consequence, the possible twisted group algebras
    for a given $\psi$ form a torsor\,%
 \footnote{~Recall that a set $S$ is a {\em torsor\/} over a group $G$
 iff there is a free transitive action of $G$ on $S$. In particular,
 $S$ and $G$ are isomorphic as sets; but the isomorphism is not canonical.}
over $H^2(H,\complexx)$.  \\[.2em]
(iii)~Setting
  \be
  \Delta(b_g) := \frac{1}{|H|} \sum_{h\in H} \frac{1}{\omega(gh^{-1},h)} \,
  b_{gh^{-1}} \oti b_h  \qquad {\rm and} \qquad
  \eps(b_g) := |H| \, \delta_{g,e} 
  \labl{eq:AH-ssFA} 
turns $A(H,\omega)$ into a haploid special Frobenius algebra in $\calc(G,\psi)$. 

\medskip\noindent
Proof: \\
The axioms can be checked by straightforward calculations.
For example, associativity of $A \,{\equiv}\, A(H,\omega)$ amounts
to $m \cir (\id_A \oti m) \cir \alpha_{A,A,A} \eq
m \cir (m \oti \id_A)$. Evaluating this relation on a basis element
$(b_{h_1} \oti b_{h_2}) \oti b_{h_3}$ gives
  \be
  \omega(h_1, h_2 h_3)\, \omega(h_2,h_3)\, \psi(h_1,h_2,h_3)^{-1}
  = \omega(h_1,h_2)\, \omega(h_1 h_2, h_3) \,, \ee
which is equivalent to $\rmd\omega(h_1,h_2,h_3)\eq\psi(h_1,h_2,h_3)$.
As another example, to establish specialness of $A$
one must check that $m \cir \Delta(b_g) \eq b_g$ and 
$\eps \cir \eta \eq \dim(A)$. This follows immediately when
substituting the definitions \erf{mCoH1} and \erf{eq:AH-ssFA}.
\qed

\dtl{Remark}{rem:algtheta}
(i)~\,\,As already pointed out,
$A(H,\omega)$ is, in general, not an associative algebra 
in the category of vector spaces, since $\omega$ is in general not closed. 
\\[.2em]
(ii)~\,There is again an equivalence relation: consider another three-co\-cycle 
$\psi'\iN Z^3(G,\complexx)$ and two-cochain $\omega'\iN C^2(H,\complexx)$ 
such that $ \rmd \omega' \eq \psi'_{|H}$. If $\psi$ and $\psi'$ are 
cohomologous, $\psi'\eq\psi \, \rmd \eta$,  
then the categories $\calc(G,\psi)$ and $\calc(G,\psi')$ are equivalent, and we
identify them. If, moreover, there is a one-cochain $\chi\iN C^1(H,\complexx)$ 
such that $\omega'\eq \omega\, \eta_{|H}\, \rmd \chi$,
then the two algebras $A(H,\omega)$ in $\calc(G,\psi)$ and $A(H,\omega')$ 
in $\calc(G,\psi')$ are isomorphic. 
\\[.2em]
(iii)~Algebras in the category $\calc(G,\psi)$ have also been considered in
\cite{ostr2}. There, the Morita classes of twisted group algebras are studied. 
It is found that $A(H_1,\omega_1)$ and $A(H_2,\omega_2)$ are Morita equivalent 
(i.e.\ have isomorphic module categories) iff the pairs $(H_1,[\omega_1])$ and 
$(H_2,[\omega_2])$ are conjugate under the action of $G$.
\\
In conformal field theory, only the Morita class of an algebra matters. 
Note, however, that in our applications pointed categories arise as Picard 
subcategories of modular tensor categories $\calc$, and that the Morita 
classes with respect to $\calc$ will in general be larger than those with 
respect to $\calpic(\calc)$. An example is provided by the critical Ising 
model, see remark \ref{warning} below.

\medskip

For the proof of the next lemma, we introduce a basis choice that we will 
also use repeatedly in the sequel.
\\[-2.2em]

\dtl{Definition}{def-adbases}
Let $\ircalc$ be a pointed sovereign tensor category equivalent to
$\calc(G,\psi)$, and $A$ a \hsfa\ in $\ircalc$.
\\[.1em]
(i)~\,\,The {\em support\/} $H(A)$ of the algebra $A$ is the subset
  \be  H(A) := \{ g\iN G \,|\, \dimc\Hom(\U_g,A)\,{>}\,0 \}  \labl{H(A)}
of $G$.
\\[.2em]
(ii)~\,An {\em adapted basis} for $A$ is the choice, for all $g\iN H(A)$, of 
morphisms $e_g\iN \Hom(\U_g,A)$ and $r_g\iN\Hom(A,\U_g)$ that form 
bases of these morphism spaces that are dual in the sense \citei\ that
$r_g\cir e_g\eq \id_{\U_g}$. 
For the unit element $\bfe\,{\equiv}\,\U_1\iN H(A)$ we take
$e_1\eq\eta$ and $r_1\eq\dim(A)^{-1}\eps$. 

\medskip

Note that for each $g\iN H(A)$, the morphism 
  \be  p_g := e_g\cir r_g ~\iN \End(A)  \ee 
is an idempotent. To make the connection with the notation in \citei, let us 
present explicitly the product and coproduct on 
$A(H,\omega)$, using the notations of (I:3.7) and (I:3.82). One finds
  \be
  m_{g,h}^{~~gh} = \omega(g,h)  \qquad {\rm and} \qquad 
  \Delta^{g,h}_{gh} = \big( \dim(A)\, \omega(g,h) \big)^{-1} \,.  \labl{ir1}
The last expression can also be obtained from (I:3.83), noting that
via $\psi|_H \eq \rmd \omega$ it is possible to express the 
$\mathsf{F}$-matrix elements in (I:3.83) in terms of $\omega$.

\dtl{Lemma}{lem}
Let $\ircalc$ be a theta-category equivalent to $\calc(G,\psi,\Omega)$ and $A$ 
a \berta\ in $\ircalc$.
\\[.1em]
(i)~\,\,For each simple object $\U_g$ of $\ircalc$ the dimension of the
    morphism space $\Hom(\U_g,A)$ is either zero or one.
\\[.2em]
(ii)~\,The support $H(A)$ is a subgroup of $G$.  \\[.2em]
(iii)~Choose an adapted basis for $A$. For all $g,h\iN H(A)$ we have 
$r_{gh} \cir m \cir (e_g \oti e_h) \,{\ne}\,0$ and
$(r_g \oti r_h) \cir \Delta \cir e_{gh} \,{\ne}\,0$.

\medskip\noindent
Proof: \\
(i)~\,\,We already know this result for every theta-category that is 
the Picard category of a modular tensor category. Here we give a proof that
does not rely on modularity.
\\
Using Frobenius reciprocity, we have
  \be  \HomA(A\oti \U_g,A\oti \U_g) \,\cong\, \Hom(\U_g,A\oti \U_g)
  \,\cong\,\complex \,, \ee
where the second equality comes from the fact that the only subobject of
$A$ that contributes is the tensor unit, which appears in a haploid algebra
with multiplicity one. This shows that all induced $A$-modules $A\oti \U_g$
are simple. Again by Frobenius reciprocity, we have
  \be  \Hom(\U_g,A) \cong \HomA(A\oti \U_g,A) \,,  \ee
which as the space of intertwiners
between simple $A$-modules has either dimension zero or dimension one.
\\[.2em]
(ii)~\,Frobenius algebras in sovereign tensor categories can only be 
defined on self-conjugate objects (see e.g.\ lemma 3.3 of \cite{fuSc16}). 
Since the conjugate of $\U_g$ is 
isomorphic to $\U_{g^{-1}}$, this implies that $H(A)$ is closed under inverses.
\\
Now choose an adapted basis for $A$. Suppose
that $g,h\iN H(A)$, but $gh\,{\not\in}\, H(A)$. Then $m\cir (e_g\oti e_h)\eq0$.
Associativity of the multiplication $m$ of $A$ then implies that 
  \be  0 = (m \cir ( m\oti \id_A)) \circ (e_g\oti e_h \oti e_{h^{-1}})
  = e_g \otimes  \llb \dim(A)^{-1} \eps\cir m \cir (e_h \oti e_{h^{-1}}) \lrb
  \,. \labl{scal}
in $\Hom(\U_g\Oti\,\U_h\Oti\,\U_{h^{-1}}, A)$. For a Frobenius algebra, 
$\eps \cir m$ is non-degenerate, and hence $\eps\cir m\cir(e_h\oti e_{h^{-1}})$ 
is non-zero. Furthermore, $e_g$ is non-zero by construction. This is a 
contradiction; hence $gh\iN H(A)$.  \\[.2em]
(iii)~By an analogous argument as in (ii), using now also directly the
Frobenius property of $A$, assuming that $m\cir (e_g\oti e_h)\eq0$ leads
to a contradiction. For the coproduct one repeats the argument using
coassociativity.
\qed

\dtl{Proposition}{prop}
(i)~\,\,Let $\ircalc$ be a theta-category equivalent to $\calc(G,\psi,\Omega)$,
and $A$ a \hsfa\ in $\ircalc$. Then $A$ is isomorphic to one of the twisted
group algebras $A(H(A),\omega)$. 
\\[.2em]
(ii)~\,A \berta\ in a modular tensor category $\calc$ is characterized by a 
subgroup $H$ in the effective center
$\Pico(\calc)$ and a trivialization of the associativity constraint on $H$.

\medskip\noindent
Proof: \\
(i)~\,\,We select bases $\bL_{g}{h}$ of $\Hom(\U_g\Oti\,\U_h,\U_{gh})$ as in
   \erf{bL-def} and define $\omega(g,h)$ by
  \be  r_{gh} \circ m \circ (e_g \oti e_h) =: \omega(g,h)\, \bL_{g}{h}
  \,.  \labl{mCoH2}
{}From lemma \ref{lem}(iii) we know that $\omega(g,h)$ is non-vanishing for 
all $g,h\iN H(A)$. Since $A$ is an algebra, we have $m \cir (\id_A \oti m) \eq 
m \cir (m \oti \id_A)$. Evaluating both sides in an adapted basis gives
   \begin{eqnarray}  &
   r_{\! h_1 h_2 h_3}^{} \cir m \cir (\id_A \oti m) \cir
   (e_{h_1} \Oti e_{h_2} \Oti e_{h_3}) =
   \omega(h_1, h_2 h_3)\, \omega(h_2,h_3)\,
   \bL_{h_1}{h_2 h_3} \cir (\id_{L_{h_1}} \Oti\, \bL_{h_2}{h_3}) \,,
   &\nonumber \\[.2em]&
   r_{\! h_1 h_2 h_3}^{} \cir m \cir (m \oti \id_A) \cir
   (e_{h_1} \Oti e_{h_2} \Oti e_{h_3}) =
   \omega(h_1 h_2, h_3)\, \omega(h_1,h_2)\, 
   \bL_{h_1 h_2}{h_3} \cir (\bL_{h_1}{h_2} \Oti\, \id_{L_{h_3}} ) \,.
   \nonumber \\[.1em]& {}
   \end{eqnarray} 
The two basis elements of 
$\Hom(L_{h_1}\Oti L_{h_2}\Oti L_{h_3},L_{h_1 h_2 h_3})$ are related by 
   \be
   \bL_{h_1\!}{h_2 h_3} \cir (\id_{L_{h_1}} \Oti\, \bL_{h_2}{h_3})
   = \psi(h_1,h_2,h_3)^{-1}\, 
   \bL_{h_1 h_2}{h_3} \cir (\bL_{h_1\!}{h_2} \oti \id_{L_{h_3}} )  \,.
   \ee
Thus the associativity of the multiplication 
implies that $\omega$ trivializes $\psi$ on the support $H(A)$.
\\[.2em]
(ii)~\,follows immediately from (i).
\qed

    \medskip

Recall from remark \ref{square}(ii) that the effective center $\Pico(\calc)$ 
is, in general, not a subgroup of $\Pic(\calc)$, but only a subset. When,
as in proposition \ref{prop}(ii) above, we talk about a subgroup in
$\Pico(\calc)$, we mean a sub{\em set\/} of $\Pico(\calc)$ that is closed under
multiplication, and as a consequence is a subgroup of $\Pic(\calc)$.

\subsection{\ksb s}

Different trivializations of the same subgroup $H$ of the effective center
that differ by an exact two-cochain give rise to isomorphic \berta s. In
order to describe (and classify) isomorphism classes of \berta s we need
at tool that allows us to take care of this. This tool is provided by the 
notion of a \ksb\ that is introduced in the present subsection; it 
constitutes an analogue of the description of ordinary twisted group 
algebras of abelian groups in terms of alternating bihomomorphisms.  

\dtl{Definition}{altbi}
For $G$ be a finite abelian group, an {\em alternating\/} 
bihomomorphism on $G$ is a bihomomorphism
  \be  \zeta: \quad  G\times G \to \complexx  \ee
such that $\zeta(g,g)\eq 1$ for all $g\iN G$. 

\smallskip

Note that $\zeta(g,g)\eq 1$ for all $g\iN G$ implies that 
$\zeta(g,h)\eq\zeta(h,g)^{-1}$ for all $g,h\iN G$, but the converse 
implication is not true. We also have \cite{hugh,BRow'}
\\[-2.3em]

\dtl{Lemma}{lem:bihom-wcomm}
The alternating bihomomorphisms on a finite {\em abelian\/} group $G$ form an 
abelian group $\AB(G,\complexx)$. The map 
  \be \begin{array}{rcl}
  H^2(G,\complexx) & \!\to\! & \AB(G,\complexx)
  \\{}\\[-1.2em]
  {}[\omega] &\!\mapsto\! & \zeta \quad{\rm with}\quad
  \zeta(g,h) := \Frac{\omega(g,h)}{\omega(h,g)} \end{array} \labl{isom}
furnishes an isomorphism of abelian groups.

\medskip

For abelian groups, this fact provides a convenient characterization of 
isomorphism classes of twisted group algebras by their commutator two-cocycles 
as defined in \erf{coco}. We would like to have an analogous characterization 
of \berta s in modular tensor
categories. The trivializing two-cochain, however, is in general not closed, 
so we cannot use an alternating bihomomorphism. 
The appropriate generalization is provided by the following 
\\[-2.3em]

\dtl{Definition}{ksb}
Let $H$ be a subgroup in the effective center $\Pico(\calc)$ of a ribbon 
category. A {\em \ksb\/} (or {\em KSB\/}, for short) on $H$ is a 
(not necessarily symmetric) bihomomorphism 
  \be  \XI:\quad H \times H \to \complexx  \ee
which on the diagonal coincides with the quadratic form $\delta$
introduced in \erf{delta},
\be  \XI(g,g) = \delta(g) \equiv \th_g \qquad {\rm for~all~~} g\iN H \,. 
\labl{ksbcond}

\dtl{Lemma}{lem:xi-xi=beta}
For any \ksb\ on a subgroup $H$ in $\Pico(\calc)$ we have
  \be  \XI(g,h)\, \XI(h,g) = \beta(g,h)  \labl{ksbcondi}
for all $g,h\iN H$, with $\beta$ the bihomomorphism \erf{beta} associated to
$\delta$.

\medskip \noindent
Proof: \\
Using first the definition of $\beta$, then the property \erf{ksbcond} and
then the bihomomorphism property of $\XI$ one finds
  \be  \begin{array}{ll}
  \beta(g,h) \!\!&
   = \delta(gh)\,\delta(g)^{-1}_{}\delta(h)^{-1}_{} 
  \\{}\\[-.8em]
  &= \XI(gh,gh)\,\XI(g,g)^{-1}_{}\,\XI(h,h)^{-1}_{}
   = \XI(g,h)\, \XI(h,g) 
  \eear \ee
for all $g,h\iN H$.
\qed

\dtl{Remark}{rems2}
(i)~\,Two KSBs on the same subgroup $H$ in the effective center
differ by an alternating bihomomorphism on $H$. Thus 
if there exist KSBs on $H$ at all (which is indeed the case, see remark 
\ref{exksb} below), then they form a torsor over $H^2(H,\complexx)$.
In particular, since $H^2(\zet_n,\complexx)\eq1$, KSBs on any
cyclic subgroup in the effective center are unique.
\\[.2em]
(ii)~\,Following the terminology in the physics literature \cite{vafa,krSc}
we also call the choice of a KSB for a given subgroup $H$ a choice of 
``discrete torsion''.  \\[.2em]
(iii)~There also exists an alternative description of KSBs, due to Kreuzer and
Schellekens \cite{krSc}, which makes use of a presentation of $H$ as a
direct product of cyclic groups. This is described in appendix \ref{ksmatrix}.

\medskip

Now let $A$ be a \berta\ in $\calc$ and fix an adapted basis for $A$. The 
following endomorphism of $\U_g$ is a multiple of the identity morphism:
  \bea \begin{picture}(300,113)(0,50)
  \put(0,0)      {\includeourbeautifulpicture{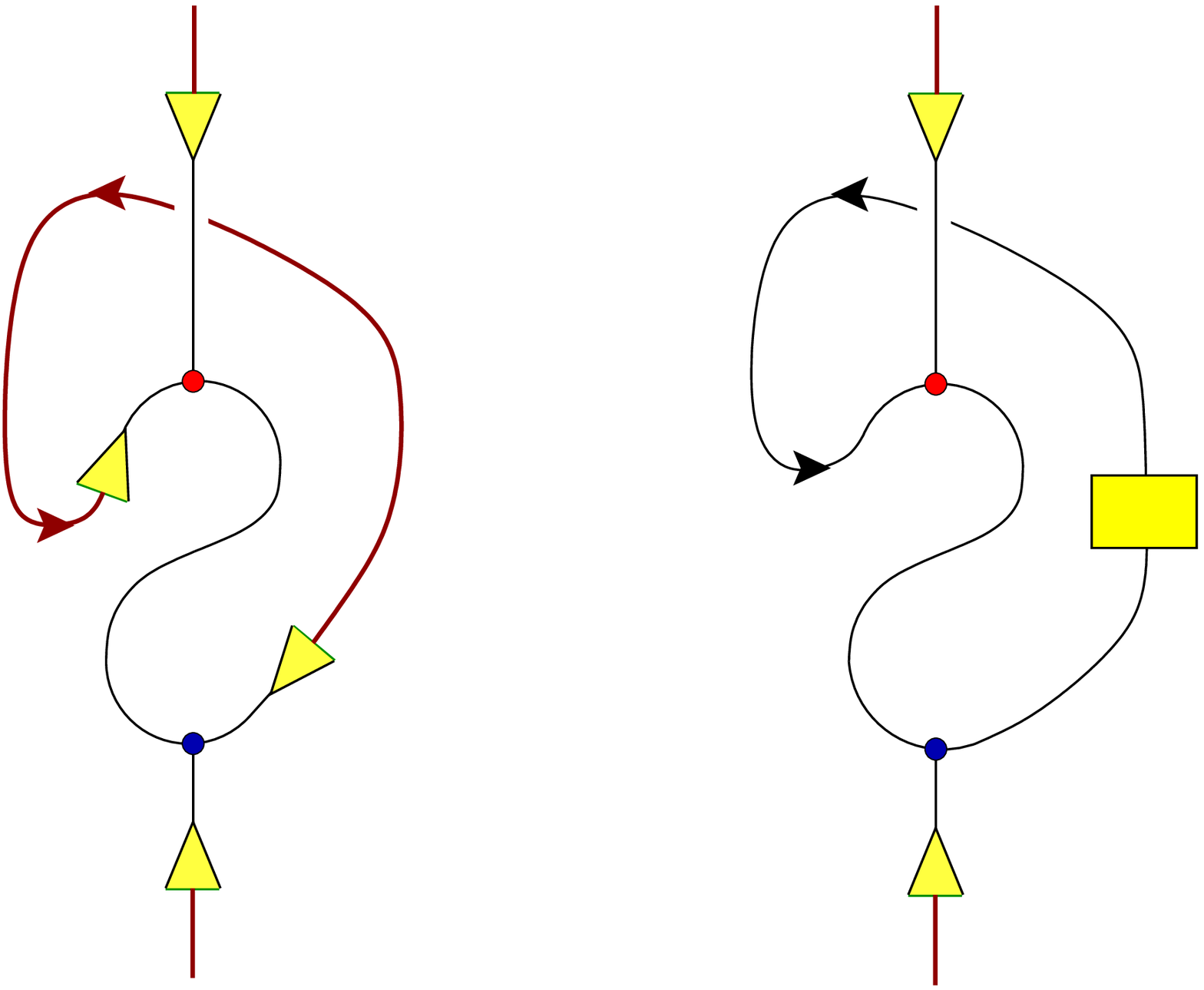}}
  \put(7.2,86)   {\scriptsize$ e_h $}
  \put(27,-8)    {\scriptsize$ \U_g $}
  \put(27,160)   {\scriptsize$ \U_g $}
  \put(30.4,73)  {\scriptsize$ A $}
  \put(34.6,20)  {\scriptsize$ e_g $}
  \put(34.6,133) {\scriptsize$ r_{\!g} $}
  \put(52.5,113) {\scriptsize$ \U_h $}
  \put(49.3,46)  {\scriptsize$ r_{\!h}^{} $}
  \put(86,75)    {$ = $}
  \put(144,-8)   {\scriptsize$ \U_g $}
  \put(144,160)  {\scriptsize$ \U_g $}
  \put(147.5,73) {\scriptsize$ A $}
  \put(170.8,112){\scriptsize$ A $}
  \put(176,74)   {\scriptsize$ p_h $}
   \put(212,75)   {$\displaystyle =\; \frac{\XI_A(h,g)}{\dim(A)}\, \id_{\U_g} \,, $}
  \epicture35 \labl{Fig1}
where $\XI_A$ is a two-cochain on $H(A)$. 
(This is essentially formula (9) of \cite{fuRs}.)
Note that by taking the trace of \erf{Fig1}, moving
the multiplication morphism past the comultiplication with the help of the
Frobenius property, and slightly deforming the resulting ribbon graph,
one arrives at the following alternative expession for $\XI_A$:
  \bea \begin{picture}(400,111)(11,31)
  \put(80,0)     {\includeourbeautifulpicture{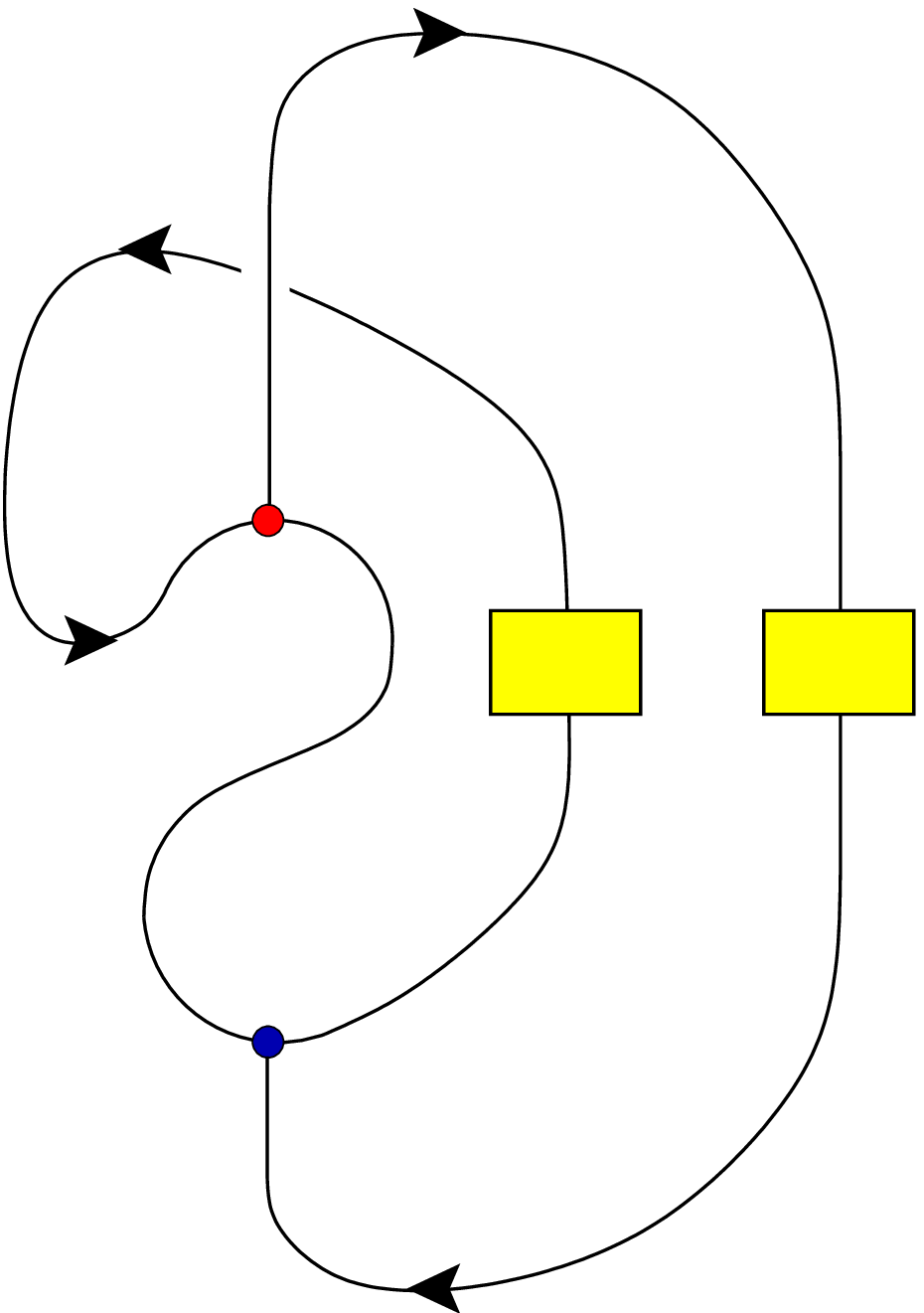}}
  \put(3,69)     {$\displaystyle \frac{\XI_A(h,g)}{\dim(A)} \;= $}
  \put(72,88)    {\scriptsize$ A $}
  \put(109,55)   {\scriptsize$ A $}
  \put(112,130)  {\scriptsize$ A $}
  \put(139,70.5) {\scriptsize$ p_h $}
  \put(169,70.5) {\scriptsize$ p_g $}
  \put(196,69)   {$ = $}
  \put(320,69)   {$ = $}
  \put(220,23)     {
    \put(0,-9)    {\includeourbeautifulpicture{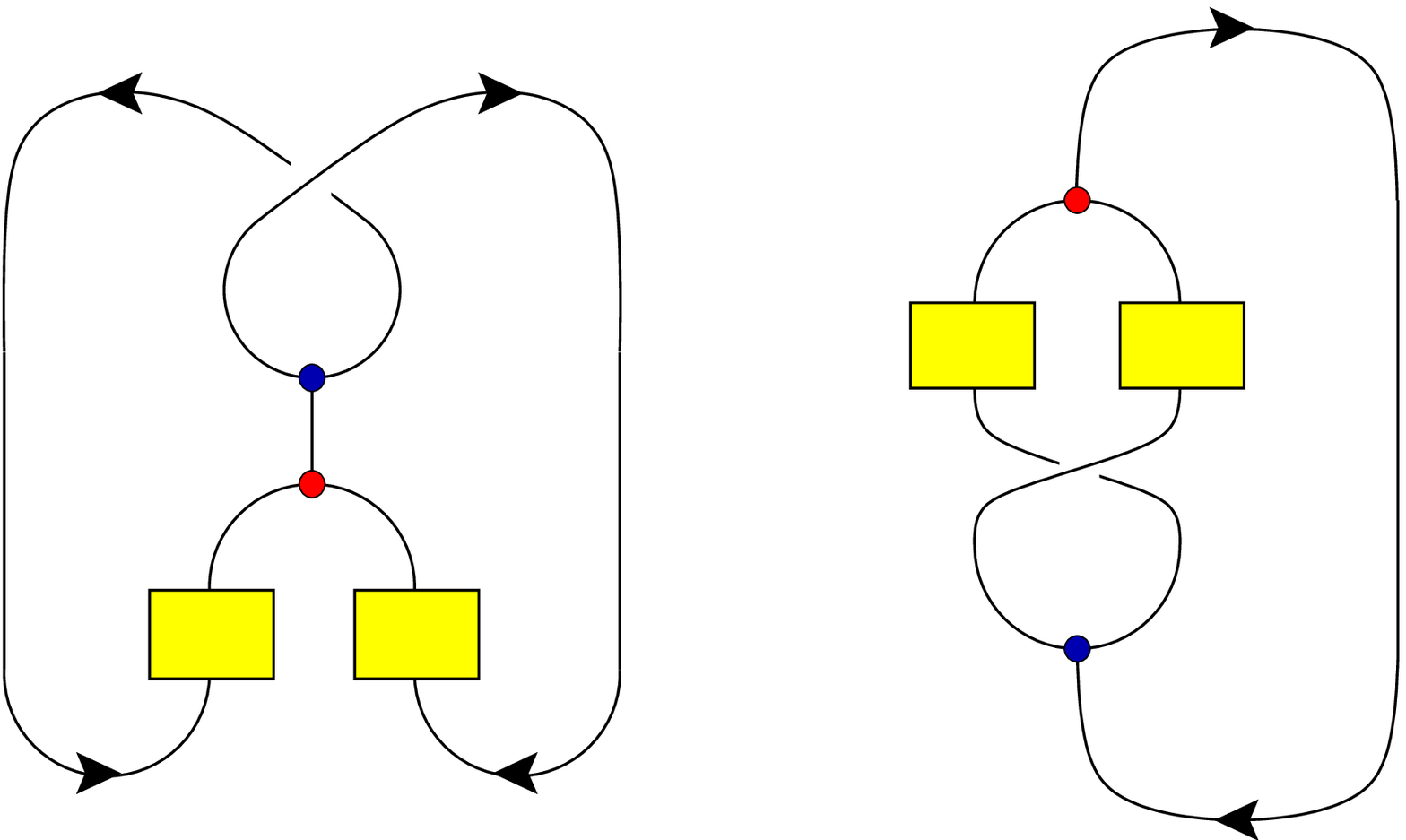}}
    \put(25,18)   {\scriptsize$ p_h $}
    \put(52,18)   {\scriptsize$ p_g $}
    \put(23,60)   {\scriptsize$ A $}
    \put(54,60)   {\scriptsize$ A $}
    \put(127,55.5){\scriptsize$ p_h $}
    \put(156,55.5){\scriptsize$ p_g $}
    \put(124.8,20){\scriptsize$ A $}
    \put(156.8,20){\scriptsize$ A $}
  } 
  \epicture11 \labl{Fig21a}
Thus we have
  \be  \XI_A(g,h) = \dim(A)\, \tr\! \big(
  m \cir (p_g \oti p_h) \cir c_{A,A} \cir \Delta \big) \,.  \labl{FigXiA}
Using the expression \erf{ir1} for the product and coproduct of 
$A$, it is easy to evaluate \erf{FigXiA}; we find
  \be
  \XI_A(g,h) = \dim(A) \, m_{g,h}^{~~gh} \, \Delta_{gh}^{h,g} \, 
  \Rs{}{h\,}g{\,gh} = \frac{1}{\Omega(g,h)} \, \frac{\omega(g,h)}{\omega(h,g)}
  \,.  \labl{XIabas}
Thus $\XI_A$ is the generalization to the braided setting of the 
concept of commutator cocycle $\omega_{\rm comm}$, which characterizes
isomorphism classes of finite abelian groups, see lemma \ref{lem:bihom-wcomm}.
Specifically, we have
  \bea \begin{picture}(170,79)(0,29)
  \put(0,0)       {\includeourbeautifulpicture{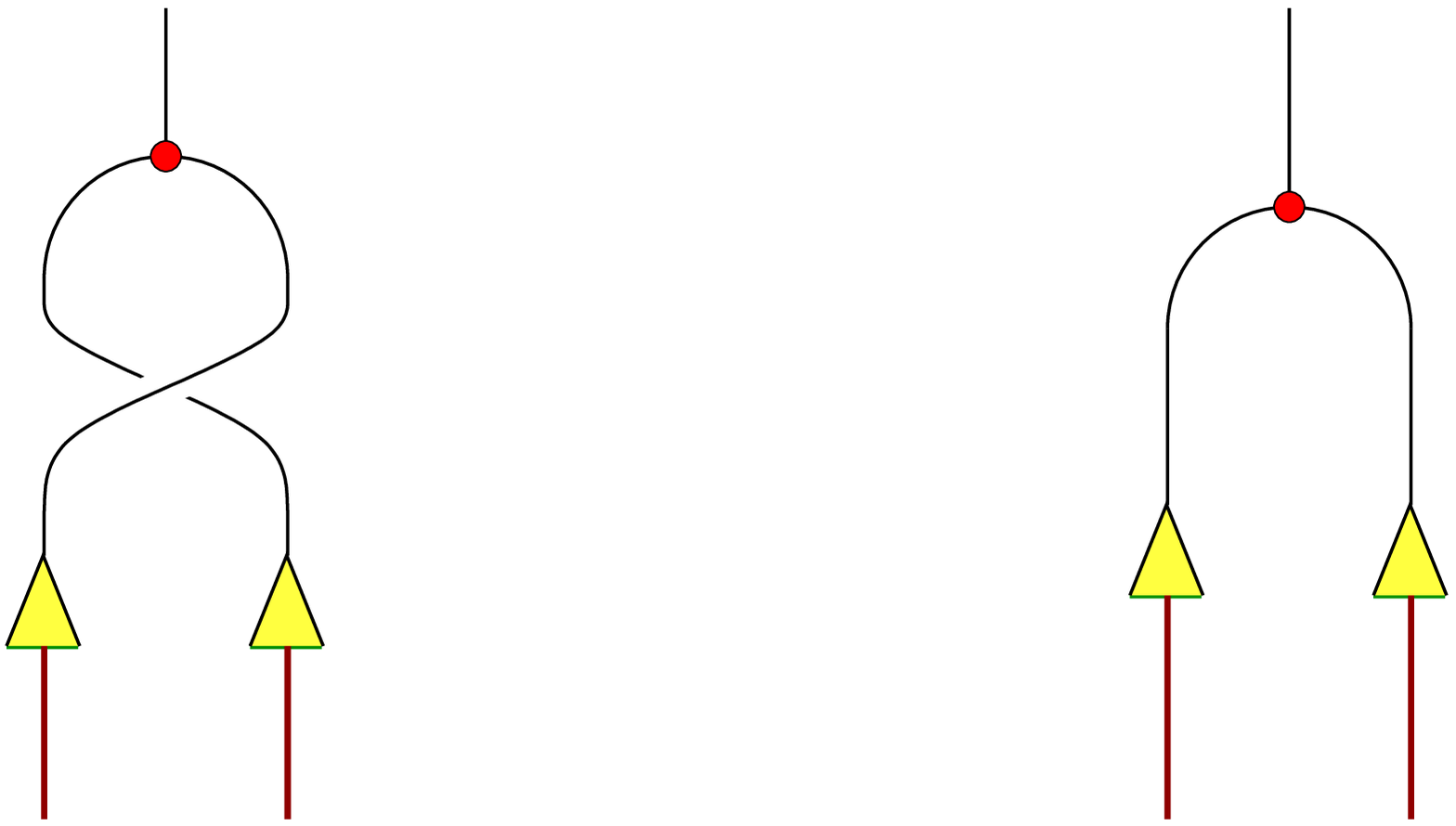}
  \setlength{\unitlength}{24810sp}
    \put( 42,264)  {\scriptsize$ A $}
    \put(  5,-20)  {\scriptsize$ \U_g $}
    \put( 80,-20)  {\scriptsize$ \U_h $}
    \put(392,264)  {\scriptsize$ A $}
    \put(353,-20)  {\scriptsize$ \U_g $}
    \put(428,-20)  {\scriptsize$ \U_h $}
  \setlength{\unitlength}{1pt}
  }
  \put(64,45)     {$ =\; \XI_A(h,g) $}
  \epicture19 \labl{Fig44}
as follows immediately from the formulas \erf{ir1} and \erf{XIabas}.

\dtl{Proposition}{prop-KSB}
(i)~\,\,\,The two-cochain $\XI_A$ defined by \erf{Fig1} does not depend on 
the choice of adapted basis.
\\[.2em]
(ii)~\,\,Isomorphic \hsfa s have identical two-cochains.
\\[.2em]
(iii)~The two-cochain $\XI_A$ is a KSB on $H(A)$.

\medskip\noindent
Proof: \\
(i)~\,and (ii)~\,are trivial to check.
\\[.2em]
(iii)~\,The defining property \erf{ksbcond} of a KSB follows from
\erf{XIabas} when stetting $h\eq g$, together with \erf{pf1}.
It remains to be shown that $\XI_A$ is a bihomomorphism. 
\\[.2em]
We first show that for a \berta, one has
  \be  m \circ \left(p_{h_1}\oti p_{h_2} \right) \circ \Delta
   = \Frac1{\dim(A)}\, p_{h_1h_2} 
  \labl{homprop}
for any two elements $h_1$ and $h_2$ in the support of $A$.
To see this, we notice that the two morphisms live in the same \onedim\
subspace of $\End(A)$ and are thus proportional. We compute the constant
of proportionality by taking the trace. For the right hand side, we find
  \be  \tr\! \left( \Frac1{\dim(A)}\, p_{h_1h_2} \right) =
  \Frac1{\dim(A)}\, \dim(\U_{h_1h_2}) = \Frac1{\dim(A)} \,.  \ee
The trace of the left hand side is computed graphically:
  \bea \begin{picture}(350,137)(0,0)
  \put(0,0)    {\includeourbeautifulpicture{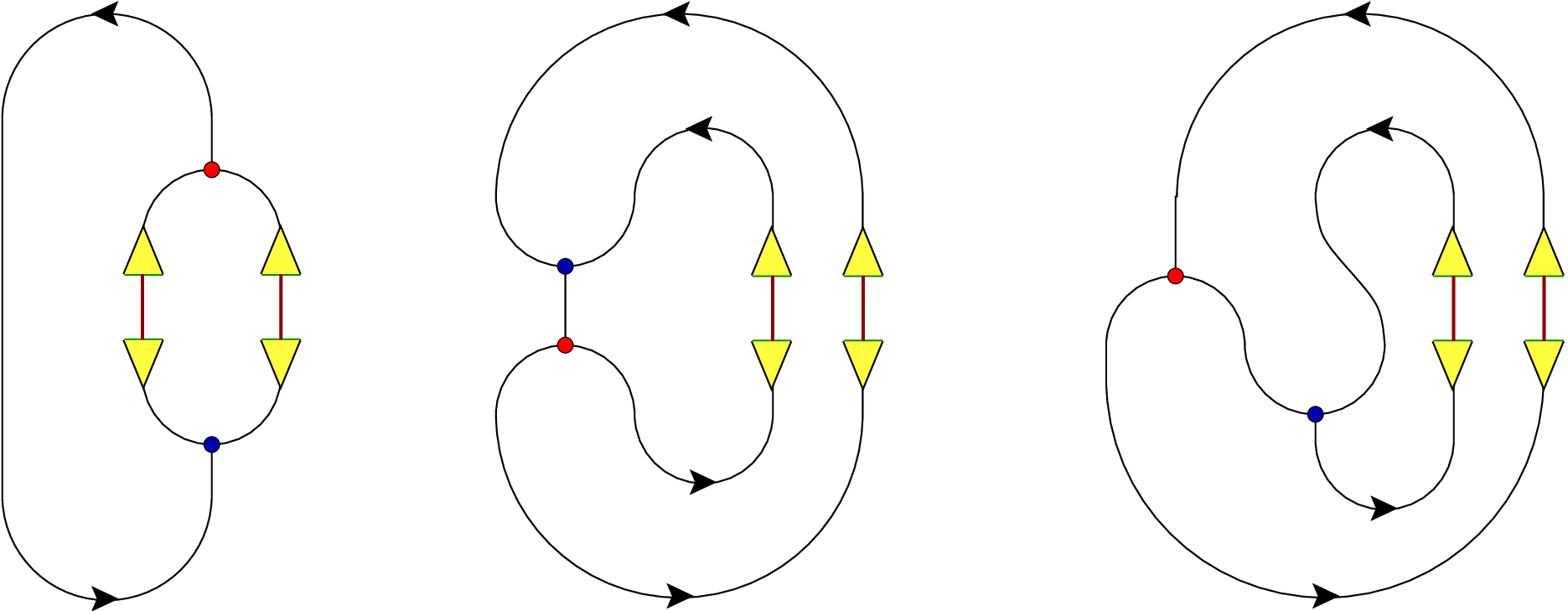}}
  \put(32.6,64.5) {\scriptsize$ \U_{h_1}^{} $}
  \put(46,118)    {\scriptsize$ A $}
  \put(63,64.5)   {\scriptsize$ \U_{h_2}^{} $}
  \put(87,63)     {$ = $}
  \put(154.5,65.5){\scriptsize$ \U_{h_1}^{} $}
  \put(180,118)   {\scriptsize$ A $}
  \put(190,64.5)  {\scriptsize$ \U_{h_2}^{} $}
  \put(214,63)    {$ = $}
  \put(319,64.5)  {\scriptsize$ \U_{h_1}^{} $}
  \put(328.7,118) {\scriptsize$ A $}
  \put(338.5,64.5){\scriptsize$ \U_{h_2}^{} $}
     \end{picture}\\[1.3em] \begin{picture}(250,133)(0,0)
  \put(66,0)      {\includeourbeautifulpicture{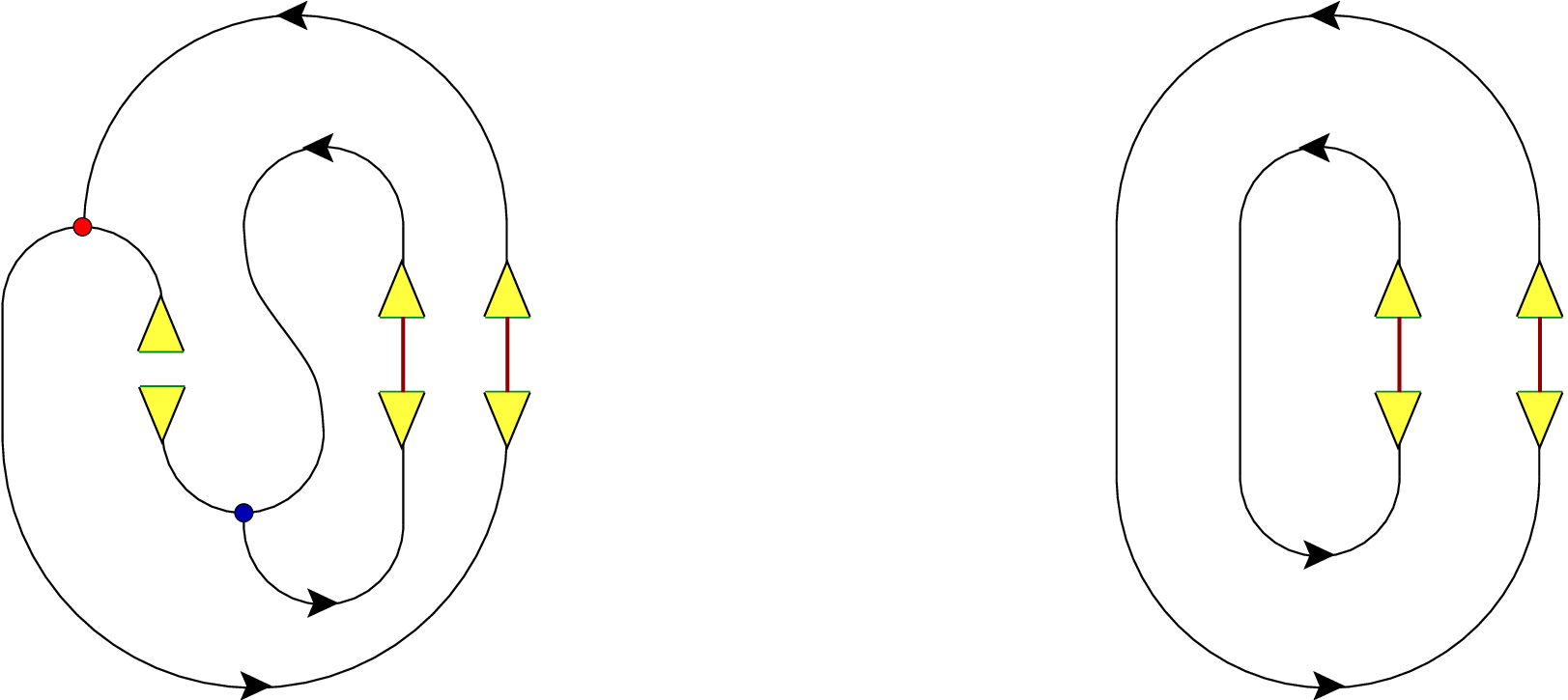}}
  \put(38,63)     {$ = $}
  \put(143.5,63.5){\scriptsize$ \U_{h_1}^{} $}
  \put(153,118)   {\scriptsize$ A $}
  \put(163.5,63.5){\scriptsize$ \U_{h_2}^{} $}
  \put(195,63)    {$ =\ \dim(A)^{-1}_{} $}
  \put(317,64.5)  {\scriptsize$ \U_{h_1}^{} $}
  \put(348.7,118) {\scriptsize$ A $}
  \put(359,63.5)  {\scriptsize$ \U_{h_2}^{} $}
  \epicture-6 \labl{Fig4}
Here the second equality employs the fact that $A$ is Frobenius. 
In the third identity it is used that in the middle $A$-ribbon only
the tensor unit propagates. Using the relation \erf{e1r1} between $e_1$ 
and $r_1$ and the unit and counit morphisms 
one obtains the last identity. The last graph is equivalent to
  \be  \Frac1{\dim(A)}\, \dim(\U_{h_1})\,  \dim(\U_{h_2}) 
  = \Frac1{\dim(A)} \,.  \ee
This shows \erf{homprop}.
\\[.2em]
We can now compute
  \begin{eqnarray} && \hsp{-5.3} \displaystyle
  \frac{ \XI_A(h_1,g)\, \XI_A(h_2,g) }{(\dim(A))^2_{}} \, \id_{\U_g}
  \nonumber \\ && \hsp{-1.9}
        \begin{picture}(313,260)(0,0)
  \put(0,0)       {\includeourbeautifulpicture{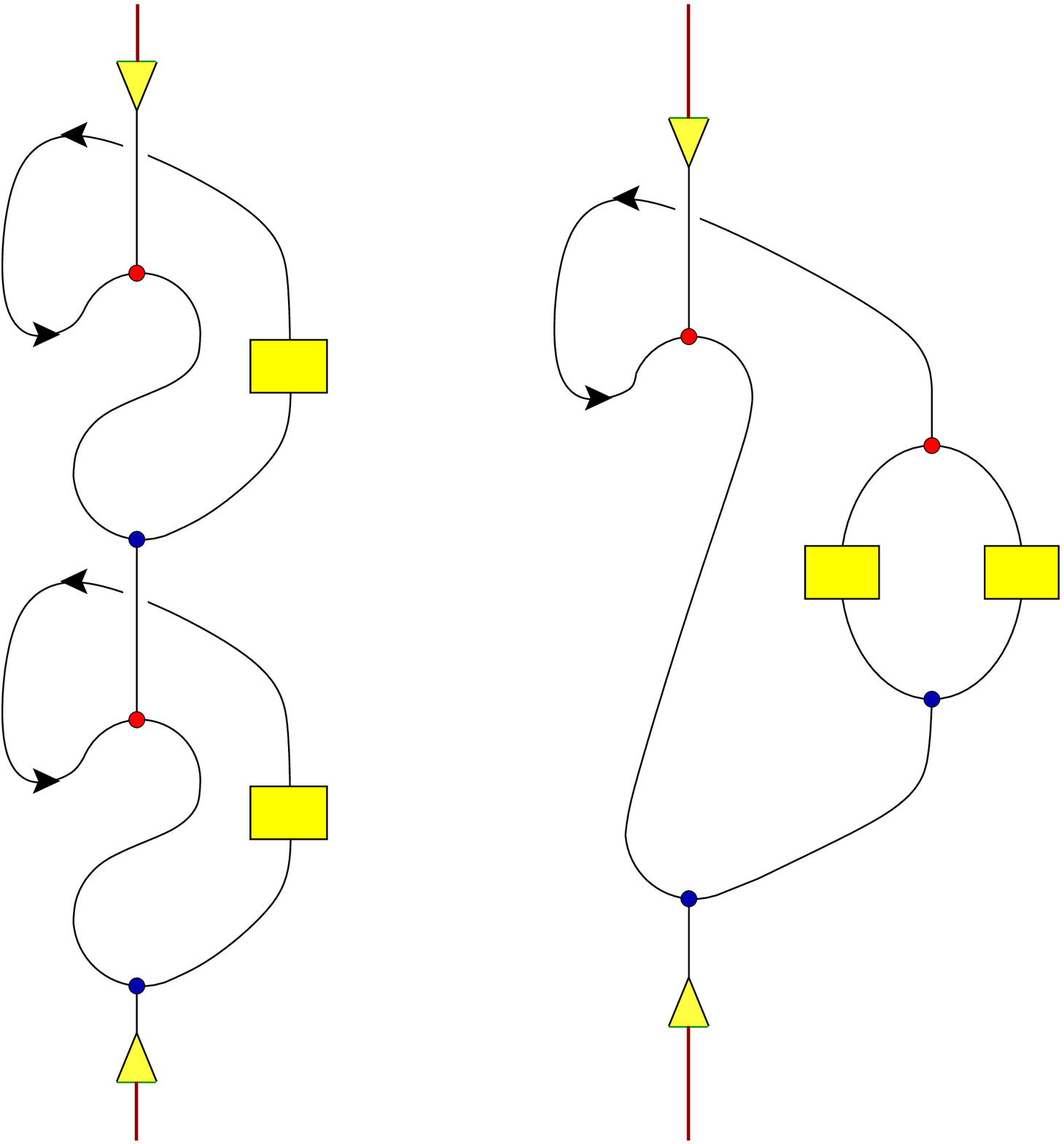}}
  \put(-32,121)   {$ = $}
  \put(26,-8)     {\scriptsize$ \U_g $}
  \put(26,251)    {\scriptsize$ \U_g $}
  \put(31,71)     {\scriptsize$ A $}
  \put(31,167)    {\scriptsize$ A $}
  \put(56.2,72)   {\scriptsize$ p_{h_2}^{} $}
  \put(56.2,168)  {\scriptsize$ p_{h_1}^{} $}
  \put(90,121)    {$ = $}
  \put(145,-8)    {\scriptsize$ \U_g $}
  \put(145,99)    {\scriptsize$ A $}
  \put(145,251)   {\scriptsize$ \U_g $}
  \put(177,123.5) {\scriptsize$ p_{h_1}^{} $}
  \put(216,123.5) {\scriptsize$ p_{h_2}^{} $}
  \put(248,121)   {$ =\; \displaystyle\frac1{\dim(A)} $}
  \put(382.5,122) {\scriptsize$ p_{h_1h_2}^{} $}
  \put(335,-8)    {\scriptsize$ \U_g $}
  \put(335,251)   {\scriptsize$ \U_g $}
  \end{picture}
  \nonumber \\[1.5em] && \hsp{22.2}
  =\, \displaystyle\frac{\XI_A(h_1 h_2,g)}{(\dim(A))^2_{}} \,,
  \label{Fig5}\end{eqnarray}
where for the second equality we used moves similar to those
in (I:5.36); the third step amounts to \erf{homprop}.
Thus $\XI_A$ is a homomorphism in the first argument.
\\[.2em]
To show that $\XI_A$ is also homomorphism in the second argument,
note that from the explicit form \erf{XIabas}, together
with \erf{d1} and \erf{pf3}, it follows that
  \be 
  \XI_A(g,h)\, \XI_A(h,g) = 
  \big( \Omega(g,h) \Omega(h,g) {\big)}^{-1} = \beta(g,h) \,.  \ee
Thus $\XI_A(g,h) \eq \beta(g,h) / \XI_A(h,g)$. Since
$\beta$ is a bihomomorphism and $\XI_A(h,g)$ a homomorphism in 
the first argument, it follows that $\XI_A(g,h)$ is a homomorphism
also in the second argument.
\qed

\dtl{Remark}{exksb}
Part (iii) of the proposition implies in particular that there does 
      exist a KSB on $H(A)$.

\dtl{Proposition}{hsfa-same}
Two \berta s with identical support are isomorphic as
Frobenius algebras if and only if they have the same KSB.

\medskip\noindent
Proof: \\
That two isomorphic \berta s have the same KSB was already established
in proposition \ref{prop-KSB}. It remains to show the converse.
We have seen in lemma \ref{def:algtheta}
that two multiplications on $A$ differ by a two-cocycle
  \be  \gamma: \quad H\times H \to\complexx  .  \ee
When the product is changed by $\gamma$, according to
\erf{XIabas} the KSB changes by the commutator two-cocycle of $\gamma$,
  \be  \XI_A(g,h) \,\mapsto\, \XI_A(g,h)\, \frac{\gamma(g,h)}
  {\gamma(h,g) } \,.  \ee
Now the isomorphism classes of \hsfa s with $H(A)\eq H$ form a torsor over 
$H^2(H,\complexx)$, while the KSBs form a torsor over the alternating 
bihomomorphisms. According to lemma \ref{lem:bihom-wcomm} the two groups are 
isomorphic, $H^2(H,\complexx)\,{\cong}\,\AB(H,\complexx)$, and equation 
\erf{isom} shows that the map $[A] \,{\mapsto}\, \XI_A$ is a morphism of 
torsors over the group. A morphism of torsors, in turn, is, as a map of sets, 
both surjective and injective.
\qed

\dt{Remark}
{}From \citeIaII\ we know that a symmetric special Frobenius algebra defines
a full CFT on {\em oriented\/} world sheets with or without boundary,
while to define a full CFT on {\em unoriented\/} and possibly non-orientable
surfaces with or without boundary we need a {\em Jandl algebra\/}. The notion 
of a Jandl algebra was introduced in definition II:2.1; it is a symmetric 
special Frobenius algebra together with a morphism $\sigma \iN \Hom(A,A)$ obeying
  \be
  \sigma \cir \sigma = \theta_A \qquad {\rm and} \qquad
  \sigma \cir m = m \cir c_{A,A} \cir (\sigma \oti \sigma) \,.
  \labl{eq:rev-def}
A morphism $\sigma$ with these properties was called a {\em reversion\/}.
\\[.2em]
Given a \berta\ $A$, we can ask what reversions, if any,
can be defined on $A$. To express \erf{eq:rev-def} in a basis,
we define $\sigma(h)\iN\complex$ for $h\iN H(A)$ via 
$\sigma(h) \id_{L_h} \,{:=}\, r_h \cir \sigma \cir e_h$. Applying
(II:2.21) we find that \erf{eq:rev-def} is equivalent to having
  \be
  \sigma(g)^2 = \theta_g \qquad{\rm and} \qquad
  \sigma(gh)\,\omega(g,h) = \sigma(g)\, \sigma(h)\, \omega(h,g)\,
  \Omega(h,g)^{-1}\,, \labl{eq:rev-berta-1}
for all $g,h\iN H(A)$. Recalling also the expression
\erf{XIabas} for the KSB, we can rewrite the second equality as
  \be
  \frac{\sigma(gh)}{\sigma(g)\,\sigma(h)} = \XI_A(h,g) = \XI_A(g,h) \,,
  \labl{eq:rev-berta-2}
where the second step follows from the fact that, by the first equality,
the KSB is symmetric. This shows in particular that a reversion can only
be defined for Schellekens algebras with {\em symmetric\/} KSB.
By the bihomomorphism property of the KSB we have
$\XI_A(g,g^{-1}) \eq \XI_A(g,g)^{-1} \eq \theta_g^{-1}$, so that when
evaluating condition \erf{eq:rev-berta-2} for $h \eq g^{-1}$ we get
$\sigma(g)\,\sigma(g^{-1})\eq\theta_g$. 
Together with the first equality in \erf{eq:rev-berta-1} this implies that 
$\sigma(g)\eq\sigma(g^{-1})$, i.e.\ $\sigma$ is a quadratic form on $H(A)$. 
\\[.2em] 
Given a Schellekens algebra $A$, the above considerations show
that the following two statements are equivalent:
\\[.1em]
i)~\,\,$\sigma$ is a reversion on $A$;
\\[.1em]
ii)~$\sigma$ is a quadratic form on $H(A)$ such that its associated 
bihomomorphism equals the KSB of $A$, $\beta_\sigma\eq\XI_A$.
\\[.2em]
Since by proposition \ref{hsfa-same} the KSB determines a Schellekens algebra
up to isomorphism, a quadratic form $\sigma \iN \QF(H,\complexx)$ with 
$\sigma(g)^2\eq\theta_g$ on some subgroup $H$ in the effective center
$\Pico(\calc)$ defines a Schellekens\hy Jandl algebra with support $H$
up to isomorphism. Distinct quadratic forms $\sigma \iN \QF(H,\complexx)$
satisfying $\sigma(g)^2\eq\theta_g$ give rise to non-isomorphic 
Schellekens\hy Jandl algebras on $H$. In summary, for a given subgroup 
$H \,{\subset}\, \Pico(\calc)$ in the effective center
of the category $\calc$ we have:
\\[.1em]
\Nxt The isomorphism classes of Schellekens algebras with support $H$ form
a torsor over $H^2(H,\complexx).$
\\[.1em]
\Nxt The isomorphism classes of Schellekens\hy Jandl algebras with support $H$ 
are in bijection with the subset 
$\{ \sigma \iN \QF(H,\complexx) \,|\, \sigma(h)^2\eq\theta_h {\rm \;for\;all\;}
h\iN H\}$.
\\[.1em]
\Nxt Different Jandl-structures on one and the same Schellekens algebra $A$ 
form a torsor over $H^1(H(A),\zet_2)$, i.e.\ differ by a character of the 
support that takes its values in $\{\pm1\}$.

\subsection{The modular invariant torus partition function}

In \citei\ it has been shown that every \hsfa\ 
provides a modular invariant torus partition function 
  \be  Z = \sum_{i,j \in \II} Z_{ij}\, \Chi_i^{} \oti \overline \Chi_j \,,  \ee
where the non-negative integers $Z_{ij}$ are defined as the invariant
of the ribbon graph in (I:5.30). 
As announced in \cite{fuRs} (and to be proven rigorously in a separate
publication), every such modular invariant is physical in the sense that 
it is part of a consistent collection of correlation functions
on oriented world sheets of arbitrary topology. 
Thus in particular every simple current modular invariant is physical.
(Recall that our notion of simple current invariant is different from
the one of \cite{gasc,gasc2}; invariants of the latter type may be
unphysical, compare remark \ref{rem}(iii) above.)

To obtain explicit expressions for the integers $Z_{ij}$, the notion of a 
monodromy charge (see remark \ref{DeltaQ}) must be extended to arbitrary 
simple objects. Note that when $U$ is a simple object of $\calc$, then so is
$\U_g\oti U$ for any $g\iN\Pic(\calc)$.

\dtl{Definition}{mono-charge}
Let $U$ be a simple object in a modular tensor category.
\\[.1em]
(i)~\,\,The {\em exponentiated monodromy charge\/} of $U$ is the one-cochain 
$\chi_U$ on $\Pic(\calc)$ defined by
  \be  c_{U,\U_g} \circ c_{\U_g,U} =: \chi_U(g) \, \id_{\U_g\otimes U} \,. 
  \labl{expmono}
(ii)~\,For any object $X$ of $\calc$ and any $g\iN G$ we abbreviate the tensor 
product of $\U_g$ and $X$ by $gX$: 
  \be  gX := \U_g\oti X \,.  \ee
In this way, the Picard group $\Pic(\calc)$ acts 
on isomorphism classes of objects (but not on objects, in general).

\dt{Remark}
Recall the description of the modular matrix S in terms of an invariant 
of the Hopf link in $S^3$, compare e.g.\ formula (I:2.22). It follows 
directly from the definition of $\chi_U$ that
  \bea \begin{picture}(350,84)(0,24)
  \put(0,50)     {$ s_{gU,V} \,=\, s_{V,gU} \,= $ }
  \put(100,0)    {\includeourbeautifulpicture{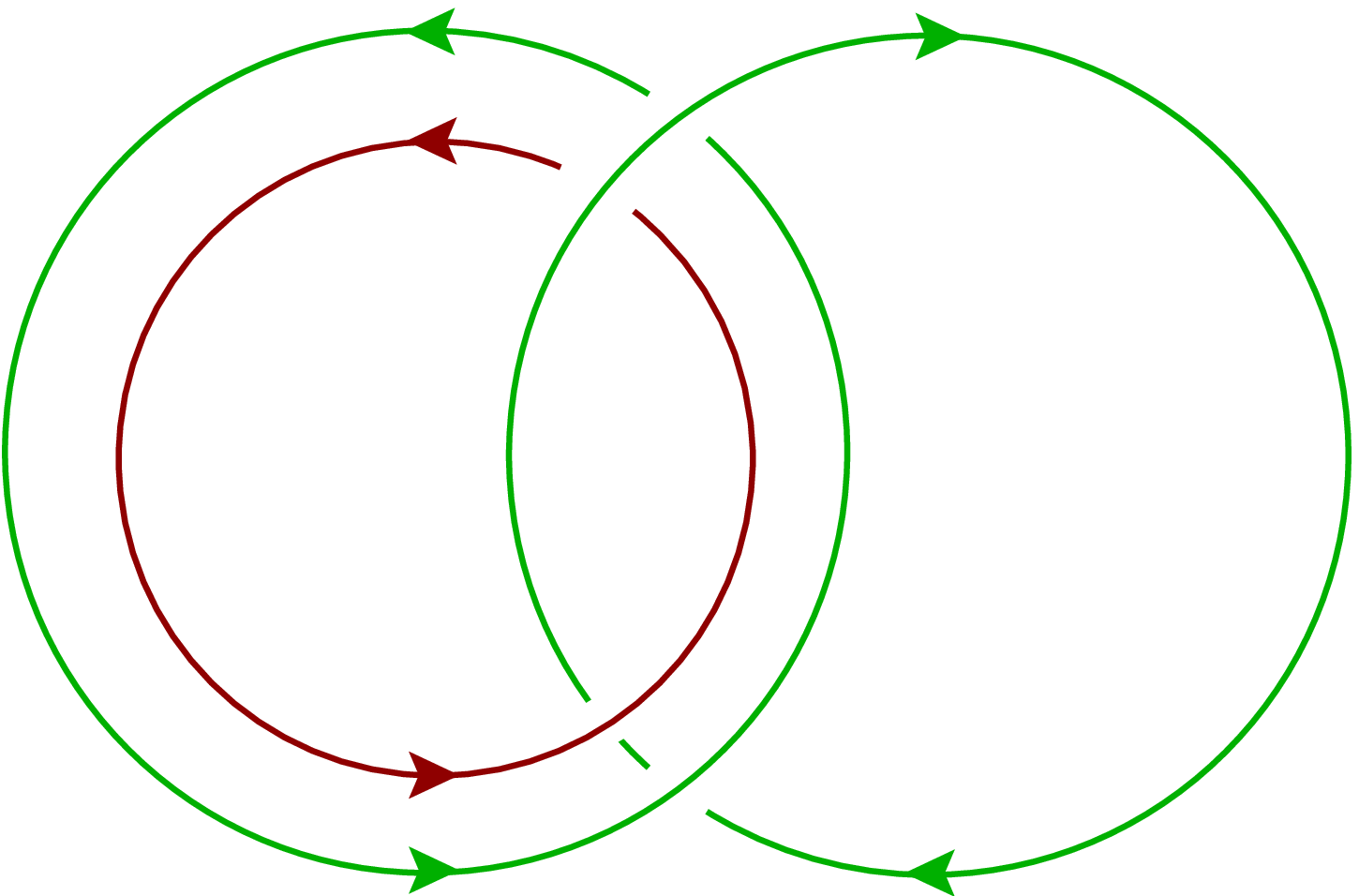}}
  \put(151.2,50) {\scriptsize$ V $}
  \put(177,50)   {\scriptsize$ \U_g $}
  \put(201,50)   {\scriptsize$ U $}
  \put(272,50)   {$ =\, \chi_V(g)\, s_{U,V} \,.$ }
  \epicture11 \labl{fundeq1}
The importance of this relation is known since the early days of simple
current theory \cite{scya,intr}.

\dtl{Proposition}{props-mono}
(i)~\,\,The exponentiated monodromy charge $\chi_U$ is a one-cocycle.
\\  Since $\Pic(\calc)$ is abelian, we can thus regard $\chi_U$ as an 
    irreducible character on the group $\Pic(\calc)$:
  \be  \chi_U \in \Pic(\calc)^*_{}  \,.  \ee
\\[.2em]
(ii)~\,The exponentiated 
monodromy charge can be expressed in terms of the twists as
  \be  \chi_{U_i}(g) = \th_{gi}^{}\, \th_g^{-1} \th_i^{-1} \, . \labl{expr}
It obeys $\chi_{U_i}(g) = \chi_{U_{\iBar}}(g^{-1})$ for all
$i \iN \cali$ and $g \iN \calpic(\calc)$.

\medskip\noindent
Proof:\\
(i)~\,The statement is a direct consequence of the definition, applied for 
$g\eq g_1g_2$, and tensoriality of the braiding, together with the fact that
$\U_{g_1g_2}$ is isomorphic to $\U_{g_1}\oti\U_{g_2}$. (At intermediate steps
an isomorphism between $\U_{g_1g_2}$ and $\U_{g_1}\oti\U_{g_2}$ must be
chosen, but the result does not depend on that choice.)
\\[.2em]
(ii)~\,The first statement is a consequence of the compatibility condition 
\erf{compat} between braiding and twist, applied to the simple objects $\U_g$ 
and $U$. That $\chi_{U_i}(g)\eq\chi_{U_{\iBar}}(g^{-1})$ then follows
from the fact that dual objects have identical balancing phases, according
to which $\th_{g^{-1}\ibar}\eq\th_{gi}$ (because 
$(U_{g^{-1} \ibar})^\vee \,{\cong}\, U_{gi}$), $\th_i\eq \th_{\ibar}$
and $\th_g \eq \th_{g^{-1}}$.
\qed

\medskip

We are now in a position to express the modular invariant partition
function given in (I:5.30) more explicitly. When writing $A$ as a direct sum, 
  \be A \,\cong \bigoplus_{g\in H(A)} \U_g \,,  \ee
we obtain (the ribbon graphs below are embedded in $S^2\,{\times}\,S^1$, with
the vertical direction being the $S^1$, so that top and bottom are identitfied)
  \begin{eqnarray}  &&
      \begin{picture}(360,195)(0,0)
  \put(-20,90)   {$ Z_{ij} \; = $}
  \put(30,0)     {\includeourbeautifulpicture{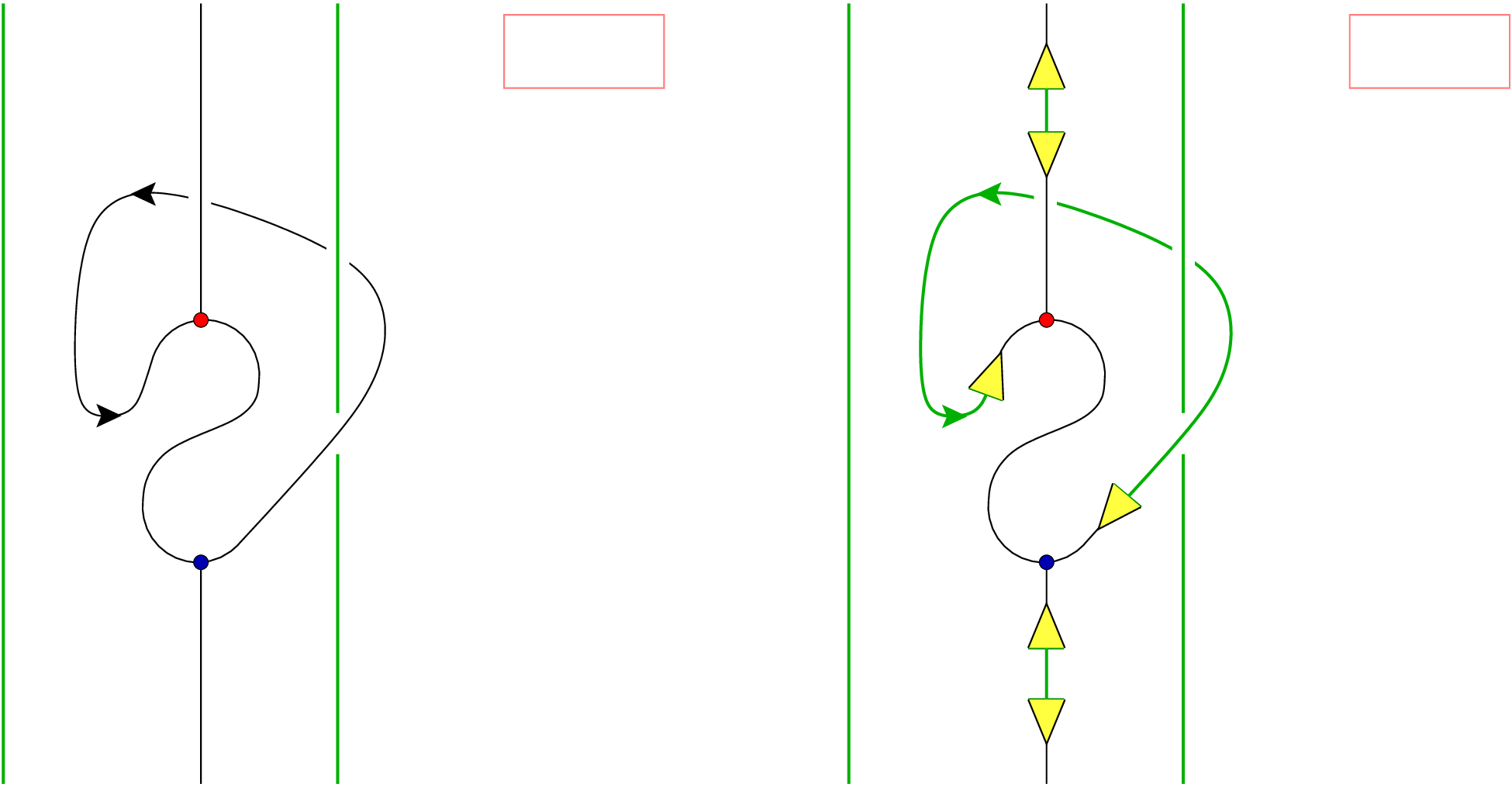}}
  \put(26.8,-8)  {\scriptsize$ U_i $}
  \put(27.6,188) {\scriptsize$ U_i $}
  \put(73.6,-8)  {\scriptsize$ A $}
  \put(74.2,188) {\scriptsize$ A $}
  \put(105.9,-8) {\scriptsize$ U_j $}
  \put(106.7,188){\scriptsize$ U_j $}
  \put(151,169)  {$ S^2_{}{\times}S^1_{} $}
  \put(162,90)   {$ =\! \displaystyle\sum_{g,h\in H(A)} $}
  \put(225.8,-8) {\scriptsize$ U_i $}
  \put(226.6,188){\scriptsize$ U_i $}
  \put(273.2,188){\scriptsize$ A $}
  \put(272.6,-8) {\scriptsize$ A $}
  \put(276.5,102){\scriptsize$ A $}
  \put(278.3,24) {\scriptsize$ \U_g $}
  \put(278.3,157){\scriptsize$ \U_g $}
  \put(304.6,-8) {\scriptsize$ U_j $}
  \put(305.4,188){\scriptsize$ U_j $}
  \put(318.4,90) {\scriptsize$ \U_h $}
  \put(350,169)  {$ S^2_{}{\times}S^1_{} $}
  \end{picture}
  \nonumber 
              \\
  &&
      \begin{picture}(0,224)(0,0)
  \put(2,90)    {$ = \; \displaystyle\sum_{g,h\in H(A)}\!\! \chi_{U_j}(h) $}
  \put(104,0)    {\includeourbeautifulpicture{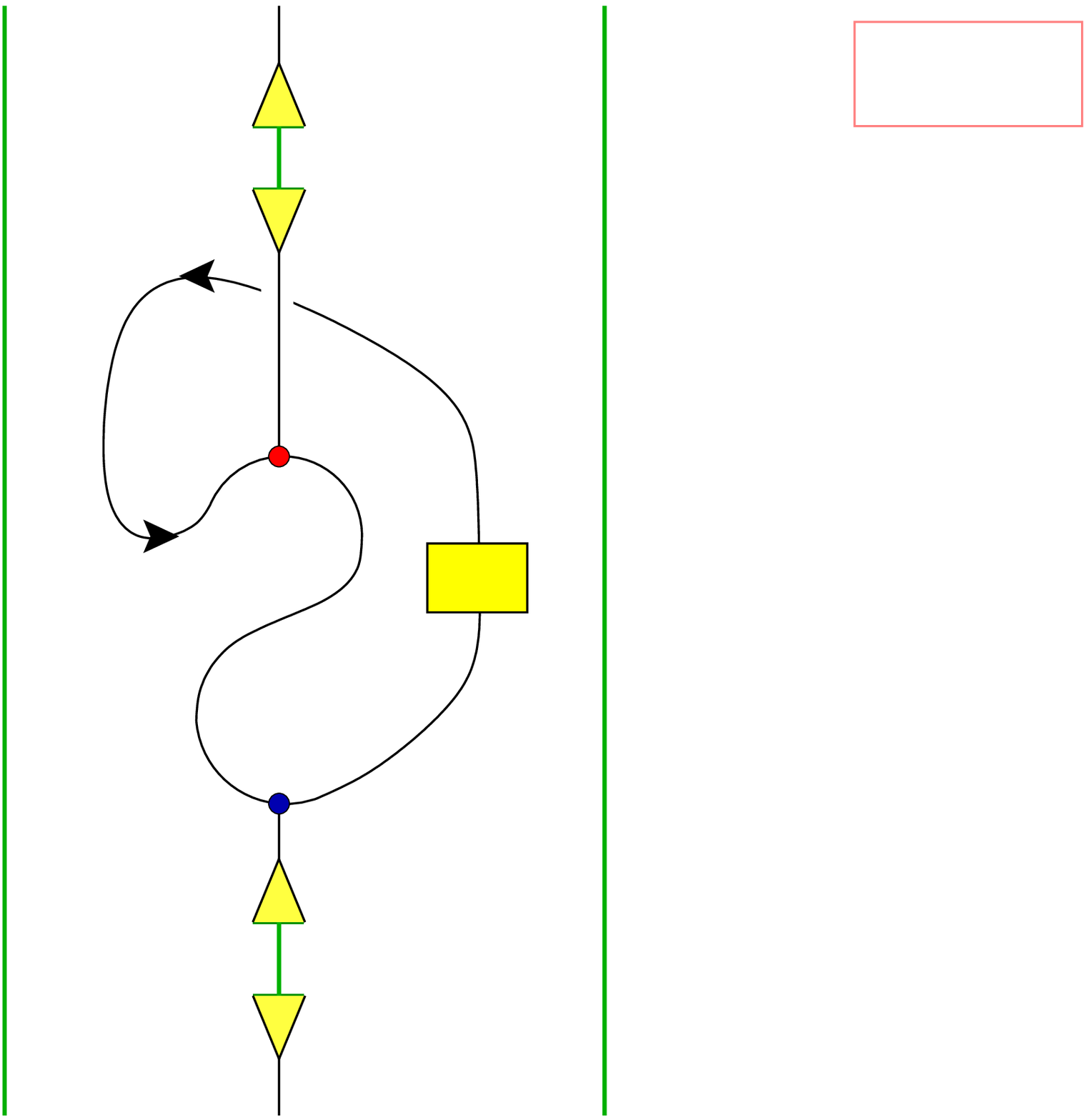}}
  \put(100.8,-8) {\scriptsize$ U_i $}
  \put(101.6,188){\scriptsize$ U_i $}
  \put(146.2,188){\scriptsize$ A $}
  \put(146.6,-8) {\scriptsize$ A $}
  \put(149.5,102){\scriptsize$ A $}
  \put(151.3,24) {\scriptsize$ \U_g $}
  \put(151.3,157){\scriptsize$ \U_g $}
  \put(179.3,89) {\scriptsize$ p_h $}
  \put(200.3,-8) {\scriptsize$ U_j $}
  \put(200.7,188){\scriptsize$ U_j $}
  \put(248,169)  {$ S^2_{}{\times}S^1_{} $}
  \end{picture}
  \nonumber \\[1.9em] && \hsp{-.2}
  =\, \Frac1{\dim(A)} \displaystyle\sum_{g,h\in H(A)}
  \chi_{U_j}(h)\, \XI_A(h,g) \,
  \dimc\Hom(U_i\Oti\, \U_g\Oti\, U_j,\one)
  \nonumber \\[.5em] && \hsp{-.2}
  =\, \Frac1{\dim(A)} \displaystyle\sum_{g,h\in H(A)}
  \chi_{U_j}(h)\, \XI_A(h,g) \, \delta_{\bar\jmath, gi} \,.
  \label{ks-form2} \end{eqnarray}
We have thus arrived at the
\\[-2.3em]

\dtl{Theorem}{thm-pf}
The modular invariant torus partition function for a \berta\ $A$ is
given by
  \be  Z_{ij}(A) =  \Frac1{|H(A)|} \sum_{g,h\in H(A)}
  \chi_{U_i}(h)\, \XI_A(h,g)\, \delta_{\bar\jmath, gi} \,.  \labl{ks-form1}

Note that
this is the charge conjugate of the partition function given in \cite{krSc}.

\dtl{Remark}{warning}
It can happen that non-isomorphic \berta s are 
Morita equivalent and thus give equivalent full conformal field theories.
In particular, they can give rise to identical torus partition functions.
A simple example appears in the Ising model: the \hsfa\ 
$A\,{\cong}\,\one\,{\oplus}\, \epsilon$
built from the primary field $\epsilon$ of conformal weight $1/2$ is 
Morita equivalent to the trivial algebra, since $A\eq\sigma^\vee\Oti\,
\sigma$ with $\sigma$ the primary field of conformal weight $1/16$.

\medskip

For an algebra in a braided tensor category, there are the notions of
a left and of a right center \cite{vazh,ostr,ffrs}. The following statements 
already appear in \cite{krSc}, but we rephrase them in our terms.
\\[-2.3em]

\dtl{Proposition}{centers}
Let $A$ be a \berta\ with support $H\,{\subset}\, \Pico(\calc)$
and KSB $\XI_A$. Consider the two subgroups 
  \be  \begin{array}{l}
  K_r(A) := \{ g\iN H\,|\,\XI_A(\,\cdot\,,g)\,\mbox{ is the trivial $H$-character}
  \}  \,, \\{}\\[-.7em]
  K_l(A) := \{ g\iN H\,|\,\XI_A(g,\cdot\,)\,\mbox{ is the trivial $H$-character}
  \} \end{array}  \ee
of $H$. Then we have:
\\[.2em]
(i)~\,\,The left and right center of $A$ are, respectively, the commutative
symmetric Frobenius subalgebras of $A$ that are defined on the subobjects
  \be  C_l(A) := \bigoplus_{g\in K_l(A)} \U_g  \qquad{\rm and}\qquad
  C_r(A) := \bigoplus_{g\in K_r(A)} \U_g \,.  \ee
(ii)~\,$A$ is commutative if and only if $\XI_A\,{\equiv}\,1$.

\medskip\noindent
Proof: \\
(i)~\,\,By proposition I:5.9, the left and the right center are
the subalgebras on the objects
  \be  C_l(A) = \bigoplus_i Z_{i0}\, U_i \qquad{\rm and}\qquad
  C_r(A) = \bigoplus_j Z_{0j}\, U_j \,.  \ee
Due to the form of the modular invariant partition function, only invertible
objects can appear. Moreover, from the forms \erf{ks-form1} and \erf{ks-form2}
of the partition function one computes that
  \be  Z_{0g} = \frac1{|H|} \sum_{h\in H} \XI_A(h,g)  \qquad{\rm and}\qquad
  Z_{g0} = \frac1{|H|} \sum_{h\in H} \XI_A(g,h) \,,  \ee
respectively. Invoking also the orthogonality of characters, this proves 
the statement.
\\[.2em]
(ii)~\,An algebra is commutative iff it coincides with both its left and its
right center. 
\qed

\dtl{Corollary}{centers3}
The structure of a commutative \berta\ can be defined precisely on those 
subgroups in $\Pico(\calc)$ on which both the monodromy charge
and the twist are identically one. Furthermore, any two commutative 
Schellekens algebras $A$ and $A'$ with $H(A) \eq H(A')$ are isomorphic.

\medskip\noindent
Proof: \\ 
The claim follows from the defining properties \erf{ksbcond} of a KSB. 
Uniqueness holds because the KSB determines the algebra up to isomorphism.
\qed

\dtl{Remark}{rem-center2}
(i)~\,\,In the physics literature, integer spin simple currents
(i.e.\ simple currents $\U_g$ with $\th_g\eq1$) such that $\beta\,{\equiv}\,1$ 
are called {\em mutually local\/}.
Thus the structure of a commutative algebra can only be defined on 
mutually local integer spin simple currents.
\\[.2em]
(ii)~\,If $\calc$ is the category of representations of a rational vertex
algebra $\cala$, the results of \cite{dolm} (see also theorem 4.3 in 
\cite{hohn6}) imply that there exists an
extension of $\cala$ as a vertex algebra whose representation category 
is equivalent to the category of {\em local\/} $A$-modules in $\calc$ 
for a suitable simple \ssfa\ $A$ in $\calc$.
(For the definition of the qualification `local', see \cite{pare23,kios}
and definition 3.15 of \cite{ffrs}.)
\\[.2em]
(iii)~On the level of partition functions, the relation between mutually
local simple currents and extensions is well-known since
the early days of simple current theory \cite{scya} (see also
\cite{intr} for the case of cyclic groups).


\sect{Modules and boundary conditions}

In this section we develop the theory of left modules over a \berta\
in a modular tensor category $\calc$. Our main motivation to study the 
representation theory of \berta s is the fact that modules correspond to 
conformally invariant boundary conditions that respect the chiral 
symmetry encoded in $\calc$. This allows
us in particular to prove the formula for boundary states that has been
presented in \cite{fhssw} and that summarizes and generalizes earlier work
\cite{fuSc5,fuSc10,fuSc11,fuSc12}. In the case of a commutative
Schellekens algebra $A$, there is an additional reason for the study of
modules: one can show that the process of forming the subcategory of local 
left modules over a commutative \berta\ -- in physical terms, a simple current 
extension \cite{scya} -- precisely corresponds to the modularisation procedure
of \cite{muge6,brug2} and thereby provides a representation theoretic origin 
for the S-matrix formula for simple current extensions that was obtained in 
\cite{fusS6}.

\subsection{Stabilizers and basis independent 6j-symbols}

We start with the
\\[-2.3em]

\dtl{Definition}{stabilizer}
Let $U$ be an object in $\calc$. \\[.2em]
(i)~\,\,The stabilizer of the action of $\Pic(\calc)$ on the isomorphism 
class of $U$ is denoted by
$\cals(U)$. Following the physics literature, we say that
$U$ is a {\em fixed point\/} of the simple current $\U_g$ iff $g\iN\cals(U)$.
\\[.2em]
(ii)~\,Given a \berta\ $A$, we denote by $\calsA(U)$ the intersection of the
stabilizer with the support of $A$, $\calsA(U)\,{:=}\,\cals(U)\,{\cap}\,H(A)$.
\\[.2em]
(iii)~Let the object $U$ be simple. For every $g\iN \cals(U)$ we fix bases 
$\bb_g(U)\iN\Hom(\U_g\oti U,U)$ and $\tildeb_g(U)\iN\Hom(U\oti \U_g,U)$. 
For $g\eq e$, we take (using strictness) $\bb_e(U)\eq \tildeb_e(U)\eq \id_U$.
We denote by $\overline{\bb_g(U})\iN\Hom(U,\U_g\Oti\, U)$ the morphism dual
to $\bb_g(U)$ (which by definition satisfies $b_g(U)\cir\overline{\bb_g(U)}
\eq \id_U$, compare equation (I:2.30)).  \\[.2em]
(iv)~Let $U$ be simple. The two-cochain $\phi_U$ on $\cals(U)$ is defined by
  \bea \begin{picture}(170,89)(0,24)
  \put(0,0)     {\includeourbeautifulpicture{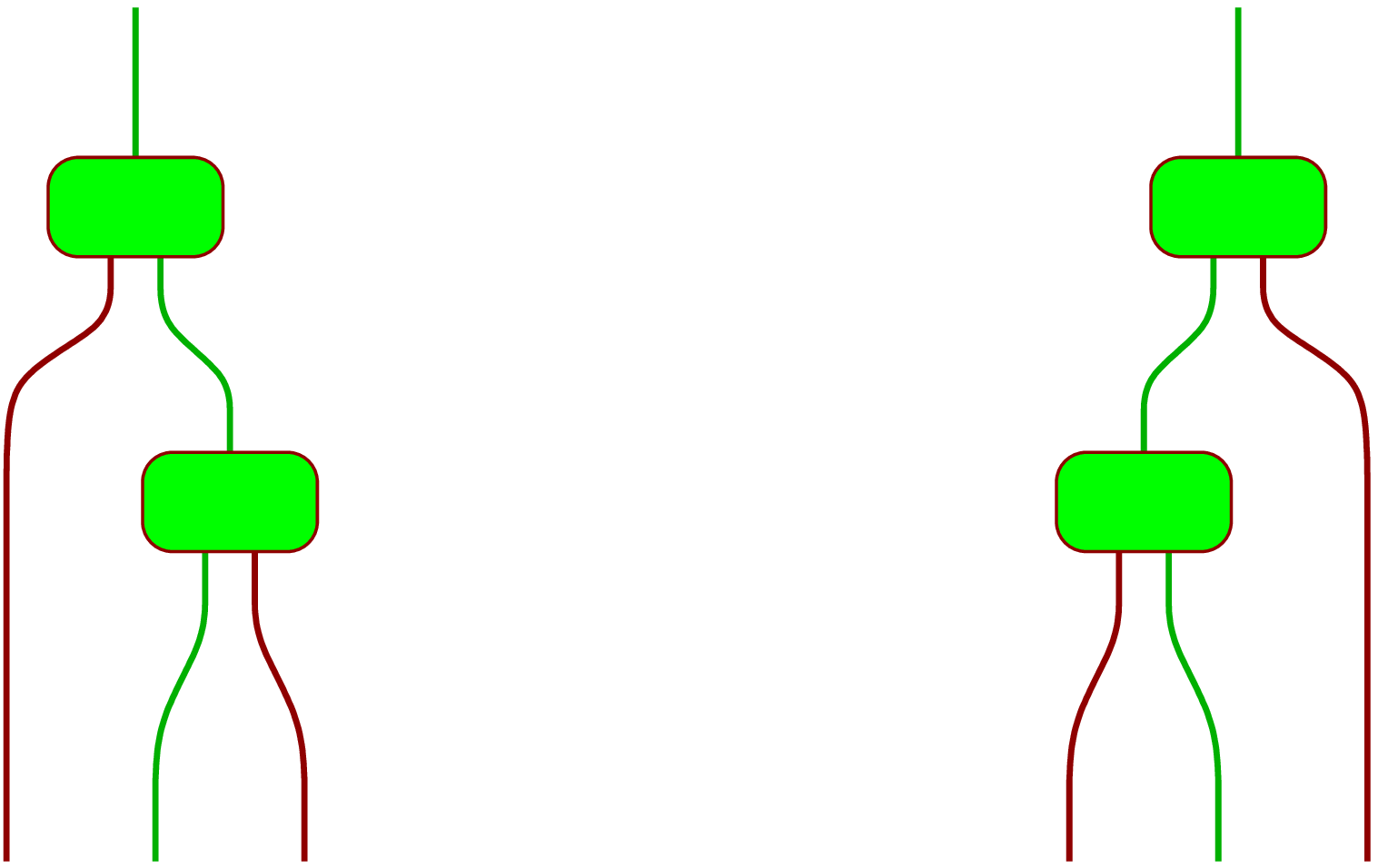}}
  \put(-4,-8)   {\scriptsize$ \U_g $}
  \put(6.6,78)  {\scriptsize$ \bb_{g\!}(U) $}
  \put(14,108)  {\scriptsize$ U $}
  \put(15,-8)   {\scriptsize$ U $}
  \put(18,42)   {\scriptsize$ \tildeb_{h\!}(U) $}
  \put(24.7,63) {\scriptsize$ U $}
  \put(32,-8)   {\scriptsize$ \U_h $}
  \put(58,47)   {$ =:\, \phi_U(g,h) $ }
  \put(124,-8)  {\scriptsize$ \U_g $}
  \put(129,42)  {\scriptsize$ \bb_{g\!}(U) $}
  \put(136,63)  {\scriptsize$ U $}
  \put(140,78)  {\scriptsize$ \tildeb_{h\!}(U) $}
  \put(146,108) {\scriptsize$ U $}
  \put(144,-8)  {\scriptsize$ U $}
  \put(160,-8)  {\scriptsize$ \U_h $}
  \epicture11 \labl{Fig7}

Once we select the morphisms $\bb_g(U)$ and $\tildeb_g(U)$ as bases for the 
respective morphism spaces, the relation \erf{Fig7} is just the definition of 
an \FF-coefficient, namely $\phi_U(g,h)\eq\Fs gUhUUU$, compare formula (I:2.36).  

\dtl{Proposition}{props-stabilizer}
Let $U$ be a simple object of $\calc$.  
\\[.2em]
(i)~\,The stabilizer $\cals(U)$ is a subgroup in
      $\Pico(\calc)$.  One has  
      $\th_g\iN\{\pm 1\}$ for all $g\iN \cals(U)$.
      The union of the stabilizers of all simple objects of $\calc$ is a
      subgroup in the effective center
      $\Pico(\calc)$.
\\[.2em]
(ii)~The two-cochain $\phi_U(g,h)$ does not depend on the choice of
     the bases $b$. It is a bihomomorphism on $\cals(U)$.

\medskip\noindent
Proof:\\ 
We start by showing that $\theta_g^2\eq 1$ for every $g\iN\cals(U)$. 
For $g\iN\cals(U)$ the equality \erf{expr} simplifies to
$\chi_U(g) \eq \th_g^{-1}$. On the other hand,
  \be  \chi_U(g)^{-1} = \chi_U(g^{-1}) = \th_{g^{-1}}^{-1} = \th_g^{-1} \,,
  \ee
where we used the fact that $\chi_U$ is a character and that
dual objects have the same twist. Multiplying the last two equations shows
$(\th_g)^2\eq1$. 
In other words, only half-integer spin simple currents can have fixed points.
\\[.2em]
Next we establish that $\cals(U)$ is a subgroup in $\Pico(\calc)$.
Clearly, $\cals(U)$ is a subgroup of the Picard group $\Pic(\calc)$.
To see that it is also a subset of $\Pico(\calc)$, by corollary 
\ref{cor:eff} it is sufficient to show that $(\theta_g)^{N_g}\eq 1$ for
all $g\iN\cals(U)$, where $N_g$ is the order of $g$.
If $N_g$ is even, then this equality follows from $\theta_g^2\eq 1$. For
$N_g$ odd we just invoke remark \ref{square}(iii), according to which 
all simple currents of odd order are in $\Pico(\calc)$. 
\\[.2em]
(ii)~That $\phi_U(g,h)$ is independent of the choice of basis holds 
because all morphism spaces in question are one-dimensional so that one
can only modify the basis elements by non-zero constants.
Since the same basis elements $\bb_{g}(U)$ and $\tildeb_{h\!}(U)$
appear on both sides of \erf{Fig7}, these constants cancel from
the definition of $\phi_U(g,h)$.
\\
To see that $\phi_U(g,h)$ is a bihomomorphism, 
introduce (basis dependent) elements of the fusing matrix by
  \bea \begin{picture}(210,103)(0,12)
  \put(0,0)     {\includeourbeautifulpicture{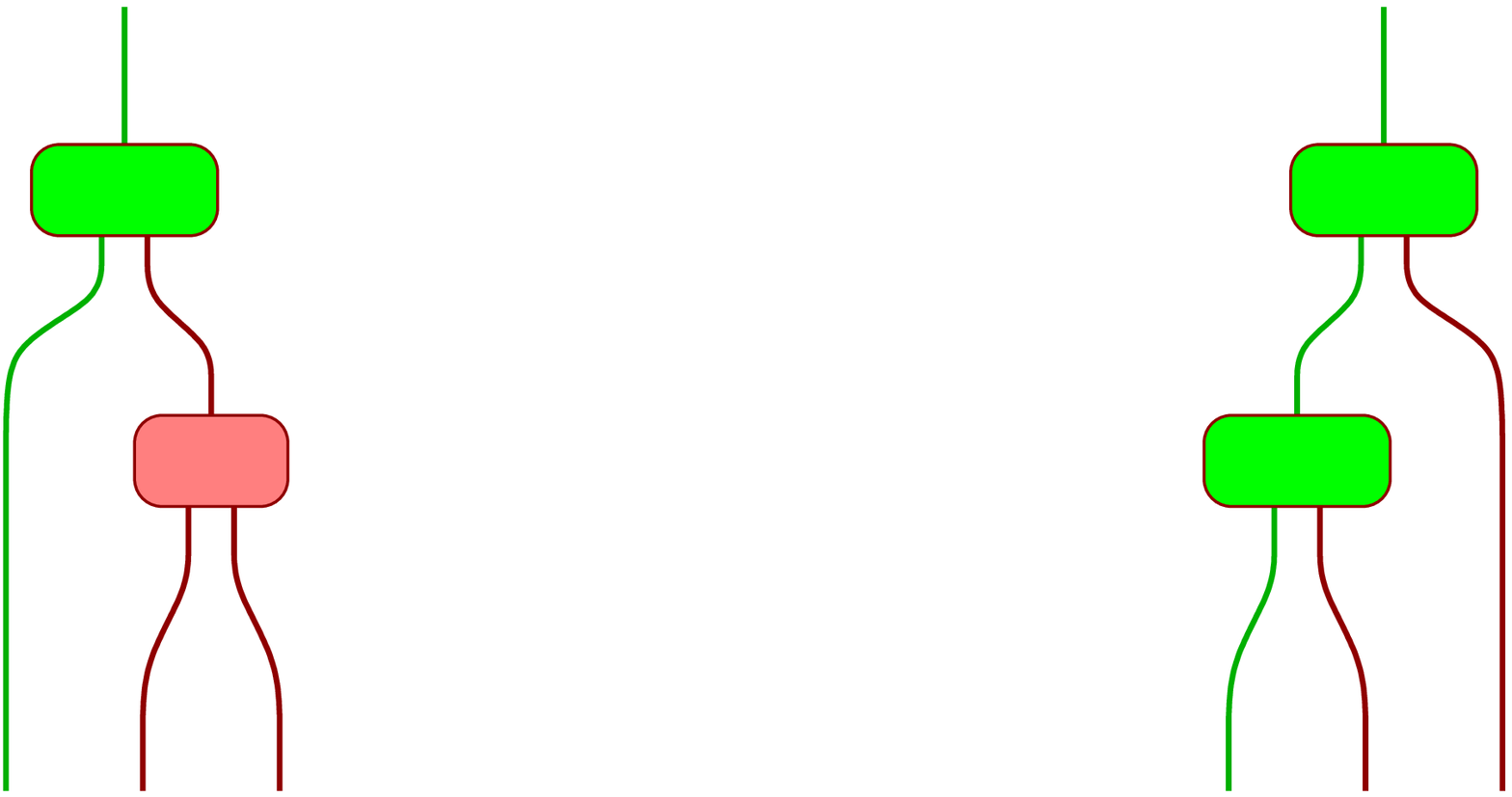}}
  \put(69,47)   {$ =:\, \psi_U^r(h_1,h_2) $ }
  \put(-2.7,-8) {\scriptsize$ U $}
  \put(6.8,78)  {\scriptsize$ \tildeb_{h_1h_2\!} $}
  \put(14,108)  {\scriptsize$ U $}
  \put(14,-8)   {\scriptsize$ \U_{h_1\!}^{} $}
  \put(18.5,42.5){\scriptsize$ \bL_{h_1\!}{h_2} $}
  \put(25.1,63) {\scriptsize$ \U_{h_1h_2}^{} $}
  \put(32,-8)   {\scriptsize$ \U_{h_2}^{} $}
  \put(158,-8)  {\scriptsize$ U $}
  \put(159,42.5){\scriptsize$ \tildeb_{h_1\!}(U) $}
  \put(168,63)  {\scriptsize$ U $}
  \put(179,108) {\scriptsize$ U $}
  \put(170,78)  {\scriptsize$ \tildeb_{h_2\!}(U) $}
  \put(174,-8)  {\scriptsize$ \U_{h_1}^{} $}
  \put(193,-8)  {\scriptsize$ \U_{h_2}^{} $}
  \epicture08 \labl{Fig9}
(where on the left side the short-hand $\tildeb_{h_1h_2}$ for
$\tildeb_{h_1h_2\!}(U)$ is used). Similarly we define
constants $\psi_U^l(g_1,g_2)$ via
  \be
  \bb_{g_1\!}(U) \cir \big( \id_{L_{g_1}} \Oti\, \bb_{g_2}(U) \big)
  = \psi_U^l(g_1,g_2) \,
  \bb_{g_1 g_2}(U) \cir \big( \bL_{g_1\!}{g_2} \oti \id_{U} \big) \,.
  \ee
In terms of $\FF$-matrix entries this amounts to
$\psi^r_U(g,h)\eq\Fs UghU{gh,}U$ and $\psi^l_U(g,h)\eq\Fs ghUU{U,}{gh}$.
Applying the pentagon identity (see appendix \ref{app:PH-ab3})
to the tensor product $L_g \oti U \oti L_{h_1} \oti L_{h_2}$ gives
  \be
  \psi^r_U(h_1,h_2) \, \phi_U(g,h_2) \, \phi_U(g,h_1) = 
  \phi_U(g,h_1 h_2) \, \psi^r_U(h_1,h_2) \,, \ee
from which it follows that $\phi_U$ is a homomorphism in the second argument.
Similarly,
evaluating the pentagon for $L_{g_1} \oti L_{g_2} \oti U \oti L_h$ results in
  \be
  \phi_U(g_2,h) \, \phi_U(g_1,h) \, \psi^l_U(g_1,g_2)
  = \psi^l_U(g_1,g_2) \, \phi_U(g_1 g_2, h) \,, \ee
thus also establishing the homomorphism property in the first argument.
\qed

\medskip

An important aspect of the bihomomorphism $\phi_U$ is that it can be
computed from matrices $\Sg_{}$ that are related to modular transformations
(and for which an explicit formula has been conjectured,
see remark \ref{rem-twining} below).
In the sequel we will discuss some properties of these matrices.

\dtl{Definition}{Sjays}
Let $g\iN\Pic(\calc)$ and $U_i,U_j$ be simple objects of $\calc$. 
Then $\Sg_{i,j}$ is the bilinear pairing
  \be  \Sg_{i,j}: \quad
  \Hom(\U_g\Oti\, U_i,U_i) \otimes_\complex \Hom(U_j,U_j\Oti\, \U_g)
  \,\to\, \complex  \ee
with
  \bea \begin{picture}(240,80)(0,28)
  \put(80,0)     {\includeourbeautifulpicture{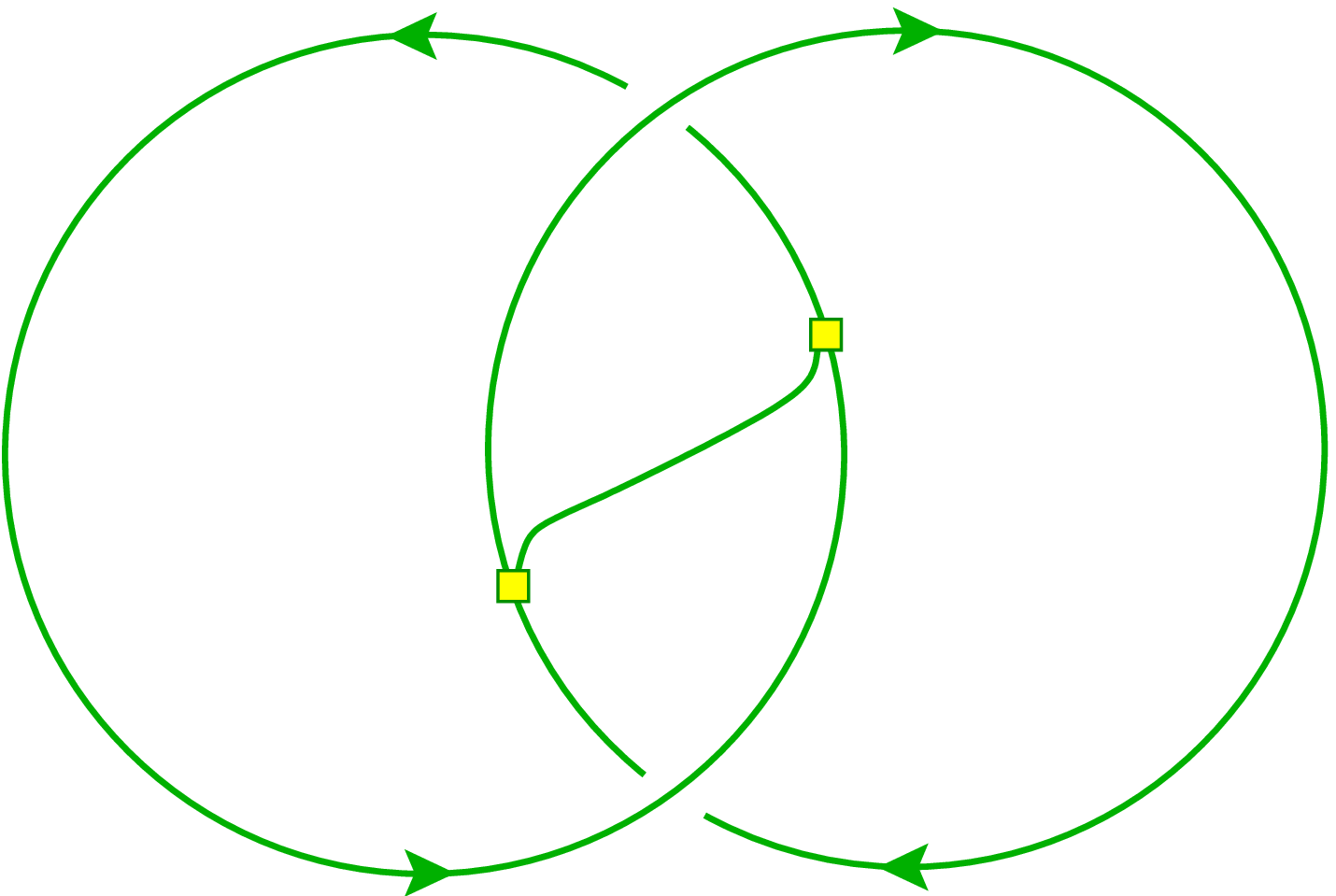}}
  \put(0,50)     {$ \Sg_{i,j}(\irbeta\Oti\iralpha) \,:= $}
  \put(128,62)   {\scriptsize$ U_j $}
  \put(131.2,35) {\scriptsize$ \iralpha $}
  \put(156,43)   {\scriptsize$ \U_g $}
  \put(177,29)   {\scriptsize$ U_i $}
  \put(180.3,63.3){\scriptsize$ \irbeta $}
  \epicture14 \labl{bilpair}
for $\irbeta\iN\Hom(\U_g\Oti\, U_i,U_i)$ and 
$\iralpha\iN\Hom(U_j,U_j\Oti\, \U_g)$.

\dtl{Proposition}{Sfix}
Let $U_i$ be a simple object of a modular tensor category
$\calc$ and let $g,h \in \cals(U_i)$.
\\[.2em]
(i)~\,There exists a $j\iN\cali$ such that $g\iN\cals(U_j)$ and
$\Sg_{i,j} \,{\neq}\, 0\,$.
\\[.2em]
(ii)~Let $j$ be any element of $\cali$ such that $\Sg_{i,j}\,{\ne}\,0\,$. Then
  \be
  \phi_{U_i}(g,h) = \chi_{U_j}(h)^{-1} .  \labl{identity}
(iii)\,We have
  \be  \phi_{U_i} (g,g) = \th_g  \in \{\pm 1\} \, .  \labl{eq:phi-gg=theta}
Thus $\phi_{U_i}$ is a KSB on $\cals(U_i)$.
In particular, if for all $g\iN\cals(U_i)$ the twist of $\U_g$ is the
identity, then $\phi_{U_i}$ is an alternating bihomomorphism. \\[.2em]
(iv)~We have 
  \be \phi_{U_i}(g,h)\, \phi_{U_i}(h,g) = 1
  \qquad {\rm and} \qquad
  \,\phi_{U_i}(g,h) = \phi_{U_{\iBar}}(h,g) \,.  \ee

\medskip\noindent
Proof:\\
(i)~\,Let $F_g$ be the subset of $\cali$ consisting of all $k\iN\cali$
such that $g\iN\cals(U_k)$. Consider the $|F_g|{\times}|F_g|\,$-matrix 
${\rm M}$ with entries
  \be
  {\rm M}_{ij} :=  
  \Sg_{i,j}(\,\bb_g(U_i) \oti \overline{\tildeb_g(U_j)}\,) \labl{Mij}
for $i,j \iN F_g$. This matrix is invertible. In fact, it is related 
to a change of basis in the space $\calh(\U_g,{\rm T})$ of conformal
blocks for a torus with one $\U_g$-insertion. We will not present the
details of the proof, but just describe the two bases.
\\[.2em]
The elements of the first basis are $|\chi_m;\U_g,{\rm T}\rangle$ for
$m\iN F_g$. The vector $|\chi_m;L_g,{\rm T}\rangle$ is given by a solid torus 
with an inscribed ribbon labeled by $U_m$ similar to (I:5.15), 
together with an additional ribbon labeled by $\U_g$ that connects
the boundary of the solid torus to the $U_m$-ribbon. The $U_m$- and
$\U_g$-ribbons are joined by the morphism
$\bb_g(U_m) \iN \Hom(\U_g \oti U_m, U_m)$. Suppose that in the solid torus 
defining $|\chi_m;L_g,{\rm T}\rangle$, the $a$-cycle of the torus ${\rm T}$ 
is contractible. Then the second basis is
of the same form as the first, but this time the solid torus is
chosen such that the $b$-cycle of ${\rm T}$ is contractible.
\\
Since the matrix ${\rm M}$ is invertible, for any $g\iN \cals(U_i)$ there
exists a $j\iN F_g$ such that ${\rm M}_{ij} \,{\ne}\, 0$. Thus for each
$g\iN \cals(U_i)$ we can find a $j\iN\cali$ such that $\Sg_{i,j} \,{\ne}\, 0$.
\\[.2em]
(ii)~\,Let $\gamma \iN\Hom(U_i, U_i\Oti \U_h)$ and $\overline\gamma\iN
\Hom(U_i\Oti\U_h,U_i)$ be any two dual isomorphisms.  Consider the move 
  \bea \begin{picture}(380,81)(18,27)
  \put(0,0)      {\includeourbeautifulpicture{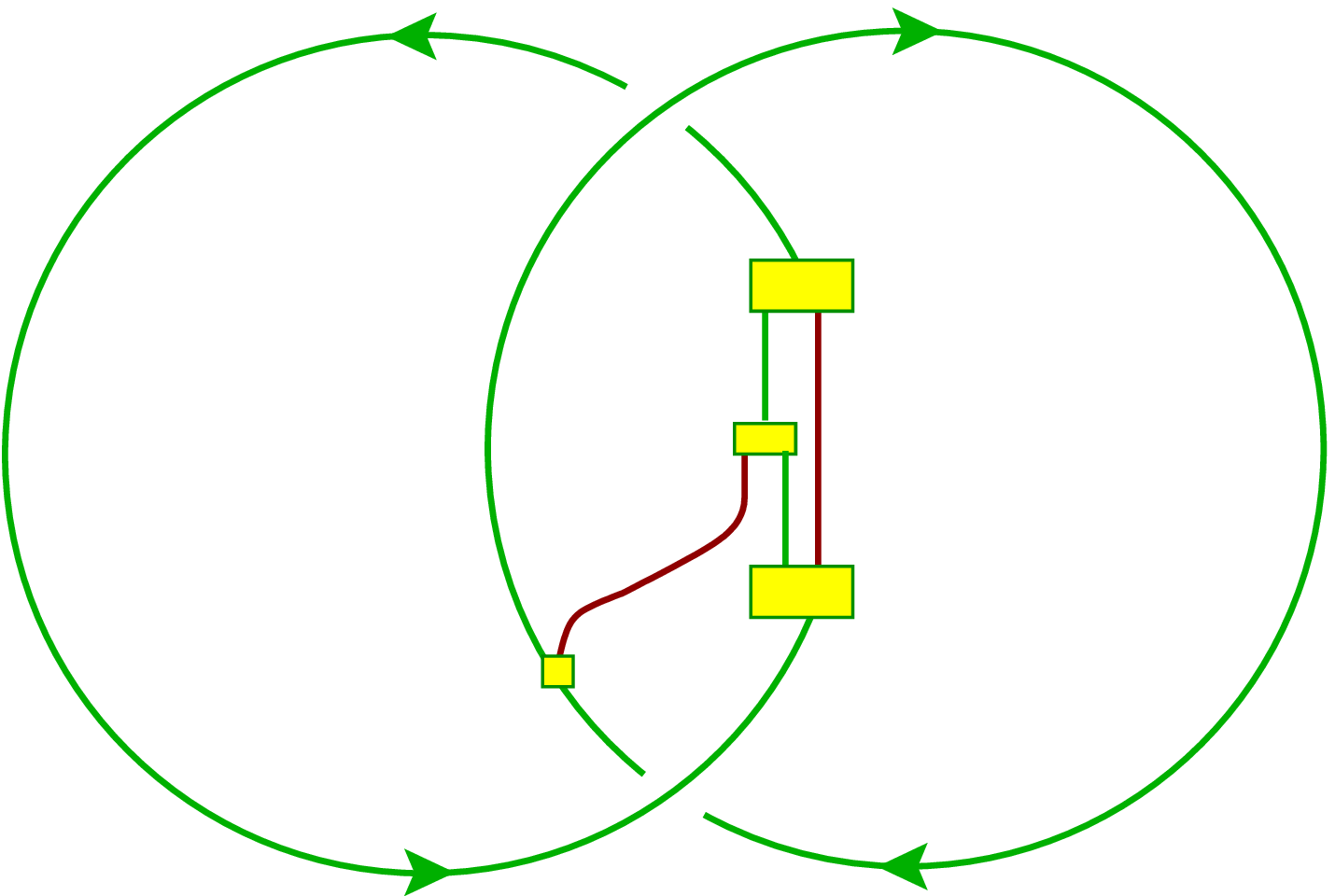}}
  \put(48,62)    {\scriptsize$ U_j $}
  \put(57,24.7)  {\scriptsize$ \iralpha $}
  \put(74.2,30)  {\scriptsize$ \U_g $}
  \put(79.7,51.4){\scriptsize$ \irbeta $}
  \put(90,17.5)  {\scriptsize$ U_i $}
  \put(97.4,51)  {\scriptsize$ \U_h $}
  \put(103,70)  {\scriptsize$ \gamma $}
  \put(103,33)  {\scriptsize$ \overline\gamma $}
  \put(174,50)   {$\displaystyle =\frac{1}{\phi_{U_i}(g,h)} $}
  \put(294,72)   {\scriptsize$ U_j $}
  \put(292.3,50.8){\scriptsize$ \iralpha $}
  \put(317.5,60) {\scriptsize$ \U_g $}
  \put(325,79.2) {\scriptsize$ \irbeta $}
  \put(336,17.5) {\scriptsize$ U_i $}
  \put(349,48)   {\scriptsize$ \U_h $}
  \put(353,62)   {\scriptsize$ \gamma $}
  \put(353,36)   {\scriptsize$ \overline\gamma $}
  \epicture11 \labl{Fig12}
which makes use of the (basis independent) 6j-symbol defined in \erf{Fig7}.
Next use the fact that $\gamma$ and $\overline\gamma$ are dual morphisms to
rewrite the left hand side of \erf{Fig12} as
  \bea \begin{picture}(380,92)(18,29)
  \put(0,0)      {\includeourbeautifulpicture{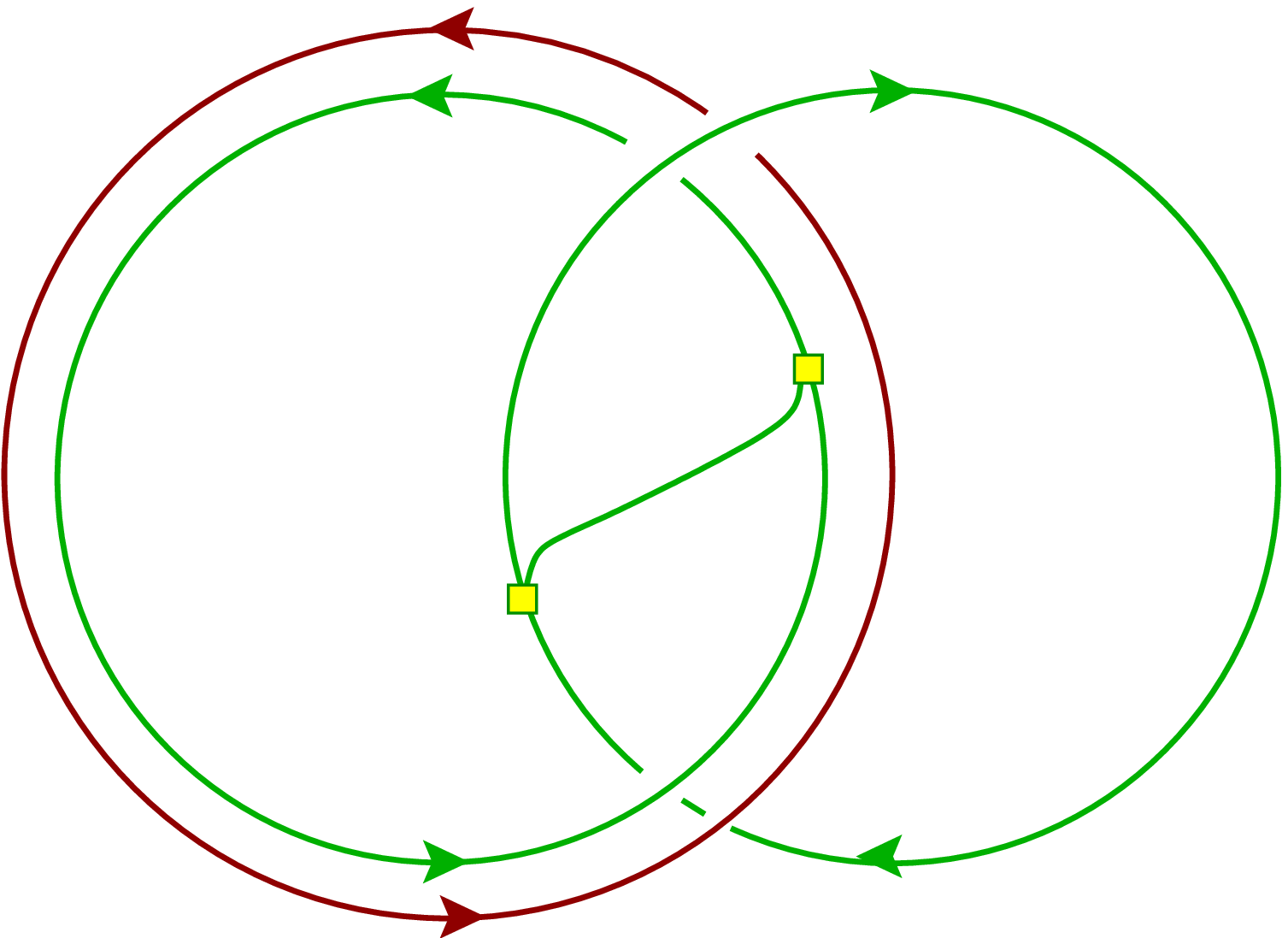}}
  \put(58,77.4)  {\scriptsize$ U_j $}
  \put(58.6,41.8){\scriptsize$ \iralpha $}
  \put(82.5,50)  {\scriptsize$ \U_g $}
  \put(88,35.3)  {\scriptsize$ U_i $}
  \put(95.4,70.4){\scriptsize$ \irbeta $}
  \put(109,88.4) {\scriptsize$ \U_h $}
  \put(178,56)   {$ =~ \chi_{U_j}(h) $}
  \put(295.3,41.8){\scriptsize$ \iralpha $}
  \put(296,82)   {\scriptsize$ U_j $}
  \put(319.8,50) {\scriptsize$ \U_g $}
  \put(344.3,70.6){\scriptsize$ \irbeta $}
  \put(337,27.2) {\scriptsize$ U_i $}
  \epicture11 \labl{Fig12a}
Here we made use of the defining property \erf{expmono} of the 
exponentiated monodromy charge. 
The equality of \erf{Fig12} and \erf{Fig12a} now amounts to
$\phi_{U_i}(g,h)^{-1} \, \Sg_{i,j}(\irbeta\Oti\iralpha) \eq
\chi_{U_j}(h)\, \Sg_{i,j}(\irbeta\Oti\iralpha)$. 
By assumption the index $j\iN\II$ has the property
that $\Sg_{i,j}(\irbeta\Oti\iralpha) \,{\ne}\, 0$ 
for non-zero $\irbeta$ and $\iralpha$;
this establishes the claim.
\\[.2em]
(iii)~By (i) we can find $j\iN\cali$ 
such that $\Sg_{i,j}\,{\ne}\,0$. By (ii) we then have
$\phi_{U_i}(g,g)\eq\chi_{U_j}(g)$. Further, since $g\iN\cals(U_j)$,
the expression \erf{expr} for the exponentiated monodromy charge reads 
$\chi_{U_j}(g)\eq \th_g^{-1}$. Finally, $\theta_g$ takes values only in 
$\{\pm 1\}$, by proposition \ref{props-stabilizer}(i). That 
$\phi_{U_i}$ is a KSB then follows by proposition \ref{props-stabilizer}(ii).
\\[.2em]
(iv)~By lemma \ref{lem:xi-xi=beta} the KSB $\phi_{U_i}$ satisfies
$\phi_{U_i}(g,h)\,\phi_{U_i}(h,g) \eq \beta(g,h) \eq 
\theta_{gh} / (\theta_g \theta_h)$ for all $g,h\iN\cals(U_i)$.
By proposition \ref{props-mono}, the monodromy charge
$\chi_{U_i}$ is a character of $\Pic(\calc)$. By \erf{expr}, for
$g,h\iN\cals(U_i)$ the property $\chi_{U_i}(gh) \eq \chi_{U_i}(g)\chi_{U_i}(h)$
of the character $\chi_{U_i}$ implies $\theta_{gh} \eq \theta_g \theta_h$.
\\
Further, the definiton of $\phi_U$ in \erf{Fig7} implies that, for simple $U$,
  \be
  \bb_g(U) \cir \big(\id_{\U_g} \oti \tildeb_h(U)\big) \cir
  \big(\,\overline{\bb_g(U)}\oti\id_{\U_h}\big) \cir \overline{\tildeb_h(U)}
  = \phi_{U}(g,h) \, \id_{U} \,.  \ee
Taking duals of both sides and invokng basis independence of the morphisms
$\phi_U$ then implies $\phi_{U^\vee}(h^{-1},g^{-1}) \eq \phi_{U}(g,h)$. Setting 
now $U\eq U_i$ and using the bihomomorphism property of $\phi_{U_{\iBar}}$
results in $\phi_{U_i}(g,h) \eq \phi_{U_{\iBar}}(h,g)$.
\qed

\dtl{Remark}{rem-twining}
(i)~\,\,It is conjectured in \cite{fusS6} that for modular tensor categories 
    arising from WZW conformal field theories,
    the bilinear pairing \erf{bilpair} can be computed using the theory of 
    twining characters \cite{fusS3,furs}. For the case of simple current
    extensions of WZW theories, see \cite{sche10}.
    The relevant numerical data have been implemented in a
    computer program {\tt kac} by Bert Schellekens \cite{kac}.
\\[.2em]
(ii)~\,The bihomomorphisms $\phi_U$ were originally introduced 
   \cite{fusS6} through the identity \erf{identity}. Implicit in the
   discussion in \cite{fusS6} is the use of a distinguished class of bases
   of the morphism spaces $\Hom(\U_g\Oti\,U,U)$; for a discussion
   of those bases see \cite{bant6}
   and the considerations preceding definition \ref{straight} below.
\\[.2em]
(iii)~By a calculation similar to (II:3.107), one computes the invariant for 
\erf{bilpair} for the basis morphisms (as in \erf{Mij}) to be
  \be
  \Sg_{i,j}\big(\,\bb_g(U_i) \oti \overline{\tildeb_g(U_j)}\,\big)
  = \sum_{m\in\cali} \sum_{\alpha=1}^{N_{ij}^{~m}}
  \frac{\theta_m}{\theta_i \theta_j} \,
  \Fs jgim{\alpha i,}{j\alpha} \dim(U_m) \,.  \labl{eq:Sg-basis}
Suppose now that there exists a choice of basis in the spaces
$\Hom(U_i \oti U_j, U_k)$ such that all $\FF$-matrix entries are real numbers. 
This is e.g.\ possible in all Virasoro minimal models (including the 
non-unitary ones) and in rational unitary WZW models.
\\
Using in addition that in a modular tensor category the twists are phases 
(compare footnote \ref{phasephootnote}),
the complex conjugate of \erf{eq:Sg-basis} is found to be
  \be
  \Sg_{i,j}{(\bb_g(U_i) \oti \overline{\tildeb_g(U_j)})}^*
  = \tSg_{i,j}(\bb_g(U_i) \oti \overline{\tildeb_g(U_j)}) \,,
  \labl{eq:Sg-conj-aux1}
where $\tSg_{i,j}$ is defined analogously as \erf{bilpair}, but with the 
two braidings replaced by inverse braidings. By rotating the $U_j$-ribbon in
\erf{bilpair} by $180^\circ$ about a horizontal axis, one then verifies that
  \be
  \tSg_{i,j}(\irbeta\oti\iralpha) = 
  \Sg_{i,\bar\jmath}(\irbeta\oti\tilde\iralpha) \labl{eq:Sg-conj-aux2}
for a suitable morphism 
$\tilde\iralpha \iN \Hom(U_{\bar\jmath},U_{\bar\jmath}\Oti\, \U_g)$.
\\ 
Combining the formulas \erf{eq:Sg-conj-aux1} and \erf{eq:Sg-conj-aux2} we 
conclude that if $\Sg_{i,j}$ is non-zero, then so is $\Sg_{i,\bar\jmath}$.
For a given $i\iN\cali$ and $g\iN\cals(U_i)$ now choose $j\iN\cali$
such that $\Sg_{i,j}$ is non-vanishing, as is possible according
proposition \ref{Sfix}(i). Thus also $\Sg_{i,\bar\jmath}$ is non-zero,
and proposition \ref{Sfix}(ii) then implies
  \be
  \phi_{U_i}(g,h) = \chi_{U_j}(h)^{-1} = 
  \chi_{U_{\overline\jmath}}(h) = \phi_{U_i}(g,h^{-1}) \,, \ee
where the second step uses proposition \ref{props-mono}(ii). Thus 
$\phi_{U_i}(g,h)^2 \eq 1$, and together with proposition \ref{Sfix}(iv) 
we find that if all $\FF$-matrix entries of $\calc$
can be chosen to be real, then
  \be
  \phi_{U_i}(g,h) = \phi_{U_i}(h,g) \in \{\pm 1\} \,.  \labl{phi+-1}

\subsection{Representation theory of \berta s} \label{bertarep}

With the help of the basis independent 6j-symbols we can now develop the
representation theory of \berta s.
For a symmetric special Frobenius algebra in a modular tensor category, 
every module appears a submodule of some induced module. 
Thus we first have a look at induced modules. 

\dtl{Lemma}{lemma-induced}
Let $A$ be a \berta\ with support $H$ and KSB $\XI_A$.
\\[.1em]
(i)~\,\,The induced $A$-modules $\IndA(U)$ and $\IndA(V)$ for two simple 
objects $U,V$ are isomorphic iff $U$ and $V$ are on the same orbit
of the action of $H$ on isomorphism classes of objects of $\calc$. 
If they are not on the same orbit, then they do not contain isomorphic submodules. 
\\[.2em]
(ii)~\,As an object of $\calc$, the induced module of a simple object $U$
decomposes as
  \be  \IndA(U) \cong\, |\calsA(U)| \!\bigoplus_{[g]\in H/\calsA(U)}\! gU \,.
  \ee
\medskip\noindent
Proof: \\
(i)~\,The statements immediately follow from reciprocity (see e.g.\ 
proposition I:4.12)
  \be \dimc\HomA(\IndA(U),\IndA(V)) = \sum_{g\in H}\dimc\Hom(U,gV) \,. 
  \labl{calc}
\\[.2em]
(ii)~\,We have
  \be
  A\oti U \,\cong\, \bigoplus_{h\in H} hU \,\cong \hspace{-1em}
  \bigoplus_{[g]\in H/\calsA(U)} \bigoplus_{h\in \calsA(U)} ghU  .  \ee
The $h$-summation amounts to an overall multiplicity $|\calsA(U)|$.
\qed

\medskip

The induced module for an object $U$ is thus, in physics terminology, 
the corresponding {\em simple current orbit\/} with a multiplicity given 
by the order of the stabilizer $\calsA(U)$. This does not, however, imply,
that the induced module decomposes into $|\calsA(U)|$ many irreducible modules. 
The remaining part of the \rep\ theory is then to decompose the induced 
modules into simple modules.  In physics terminology, this step is known as
{\em fixed point resolution\/}. The essential information is contained in 
\\[-2.3em]

\dtl{Proposition}{prop-ends}
Let $U \iN \obj(\calc)$ be simple.
The algebra of module endomorphisms of the induced module $\IndA(U)$
is a twisted group algebra over the stabilizer $\calsA(U)$, 
  \be  \EndA(\IndA(U)) \cong \complex_{\beta_{U,A}} \calsA(U) \, ,  \ee
for some two-cocycle $\beta_{U,A}$. 
The cohomology class $[\beta_{U,A}]\in H^2(\calsA(U),\complexx)$ is
described by the alternating bihomomorphism 
   \be  \eps_{U,A}(g,h) := \frac{\beta_{U,A}(g,h)}{\beta_{U,A}(h,g)} =
   \phi_{U}(g,h) \, \XI_A(h,g) \,,  \labl{epsUA}
with $\XI_A$ as defined in \erf{Fig1}.

\medskip\noindent
Proof: \\
By reciprocity, the algebra $\EndA(\IndA(U))$ is isomorphic as a vector
space to a morphism space in $\calc$:
  \be  \EndA(\IndA(U)) \cong \Hom(U,A\otimes U) \,.  \ee
The associative product on the latter space inherited from the product
on $\EndA(\IndA(U))$ is as follows: for 
$\phi_1,\phi_2\iN\Hom(U,A\oti U) $ we have
  \bea \begin{picture}(100,88)(0,28)
  \put(0,52) {$ \phi_1 \star \phi_2 \,= $}
  \put(70,0)     {\includeourbeautifulpicture{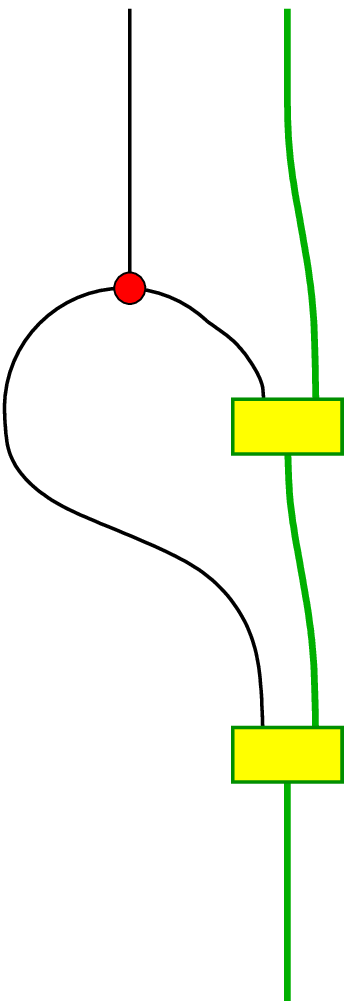}}
  \put(81,115)   {\scriptsize$ A $}
  \put(99,-8)    {\scriptsize$ U $}
  \put(99,115)   {\scriptsize$ U $}
  \put(109.5,26) {\scriptsize$ \phi_2 $}
  \put(109.5,62) {\scriptsize$ \phi_1 $}
  \epicture16 \labl{Fig13}
Note that this product depends on the product on the Frobenius algebra $A$.
Together with the morphism $e_g$ of the adapted basis we get the basis 
  \be  \phi_g := (e_g\oti\id_U) \circ \overline{\bb_g(U}) \, .  \labl{phig}
for the algebra. {}From the form 
  \bea \begin{picture}(170,125)(0,22)
  \put(0,0)     {\includeourbeautifulpicture{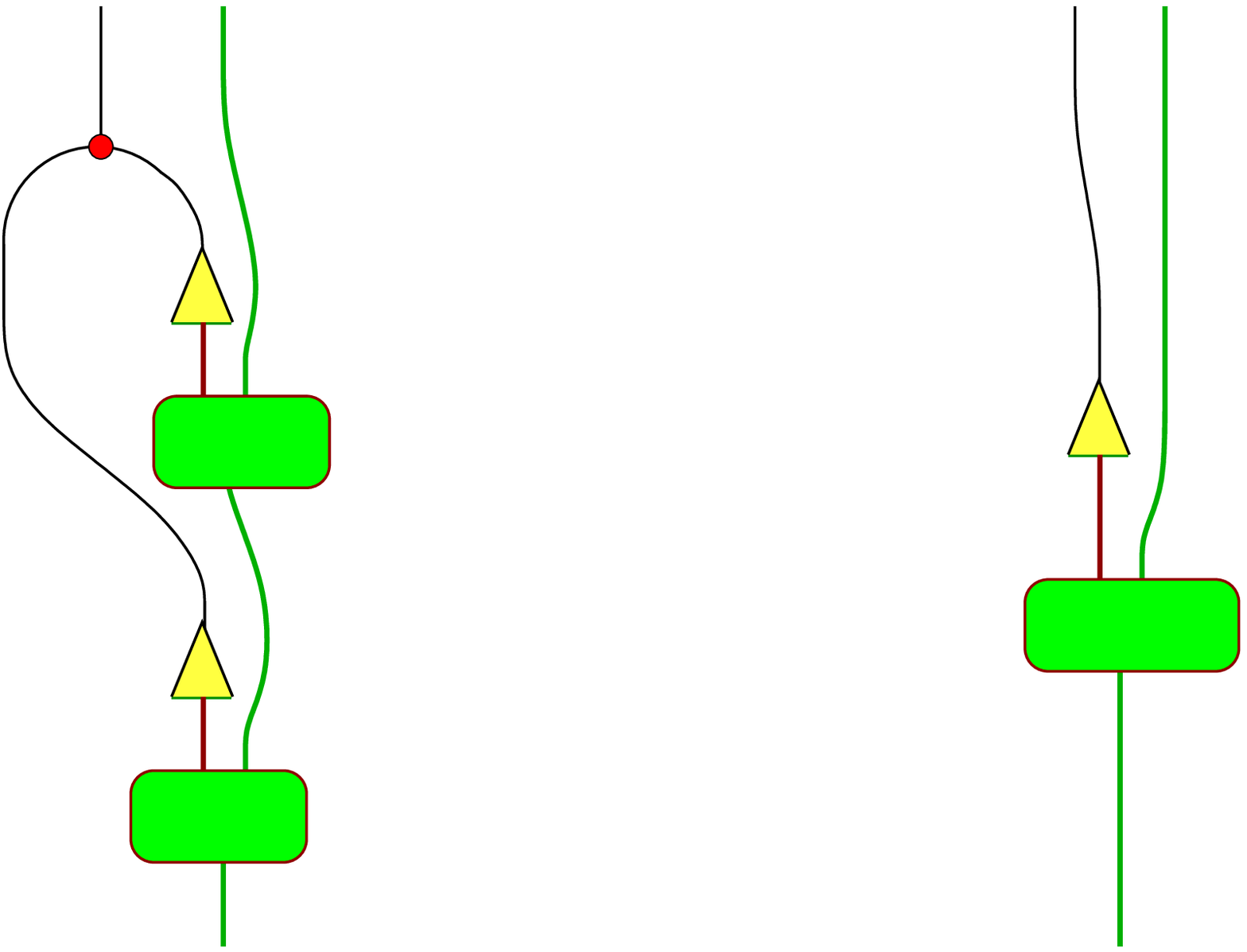}}
  \put(11,137)   {\scriptsize$ A $}
  \put(14.3,30){\scriptsize$ \U_{g_2} $}
  \put(14.3,83)  {\scriptsize$ \U_{g_1} $}
  \put(27.5,-8)  {\scriptsize$ U $}
  \put(28,137)   {\scriptsize$ U $}
  \put(21,17){\scriptsize$ \overline{\bb_{g_2\!}(\!U\!}) $}
  \put(25,70){\scriptsize$ \overline{\bb_{g_1\!}(\!U\!}) $}
  \put(57,66)    {$ =\ \beta_{U,A}(g_1,g_2) $}
  \put(133.4,57) {\scriptsize$ \U_{g_1g_2} $}
  \put(149,137)  {\scriptsize$ A $}
  \put(154.5,-8) {\scriptsize$ U $}
  \put(162.5,137){\scriptsize$ U $}
  \put(147,44){\scriptsize$ \overline{\bb_{g_1g_2\!}(\!U\!}) $}
  \epicture13 \labl{Fig14}
of the product we see that the algebra is graded by the stabilizer
$\calsA(U)$: for $g_1,g_2\iN\cals(U)$, we have
  \be  \phi_{g_1} \star\phi_{g_2} = \beta_{U,A}(g_1,g_2)\,\phi_{g_1g_2}  \labl{a1}
for some two-cochain $\beta_{U,A}$ on $\calsA(U)$. The multiplication on the
$\complex$-algebra $\EndA(\IndA(U))$ is associative, so this cochain is closed.
A change of basis will not change its cohomology class, which is 
the object we are interested in. We characterize it by computing the
commutator cocycle $\eps_{U,A}$ of $\beta_{U,A}$, which appears in
  \bea \begin{picture}(220,118)(0,28)
  \put(0,0)     {\includeourbeautifulpicture{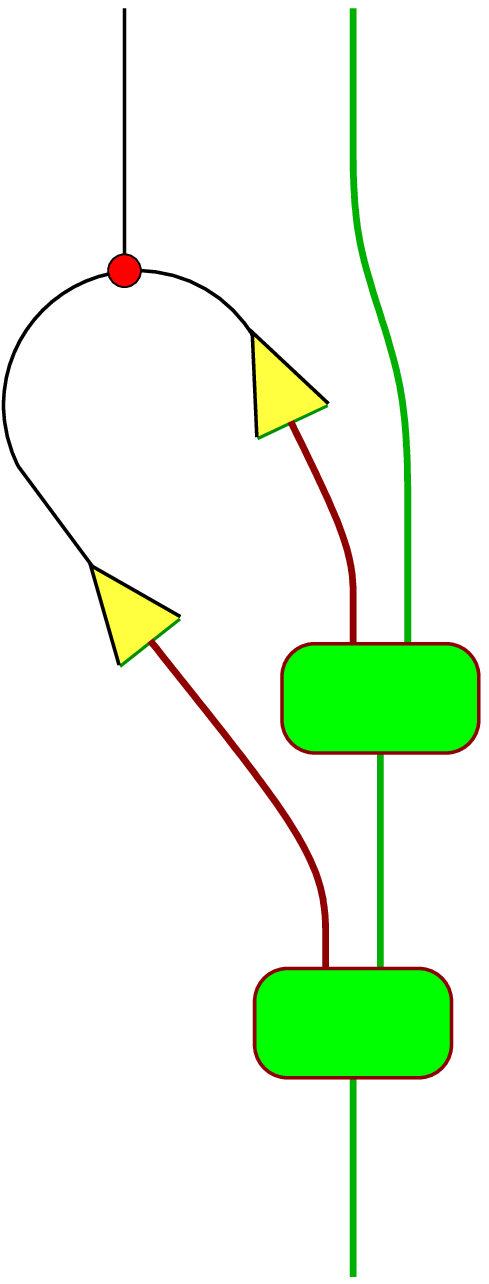}}
  \put(155,0)   {\includeourbeautifulpicture{Fig15}}
  \put(10,144)  {\scriptsize$ A $} 
  \put(21,47)   {\scriptsize$ \U_g $}
  \put(26.2,80) {\scriptsize$ \U_h $}
  \put(29,26)   {\scriptsize$ \overline{\bb_g(U}) $}
  \put(32,62)   {\scriptsize$ \overline{\bb_{h\!}(U}) $}
  \put(35.3,-8) {\scriptsize$ U $}
  \put(35.8,144){\scriptsize$ U $}
  \put(78,67)   {$ =\ \eps_{U,A}(h,g) $}
  \put(165,144) {\scriptsize$ A $} 
  \put(176,47)  {\scriptsize$ \U_h $}
  \put(181.5,80){\scriptsize$ \U_g $}
  \put(184,26)  {\scriptsize$ \overline{\bb_h(U}) $}
  \put(187,62)  {\scriptsize$ \overline{\bb_g(U}) $}
  \put(190.2,-8){\scriptsize$ U $}
  \put(190.7,144){\scriptsize$ U $}
  \epicture16 \labl{Fig15}
To compute $\eps_{U,A}(h,g)$, we compose both sides of \erf{Fig15} with the 
morphism
  \bea \begin{picture}(60,118)(0,28)
  \put(0,0)     {\includeourbeautifulpicture{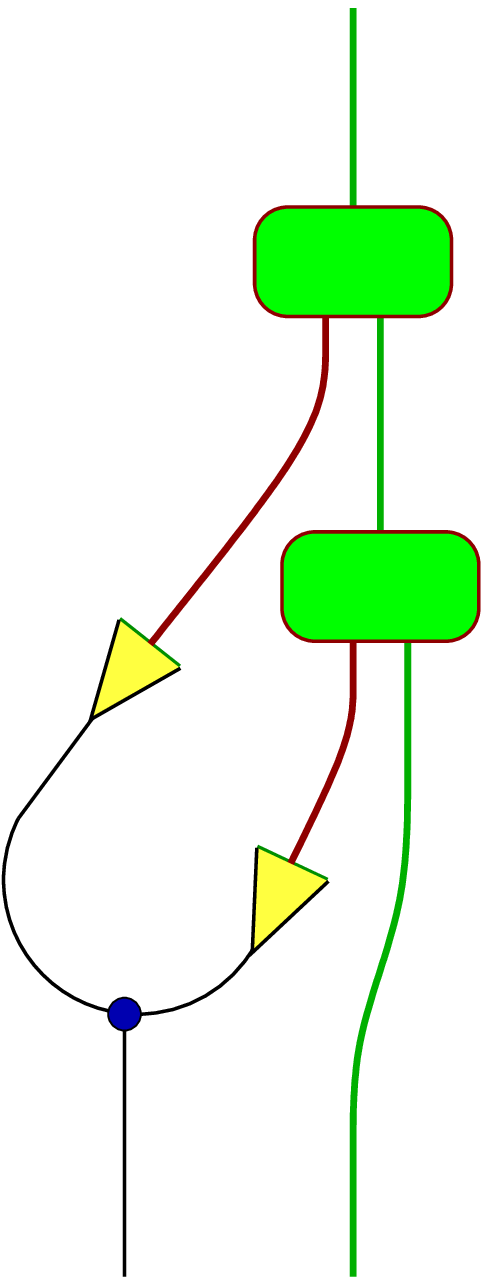}}
  \put(9.2,-8)  {\scriptsize$ A $} 
  \put(18,87.8) {\scriptsize$ \U_h $}
  \put(26,57.2) {\scriptsize$ \U_g $}
  \put(29,110)  {\scriptsize$ \bb_{h\!}(U) $}
  \put(32,74)   {\scriptsize$ \bb_g(U) $}
  \put(35,-8)   {\scriptsize$ U $}
  \put(35.8,144){\scriptsize$ U $}
  \epicture15 \labl{Fig16}
Then on the \rhs, after using the Frobenius property, only the tensor unit 
can propagate in the middle $A$-ribbon, so that we obtain a multiple of $\id_U$:
  \bea \begin{picture}(400,203)(10,30)
  \put(0,0)     {\includeourbeautifulpicture{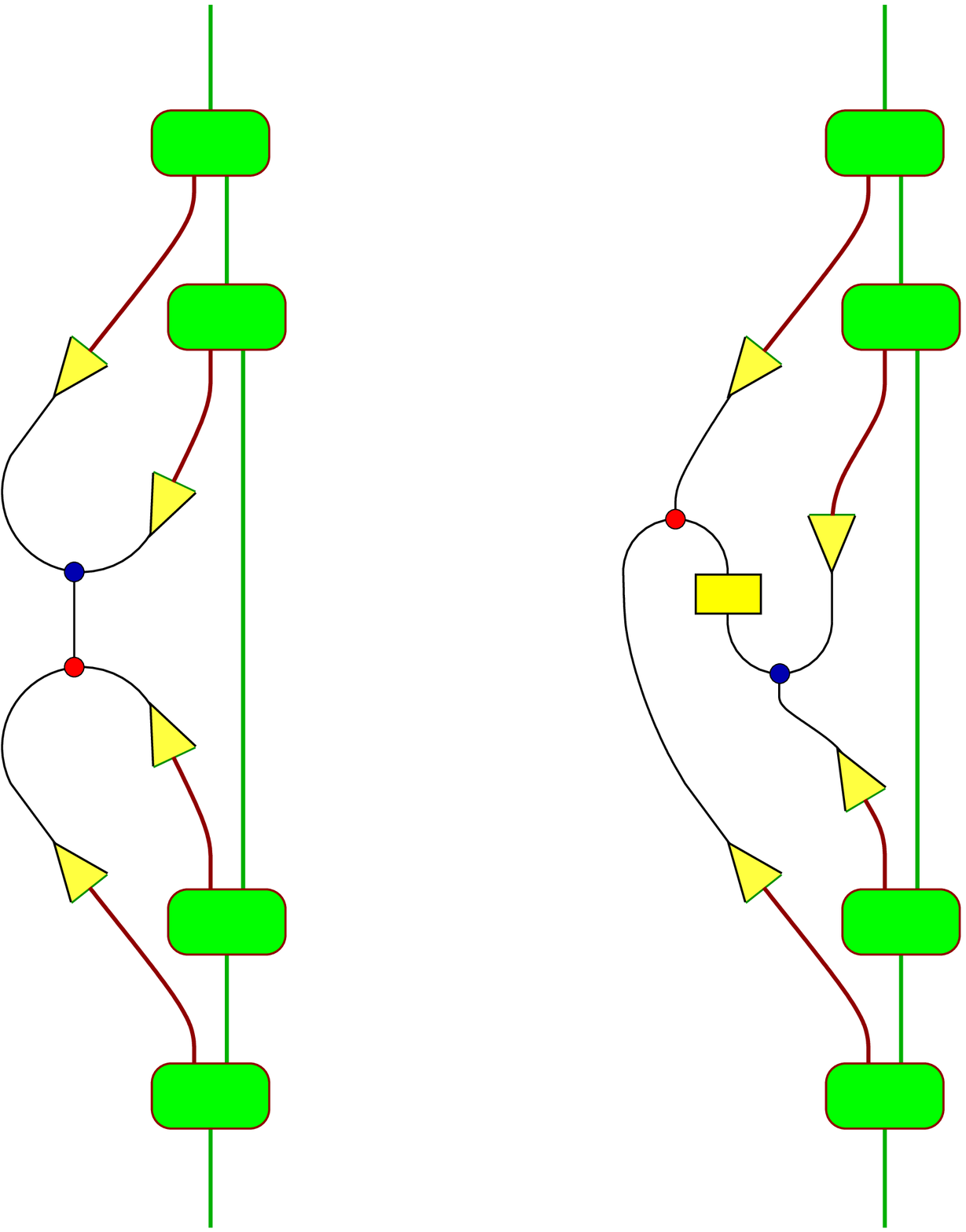}}
  \setlength{\unitlength}{24810sp}
    \put(43,473)  {\scriptsize$ \U_h $}
    \put(65,383)  {\scriptsize$ \U_g $}
    \put(50,107)  {\scriptsize$ \U_h $}
    \put(72,179)  {\scriptsize$ \U_g $}
    \put(15,285)  {\scriptsize$ A $}
    \put(80,529)  {\scriptsize$ \bb_h(\!U\!) $}
    \put(88,442)  {\scriptsize$ \bb_g(\!U\!) $}
    \put(88,145)  {\scriptsize$ \overline{\bb_g(\!U\!}) $}
    \put(78, 59)  {\scriptsize$ \overline{\bb_h(\!U\!}) $}
    \put(93,-21)  {\scriptsize$ U $}
    \put(93,603)  {\scriptsize$ U $}
    \put(374,469)  {\scriptsize$ \U_h $}
    \put(385,376)  {\scriptsize$ \U_g $}
    \put(402,178)  {\scriptsize$ \U_g $}
    \put(373,113)  {\scriptsize$ \U_h $}
    \put(310,367)  {\scriptsize$ A $}
    \put(366,230)  {\scriptsize$ A $}
    \put(410,529)  {\scriptsize$ \bb_h(\!U\!) $}
    \put(415,442)  {\scriptsize$ \bb_g(\!U\!) $}
    \put(415,144)  {\scriptsize$ \overline{\bb_g(\!U\!}) $}
    \put(406, 59)  {\scriptsize$ \overline{\bb_h(\!U\!}) $}
    \put(421,-21)  {\scriptsize$ U $}
    \put(418,603)  {\scriptsize$ U $}
    \put(749,466)  {\scriptsize$ \U_h $}
    \put(771,382)  {\scriptsize$ \U_g $}
    \put(777,187)  {\scriptsize$ \U_g $}
    \put(761,103)  {\scriptsize$ \U_h $}
    \put(676,322)  {\scriptsize$ A $}
    \put(746,295)  {\scriptsize$ A $}
    \put(787,529)  {\scriptsize$ \bb_h(\!U\!) $}
    \put(794,442)  {\scriptsize$ \bb_g(\!U\!) $}
    \put(792,145)  {\scriptsize$ \overline{\bb_g(\!U\!}) $}
    \put(785, 61)  {\scriptsize$ \overline{\bb_h(\!U\!}) $}
    \put(799,-21)  {\scriptsize$ U $}
    \put(799,603)  {\scriptsize$ U $}
  \setlength{\unitlength}{1pt}
  \put(79,110)  {$ = $}
  \put(130.5,116.5) {\scriptsize$ p_e $}
  \put(210,110) {$ =\ \displaystyle\frac1{|H|} $}
  \put(344,110) {$ =\ \displaystyle\frac1{|H|}\;\id_U \;. $}
  \epicture18 \labl{Fig17}
To compute the left hand side, we note that, since all coupling spaces
involved are one-dimensional, there exist non-zero scalars $\alpha_g(U)$
such that
  \bea \begin{picture}(140,68)(0,28)
  \put(0,0)     {\includeourbeautifulpicture{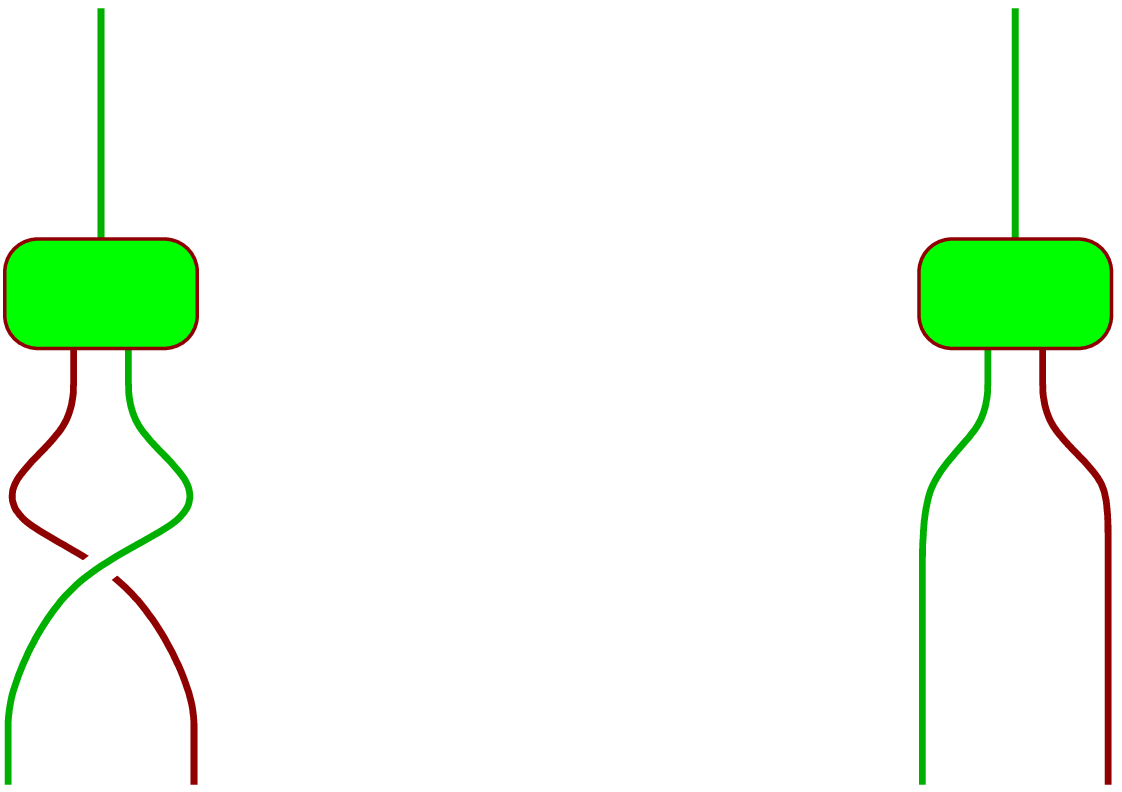}}
  \setlength{\unitlength}{24810sp}
    \put(20,235)  {\scriptsize$ U $}
    \put(-5,-20)  {\scriptsize$ U $}
    \put(48,-20)  {\scriptsize$ \U_g $}
    \put( 6,140)  {\scriptsize$ \bb_g(\!U\!) $}
    \put(283,235)  {\scriptsize$ U $}
    \put(257,-20)  {\scriptsize$ U $}
    \put(310,-20)  {\scriptsize$ \U_g $}
    \put(270,140)  {\scriptsize$ \tildeb_g(\!U\!) $}
  \setlength{\unitlength}{1pt}
  \put(40,40)   {$ =\ \alpha_g(U) $}
  \epicture13 \labl{def-alpha}
We now compute
  \begin{eqnarray} \begin{picture}(400,232)(17,0)
  \put(0,0)     {\includeourbeautifulpicture{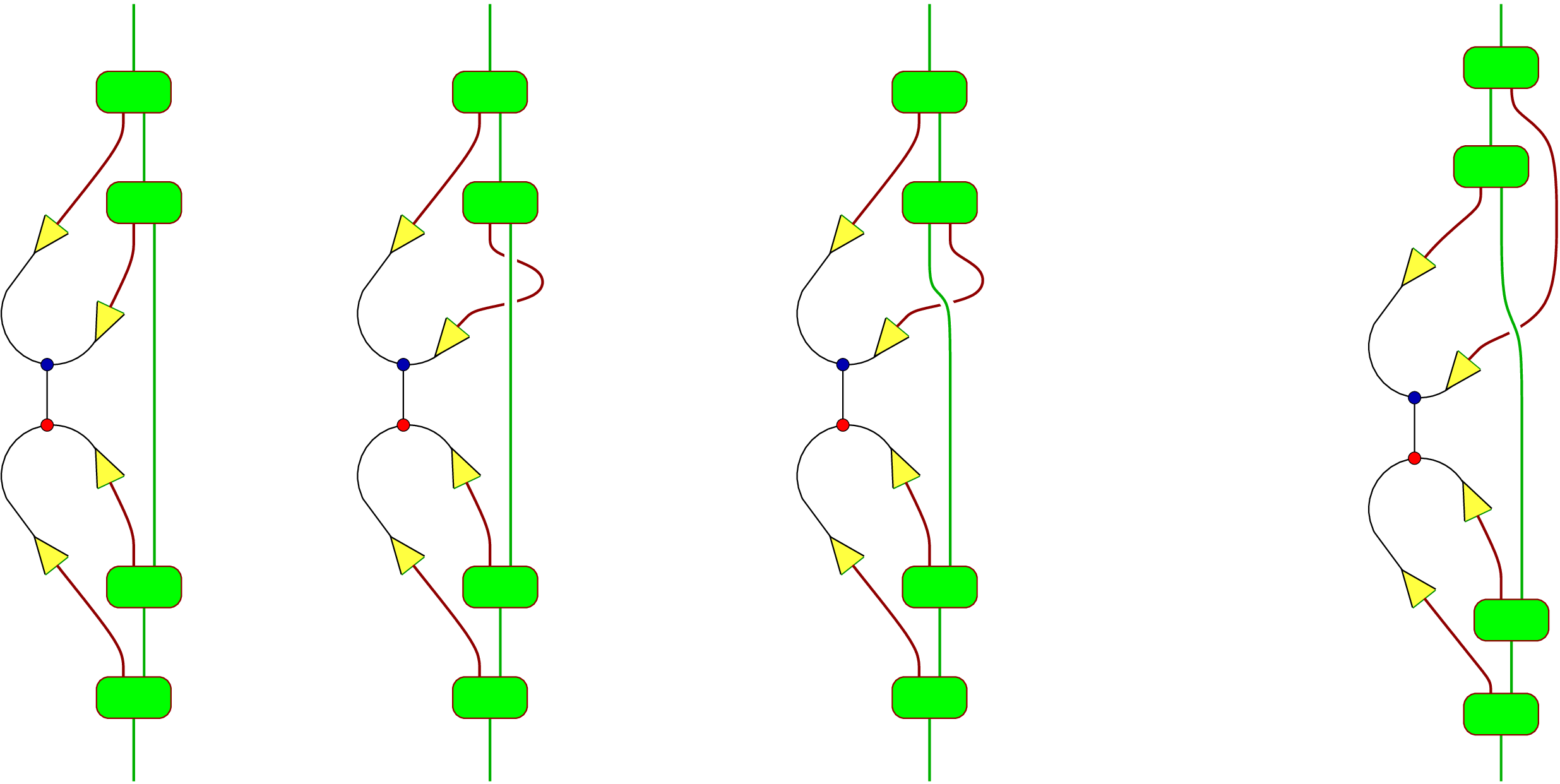}}
  \setlength{\unitlength}{24810sp}
    \put(43,469)  {\scriptsize$ \U_h $}
    \put(65,383)  {\scriptsize$ \U_g $}
    \put(50,107)  {\scriptsize$ \U_g $}
    \put(72,179)  {\scriptsize$ \U_h $}
    \put(15,285)  {\scriptsize$ A $}
    \put(78,524)  {\scriptsize$ \bb_h(\!U\!) $}
    \put(88,441)  {\scriptsize$ \bb_g(\!U\!) $}
    \put(88,145)  {\scriptsize$ \overline{\bb_h(\!U\!}) $}
    \put(80, 60)  {\scriptsize$ \overline{\bb_g(\!U\!}) $}
    \put(93,-18)  {\scriptsize$ U $}
    \put(93,603)  {\scriptsize$ U $}
    \put(313,467)  {\scriptsize$ \U_h $}
    \put(420,371)  {\scriptsize$ \U_g $}
    \put(321,113)  {\scriptsize$ \U_g $}
    \put(345,185)  {\scriptsize$ \U_h $}
    \put(288,285)  {\scriptsize$ A $}
    \put(350,524)  {\scriptsize$ \bb_h(\!U\!) $}
    \put(360,441)  {\scriptsize$ \bb_g(\!U\!) $}
    \put(360,144)  {\scriptsize$ \overline{\bb_h(\!U\!}) $}
    \put(352, 60)  {\scriptsize$ \overline{\bb_g(\!U\!}) $}
    \put(370,-18)  {\scriptsize$ U $}
    \put(370,603)  {\scriptsize$ U $}
    \put(335,0){
      \put(313,467)  {\scriptsize$ \U_h $}
      \put(420,371)  {\scriptsize$ \U_g $}
      \put(321,113)  {\scriptsize$ \U_g $}
      \put(345,185)  {\scriptsize$ \U_h $}
      \put(288,285)  {\scriptsize$ A $}
      \put(350,524)  {\scriptsize$ \bb_h(\!U\!) $}
      \put(360,441)  {\scriptsize$ \bb_g(\!U\!) $}
      \put(360,144)  {\scriptsize$ \overline{\bb_h(\!U\!}) $}
      \put(352, 60)  {\scriptsize$ \overline{\bb_g(\!U\!}) $}
      \put(370,-18)  {\scriptsize$ U $}
      \put(370,603)  {\scriptsize$ U $}
    }
    \put(1081,429)  {\scriptsize$ \U_h $}
    \put(1195,426)  {\scriptsize$ \U_g $}
    \put(1090, 87)  {\scriptsize$ \U_g $}
    \put(1114,162)  {\scriptsize$ \U_h $}
    \put(1059,259)  {\scriptsize$ A $}
    \put(1118,467)  {\scriptsize$ \bb_h(\!U\!) $}
    \put(1124,543)  {\scriptsize$ \bb_g(\!U\!) $}
    \put(1133,119)  {\scriptsize$ \overline{\bb_h(\!U\!}) $}
    \put(1125,48)   {\scriptsize$ \overline{\bb_g(\!U\!}) $}
    \put(1140,-18)  {\scriptsize$ U $}
    \put(1140,603)  {\scriptsize$ U $}
  \setlength{\unitlength}{1pt}
  \put(72,106)  {$ = $}
  \put(174,106) {$ =\ \alpha_g(U) $}
  \put(298,102) {$ =\ \alpha_g(U)\,\phi_U(h,g) $}
  \end{picture} \nonumber\\[.7em]{} \label{Fig19}
  \\[-1.7em]{}\nonumber\end{eqnarray}
where the basis independent $6j$-symbols $\phi_U(h,g)$ are used to interchange 
the order of the $\U_h$- and $\U_g$-ribbons. After a slight deformation of the
$\U_h$-ribbons we can now invoke the relation
\erf{sc-rel} that is valid for invertible objects in order to tie them together
in a different order. The dual morphisms $\bbb_h(U)$ and $\overline{\bbb_h(U})$ 
cancel each other, so that we arrive at
  \bea \begin{picture}(390,203)(0,32)
  \put(0,0)      {\includeourbeautifulpicture{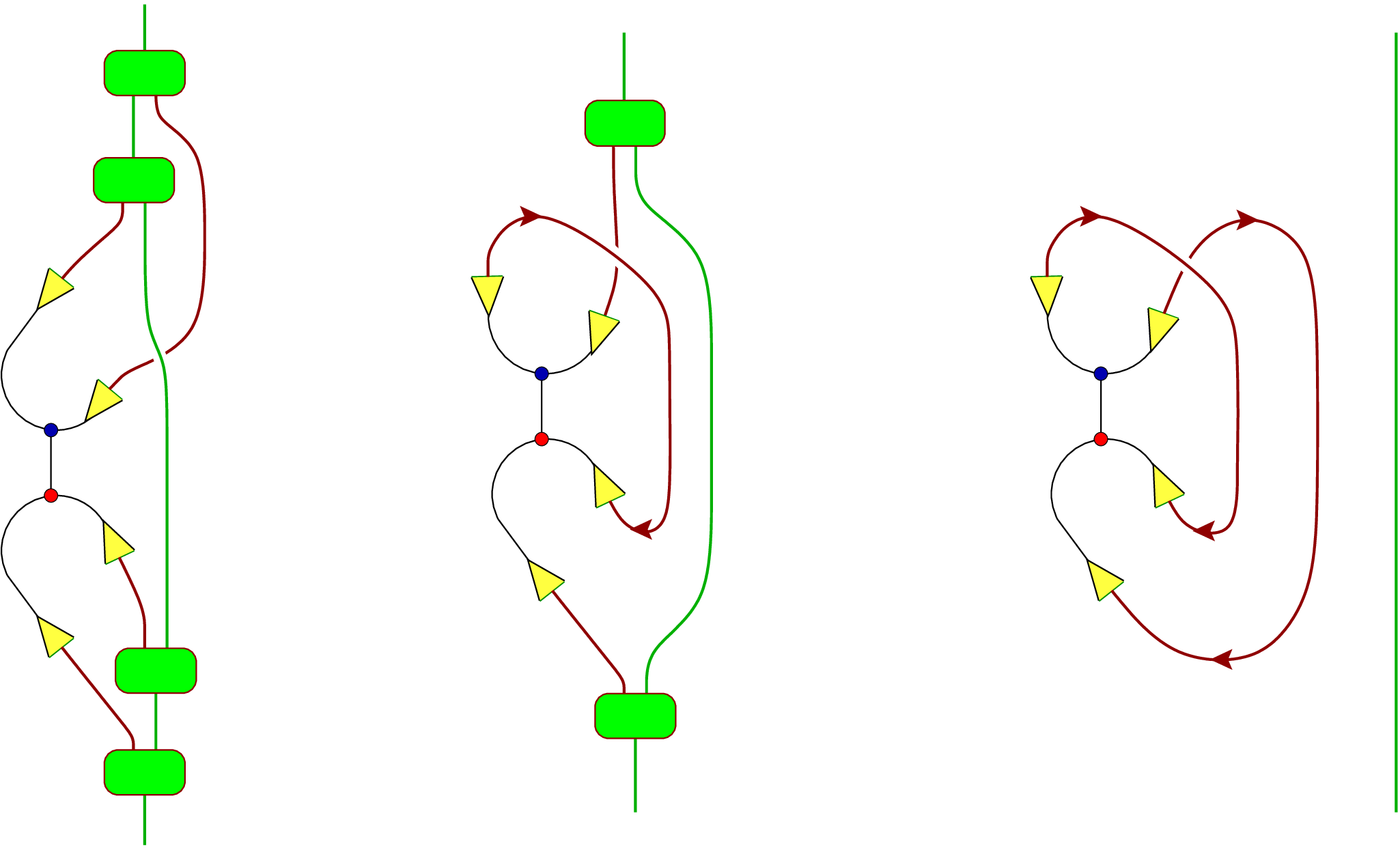}}
  \setlength{\unitlength}{24810sp}
  \put(-1044,0){
    \put(1081,429)  {\scriptsize$ \U_h $}
    \put(1195,426)  {\scriptsize$ \U_g $}
    \put(1090, 87)  {\scriptsize$ \U_g $}
    \put(1114,162)  {\scriptsize$ \U_h $}
    \put(1059,259)  {\scriptsize$ A $}
    \put(1116,467)  {\scriptsize$ \bb_h(\!U\!) $}
    \put(1124,543)  {\scriptsize$ \bb_g(\!U\!) $}
    \put(1131,119)  {\scriptsize$ \overline{\bb_h(\!U\!}) $}
    \put(1125,48)   {\scriptsize$ \overline{\bb_g(\!U\!}) $}
    \put(1140,-18)  {\scriptsize$ U $}
    \put(1140,603)  {\scriptsize$ U $}
  }
    \put(444,296)  {\scriptsize$ \U_h $}
    \put(408,459)  {\scriptsize$ \U_g $}
    \put(390,132)  {\scriptsize$ \U_g $}
    \put(361,302)  {\scriptsize$ A $}
    \put(418,508)  {\scriptsize$ \bb_g(\!U\!) $}
    \put(425,88)  {\scriptsize$ \overline{\bb_g(\!U\!}) $}
    \put(439,584)  {\scriptsize$ U $}
    \put(439,  3)  {\scriptsize$ U $}
    \put(719,440)  {\scriptsize$ \U_h $}
    \put(792,129)  {\scriptsize$ \U_g $}
    \put(759,303)  {\scriptsize$ A $}
    \put(980,584)  {\scriptsize$ U $}
    \put(980,  3)  {\scriptsize$ U $}
  \setlength{\unitlength}{1pt}
  \put(65,106)   {$\displaystyle  =\; \frac{1}{\alpha_g(U)}$}
  \put(218,106)  {$\displaystyle  =\; \frac{1}{\alpha_g(U)}$}
  \epicture27 \labl{Fig20}
where in the last step we have used the relation \erf{sc-rel} for $\U_g$.
Comparing to \erf{Fig21a} we see that the last morphism is equal to 
$\XI_A(g,h) / (\alpha_g(U) \dim(A)) \, \id_U$. Combining with \erf{Fig17}
and \erf{Fig19}, and noting that $\dim(A)\eq |H(A)|$, gives
  \be  \eps_{U,A} (h,g) = \phi_U(h,g)\, \XI_A(g,h) \,,  \ee
in agreement with \erf{epsUA}.
\qed

\medskip

The general theory of twisted group algebras of finite abelian groups
\cite{KArp4} motivates the
\\[-2.3em]

\dtl{Definition}{ustab}
Given a \berta\ $A$, the {\em untwisted stabilizer\/}, or {\em central 
stabilizer\/}, of a simple object $U$ is the subgroup 
  \be  \calu_A(U) := \{ g \iN \calsA(U) \,|\, \eps_{U,A} (g,h) \eq 1
  \;{\rm for}\;{\rm all}\;h\iN\calsA(U) \}  \ee 
of the stabilizer $\calsA(U)$.

\medskip

The stabilizer $\calsA(U_i)$ only depends on the support $H(A)$ of the \berta\
$A$; the central stabilizer $\calu_A(U_i)$, however, also depends on the 
algebra structure chosen on the object $A$.
The following property of the central stabilizer
relates it to a twisted group algebra (compare e.g.\ \cite{fuSc11}):
\\[-2.3em]

\dtl{Lemma}{lem:CU-cent}
The group algebra of $\calu_A(U)\,{\le}\,\calsA(U)$ is the center of the 
$\beta_{U,A}$-twisted group algebra of $\calsA(U)$,
  \be  \complex\calu_A(U) = \calz(\complex_{\beta_{U,A}}\calsA(U))\,.  \ee

\dtl{Remark}{rem-ustab}
(i)~\,\,The modules over the twisted group algebra are in bijection with
        the representations of the \central\ stabilizer. 
\\[.2em]
(ii)~\,The \central\ stabilizer, rather than the full stabilizer, is 
thus the group whose characters label the ``resolved fixed points''. This
holds true even in the absence of discrete torsion, i.e.\ for 
$\XI_A\,{\equiv}\, 1$.
\\[.2em]
(iii)~For a certain class of simple objects $U_i$, a simple formula 
for the order $|\calu_A(U_i)|$ of the \central\ stabilizer has been derived in 
\cite{bant7}. It states that, provided that the number
  \be  Z_H(U_i) := \Frac1{|\calsA(U_i)|} \sum_{g\in H}
  \th_g^{1/2} \sum_{j,k\in \II} \caln_{jk}^{~i}\, S_{g,j}\,S_{0,k}\,
  \frac{\th_j^2}{\th_k^2} \,,  \ee
defined for a simple current group $H$ and a simple object $U_i$,
is non-zero, then
  \be  |\calu_A(U_i)| = |\calsA(U_i)|\, (Z_H(U_i))^2_{} \, .  \ee
(iv)~It is instructive to look at the situation in the case of
theta-categories. Let $\calc(G,\psi,\omega)$ be a theta-category and $A$ a 
commutative algebra in it. Then the simple $A$-modules are in correspondence 
with elements of the quotient group $G/H$; the {\em local\/} modules 
are those classes $[q]\iN G/H$ such that
$\beta(g,q)\eq 1$ for all $g\iN H$. They form a subgroup $(G/H)^0$
of $G/H$. The category of local modules is again a theta-category;
it is characterized by the quadratic form $\bar q ([g])\eq q(g)$, which
is well-defined.

\medskip

As already mentioned at the beginning of this section, every simple $A$-module 
appears as a submodule of an induced module. Suppose a given induced module 
$\IndA(U)$ over a Schellekens algebra $A$ decomposes according to
  \be
  \IndA(U) \cong \bigoplus_{\kappa\in\JJ} M_{\kappa}^{\oplus n_\kappa}
  \labl{eq:IndA-decomp}
into simple modules $M_\kappa$. The algebra $\EndA(\IndA(U))$ is then 
isomorphic to a direct sum of matrix algebras $\EndA(\IndA(U)) \,{\cong}\, 
\bigoplus_{\kappa\in\JJ}\! {\rm Mat}_{n_\kappa} (\complex)$. 
The unit matrices of the individual blocks then form a basis of the center
of $\EndA(\IndA(U))$ consisting of idempotents. Each such idempotent
gives one isotypical component $M_\kappa^{\oplus n_\kappa}$
in the decomposition \erf{eq:IndA-decomp}.

To construct these projectors, let us start by specifying a basis
$\{\varphi_g \,|\, g\iN\calu_A(U)\}$ of the $\complex$-algebra
$\calz(\EndA(\IndA(U)))$ for a simple object $U$. It is given by
  \bea \begin{picture}(230,79)(0,22)
  \put(180,0)       {\includeourbeautifulpicture{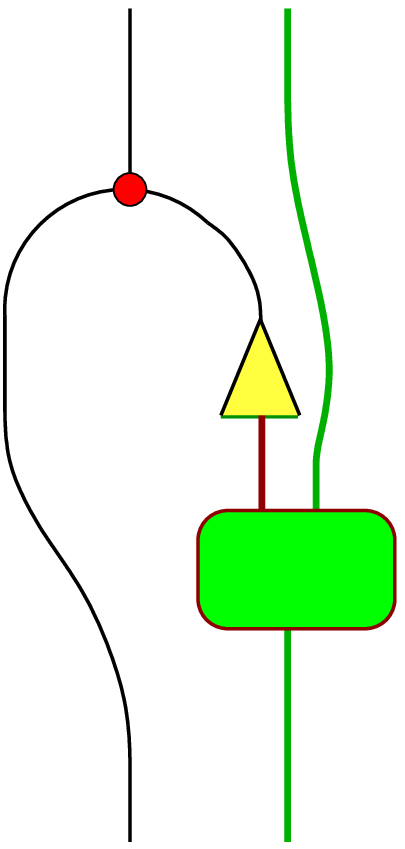}
  \setlength{\unitlength}{24810sp}
    \put(29,253)  {\scriptsize$ A $}
    \put(29,-21)  {\scriptsize$ A $}
    \put(73,253)  {\scriptsize$ U $}
    \put(73,-21)  {\scriptsize$ U $}
    \put(47,108)  {\scriptsize$ \U_g $}
    \put(61,75)  {\scriptsize$ \overline{\bb_g(\!U\!)} $}
  \setlength{\unitlength}{1pt}
  }
  \put(0,42)     {$ \varphi_g\,:=\,(m\oti\id_U) \cir (\id_A\oti\phi_g) \,= $}
  \epicture11 \labl{Fig27} 
for $g\iN\calu_A(U)$, where $\phi_g$ is the morphism introduced in \erf{phig}. 
That this is indeed a basis is implied by lemma \ref{lem:CU-cent}.

When making the choice arbitrarily, under composition 
this set of morphism closes up to scalar factors:
  \bea \begin{picture}(350,126)(0,31)
  \put(75,0)      {\includeourbeautifulpicture{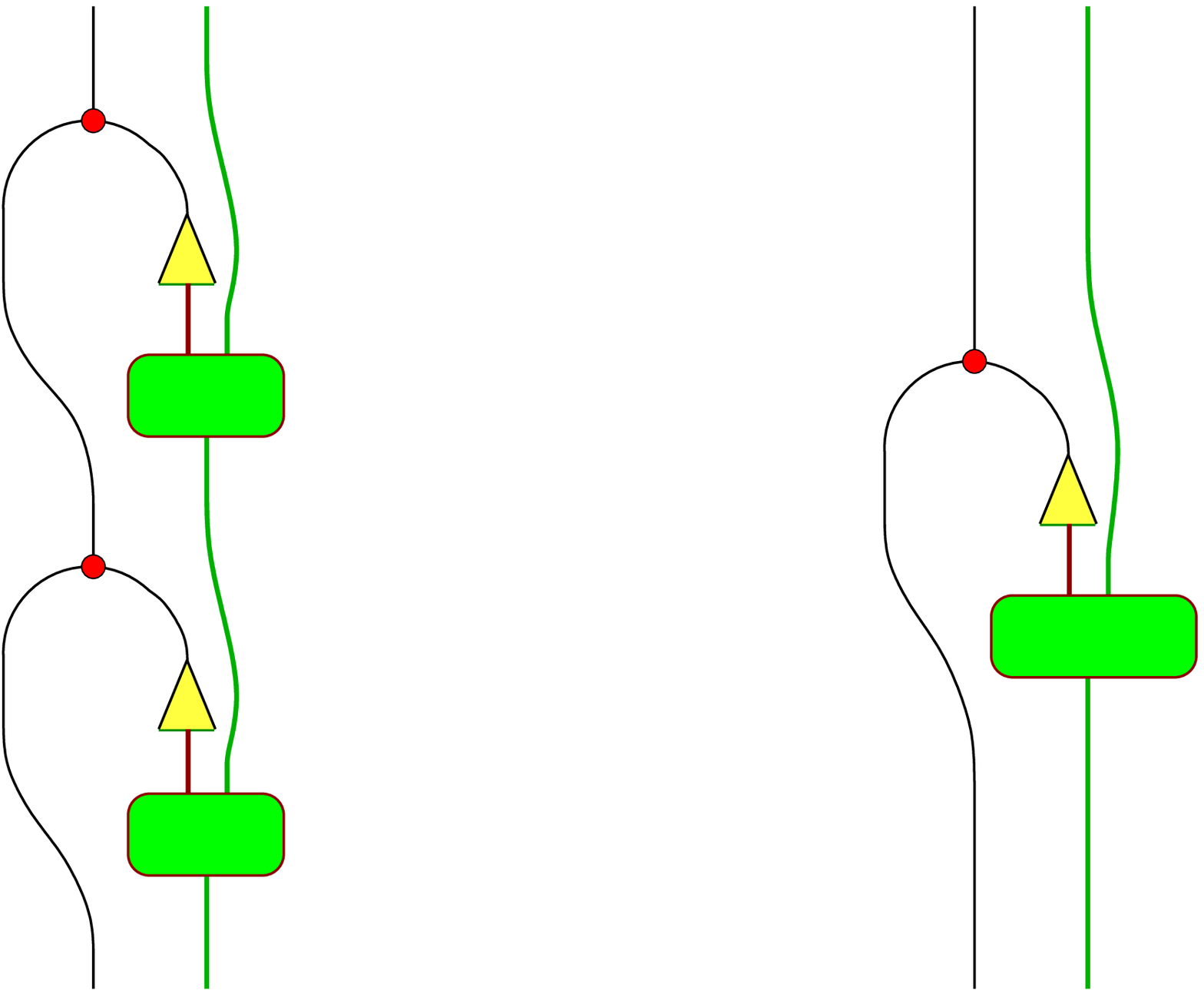}
  \setlength{\unitlength}{24810sp}
    \put(29,407)  {\scriptsize$ A $}
    \put(29,-21)  {\scriptsize$ A $}
    \put(73,407)  {\scriptsize$ U $}
    \put(73,-21)  {\scriptsize$ U $}
    \put(47,269)  {\scriptsize$ \U_g $}
    \put(55,235)  {\scriptsize$ \overline{\bb_g(U}) $}
    \put(47, 87)  {\scriptsize$ \U_h $}
    \put(54,59)  {\scriptsize$ \overline{\bb_h(U}) $}
    \put(355,0){
      \put(29,407)  {\scriptsize$ A $}
      \put(29,-21)  {\scriptsize$ A $}
      \put(73,407)  {\scriptsize$ U $}
      \put(73,-21)  {\scriptsize$ U $}
    }
    \put(390,172)  {\scriptsize$ \U_{gh} $}
    \put(410,137)  {\scriptsize$ \overline{\bb_{gh}(U}) $}
  \setlength{\unitlength}{1pt}
  }
  \put(0,72)     {$ \varphi_g \cir \varphi_h \;= $}
  \put(137,72)   {$ =\ \beta_{U,A}(g,h) $}
  \put(270,72)   {$ =\ \beta_{U,A}(g,h)\,\varphi_{gh} \;, $}
  \epicture19 \labl{Fig29} 
where $\beta_{U,A}$ was defined in \erf{Fig14}.
Note that since $g,h\iN\calu_A(U)$, by definition the commutator two-cocycle
$\eps_{U,A}(g,h)$ of $\beta_{U,A}(g,h)$ is equal to one, and thus
$\beta_{U,A} \eq {\rm d}\eta$ for some one-cochain $\eta$ on $\calu(U)$.
This implies 
that it is possible to modify our initial choice of basis by suitable scalar 
factors in such a way that with the new choice we simply have
  \be
  \varphi_g \cir \varphi_h = \varphi_{gh} \,.  \ee
This choice is unique up to a character $\psi_U $ of $\calu_A(U)$.
The isotypical components are labeled by characters of $\calu_A(U)$, but without 
distinguishing the trivial character, i.e.\
more precisely, they form a torsor over $\calu_A(U)^*$.
This phenomenon has been called fixed point homogeneity in \cite{fusS6}.

Note that with this choice of scalar factors the product in \erf{a1} takes the 
simple form
  \be  \phi_{g_1}
  \star \phi_{g_2} = \phi_{g_1 g_2} = \phi_{g_2} \star \phi_{g_1}
  \qquad\mbox{for all}~~ g_1,g_2\iN \calu_A(U) \,.  \labl{pp=p}
Such a choice is indeed possible: Fix an adapted basis for the algebra
$A$; we may choose, for every simple object $U_i$ separately, the morphisms
$\bb_g(U_i)$ in such a way that the morphism $\phi_g$ constructed from these 
$\bb_g(U_i)$ and from the adapted basis obey \erf{pp=p}. We therefore make the
\\[-2.3em]

\dtl{Definition}{straight}
Given a \berta\ $A$ in $\calc$, we fix an adapted basis for $A$. A family of 
basis choices $\bb_g(U)\iN\Hom(\U_g\oti U,U)$ for all simple objects $U$
and for $g\iN\calu_A(U)$, such that \erf{pp=p} holds for the morphisms
$\phi_g$ constructed with these choices is called
an {\em $A$-straight family\/} of bases. The corresponding family of
morphisms $\phi_g$ is called $A$-straight as well. Whenever the algebra 
evident, we drop it from our notation.

\medskip

Clearly, an $A$-straight family is not uniquely determined.
Rather, for each simple object $U$, the possible choices form a torsor
over $\calu_A(U)^*_{}$. Note that the choices for non-isomorphic simple
objects are not correlated.

Let now $\phi_g$ be straight, and introduce
for every $\psi\iN\calu_A(U)^*$ the morphism
  \be
  \tP_{U,\psi} := \Frac1{|\calu_A(U)|} \sum_{g\in\calu_A(U)} 
  \psi^*(g)\, \varphi_g \,. \labl{tildeP}
By orthogonality of the characters we have $\tP_{U,\psi_1} \cir \tP_{U,\psi_2}
\eq \delta_{\psi_1,\psi_2}\, \tP_{U,\psi_1}$, for $\psi_1,\psi_2\iN
\calu_A(U)^*_{}$. Indeed,
the $\tP_{U,\psi}$ with $\psi\iN\calu_A(U)^*_{}$
form a basis of $\calz(\EndA(\IndA(U)))$ consisting of idempotents. Comparing 
with \erf{eq:IndA-decomp} we see that the images $\tM_{U,\psi}$ of the
morphisms $\tP_{U,\psi}$ are the isotypical components in the decomposition
of $\IndA(U)$. Note that, for $\calu_A(U)\,{\ne}\,\calsA(U)$, $\tM_{U,\psi}$
is not simple, but rather the direct sum of 
  \be
  d_A(U) := \sqrt{ \Frac{ |\calsA(U)| }{ |\calu_A(U)| }}  \labl{eq:da-def}
isomorphic simple modules.

The arguments in this section have established the following
algorithm to determine the list of (isomorphy classes of) simple modules
over a \berta\ and thus of boundary conditions in a CFT of simple current type:
\begin{itemize}
\item[\NXT] 
The set $\mathcal{I}$ of isomorphism classes of simple objects
in $\calc$ admits a partition in $r$ orbits under the group $H(A)$. 
We denote by $U_{m_1},...\,,U_{m_r}$ a set of representatives of these orbits.
\item[\NXT] 
Lemma \ref{lemma-induced} implies that for $m_i\,{\neq}\, m_j$
the induced modules $\IndA(U_{m_i})$ and $\IndA(U_{m_j})$ do not contain 
any common simple submodule. Moreover, each simple module appears as the 
submodule of an induced module. The decomposition of the induced
modules $\IndA(U_{m_i})$ into their isotypical components thus
provides us with the complete list of simple modules, without overcounting.
\item[\NXT] 
The isotypical components $M_\kappa^{\oplus n(i)}$ in the decomposition 
$\IndA(U_{m_i})\eq\bigoplus_{\kappa\in\JJ}\!  M_\kappa^{\oplus n(i)}$ 
are given as images of the projectors
$\tP_{U,\psi}$ defined in \erf{tildeP}. The projector is labeled
by a character $\psi\iN \calu_A(U)^*_{}$, once an $A$-straight basis
has been chosen. The multiplicites $n(i) \eq d_A(U_{m_i})$ in an 
isotypical component depend only on $i$, not on $\kappa$.
\item[\NXT] With a given choice of straight basis, the simple $A$-modules are 
labeled by pairs $(n,\psi)$ where $n\eq1,2,...\,,r$ labels an $H(A)$-orbit
and $\psi\iN\calu_A(U_{m_n})^*$ is a character of the central stabilizer
$\calu_A(U_{m_n})$ of $U_{m_n}$.
\end{itemize}

\dt{Remark}
(i)~\,We have developed the representation theory of
left $A$-modules. As mentioned, they describe the conformal boundary
conditions of the CFT associated to $A$. However, the fact that we are dealing
with left, rather than with right, modules over $A$ is merely a matter of
convention. One can equivalently describe all conformal boundary conditions
by right $A$-modules. 
\\[.2em]
(ii)~Right $A$-modules over a Schellekens algebra $A$ can be
studied by the same methods as were used above. Every right $A$-module
is submodule of an induced right module
${\rm Ind}(U)_{\!A} \,{:=}\, (U \oti A , \id_U \oti m)$. To obtain the
isotypical components in the decomposition of ${\rm Ind}(U)_{\!A}$
we study the endomorphism algebra $\EndA({\rm Ind}(U)_{\!A})$ of right 
module endomorphisms of ${\rm Ind}(U)_{\!A}$. Let now $U$ be a simple object
of $\calc$; a basis of $\EndA({\rm Ind}(U)_{\!A})$ is given by
  \be
  \tilde\varphi_g := (\id_U \oti m) \cir 
  (\id_U \oti e_g \oti \id_A)\cir (\overline{\tildeb_g(U)} \oti \id_A) \,.
  \ee
The composition of two basis elements yields
$\tilde\varphi_g \cir \tilde\varphi_h \eq
\tilde\beta_{U,A}(g,h)\, \tilde\varphi_{gh}$ for some two-cocycle
$\tilde\beta_{U,A}$. The endomorphism algebra is thus isomorphic
to a twisted group algebra,
  \be
  \EndA({\rm Ind}(U)_{\!A}) \cong \complex_{\tilde\beta_{U,A}} \calsA(U) \,.
  \ee
The commutator two-cocycle $\tilde\eps_{U,A}$ of $\tilde\beta_{U,A}$ can be 
worked out analogously as for left modules; we find
  \be
  \tilde\eps_{U,A}(g,h) =
  \frac{\tilde\beta_{U,A}(g,h)}{\tilde\beta_{U,A}(h,g)}
  = \phi_U(g,h)\, \XI_A(g,h) \,.  \ee
Comparison with \erf{epsUA} shows that $\EndA({\rm Ind}(U)_{\!A})$ and 
$\EndA(\IndA(U))$ are isomorphic if the KSB $\XI_A$ is symmetric.
The situation becomes particularly transparent for categories in which all 
$\FF$-matrix entries are real. The result \erf{phi+-1} implies that in this 
case, for any \berta, the $\complex$-algebra $\EndA({\rm Ind}(U)_{\!A})$ 
for induced right modules is just the opposed algebra of $\EndA(\IndA(U))$ for
induced left modules.
\\[.2em]
(iii)~As in the case of left modules, one introduces for a given
adapted basis of $A$ the notion of $A$-straightness for a family of morphisms 
in $\Hom(U_i,U_i\oti A)$. By an appropriate choice of the morphisms
$\tildeb_g(U)$ one shows that straight families exist and that they form a 
torsor over the character group of the (right) central stabilizer.

   \medskip

Using the coproduct of $A$ rather then the product, one can endow
also the vector space $\Hom(A\oti U,U)$ with the structure
of a $\complex$-algebra. This product is inherited from the endomorphism
algebra $\EndA(A\oti U,A\oti U)$ as well, this time using Frobenius
reciprocity. One checks that if a choice of morphisms $\overline{\bb_g(U)}$
straightens the product on $\Hom(U,A\oti U)$, then the dual morphisms
$\bb_g(U)$ straighten the product on $\Hom(A\oti U,U)$. An analogous
statement holds for right modules. All straight choices are unique up
to characters of the relevant central stabilizers.

\subsection{The boundary state}

Our next goal is to determine the {\em boundary states\/} for the full
\cft\ that corresponds to some given \berta\ in a modular tensor category. 
The boundary state for a given boundary condition is, by definition, the 
information about the collection of 
all one-point functions of bulk fields on a disk with that boundary condition.
We have already determined the possible boundary conditions: they correspond
to left $A$-modules as described in the previous subsection. 

The bulk fields that can have a non-vanishing one-point function on the disk
are described by the vectors $\alph$ in the spaces 
  \be  \HomAA(U_i \,{\otimes^+} A \,{\otimes^-}\, U_{\ibar}, A) 
  \labl{bulkspace}
of morphisms of bimodules.
Here the superscripts at the tensor symbols remind us that the object
$U_i \,{\otimes} A \,{\otimes}\, U_{\ibar}$ of $\calc$ has been
endowed with the following structure of an $A$-bimodule: the left action
of $A$ is $\rho_l \,{:=}\, (\id_{U_i} \oti m \oti \id_{U_{\iBar}}) \cir
(c_{U_i,A}^{-1}\oti \id_A \oti \id_{U_{\iBar}})$, while the right action is given by
$\rho_r \,{:=}\, (\id_{U_i} \oti m \oti \id_{U_{\ibar}}) \cir
(\id_{U_i} \oti \id_A \oti c_{A,U_{\ibar}}^{-1})$. In other words, for the left
action the $A$-ribbon is braided under the $U_i$-ribbon, while for
the right action the $A$-ribbon is braided over the $U_{\ibar}$-ribbon.

Only bulk fields with conjugate left and right chiral labels 
$i$ and $\ibar$ contribute. Thus for every $i\iN \II$ there are $Z_{i\ibar}$ 
bulk fields that can have a non-vanishing one-point function on the disk.  
Correlators on the disk have already been presented schematically in the
figure (21) of \cite{fuRs} as ribbon graphs in the three-ball $B$.
With the conventions chosen here, the ribbon graph for the
correlator of a bulk field $\Phi_{i,\ibar}^{(\alph)}$ labeled by the
morphism $\alph$ in the space \erf{bulkspace} on a disk with boundary 
condition given by the $A$-module $M$ looks as follows \citeiv:
  \bea \begin{picture}(150,141)(8,41)
  \put(0,0)       {\includeourbeautifulpicture{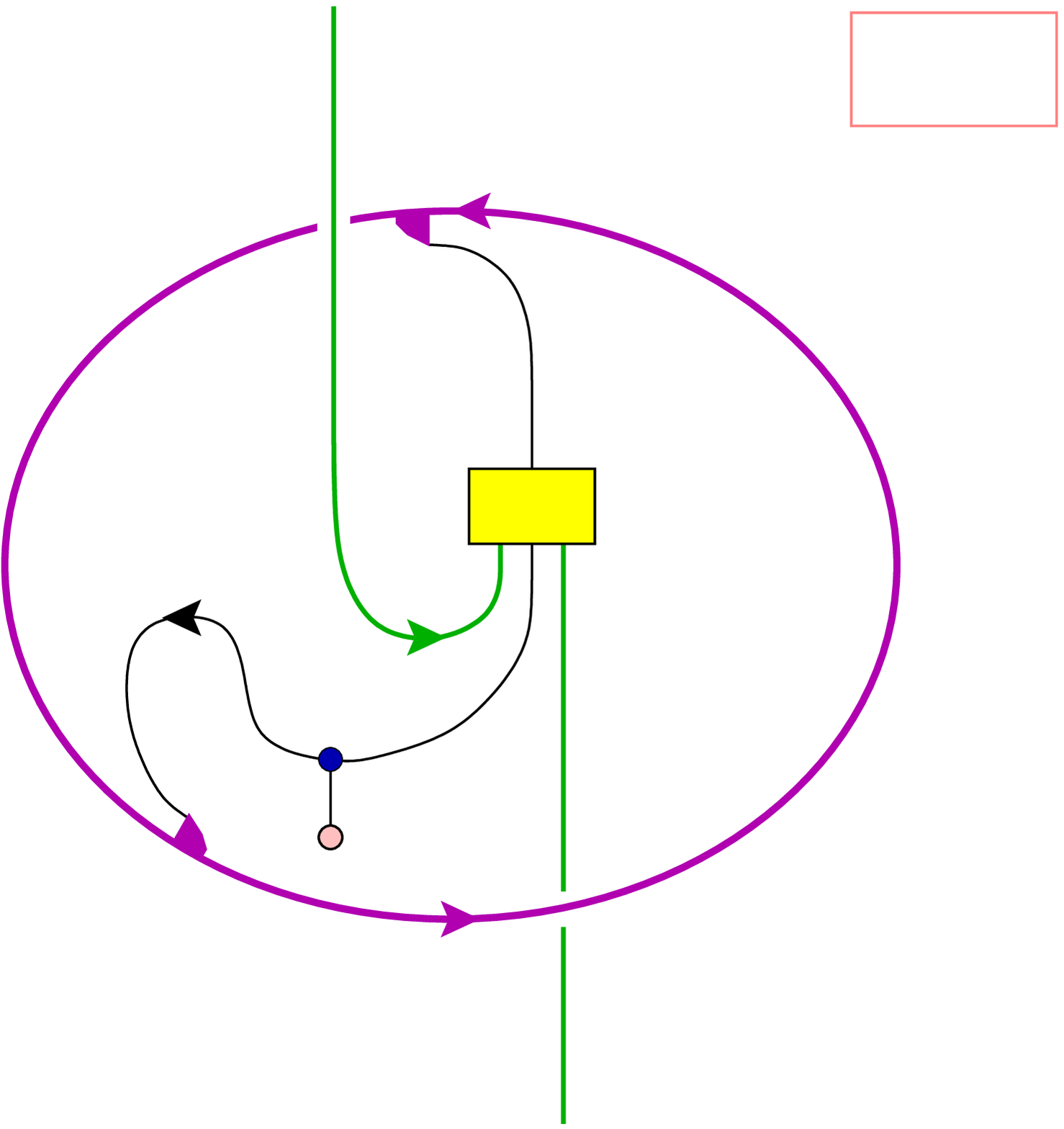}}
  \put(39,64.5)   {\scriptsize$ A $}
  \put(46,176)    {\scriptsize$ U_i^\vee $}
  \put(58,131)    {\scriptsize$ \rho_{\!M}^{} $}
  \put(78.4,93.3) {\scriptsize$ \alph $}
  \put(82,111)    {\scriptsize$ A $}
  \put(83,-8)     {\scriptsize$ U_\ibar $}
  \put(135.7,68)  {\scriptsize$ \dot M $}
  \put(141,157)   {$ B $}
  \epicture27 \labl{Fig23}
Here the morphisms connecting the $A$-ribbons to the $M$-ribbon are
representation morphisms $\rho_M$. The graph \erf{Fig23} can be simplified by 
first dragging one of the representation
morphisms along the $\dot M$-ribbon, then using the representation property,
then the fact that $A$ is symmetric Frobenius, then that $\alpha$ is
a bimodule morphism, and finally again that $A$ is symmetric Frobenius.
Depending on whether one applies the bimodule morphism property of $\alpha$
to the incoming or outgoing $A$-ribbon, one thereby arrives at one of the
two following equivalent graphs:
  \bea \begin{picture}(350,143)(8,40)
  \put(0,0)       {\includeourbeautifulpicture{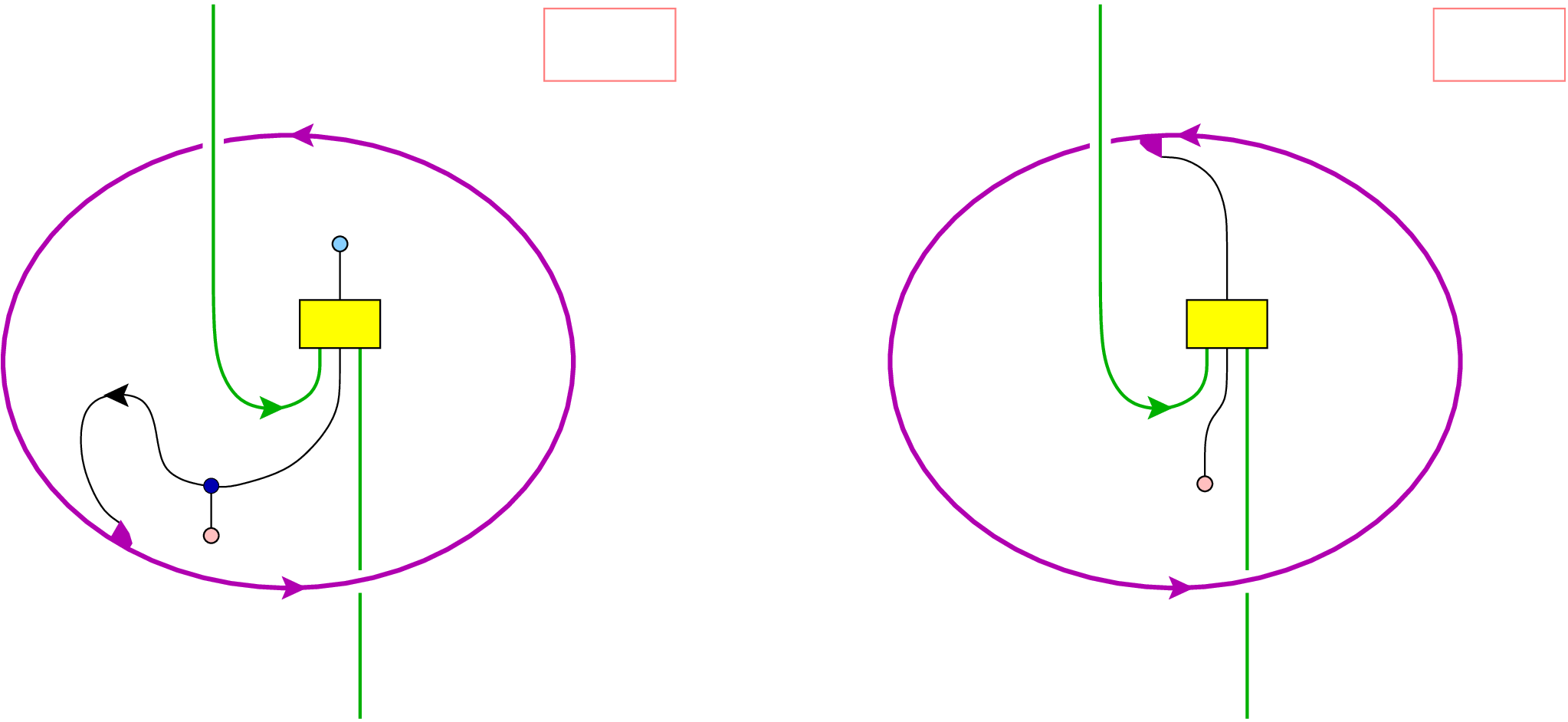}}
  \put(39,64.5)   {\scriptsize$ A $}
  \put(46,176)    {\scriptsize$ U_i^\vee $}
  \put(78.7,92.8) {\scriptsize$ \alph $}
  \put(83,-8)     {\scriptsize$ U_\ibar $}
  \put(135.7,68)  {\scriptsize$ \dot M $}
  \put(141,157)   {$ B $}
  \put(172,84)    {$ = $} 
  \put(257,176)   {\scriptsize$ U_i^\vee $}
  \put(290.1,92.8){\scriptsize$ \alph $}
  \put(294,111)   {\scriptsize$ A $}
  \put(294,-8)    {\scriptsize$ U_\ibar $}
  \put(347.6,68)  {\scriptsize$ \dot M $}
  \put(352,157)   {$ B $}
  \epicture27 \labl{Fig24}

Given a bulk field and a boundary condition, we would like to express
the one-point function in terms of a conformal two-point block on the
Riemann sphere. These conformal blocks form, for every label $i\iN\II$, a 
one-dimensional complex vector space
  \be  V_{i\ibar } = \Hom(U_i\oti U_{\ibar},\one) \,.  \labl{Viibar}
Since this vector space does not have any preferred
basis, there is no canonical identification with the ground field $\complex$. 
Such an identification becomes only possible once we have chosen a basis 
$\Beta_{i\ibar}$ in $V_{i\ibar }$.\,%
  \footnote{~In \citei\ a basis of this space was denoted by
  $\lambda_{(i,\ibar)0}$, see formula (I:2.29). Here we use the short-hand
  $\lambda_{i \ibar}$.}
Such a basis vector in the space of two-point blocks is, in this context, 
also known as an ``Ishibashi state''. The Ishibashi state for the label $i$ 
can be depicted as the following ribbon graph in the three-ball:
  \bea \begin{picture}(50,101)(8,44)
  \put(0,0)       {\includeourbeautifulpicture{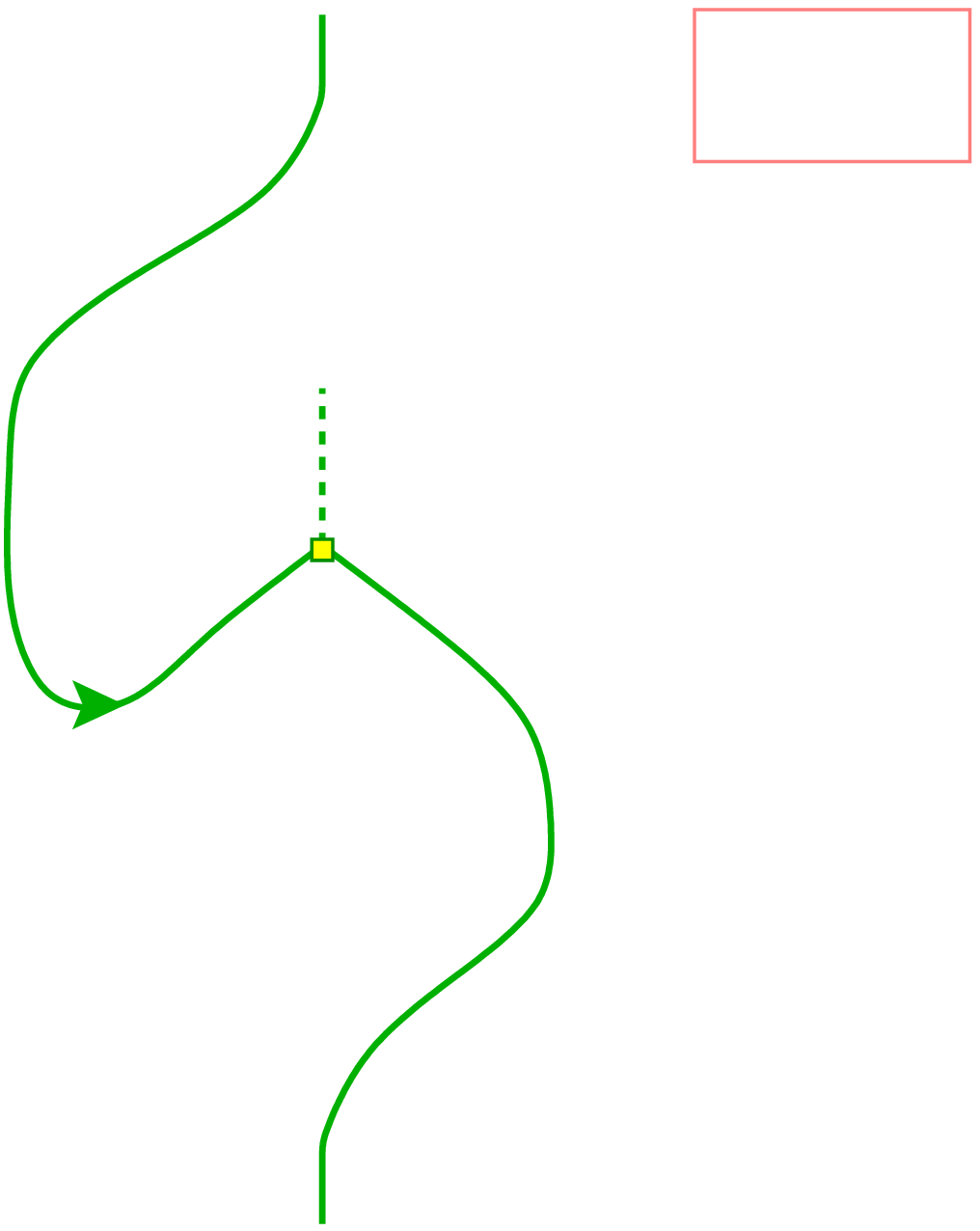}}
  \put(33,-8)     {\scriptsize$ U_\ibar $}
  \put(33,143)    {\scriptsize$ U_i^\vee $}
  \put(39,79)     {\scriptsize$ \Beta_{i\ibar} $}
  \put(89.5,127)  {$ B $}
  \epicture28 \labl{Fig38}

The one-point functions of bulk fields on a disk are then encoded in
$\complex$-linear maps
  \be  \tilde\Phi_M : \quad
  \HomAA(U_i \,{\otimes^+}\, A \,{\otimes^-}\, U_{\ibar}, A) \,\to\, V_{i\ibar}
  \ee
that send the morphism $\alph$ to the morphism in $V_{i\ibar}$ that
appears in figure \erf{Fig24}. 
Only once we have selected a basis $\Beta_{i\ibar}$ in $V_{i\ibar}$ we
can encode the same information in a linear form $\Phi_M$ on the space of
bulk fields,
  \be  \Phi_M \,\in\, \big( \HomAA(U_i\,{\otimes^+}\,A\,{\otimes^-}\,U_{\ibar},A)
  {\big)}^*_{} \,,  \ee
according to
  \be  \tilde\Phi_M (\alph) = \Phi_M(\alph) \, \Beta_{i\ibar} \,.  \ee
The linear form $\Phi_M$ can be thought of as providing the coefficients
of the boundary state for the boundary condition $M$, written in terms of a
fixed set of Ishibashi states. The values of this linear form are given by
the invariant of the following graph in $S^3$:
  \bea \begin{picture}(60,110)(8,58)
  \put(0,0)       {\includeourbeautifulpicture{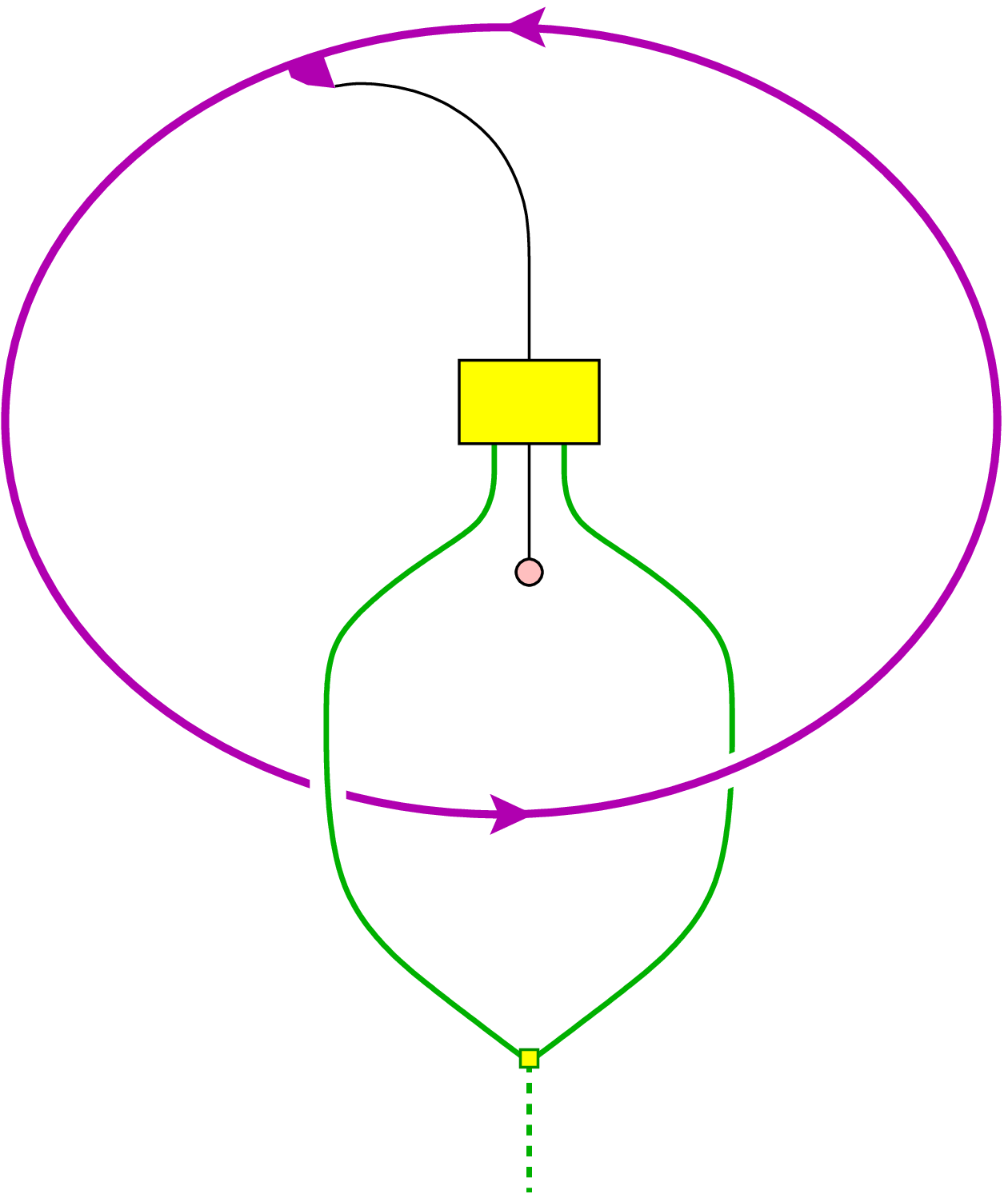}}
  \put(-73,86)    {$ \Phi_M (\alph) \;= $}
  \put(46,26)     {\scriptsize$ U_i $}
  \put(69.7,107.5){\scriptsize$ \alph $}
  \put(74,130)    {\scriptsize$ A $}
  \put(76,14)     {\scriptsize$ \Betabar_{i\ibar} $}
  \put(91,26)     {\scriptsize$ U_\ibar $}
  \put(128.7,74)  {\scriptsize$ \dot M $}
  \epicture36 \labl{Fig25}
Here the morphism $\Betabar_{i\ibar}$ that is dual to $\Beta_{i\ibar}$ appears.

\medskip

In the rest of this subsection, we will make these general expressions more
concrete for the particular case of Schellekens algebras. We start with a 
discussion of bulk fields that can have a non-vanishing one-point function 
on the disk, i.e.\ that have conjugate left and right chiral labels.

\dtl{Lemma}{bashi}
There is a bijection between bulk fields with conjugate left and right chiral 
labels $i$ and $\ibar$ and pairs $(U_i,g)$, with $U_i$ a simple object and 
$g\iN\calsA(U_i)$ such that
  \be  \XI_A(\,\cdot,g)\, \chi_{U_i}(\cdot)  \ee
is the trivial character on $H(A)$. 
\\
In terms of the morphisms we have already chosen (see appendix C), a basis 
of bulk fields is given by the morphisms 
  \bea \begin{picture}(160,110)(0,30)
  \put(70,0)       {\includeourbeautifulpicture{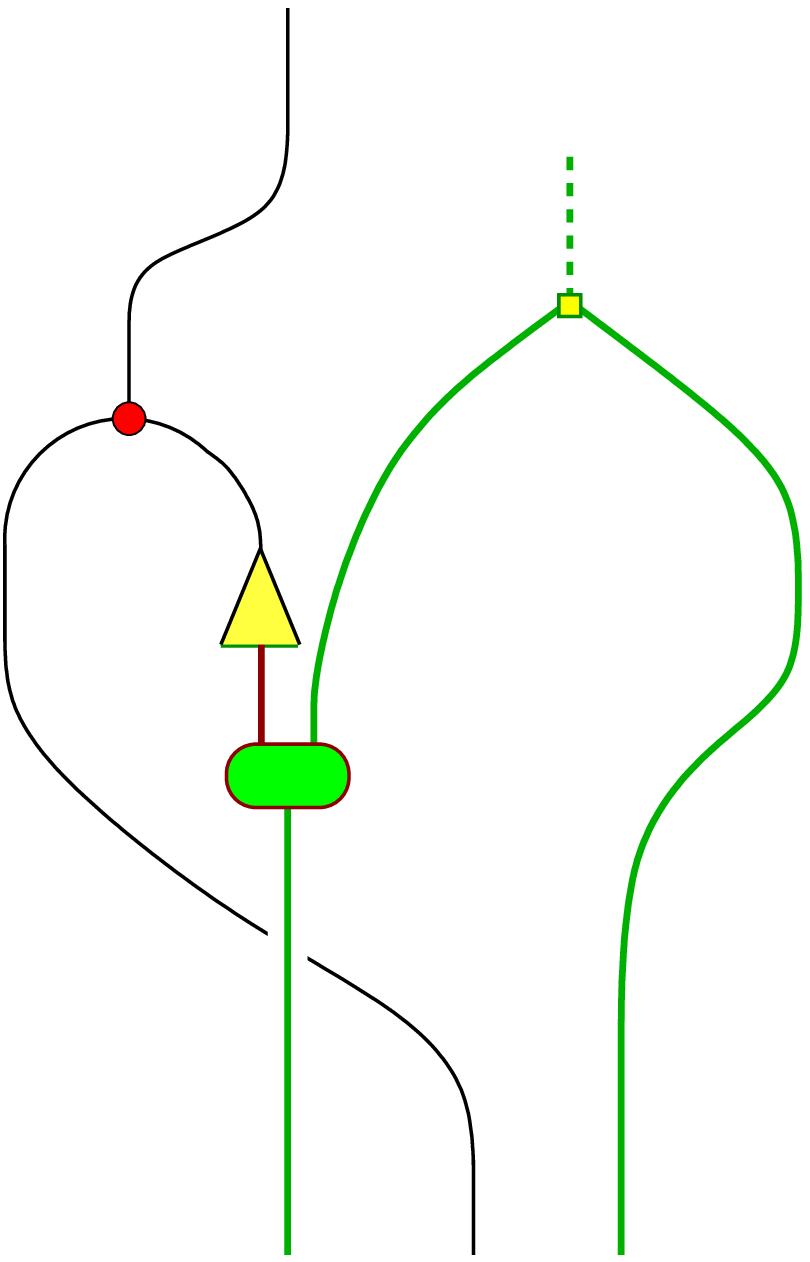} 
  \setlength{\unitlength}{24810sp}
    \put( 48,159)  {\scriptsize$ \U_g $}
    \put( 73,-20)  {\scriptsize$ U_i $}
    \put( 75,372)  {\scriptsize$ A $}
    \put(106,135)  {\scriptsize$ \overline{\bb_g(U_i)} $}
    \put(127,-20)  {\scriptsize$ A $}
    \put(171,-20)  {\scriptsize$ U_{\ibar} $}
  \setlength{\unitlength}{1pt}
  }
  \put(86.5,94)    {\scriptsize$ m $}
  \put(135.5,104.5){\scriptsize$ \Beta_{i\ibar} $}
  \put(0,63)       {$ \Psi_g (U_i) := $}
  \epicture06 \labl{Fig41}

\medskip\noindent
Proof: \\
The space of bimodule morphisms 
$\HomAA(U_i \,{\otimes^+} A \,{\otimes^-}\, U_{\ibar},A)$
is trivially a subspace of the space of left-module morphisms
$\HomA(U_i\,{\otimes^+}A\,{\otimes}\,U_{\ibar},A)$. By reciprocity, one easily
sees that the latter space is graded over the stabilizer $\calsA(U_i)$, with
one-dimensional homogeneous subspaces. A basis of each homogeneous subspace is
given by the morphisms \erf{Fig41}. We must determine which of these morphisms 
are even morphisms of bimodules. In other words, we are looking for the linear 
subspace consisting of linear combinations of the morphisms $\Psi_g(U_i)$ with 
coefficients $\xi_g\iN\complex$ that obey
  \be
  \sum_{g'\in \calsA(U_i)} \xi_{g'} \,
  \rho_r \cir (\Psi_{g'}(U_i) \oti \id_A) =
  \sum_{g'\in \calsA(U_i)} \xi_{g'} \,\Psi_{g'} (U_i) \cir \rho_r \,.  \ee
Composing both sides with $r_{gh}$ from the left and with
$\id_{U_i}\oti\eta\oti\id_{U_{\ibar}} \oti e_h$ from the right isolates the term 
$g'\eq g$ in the sum. The resulting condition must be valid for all $h\iN H(A)$, 
which shows that $\xi_g$ is allowed to be non-zero if and only if the equality
  \bea \begin{picture}(280,110)(0,43)
  \put(0,0)       {\includeourbeautifulpicture{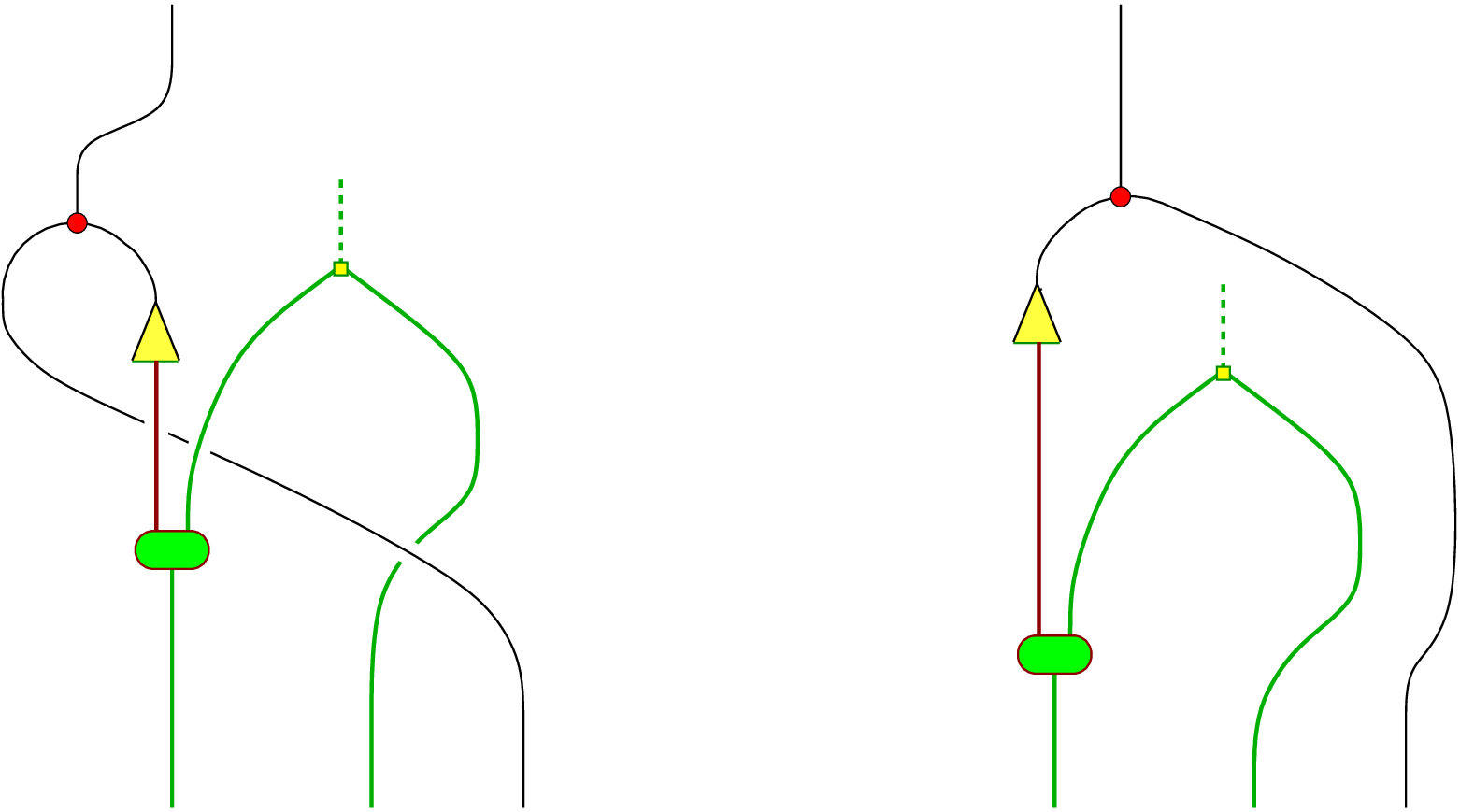}
  \setlength{\unitlength}{24810sp}
    \put( 75,398)  {\scriptsize$ A $}
    \put(244,-20)  {\scriptsize$ A $}
    \put( 73,-20)  {\scriptsize$ U_i $}
    \put(171,-20)  {\scriptsize$ U_{\ibar} $}
    \put( 45,159)  {\scriptsize$ \U_g $}
    \put(106,118)  {\scriptsize$ \overline{\bb_g(U_i)} $}
    \put(535,398)  {\scriptsize$ A $}
    \put(669,-20)  {\scriptsize$ A $}
    \put(499,-20)  {\scriptsize$ U_i $}
    \put(595,-20)  {\scriptsize$ U_{\ibar} $}
    \put(473,151)  {\scriptsize$ \U_g $}
    \put(533, 70)  {\scriptsize$ \overline{\bb_g(U_i)} $}
  \setlength{\unitlength}{1pt}
  }
  \put(132,72)     {$ = $}
  \epicture28 \labl{Fig42}
is satisfied. 
\\
To compute the graph on the left hand side of \erf{Fig42}, we insert the identity
$ \id_A \,{=} \sum_{h\in H(A)} p_h $ that is valid for \berta s 
together with the definition \erf{expmono} of the exponentiated monodromy charge
to obtain 
  \bea \begin{picture}(400,115)(12,40)
  \put(75,0)       {\includeourbeautifulpicture{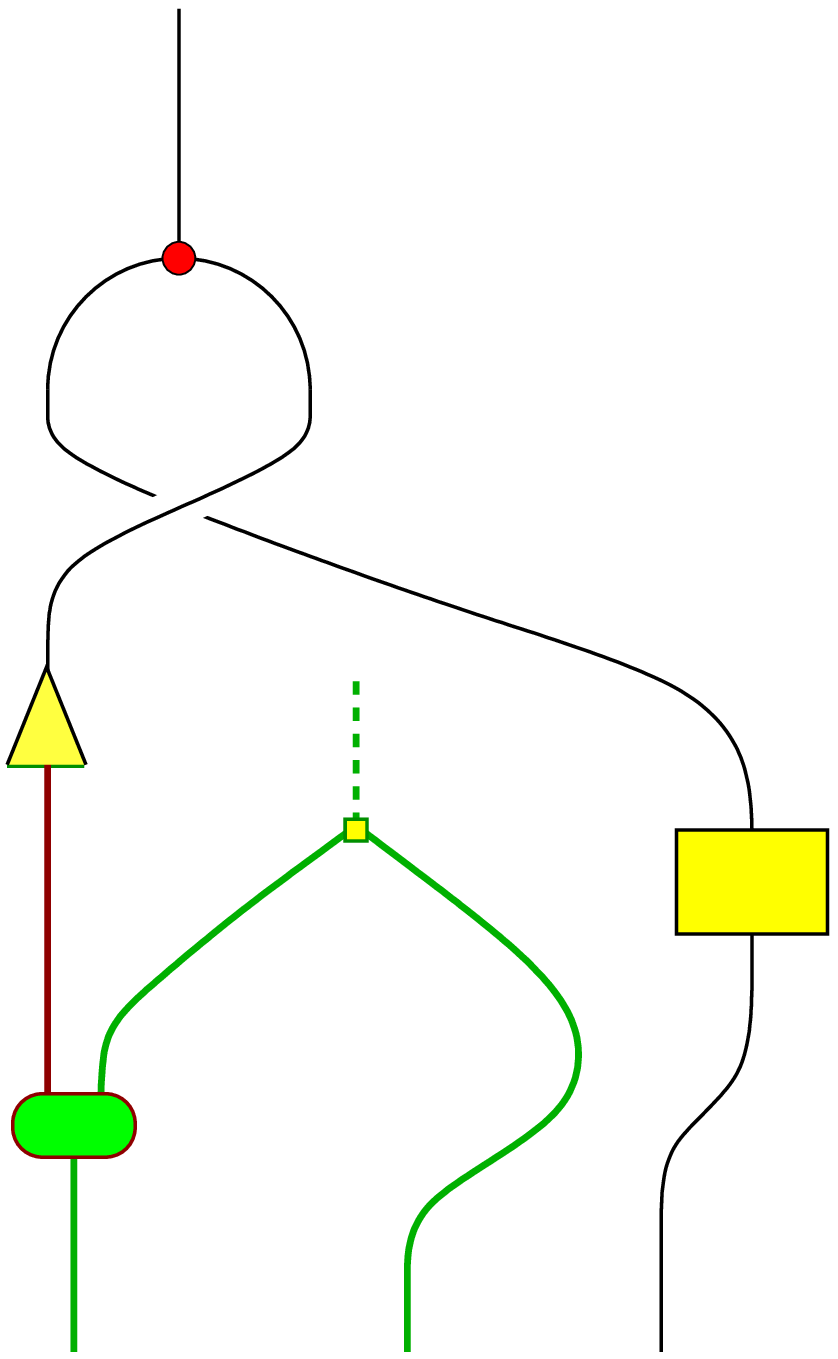}
  \setlength{\unitlength}{24810sp}
    \put( 42,399)  {\scriptsize$ A $}
    \put(180,-20)  {\scriptsize$ A $}
    \put( 12,-20)  {\scriptsize$ U_i $}
    \put(108,-20)  {\scriptsize$ U_{\ibar} $}
    \put(-17,139)  {\scriptsize$ \U_g $}
    \put( 45, 58)  {\scriptsize$ \overline{\bb_g(U_i)} $}
    \put(207,135)  {\scriptsize$ p_h $}
  \setlength{\unitlength}{1pt}
  }
  \put(9,66)      {$ \displaystyle\sum_{h\in H} \chi_{U_i}(h) $}
  \put(185,66)    {$ =\;\displaystyle\sum_{h\in H} \chi_{U_i}(h)\,\XI_A(h,g) $}
         \put(200,0){\begin{picture}(0,0)
  \put(115,0)       {\includeourbeautifulpicture{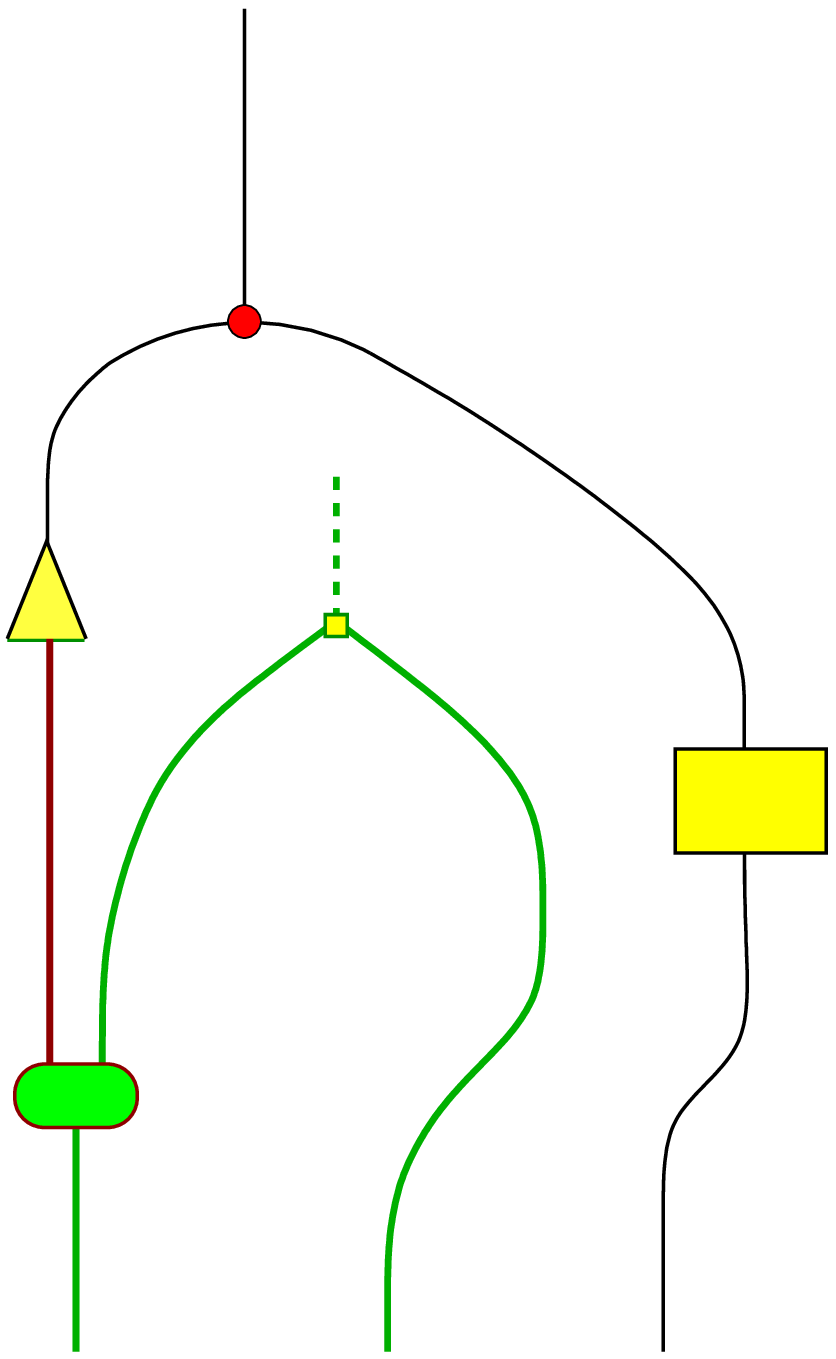}
  \setlength{\unitlength}{24810sp}
    \put( 62,399)  {\scriptsize$ A $}
    \put(180,-20)  {\scriptsize$ A $}
    \put( 12,-20)  {\scriptsize$ U_i $}
    \put(104,-20)  {\scriptsize$ U_{\ibar} $}
    \put(-17,139)  {\scriptsize$ \U_g $}
    \put( 45, 67)  {\scriptsize$ \overline{\bb_g(U_i)} $}
    \put(206,158)  {\scriptsize$ p_h $}
  \setlength{\unitlength}{1pt}
  }
       \end{picture} }
  \epicture23 \labl{Fig43+45}
Here in the second step we have inserted \erf{Fig44}.
Thus we have equality in \erf{Fig42} if and only if $\chi_{U_i}(\cdot)\, 
\XI_A(\,\cdot,g)$ is the trivial character on $H$.
(This is precisely the condition found in \cite{fhssw}.)
\qed 

\medskip

Next we evaluate the invariant of the graph \erf{Fig25} for $\Phi_M(\alpha)$
in the case when $A$ is a \berta\ and $M\eq \tM_{U_j,\psi}$ is the 
$A$-module that corresponds to the idempotent \erf{tildeP}. (Recall from 
\erf{eq:da-def} that if the central stabilizer is strictly smaller than the 
stabilizer, then this module is still reducible.) The graph for 
$\Phi_{\tM_{(U_j,\psi)}}(\alpha)$ looks as follows:
  \begin{eqnarray} \begin{picture}(400,154)(8,0)
  \put(0,0)        {\includeourbeautifulpicture{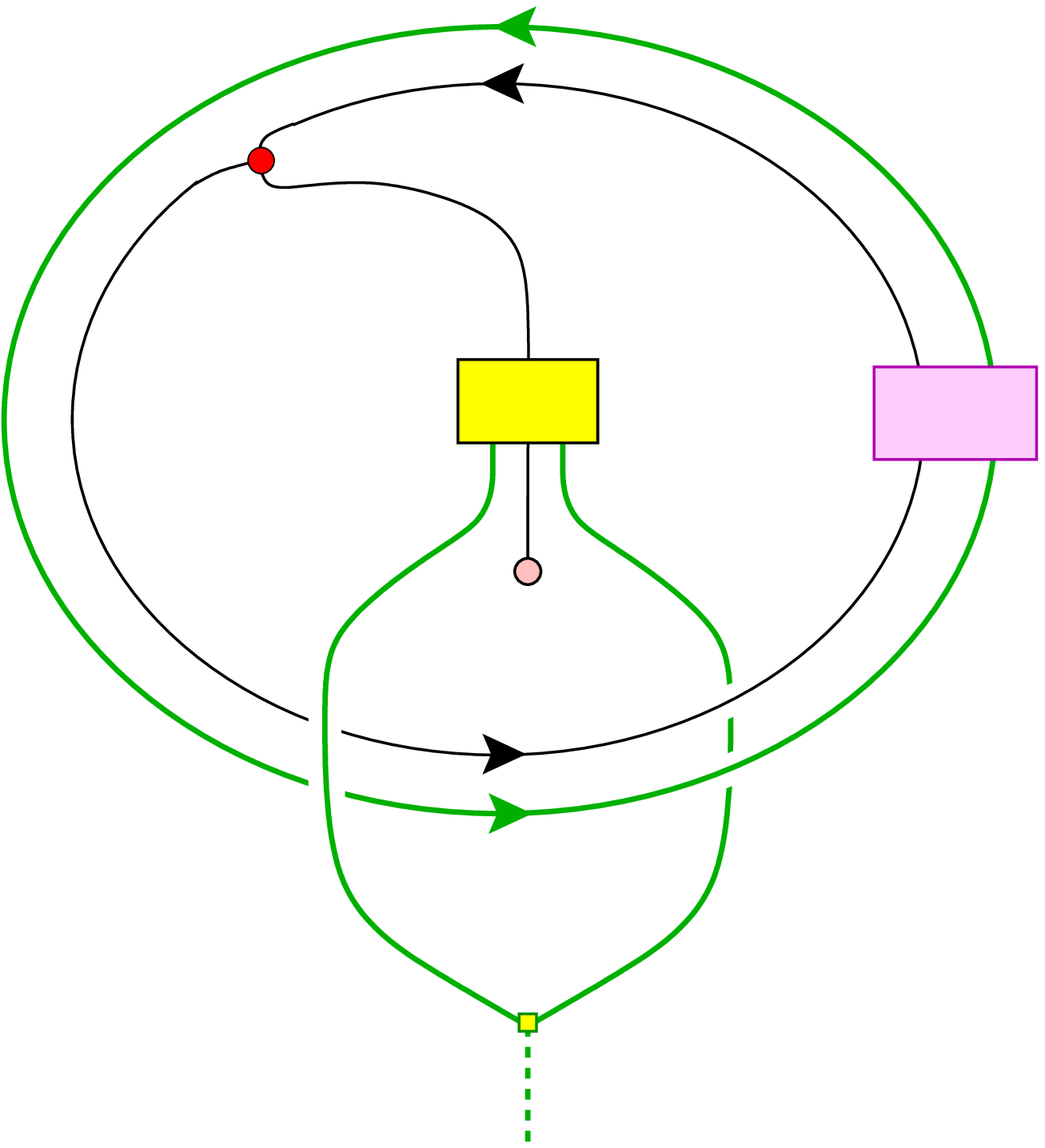}}
  \put(50,19)      {\scriptsize$ U_i $}
  \put(69.8,101.7) {\scriptsize$ \alph $}
  \put(38.8,135.6) {\scriptsize$ A $}
  \put(85.9,19)    {\scriptsize$ U_\ibar $}
  \put(129.7,68)   {\scriptsize$ U_j $}
  \put(120.7,100.3){\scriptsize$ \tP_{U_j,\psi} $}
  \put(158,93)     {$ =\! \displaystyle\sum_{g\in\calsA(U_j)}
                        \frac {\psi^*(g)} {|\calu_A(U_j)|} $}
  \put(310,19)     {\scriptsize$ U_i $}
  \put(329.5,101.7){\scriptsize$ \alph $}
  \put(288.4,129.2){\scriptsize$ A $}
  \put(345.9,19)   {\scriptsize$ U_\ibar $}
  \put(403.5,108)  {\scriptsize$ U_j $}
  \put(404.5,78.3) {\scriptsize$ \overline{\bb_g(U_j}) $} 
  \end{picture} \nonumber\\[-1.6em]{} \label{Fig26}
  \\[-1.8em]{}\nonumber\end{eqnarray}
Using the associativity of $A$, one can drag the product morphism coming from 
$\tP_{U_j,\psi}$ along the dashed line indicated on the \rhs. Afterwards one 
can use successively the bimodule morphism property of $\alpha$, the unit 
property of $\eta$, and then again the bimodule morphism property so as to 
arrive at the graph
  \bea \begin{picture}(400,112)(8,46)
  \put(0,0)        {\includeourbeautifulpicture{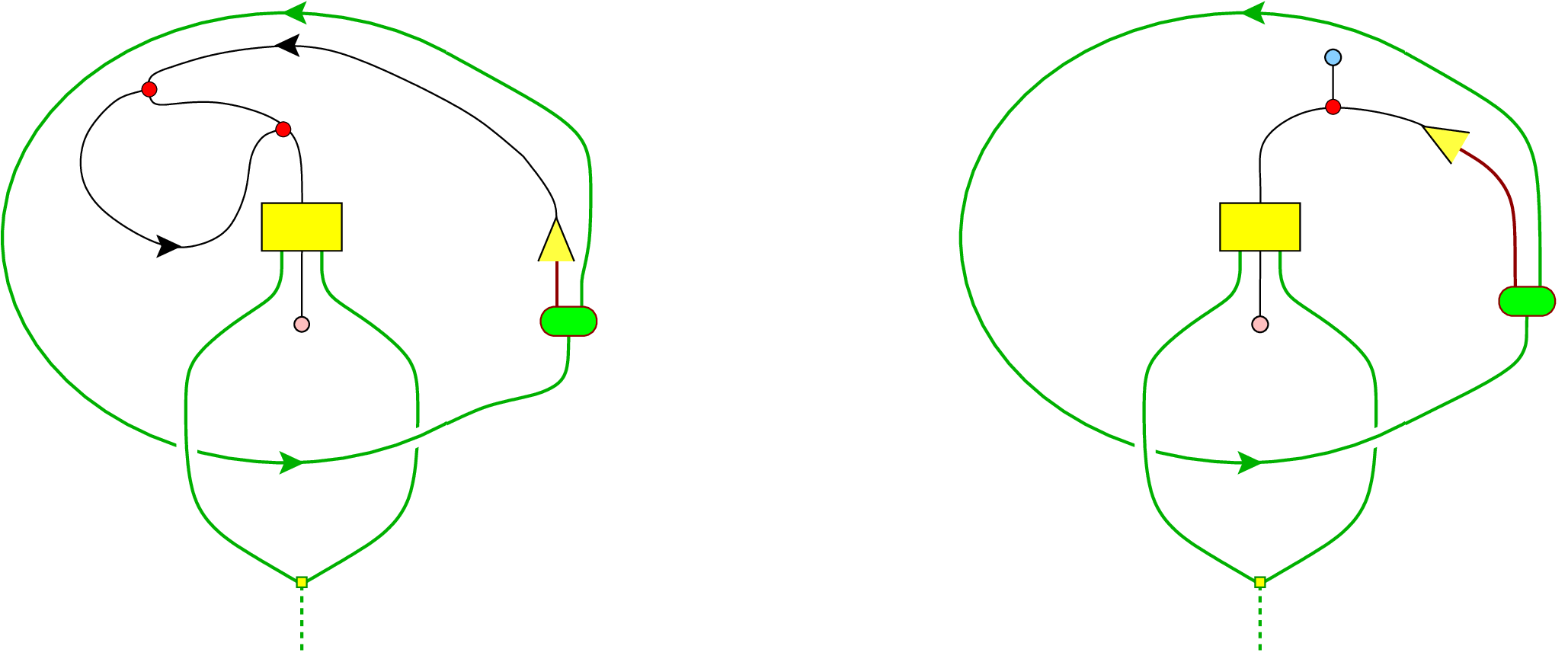}}
  \put(28.7,129.2) {\scriptsize$ A $}
  \put(51,19)      {\scriptsize$ U_i $}
  \put(70.2,101.7) {\scriptsize$ \alph $}
  \put(86.6,19)    {\scriptsize$ U_\ibar $}
  \put(144.1,108)  {\scriptsize$ U_j $}
  \put(144.8,78.3) {\scriptsize$ \overline{\bb_g(U_j}) $}
  \put(196,84)     {$ = $}
  \put(281,19)     {\scriptsize$ U_i $}
  \put(300.2,101.7){\scriptsize$ \alph $}
  \put(304.2,119.5){\scriptsize$ A $}
  \put(316.9,19)   {\scriptsize$ U_\ibar $}
  \put(353,95)     {\scriptsize$ \U_g $}
  \put(372.5,108)  {\scriptsize$ U_j $}
  \put(375.2,83.3) {\scriptsize$ \overline{\bb_g(U_j}) $} 
  \epicture21 \labl{Fig28}

To proceed, we must specify the morphism 
$\alph\iN\HomAA(U_i \,{\otimes^+} A \,{\otimes^-}\, U_{\ibar},A)$.
In view of lemma \ref{bashi} it is natural to take one of the basis morphisms 
\erf{Fig41}. Setting thus $\alpha\eq\Psi_h(U_i)$, with $\Psi_h(U_i)$ as in 
\erf{Fig41}, on the \rhs\ of \erf{Fig28}, we obtain
  \bea \begin{picture}(400,124)(0,35)
  \put(0,0)        {\includeourbeautifulpicture{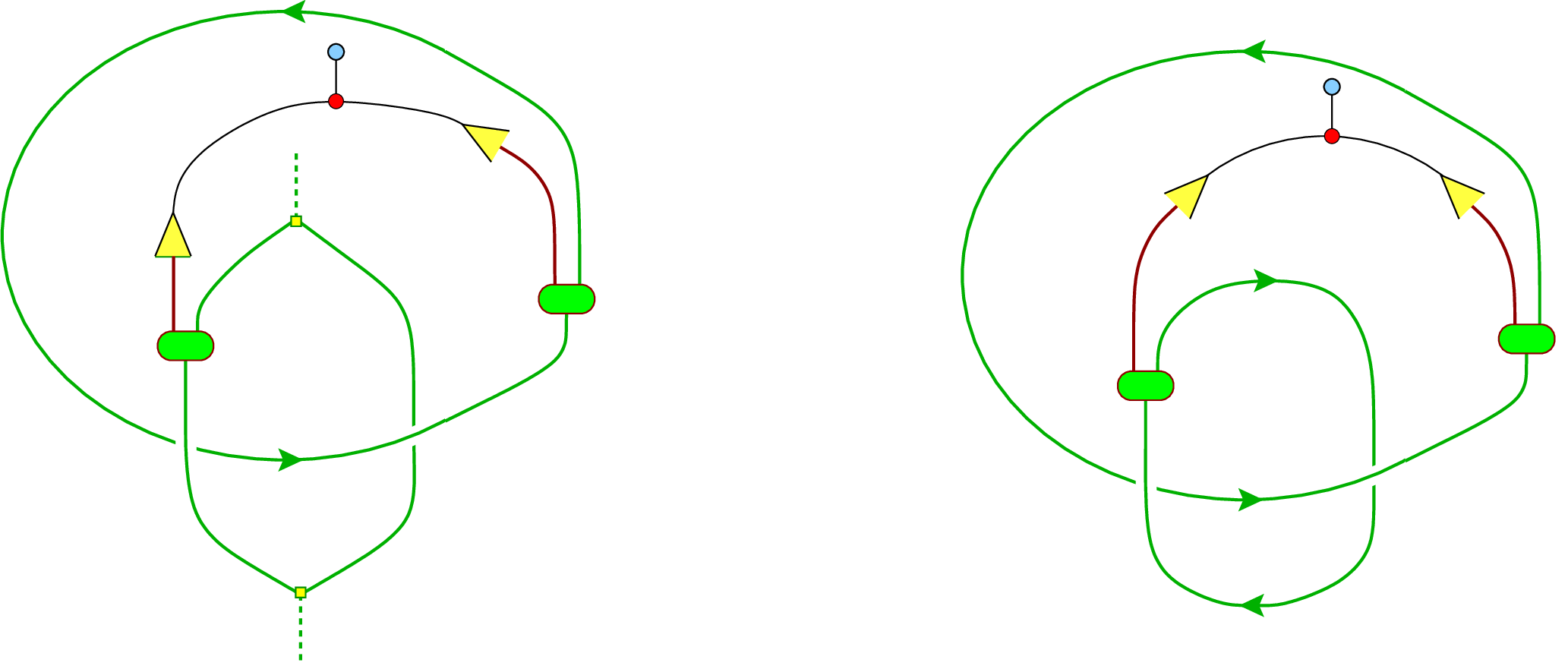}}
  \put(30,89)      {\scriptsize$ \U_h $}
  \put(51,19)      {\scriptsize$ U_i $}
  \put(52.8,75.3)  {\tiny$ \overline{\bb_{h\!}(U_i}) $}
  \put(58.7,125.5) {\scriptsize$ A $}
  \put(86.6,19)    {\scriptsize$ U_\ibar $}
  \put(107.5,87.3) {\tiny$ \overline{\bb_g(U_j}) $}
  \put(122,103)    {\scriptsize$ \U_g $}
  \put(141.1,108)  {\scriptsize$ U_j $} 
  \put(160,82)     {$ =\; \displaystyle\frac 1{\dim(U_i)} $}
  \put(262,89)     {\scriptsize$ \U_h $}
  \put(267,27)     {\scriptsize$ U_i $}
  \put(284.4,66.3) {\tiny$ \overline{\bb_{h\!}(U_i}) $}
  \put(303,121.5)  {\scriptsize$ A $}
  \put(338.2,78.3) {\tiny$ \overline{\bb_g(U_j}) $}
  \put(352.4,95)   {\scriptsize$ \U_g $}
  \put(373.3,101)  {\scriptsize$ U_j $}
  \epicture09 \labl{Fig46} 

This graph is non-zero only for $h\eq g^{-1}$ so that the summation
in \erf{Fig26} can be carried out explicitly. Pulling now the morphism
$\overline{\bb_g(U_j)}$ along the $U_j$-ribbon and 
slightly deforming the ribbons, using also sovereignty on the $U_i$-ribbon,
the graph becomes
  \bea \begin{picture}(400,144)(5,40)
  \put(0,0)        {\includeourbeautifulpicture{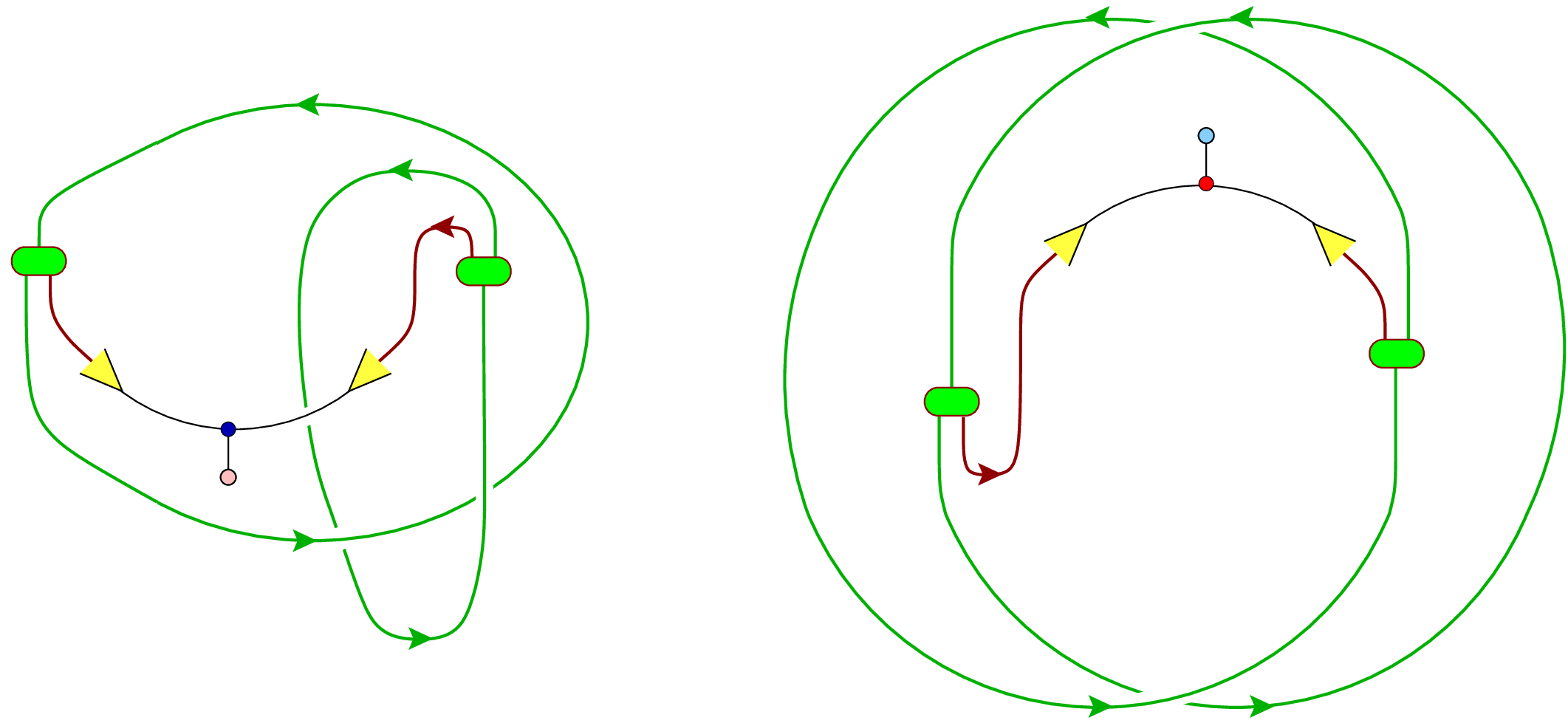}
  \setlength{\unitlength}{24810sp}
    \put(116,205)  {\scriptsize$ A $}
    \put(337,365)  {\scriptsize$ U_j $}
    \put(321,218)  {\scriptsize$ U_i $}
    \put( 46,258)  {\scriptsize$ \U_g $}
    \put(242,307)  {\scriptsize$ \U_h $}
    \put(322,262)  {\scriptsize$ \overline{\bb_h(U_i\!}) $}
    \put( 49,294)  {\scriptsize$ \overline{\bb_g(U_j)} $}
    \put(435,245)  {$ = $}
    \put(721,349)  {\scriptsize$ A $}
    \put(982, 82)  {\scriptsize$ U_j $}
    \put(868,117)  {\scriptsize$ U_i $}
    \put(670,229)  {\scriptsize$ \U_g $}
    \put(875,265)  {\scriptsize$ \U_h $}
    \put(935,233)  {\scriptsize$ \overline{\bb_h(U_i\!}) $}
    \put(537,205)  {\scriptsize$ \overline{\bb_g(U_j)} $}
  \setlength{\unitlength}{1pt}
  }
  \epicture23 \labl{Fig47}
The equality between the left and right hand sides uses that $A$ is symmetric.
Comparison with the bilinear pairing \erf{bilpair} shows that the \rhs\ is 
nothing but 
  \be  \Sg_{i,\bar\jmath}(\irbeta\Oti\iralpha) =: s(A)_{i,\bar\jmath}^g  
  \labl{def-sg}
with $\irbeta\iN\Hom(\U_g\Oti\, U_i,U_i)$ 
and $\iralpha\iN\Hom(U_{\bar\jmath},U_{\bar\jmath} \Oti\, \U_g)$ given by 
  \bea \begin{picture}(300,103)(5,29)
  \put(0,0)        {\includeourbeautifulpicture{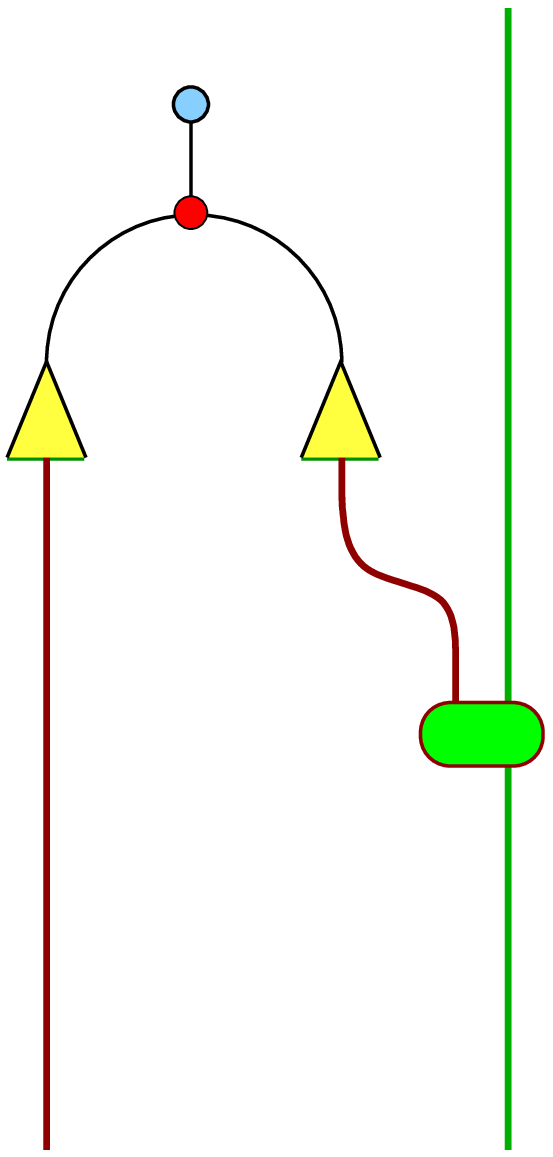}
  \setlength{\unitlength}{24810sp}
    \put(  1,-20)  {\scriptsize$ \U_g $}
    \put( 70,155)  {\scriptsize$ \U_{g^{-1}} $}
    \put(139,-20)  {\scriptsize$ U_i $}
    \put(139,342)  {\scriptsize$ U_i $}
    \put( 73,274)  {\scriptsize$ A $}
    \put(164,117)  {\scriptsize$ \overline{\bb_g(U_i)} $}
    \put(-110,140)  {$ \irbeta\;= $}
  \setlength{\unitlength}{1pt}
  }
  \put(0,0){ \put(200,0){  {\includeourbeautifulpicture{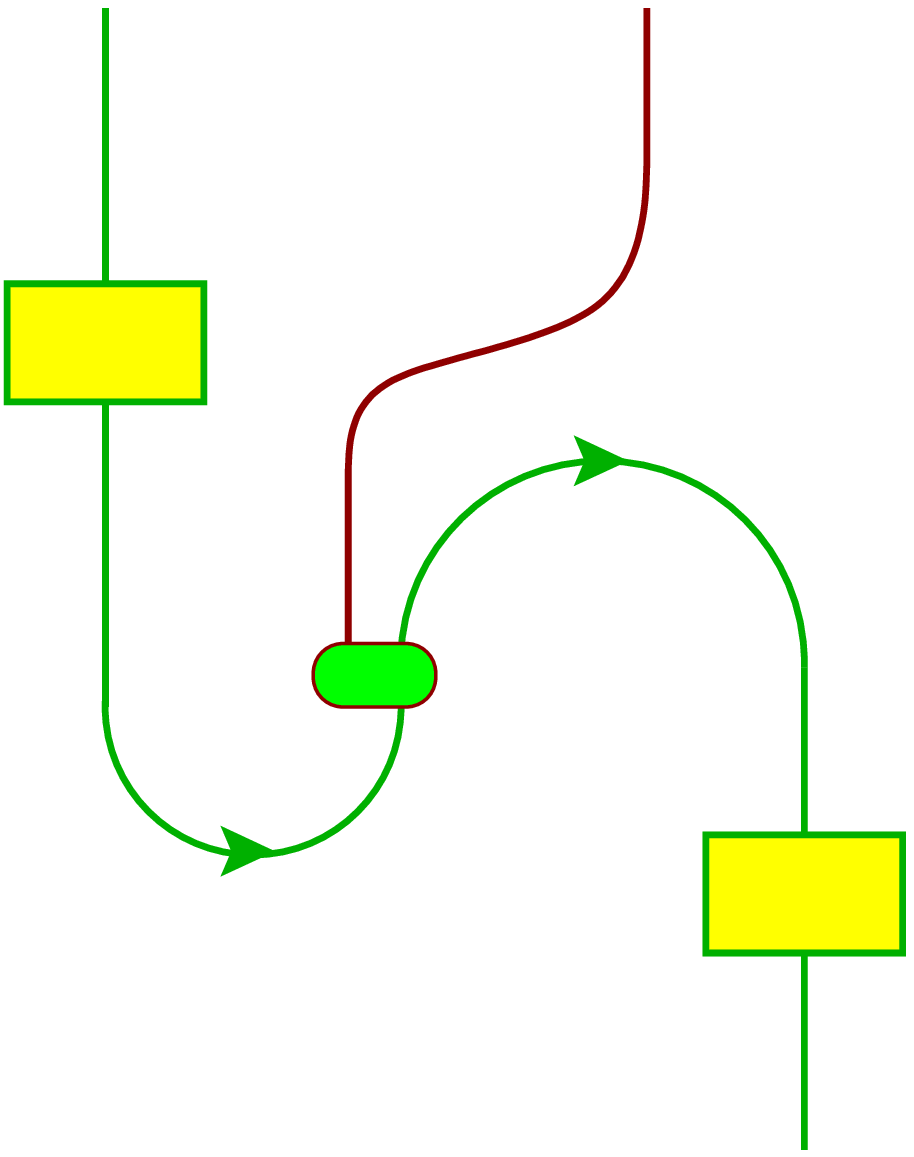} 
  \setlength{\unitlength}{24810sp}
    \put(177,342)  {\scriptsize$ \U_g $}
    \put( 22,342)  {\scriptsize$ U_{\bar\jmath} $}
    \put(225,-20)  {\scriptsize$ U_{\bar\jmath} $}
    \put(110, 88)  {\scriptsize$ U_j $}
    \put(139,169)  {\scriptsize$ U_j $}
    \put( 12,230)  {\scriptsize$ \pi_{\bar\jmath}^{-1} $}
    \put(220, 72)  {\scriptsize$ \pi_{\bar\jmath} $}
    \put(131,131)  {\scriptsize$ \overline{\bb_g(U_j)} $}
    \put(-215,140) {and}
    \put(-106,140)  {$ \iralpha\;= $}
  \setlength{\unitlength}{1pt}
  }}}
  \epicture15 \labl{eq:al-be}

\dtl{Remark}{rem:sg-props}
(i)~\,\,The scalar matrix $s(A)_{i,\bar\jmath}^g$ defined in \erf{def-sg} 
depends both on the algebra $A$ and on a choice of $A$-straight bases. 
However, choosing different bases changes the value of $s(A)_{i,\bar\jmath}^g$ 
at most up to characters of the central stabilizers. 
\\[.2em]
(ii)~\,In the following we list some properties of the matrices $s(A)^g$,
without a detailed proof. First, by setting $g\eq e$ in \erf{Fig47} we
get immediately
  \be  s(A)^e_{i,j} = \dim(A) \, s_{i,j} \,,  \ee
where $s$ is $S_{0,0}^{-1}$ times the ordinary unitary
modular S-matrix. 
\\
Next, let $\tilde s(A)^g_{i,j}$ be defined as $s(A)^g_{i,j}$, but with the 
inverse braiding instead of the braiding.  Then we have
  \be
  \sum_{k\in F(g)} s(A)^g_{i,k}\, \tilde s(A)^g_{\bar k,\bar\jmath} = 
  \Frac{\dim(A)^2}{(S_{00}^{}{)}^2_{}} \, \delta_{i,j} \,,
  \labl{eq:sg-prop-b}
the sum being over the set
$F(g)$ of all $k\iN\cali$ such that $g\iN\cals(U_k)$, compare 
part (i) of the proof of proposition \ref{Sfix}. To show the validity
of \erf{eq:sg-prop-b}, one uses that $A$-straight bases have been chosen.
\\
We also have
  \be  s(A)^g_{i,j} = s(A)^{g^{-1}}_{\bar\jmath,\bar\imath} . \ee      
This is seen by turning the graph \erf{Fig47}
upside down and using that the algebra $A$ is symmetric.
\\
Finally, if there exists a basis of the coupling spaces in which all 
\FF-matrix entries are real (compare remark \ref{rem-twining}(iii)), then 
we also have
  \be
  \big( s(A)^g_{i,j} \big)^{\!*} = \tilde s(A)^g_{i,j} \,.  \ee

   \vskip .2em 
\noindent
(iii)~The matrices $s(A)^g$ should be compared to the corresponding
quantities $S^J$ in \cite{fusS6}. The two quantities can be identified via
$S^J_{i,j} \eq \big({S_{00}}/{\dim(A)}\big)\,\big(s(A)^g_{i,\bar\jmath}\big)^*$, 
where $J\eq g$.

\medskip

Collecting the information in the relations \erf{Fig26}\,--\,\erf{def-sg},
we arrive at the following expression for
the one-point function of the bulk field $\Psi_h(U_i)$ on a disk
with boundary condition $\tM_{(U_j,\psi)}$:
  \be
  \Phi_{\tM_{(U_j,\psi)}\!}(\Psi_h(U_i)) = \frac{1}{\dim(U_i)} \,
  \frac{\psi^*(h^{-1})}{|\calu_A(U_j)|} \, s(A)^{h^{-1}}_{i,\bar\jmath} \,.  \ee
Now recall from \erf{eq:da-def} that the $A$-module $\tM_{(U_j,\psi)}$ is 
the direct sum of $d_A(U_j)$ isomorphic simple modules. Accordingly, the 
one-point function on a disk with boundary condition labeled by the simple 
$A$-module $M_{(n,\psi)}$ is given by
  \be
  \Phi_{M_{(n,\psi)}\!}(\Psi_{h^{-1}\!}(U_i))
  = \frac{1}{\dim(U_i)} \, \frac{\psi^*(h)}{
  \sqrt{|\calsA(U_{m_n})|\,|\calu_A(U_{m_n})|}}\, s(A)^h_{i,\overline{m_n}} \,.
  \labl{therealthing}
This coincides with formula (11) of \cite{fhssw}, except for an overall factor.
This factor comes from the normalisation of the bulk fields and of the
matrix $s(A)^g$. For the latter, compare formula \erf{eq:sg-prop-b};
also note that the normalisation of bulk fields differs from the one 
used in \cite{fhssw} already when $A\eq\bfe$ \cite{fffs2}.
That the prefactor in \erf{therealthing} precisely stems from these
normalisations can e.g.\ be seen from the fact that both the formula
given in \cite{fhssw} and our result \erf{therealthing} lead to
annulus coefficients that are non-negative integers.


\sect{Bimodules and defects}\label{Bimodules}

$A$-bimodules, which in the application to CFT correspond to defect lines
\citei\ can be treated in a manner that is quite analogous to the description 
of $A$-modules in section \ref{bertarep}.  
Again the basic ingredient is a notion of induction.
As it turns out, it is convenient not to use $\alpha$-in\-duc\-tion. Rather,
we proceed as follows. Consider the object $A\oti U\oti A$; the 
multiplication in $A$ endows it with a natural structure of an 
$A$-bimodule that we call the {\em induced bimodule\/} $\IndAA(U)$:
  \be  \IndAA(U) = (A\oti U\oti A, m\oti\id_U\oti\id_A, \id_A\oti\id_U\oti m)
  \,.  \ee
Obviously, such bimodules can be studied for any \ssfa\ $A$, not only
for Schellekens algebras. Moreover, there is no need to take the same
\ssfa\ for the left action and the right action. This leads us to the
\\[-2.3em]

\dtl{Definition}{inducedbimodules}
Let $A_1$ and $A_2$ be two \ssfa s in a tensor category $\calc$ satisfying
the assumptions in convention \ref{categories}. The {\em induced\/} 
$A_1$-$A_2$-{\em bimodule\/} associated to an object $U$ of $\calc$ is 
the bimodule
  \be  \IndAAv(U) := (A_1\oti U\oti A_2, m_1\oti\id_U\oti\id_{A_2}, 
\id_{A_1}\oti\id_U\oti m_2) \,.  \ee

The following statement is immediate:
\\[-2.3em]

\dtl{Lemma}{getall}
Every $A_1$-$A_2$-bimodule $B$ over special Frobenius algebras $A_1$ and 
$A_2$ is a bimodule retract of some induced $A_1$-$A_2$-bimodule.

\medskip\noindent
Proof: \\
Indeed, $B$ is the bimodule retract of $\IndAAv(\dot B)$ that is defined by the
embedding and retraction morphisms
  \bea \begin{picture}(400,85)(0,25)
  \put(92,0)     {\includeourbeautifulpicture{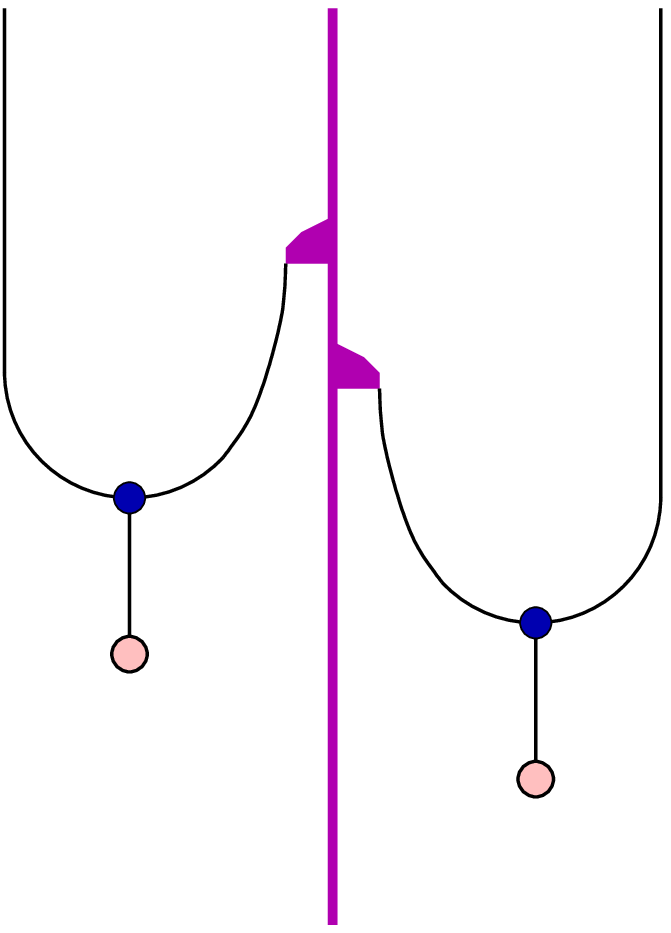}}
  \put(-4,54)    {$ e^{}_{B\prec\IndAAv(\dot B)} \;:= $}
  \put(89,107)   {\scriptsize$ A_1 $}
  \put(124.5,-9.8){\scriptsize$ \dot B $}
  \put(125,107)  {\scriptsize$ \dot B $}
  \put(160,107)  {\scriptsize$ A_2 $}
  \put(195,54)   {and $\quad\ r^{}_{\IndAAv(\dot B)\succ B} \;:= $}
  \put(324.3,-8) {\scriptsize$ A_1 $}
  \put(356.9,-9.8){\scriptsize$ \dot B $}
  \put(357.5,107){\scriptsize$ \dot B $}
  \put(389,-8)   {\scriptsize$ A_2 $}
  \epicture13 \labl{Fig30}
By the Frobenius and associativity properties of $A_1$ and $A_2$,
these are morphisms of bimodules, and by specialness we have 
$r_{\IndAAv(\dot B)\succ B} \cir e_{B\prec\IndAAv(\dot B)} \eq \id_B$.

\medskip

The central task in developing the bimodule theory of \berta s is thus 
to decompose induced bimodules. To this end,
we must describe bimodule morphisms between induced bimodules.
A `double-sided reciprocity' turns out to be a convenient tool:
\\[-2.3em]

\dtl{Lemma}{doublrec}
For any two $A_1$ and $A_2$ and any two objects $U$ and $V$ of $\calc$ 
there are canonical isomorphisms
  \be  f_{U,V}:\quad \HomAAv(\IndAAv(U),\IndAAv(V)) \,\stackrel\cong\to\, 
  \Hom(U,A_1\oti V\oti A_2)  \ee
of complex vector spaces. They are given by
  \bea \begin{picture}(380,69)(0,30)
  \put(0,0)     {\includeourbeautifulpicture{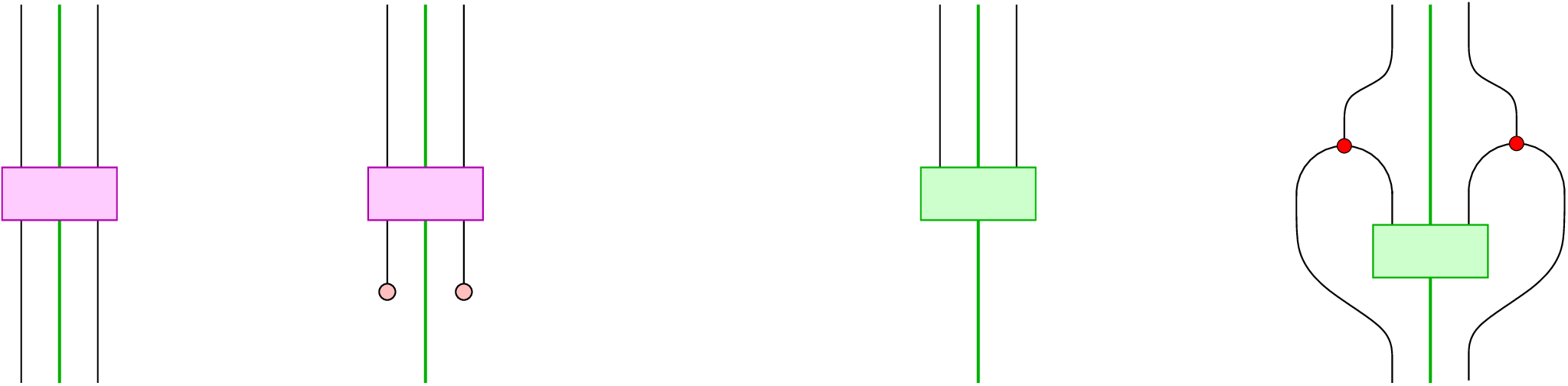}}
  \put(-1.8,-8)  {\scriptsize$ A_1 $}
  \put(-1,95)    {\scriptsize$ A_1 $}
  \put(11,-8)    {\scriptsize$ U $}
  \put(11.4,95)  {\scriptsize$ V $}
  \put(11.7,44)  {\scriptsize$ \alpha $}
  \put(20,-8)    {\scriptsize$ A_2 $}
  \put(20,94)    {\scriptsize$ A_2 $}
  \put(50,43)    {$ \stackrel{f_{U,V}^{}}\longmapsto $}
  \put(87,95)    {\scriptsize$ A_1 $}
  \put(98,-8)    {\scriptsize$ U $} 
  \put(99,95)    {\scriptsize$ V $}
  \put(99.7,44)  {\scriptsize$ \alpha $}
  \put(108,95)   {\scriptsize$ A_2 $}
  \put(159,43)   {and}
  \put(219,95)   {\scriptsize$ A_1 $}
  \put(231,-8)   {\scriptsize$ U $}
  \put(231,95)   {\scriptsize$ V $}
  \put(231.7,44) {\scriptsize$ \kapa $}
  \put(240,95)   {\scriptsize$ A_2 $}
  \put(271,43)   {$ \stackrel{f_{U,V}^{-1}}\longmapsto $}
  \put(326,-8)   {\scriptsize$ A_1 $}
  \put(326,95)   {\scriptsize$ A_1 $}
  \put(338.3,-8) {\scriptsize$ U $}
  \put(338.8,95) {\scriptsize$ V $}
  \put(339.4,30.6){\scriptsize$ \kapa $}
  \put(348,-8)   {\scriptsize$ A_2 $}
  \put(348,95)   {\scriptsize$ A_2 $}
  \epicture15 \labl{Fig31}
for $\alpha\iN\HomAAv(\IndAAv(U),\IndAAv(V))$ and 
$\kapa\iN\Hom(U,A_1\oti V\oti A_2)$.

\medskip\noindent
Proof:\\
That the two mappings are inverse to each other follows immediately by
the associativity and unit properties.
\qed
 
Note that the statement holds for any pair $A_1,A_2$ of unital associative 
algebras; the Frobenius property is not needed.

The composition of bimodule morphisms induces a concatenation on the
morphism spaces $\Hom(U,A_1\oti V\oti A_2)$:
  \bea \begin{picture}(180,95)(0,26)
  \put(74,0)     {\includeourbeautifulpicture{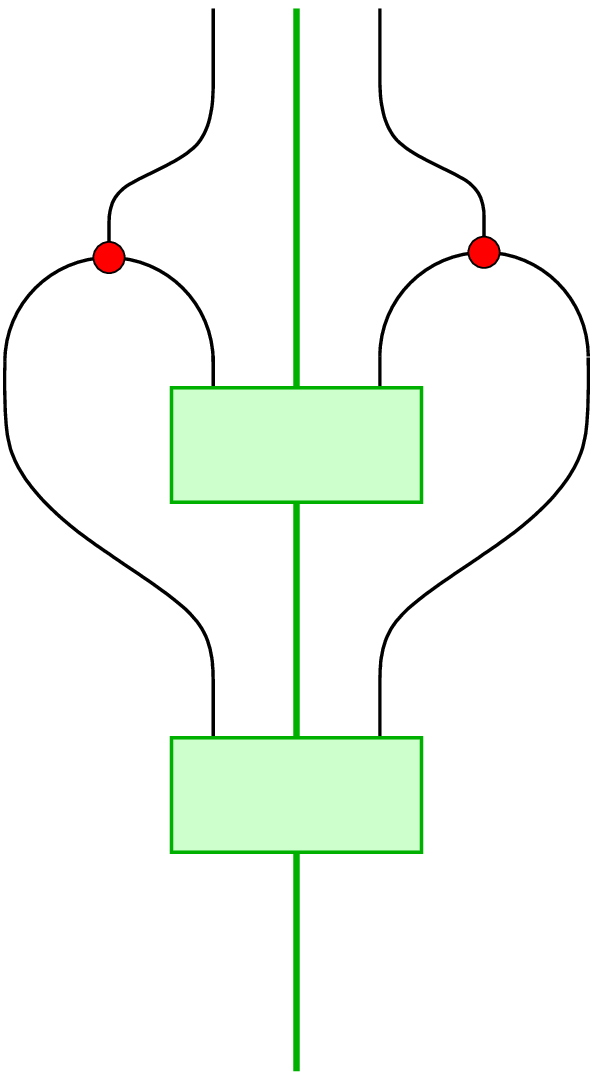}}
  \put(0,57)     {$ \kapa_1 \star \kapa_2 \;:= $}
  \put(82,82.7)  {\scriptsize$ m_1^{} $}
  \put(91,121)   {\scriptsize$ A_1 $}
  \put(103,29)   {\scriptsize$ \kapa_2 $}
  \put(103,68)   {\scriptsize$ \kapa_1 $}
  \put(103,-8)   {\scriptsize$ U $}
  \put(103,121)  {\scriptsize$ W $}
  \put(107.6,52) {\scriptsize$ V $}
  \put(113,121)  {\scriptsize$ A_2 $}
  \put(123,82.7) {\scriptsize$ m_2^{} $}
  \epicture10 \labl{Fig32}

We now consider more specifically the situation that $A_1$ and $A_2$ are
\berta s, $A_i\,{\cong}\,\bigoplus_{h_i\in H_i}\!\U_{h_i}$, with two not
necessarily equal subgroups $H_1$ and $H_2$ in the effective center
$\Pico(\calc)$. In this case, the morphism space $\Hom(U,A_1\oti V\oti A_2)$ is 
non-zero only if $U$ and $V$ are on the same orbit of the subgroup 
$H \,{\le}\, \Pic(\calc)$ that is generated by $H_1$ and $H_2$. Moreover, for 
$U\eq V$, this morphism space is naturally isomorphic to the endomorphism space
  \be  \bearll  \EndAAv(\IndAAv(U)) \!\!& \cong\, \Hom(U,A_1\oti U\oti A_2)
  \\{}\\[-.8em]& \cong\! \displaystyle
  \bigoplus_{h_1\in H_1,\,h_2\in H_2}\!\! \Hom(U,\U_{h_1}\oti U\oti \U_{h_2})
  \,.  \eear \ee
It is thus a graded vector space over
  \be  \bicals_{\!A_1|A_2}(U) := \{ 
  (h_1,h_2)\iN H_1{\times H_2} \,|\, h_1h_2\iN\cals(U) \} \,,  \ee
where the stabilizer $\cals(U)$ of $U$ is defined with respect to the whole 
effective center $\Pico(\calc)$.
This subset of $H_1\Times H_2$ is a subgroup (since $H$ is abelian); it will be
called the {\em bi-stabilizer\/} of $U$ with respect to $A_1$ and $A_2$. 
Clearly we have
  \be  
  \cals_{\!A_1}(U) \Times \cals_{\!A_2}(U) \le
  \bicals_{\!A_1|A_2}(U) \le H_1\Times H_2 \,.  \ee
Next we analyze the structure of the concatenation product introduced in
\erf{Fig32} on the space $\EndAAv(\IndAAv(U))$. To this end we define, for 
$(g_1,g_2) \iN \bicals_{\!A_1|A_2}(U)$, the morphisms 
  \bea \begin{picture}(170,115)(0,32)
  \put(81,0)     {\includeourbeautifulpicture{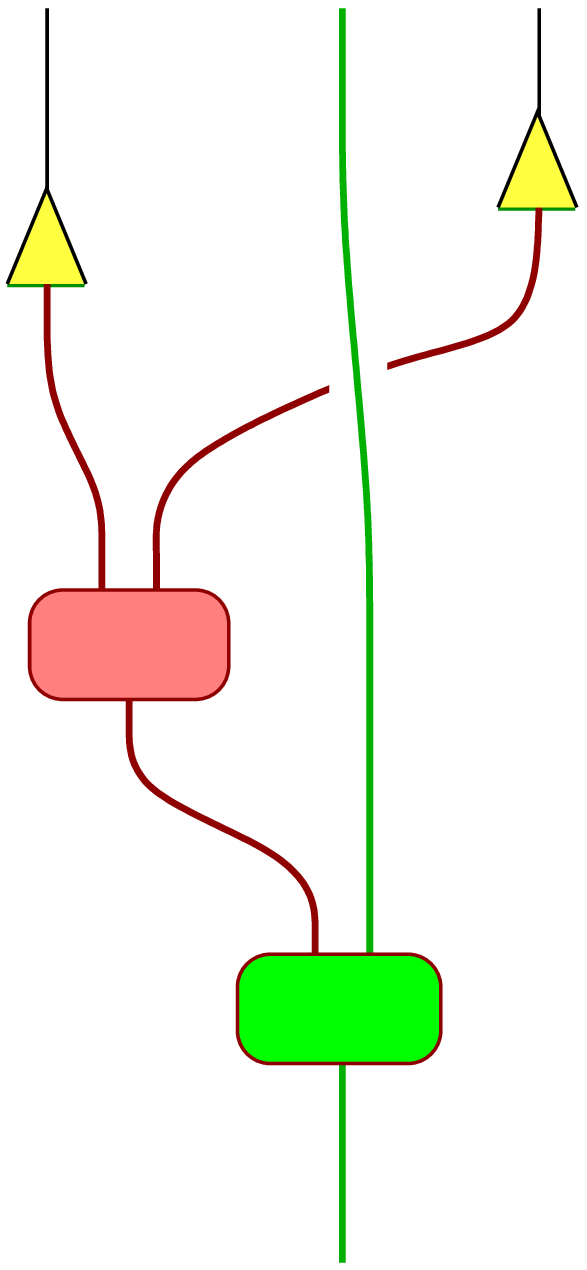}}
  \put(0,67)     {$ \biphi{g_1}{g_2}(U) \;:= $}
  \put(78,83)    {\scriptsize$ \U_{g_1} $}
  \put(90,43)    {\scriptsize$ \U_{g_1g_2} $}
  \put(86.2,66.3){\scriptsize$ \overline{\bL_{g_1}{g_2}} $}
  \put(100,80)   {\scriptsize$ \U_{g_2} $}
  \put(109.2,26.3){\scriptsize$ \overline{\bb_{g_1g_2}} $}
  \put(81,143)   {\scriptsize$ A_1 $}
  \put(115.3,143){\scriptsize$ U $}
  \put(115,-8)   {\scriptsize$ U $}
  \put(135,143)  {\scriptsize$ A_2 $}
  \epicture16 \labl{Fig33} 
where we have suppressed the argument $U$ in the morphism 
$\overline{\bb_{g_1g_2}(U})\iN\Hom(U,\U_{g_1g_2}\Oti U)$ (as introduced in 
definition \ref{stabilizer}(iii)). These form a basis of 
$\Hom(U,A_1\oti U\oti A_2)$.

\dtl{Proposition}{uustab}
Let $U\iN {\rm Obj}(\calc)$ be simple and $A_1$, $A_2$ be two Schellekens 
algebras in $\calc$.
The $\complex$-algebra $\Hom(U,A_1\oti U\oti A_2)$ is a twisted group
algebra over the group $\bicals_{\!A_1|A_2}(U)$. The cohomology class of the
corresponding two-cocycle is described by the alternating bihomomorphism
  \be  \eps_{U,A_1,A_2} (g_1,g_2,h_1,h_2) = \phi_U(g_1g_2,h_1h_2)\,
  \beta(h_1,g_2) \, \XI_{A_1}(h_1,g_1) \, \XI_{A_2}(g_2,h_2)
  \,.  \labl{prop-epsUA1A2}

\noindent
Proof: \\
We first notice that the algebra is also graded 
over the group $\bicals_{\!A_1|A_2}(U)$ as an algebra, i.e.\ we have
  \be  \biphi{g_1}{g_2}(U) \star \biphi{h_1}{h_2}(U)
  = \beta_{U,A_1,A_2}(g_1,g_2,h_1,h_2)\, \biphi{g_1 h_1}{g_2 h_2}(U)
  \labl{eq:biphi-mult}
for some two-cocyle $\beta_{U,A_1,A_2}$ on $\bicals_{\!A_1|A_2}(U)$.
This follows from the fact that the morphism space 
$\Hom(U,\U_{g_1h_1}\Oti\U_{g_2h_2}\Oti U)$ is one-dimen\-si\-onal.
\\[.2em]
Again, we determine the isomorphy type of the algebra by computing the
commutator two-cocyle $\eps_{U,A_1,A_2}(g_1,g_2,h_1,h_2)$ appearing in
  \be
  \biphi{g_1}{g_2}(U) \star \biphi{h_1}{h_2}(U)
  = \eps_{U,A_1,A_2}(g_1,g_2,h_1,h_2)\,
  \biphi{h_1}{h_2}(U) \star \biphi{g_1}{g_2}(U) \,.  \labl{epsUA1A2} 
To determine the scalars $\eps_{U,A_1,A_2}(g_1,g_2,h_1,h_2)$, we
compose the equality \erf{epsUA1A2} with the concatenation
$\overline{\biphi{g_1}{g_2}(U)}\,{\star}\,\overline{\biphi{h_1}{h_2}(U)}$
of dual basis morphisms.
Then on the \rhs, after using the Frobenius property for both $A_1$ and
$A_2$, only the tensor unit can propagate in the middle $A_i$-ribbons, so 
that analogously as in \erf{Fig19} one obtains a multiple of $\id_U$: 
  \bea \begin{picture}(380,189)(0,44)
  \put(0,0)     {\includeourbeautifulpicture{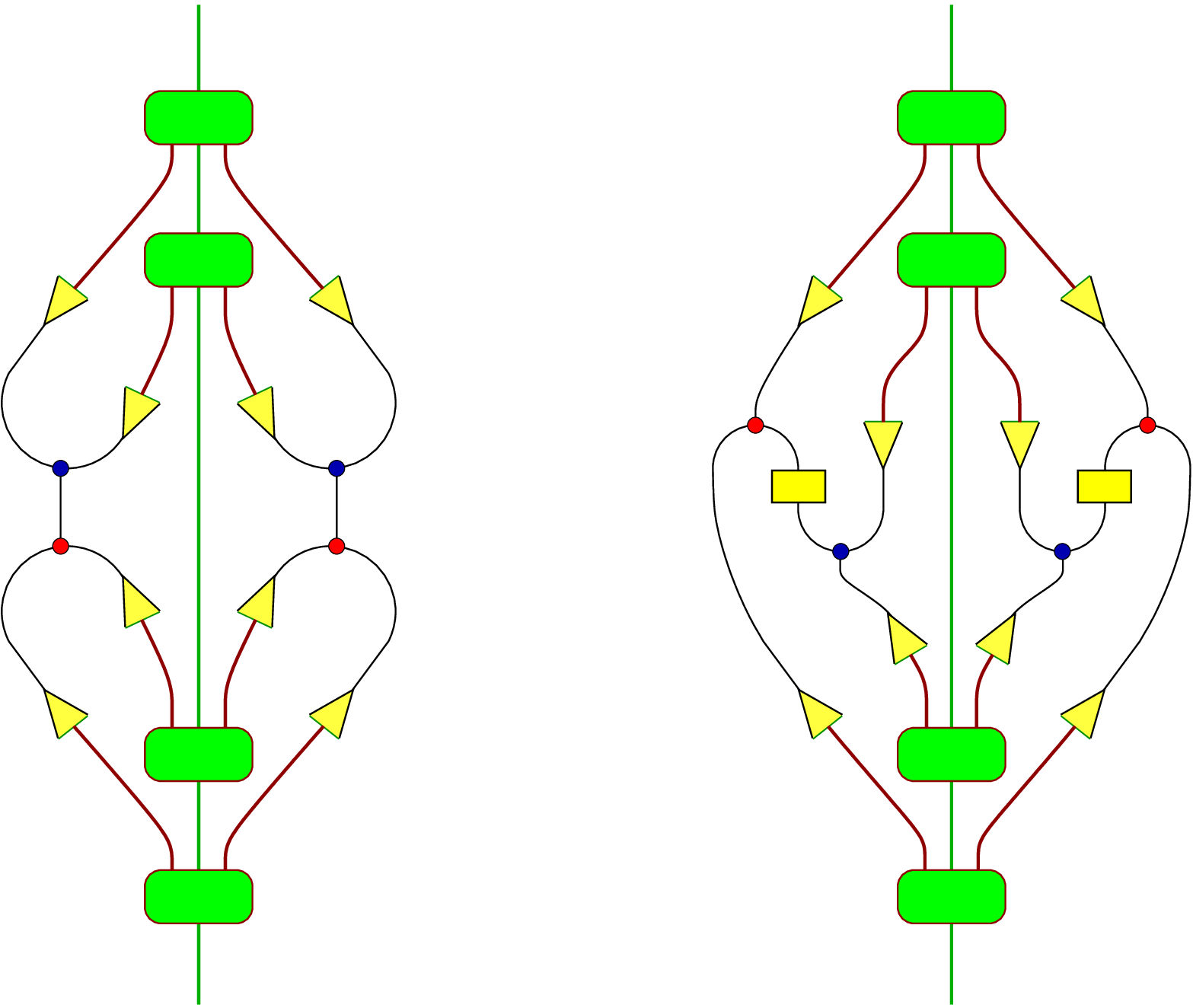}}
  \put(14.4,110)    {\scriptsize$ A_1 $}
  \put(19.1,184)    {\scriptsize$ \U_{g_1}^{} $}
  \put(22.5,155)    {\scriptsize$ \U_{h_1}^{} $}
  \put(24.5,37)     {\scriptsize$ \U_{g_1}^{} $}
  \put(24.5,72)     {\scriptsize$ \U_{h_1}^{} $}
  \put(35.3,24.2)   {\scriptsize$ \biphi{g_1}{g_2} $}
  \put(35.3,55.9)   {\scriptsize$ \biphi{h_1}{h_2} $}
  \put(35.3,166.5)  {\scriptsize$ \overline{\biphi{h_1}{h_2}} $}
  \put(35.3,197.9)  {\scriptsize$ \overline{\biphi{g_1}{g_2}} $}
  \put(41.3,-8)     {\scriptsize$ U $}
  \put(41.8,230)    {\scriptsize$ U $}
  \put(52.2,155)    {\scriptsize$ \U_{h_2}^{} $}
  \put(53.3,72)     {\scriptsize$ \U_{h_2}^{} $}
  \put(55.3,37)     {\scriptsize$ \U_{g_2}^{} $}
  \put(56.1,184)    {\scriptsize$ \U_{g_2}^{} $}
  \put(76.7,110)    {\scriptsize$ A_2 $}
  \put(119,110)     {$ = $}
  \put(176.5,116.5) {\scriptsize$ p_e $}
  \put(188.6,184)   {\scriptsize$ \U_{g_1}^{} $}
  \put(192.2,155)   {\scriptsize$ \U_{h_1}^{} $}
  \put(193.7,37)    {\scriptsize$ \U_{g_1}^{} $}
  \put(194.2,70)    {\scriptsize$ \U_{h_1}^{} $}
  \put(204.3,24.2)  {\scriptsize$ \biphi{g_1}{g_2} $}
  \put(204.3,55.9)  {\scriptsize$ \biphi{h_1}{h_2} $}
  \put(204.3,166.5) {\scriptsize$ \overline{\biphi{h_1}{h_2}} $}
  \put(204.3,197.9) {\scriptsize$ \overline{\biphi{g_1}{g_2}} $}
  \put(221.6,155)   {\scriptsize$ \U_{h_2}^{} $}
  \put(221.8,70)    {\scriptsize$ \U_{h_2}^{} $}
  \put(224.5,37)    {\scriptsize$ \U_{g_2}^{} $}
  \put(225.3,184)   {\scriptsize$ \U_{g_2}^{} $}
  \put(211.3,-8)    {\scriptsize$ U $}
  \put(211.6,230)   {\scriptsize$ U $}
  \put(245.5,116.5) {\scriptsize$ p_e $}
  \put(290,110)     {$ =\ \displaystyle\frac{1}{|H_1|\,|H_2|} \, \id_U \;. $} 
  \epicture26 \labl{Fig34} 
On the left hand side one gets, after inserting the explicit form 
\erf{Fig33} of the morphisms, and by analogous manipulations as in \erf{Fig19} 
  \bea \begin{picture}(400,296)(5,90)
  \put(0,0)         {\includeourbeautifulpicture{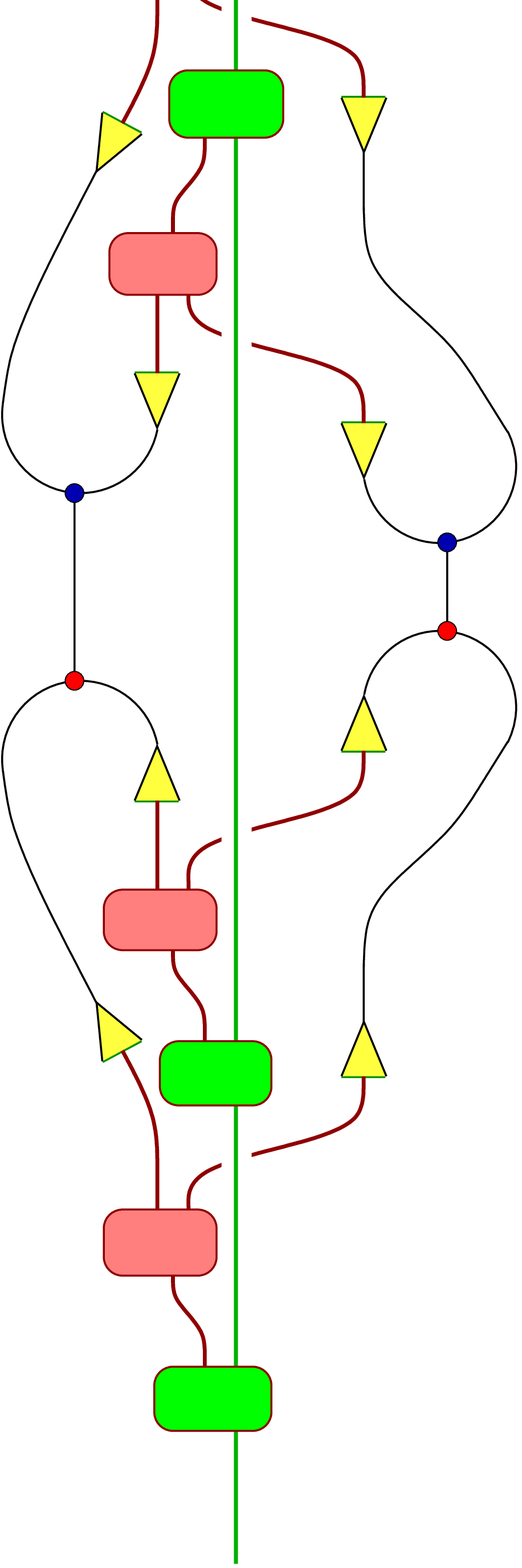}}
  \setlength{\unitlength}{24810sp}
    \put( 80,907)  {\scriptsize$ \bb_{g_1 g_2} $}
    \put( 88,741)  {\scriptsize$ \bb_{h_1 h_2} $}
    \put( 57,824)  {\scriptsize$ \bL_{g_1\!}{g_2} $}
    \put( 57,660)  {\scriptsize$ \bL_{h_1\!}{h_2} $}
    \put( 83,244)  {\scriptsize$ \overline{\bb_{g_1 g_2}} $}
    \put( 83, 80)  {\scriptsize$ \overline{\bb_{h_1 h_2}} $}
    \put( 54,324)  {\scriptsize$ \overline{\bL_{g_1\!}{g_2}} $}
    \put( 56,161)  {\scriptsize$ \overline{\bL_{h_1\!}{h_2}} $}
    \put(145,356)  {\scriptsize$ \U_{g_2}^{} $}
    \put( 70,283)  {\scriptsize$ \U_{g_1 g_2}^{} $}
    \put(153,196)  {\scriptsize$ \U_{h_2}^{} $}
    \put( 12,480)  {\scriptsize$ A_1 $}
    \put(231,488)  {\scriptsize$ A_2 $}
    \put(788,843)  {\scriptsize$ \bb_{g_1 g_2} $}
    \put(810,916)  {\scriptsize$ \tildeb_{h_1 h_2} $}
    \put(758,766)  {\scriptsize$ \bL_{g_1}{g_2} $}
    \put(884,796)  {\scriptsize$ \bL_{h_1}{h_2} $}
    \put(788,247)  {\scriptsize$ \overline{\bb_{g_1 g_2}} $}
    \put(788, 80)  {\scriptsize$ \overline{\bb_{h_1 h_2}} $}
    \put(759,324)  {\scriptsize$ \overline{\bL_{g_1}{g_2}} $}
    \put(761,160)  {\scriptsize$ \overline{\bL_{h_1}{h_2}} $}
    \put(743,722)  {\scriptsize$ \U_{g_1}^{} $}
    \put(945,682)  {\scriptsize$ \U_{g_2}^{} $}
    \put(866,760)  {\scriptsize$ \U_{h_1}^{} $}
    \put(923,734)  {\scriptsize$ \U_{h_2}^{} $}
    \put(752,364)  {\scriptsize$ \U_{g_1}^{} $}
    \put(856,355)  {\scriptsize$ \U_{g_2}^{} $}
    \put(750,205)  {\scriptsize$ \U_{h_1}^{} $}
    \put(866,193)  {\scriptsize$ \U_{h_2}^{} $}
    \put(735,805)  {\scriptsize$ \U_{g_1 g_2}^{} $}
    \put(881,864)  {\scriptsize$ \U_{h_1 h_2}^{} $}
    \put(776,285)  {\scriptsize$ \U_{g_1 g_2}^{} $}
    \put(750,118)  {\scriptsize$ \U_{h_1 h_2}^{} $}
    \put(720,480)  {\scriptsize$ A_1 $}
    \put(939,488)  {\scriptsize$ A_2 $}
  \setlength{\unitlength}{1pt}
  \put(12.5,333)    {\scriptsize$ \U_{g_1g_2}^{} $}
  \put(15.5,138)    {\scriptsize$ \U_{g_1}^{} $}
  \put(16.2,78)     {\scriptsize$ \U_{h_1}^{} $}
  \put(16.2,238)    {\scriptsize$ \U_{h_1}^{} $}
  \put(15.5,298)    {\scriptsize$ \U_{g_1}^{} $}
  \put(16.5,45)     {\scriptsize$ \U_{h_1h_2}^{} $}
  \put(23.5,267)    {\scriptsize$ \U_{h_1h_2}^{} $}
  \put(42.3,-8)     {\scriptsize$ U $}
  \put(43,384)      {\scriptsize$ U $}
  \put(51.5,238)    {\scriptsize$ \U_{h_2}^{} $}
  \put(51.5,300)    {\scriptsize$ \U_{g_2}^{} $}
  \put(122,186)     {$ =\ \alpha_{h_1h_2}(U)\,\phi_U(g_1g_2,h_1h_2) $}
  \put(310.3,-8)    {\scriptsize$ U $}
  \put(311,384)     {\scriptsize$ U $} 
  \epicture68 \labl{Fig35}
Here the basis indepedent $6j$-symbol $\phi_U(h_1h_2,g_1g_2)$ as defined in
\erf{Fig7} (note that both products $h_1h_2$ and $g_1g_2$ are in $\cals(U)$,
so that this definition is applicable) appears when interchanging the order 
of the $\U_{h_1h_2}$- and $\U_{g_1g_2}$-ribbons. Next, we invert the braiding
on the \rhs\ of \erf{Fig35} that is marked with a circle;
this gives rise to a factor $\beta(h_1,g_2)$, compare to \erf{pf3}.
Similarly like in the first step of \erf{Fig20} one can now use
the relation \erf{sc-rel} and the fact that $\bb_{g_1 g_2}(U)$ and 
$\overline{\bb_{g_1 g_2}(U)}$ are dual bases so as to obtain
  \begin{eqnarray}  \begin{picture}(400,342)(18,0)
  \put(35,0)     {\includeourbeautifulpicture{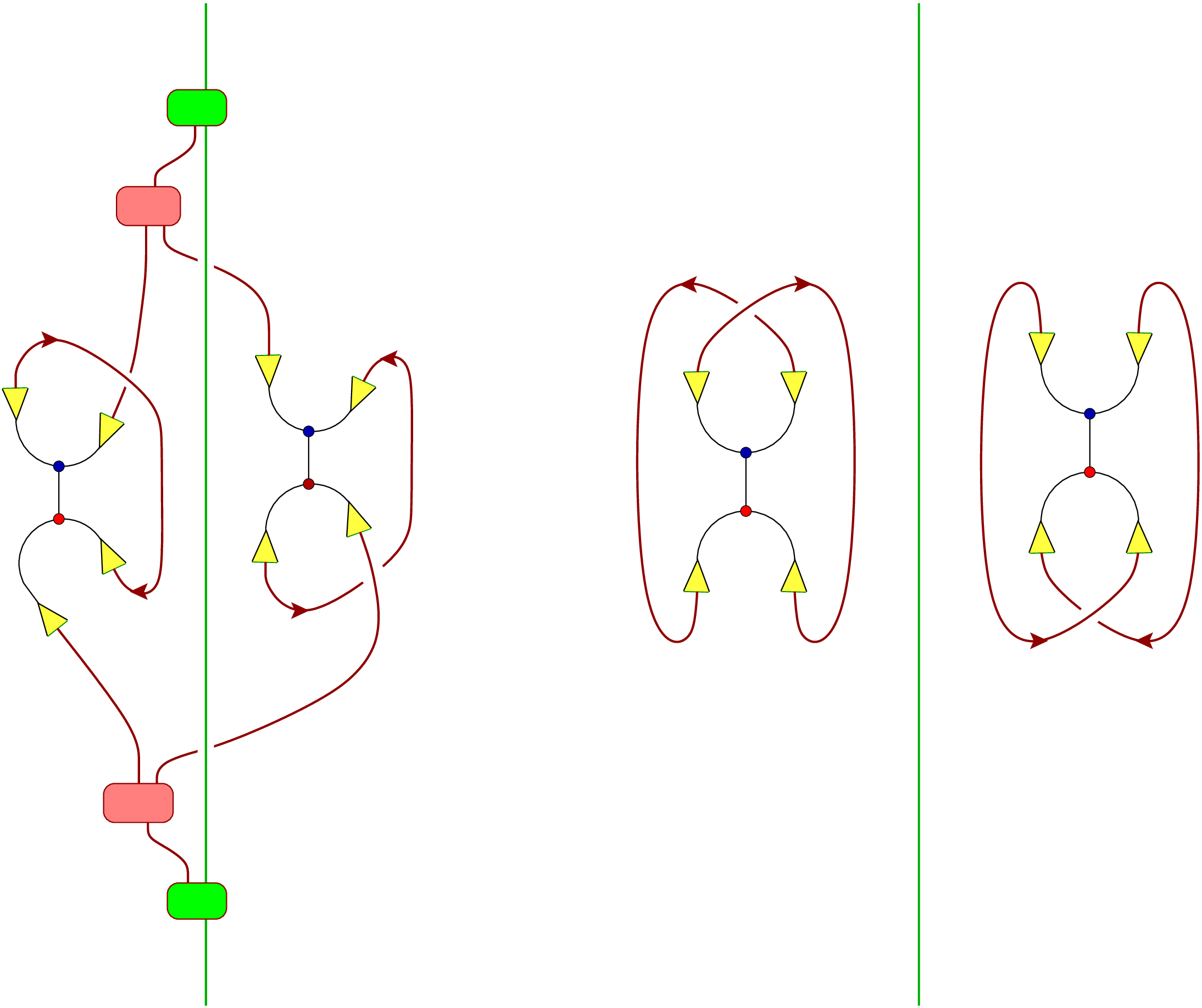}
  \setlength{\unitlength}{24810sp}
    \put(147,786)  {\scriptsize$ \bb_{h_1 h_2} $}
    \put(103,700)  {\scriptsize$ \bL_{h_1}{h_2} $}
    \put(147, 89)  {\scriptsize$ \overline{\bb_{h_1 h_2}} $}
    \put( 95,174)  {\scriptsize$ \overline{\bL_{h_1}{h_2}} $}
    \put( 90,756)  {\scriptsize$ \U_{h_1 h_2}^{} $}
    \put( 90,131)  {\scriptsize$ \U_{h_1 h_2}^{} $}
    \put( 20,445)  {\scriptsize$ A_1 $}
    \put(240,473)  {\scriptsize$ A_2 $}
    \put(612,323)  {\scriptsize$ \U_{h_1}^{} $}
    \put(670,323)  {\scriptsize$ \U_{g_1}^{} $}
    \put(918,613)  {\scriptsize$ \U_{h_2}^{} $}
    \put(965,613)  {\scriptsize$ \U_{g_2}^{} $}
    \put(625,455)  {\scriptsize$ A_1 $}
    \put(928,483)  {\scriptsize$ A_2 $}
  \setlength{\unitlength}{1pt}
  }
  \put(-6,169)    {$ =\; F $}
  \put(59.4,99.5){\scriptsize$ \U_{h_1} $}
  \put(67.8,235) {\scriptsize$ \U_{h_1} $}
  \put(74.6,169) {\scriptsize$ \U_{g_1} $}
  \put(99.3,-8.8){\scriptsize$ U $}
  \put(100.5,339){\scriptsize$ U $}
  \put(124.8,235){\scriptsize$ \U_{h_2} $}
  \put(145.8,99) {\scriptsize$ \U_{h_2} $}
  \put(157.8,182){\scriptsize$ \U_{g_2} $}
  \put(199,169)  {$ =\ F $}
  \put(338,-8.8) {\scriptsize$ U $}
  \put(338.5,339)  {\scriptsize$ U $}
  \end{picture}  \nonumber\\[-.7em]{}  \label{Fig36}
  \\[-1.6em]{}\nonumber\end{eqnarray}
where we use the abbreviation $F\eq\phi_U(g_1g_2,h_1h_2)\, \beta(h_1,g_2)$. 
Using \erf{Fig21a} we thus arrive at
  \be  \eps_{U,A_1,A_2}(g_1,g_2,h_1,h_2)  = \phi_U(g_1g_2,h_1h_2)\,
  \beta(h_1,g_2) \, \XI_{A_1}(h_1,g_1) \, \XI_{A_2}(g_2,h_2)
  \,, \labl{epsUA1A2'}
Thus we have established \erf{prop-epsUA1A2}. 
\qed

\medskip

After these preliminaries, the determination of the simple 
$A_1$-$A_2$-bimodules continues along the same lines as in the case of 
left modules that was treated in section \ref{bertarep}.

Denote by $\{ B_\kappa \,|\, \kappa\iN\calk \}$ a set of representatives 
of isomorphism classes of simple $A_1$-$A_2$-bimodules, and let $U$ be a 
simple object in $\calc$. As for left modules, the isotypical components
$B_\kappa^{\oplus n_\kappa}$ in the decomposition
  \be
  \IndAAv(U) \cong \bigoplus_{\kappa \in \mathcal{K}} 
  B_\kappa^{\oplus n_\kappa} \ee
of an induced bimodule can be extracted as the images of idempotents in the
center of the endomorphism algebra $\EndAAv(\IndAAv(U)) 
\,{\cong}\, \complex_{\beta_{U,A_1,A_2}}\bicals_{\!A_1|A_2}(U)$.

To describe this center we introduce the {\em central bi-stabilizer\/}
  \be \bearll
  \bicalu_{A_1|A_2}(U) := \big\{ (g_1,g_2) \iN \bicals_{\!A_1|A_2}(U)
  \,\big| \hsp{-.6} & 
  \eps_{U,A_1,A_2}(g_1,g_2,h_1,h_2) \eq 1
  \\&\hsp{3.4}
  {\rm~for\;all~} (h_1,h_2) \iN \bicals_{\!A_1|A_2}(U) \big\}  \,.  \eear \ee
As for modules we then have
  \be
  \calz \equiv \calz(\EndAAv(\IndAAv(U)) ) \cong 
  \calz( \complex_{\beta_{U,A_1,A_2}} \bicals_{\!A_1|A_2}(U) )
  = \complex\bicalu_{A_1|A_2}(U) \,.  \ee
Note that, by construction, a basis of the morphism space $\EndAAv(\IndAAv(U))$ 
is given by $\{ \bivarphi{g_1}{g_2}(U)\,|\,(g_1,g_2)\iN\bicals_{\!A_1|A_2}(U) \}$
with $\bivarphi{g_1}{g_2}(U) \,{:=}\, f_{U,U}^{-1}(\biphi{g_1}{g_2}(U))$,
and with the isomorphism $f_{U,U}$ the one defined in lemma \ref{doublrec}.
As already stated, the multiplication is given by concatenation;
following \erf{eq:biphi-mult} it takes the form
  \be
  \bivarphi{g_1}{g_2}(U) \cir
  \bivarphi{h_1}{h_2}(U) = \beta_{U,A_1,A_2}(g_1,g_2,h_1,h_2)\,
  \bivarphi{g_1 h_1}{g_2 h_2}(U) \,.  \ee
The definition of the central bi-stabilizer implies that 
a basis of $\calz$ is, in turn, given by
$\{ \bivarphi{g_1}{g_2}(U) \,|\, (g_1,g_2) \iN \bicalu_{A_1|A_2}(U) \}$.
On $\calz$ the commutator two-cocycle of $\beta_{U,A_1,A_2}$ is trivial,
and hence we can modify the original choice of basis by suitable
scalars so that in the new basis
we have, for  $(g_1,g_2)$ and $(h_1,h_2)$ in $\bicalu_{A_1|A_2}(U)$,
  \be
  \bivarphi{g_1}{g_2}(U) \cir \bivarphi{h_1}{h_2}(U) =
  \bivarphi{g_1 h_1}{g_2 h_2}(U) \,.  \labl{eq:better-basis}
The primitive idempotents in $\calz$ can then be given explicitly as
  \be
  \widehat P_{U,\psi} = 
  \frac{1}{|\bicalu_{A_1|A_2}(U)|}
  \sum_{(g,h)\in\bicalu_{A_1|A_2}(U)} \psi^*(g,h)\, \bivarphi{g}{h}(U) \,,
  \labl{eq:bimod-proj}
where $\psi \iN \bicalu_{A_1|A_2}(U)^*$ is a character of
the central bi-stabilizer. All simple bimodules in the
decomposition of $\IndAAv(U)$ occur with the same multiplicity
  \be
  d_{A_1|A_2}(U) = \sqrt{\frac{|\bicals_{\!A_1|A_2}(U)|}{
  |\bicalu_{A_1|A_2}(U)|}} \,.  \labl{eq:isotyp-mult}

The inequivalent simple bimodules in a given induced bimodule
$\IndAAv(U)$ are thus labeled by characters of $\bicalu_{A_1|A_2}(U)$, 
once a basis satisfying \erf{eq:better-basis}
has been chosen. Furthermore, two induced bimodules
$\IndAAv(U_i)$ and $\IndAAv(U_j)$ for simple objects $U_i$, $U_j$
are isomorphic if $U_i$ and $U_j$ lie on the same $H$-orbit,
and do not contain any common submodules otherwise.  
This establishes the 
\\[-2.3em]

\dtl{Proposition}{prop:bimod}
Let $A_1$ and $A_2$ be two Schellekens algebras in a modular
tensor category $\calc$. Denote by $H$ the subgroup of $\Pic(\calc)$ 
generated by $H(A_1)$ and $H(A_2)$. Given a choice of basis satisfying 
\erf{eq:better-basis}, the isomorphism classes of simple 
$A_1$-$A_2$-bimodules are uniquely labeled by pairs $(n,\psi)$, where 
$n$ labels an $H$-orbit in the set $\{U_i\,|\, i\iN\II \}$ of simple objects
of $\calc$ and $\psi$ is a character of
the central bi-stabilizer $\bicalu_{A_1|A_2}(U_{m_n})$, for an arbitrary 
fixed choice of representative $U_{m_n}$ of the $n$th $H$-orbit.

\dt{Remark}
(i)~\,The total number of isomorphism classes of simple $A_1$-$A_2$-bimodules
is given by the formula
$\tr\big( Z(A_1) Z(A_2)^t \big)$, see remark I:5.19(ii).
\\[.2em]
(ii)~For $A_1\eq A_2\eq A$, the category $\calcAA$ of
$A$-$A$-bimodules is itself a tensor category. 
It thus makes sense to consider the Grothendieck ring (or fusion
ring) of $\calcAA$. This ring is in general non-commutative. Denoting 
again by $\{ B_\kappa \,|\, \kappa \iN \calk \}$ a set of representatives
of isomorphism classes of simple bimodules, the 
structure constants of the fusion ring of $\calcAA$ are defined via
  \be
  B_\mu \,{\otimes}_A\, B_\nu \,\cong\,
  \bigoplus_{\kappa\in\calk} \widehat N_{\mu \nu}^{~~\kappa} B_\kappa \,.
  \labl{eq:bi-fusion}
The $\widehat N_{\mu \nu}^{~~\kappa}$ are non-negative integers by construction.
One can compute these numbers as invariants of a
ribbon graph in $S^2 \Times S^1$. We have
  \be
  \widehat N_{\mu \nu}^{~~\kappa} = 
  \dim \HomAA(B_\mu \,{\otimes}_A\, B_\nu , B_\kappa) = Z^{X|Y}_{00} \,,
  \ee
where we set $X \eq B_\mu \,{\otimes_A}\, B_\nu$, $Y \eq B_\kappa$, and
where $Z^{X|Y}_{00}$ is the ribbon graph given in (I:5.151). 
Note that the knowledge
of the projectors \erf{eq:bimod-proj} is sufficient to compute the 
$\widehat N_{\mu \nu}^{~~\kappa}$. However, since the image of these
projectors is a direct sum of isomorphic simple bimodules, one must
divide the resulting invariant by the corresponding multiplicities
\erf{eq:isotyp-mult}.
\\[.2em]
(iii)~Let $A$ be a simple symmetric special Frobenius algebra.
The fusion algebra of $A$-$A$-bimodules has a direct physical
interpretation. It describes the fusion of conformal defects,
i.e.\ the process of joining two defect lines and decomposing
the resulting defect line again into simple defects.
As a consequence of \erf{eq:bi-fusion}, the matrices
$Z^{B_\mu|B_\nu}_{ij}$ obey the condition
  \be
  Z^{B_\mu|B_\kappa}_{ij} = \sum_{\nu\in\calk} \widehat N_{\mu \nu}^{~~\kappa} 
  Z^{A|B_\nu}_{ij} \,. \ee
This is the way the fusion algebra of defect lines was
introduced in \cite{pezu5}.


\newpage
\appendix

\sect{Some notions from group cohomology} \label{appcoho}

In this appendix we summarize a few notions from the cohomology of
finite groups. In the applications of our interest, the group is abelian.
Still, we write the group operation multiplicatively, since the groups we 
are interested in arise from products in the Grothendieck group
of a tensor category and thus from a tensor {\em product}.

\subsection{Ordinary cohomology}

We start with ordinary group cohomology with values in $\complexx$. 
Let $G$ be a (not necessarily abelian) finite group.
A $k$-cochain on $G$ with values in $\complexx$ is a function 
  \be  \kappa:\quad  G^k \equiv G\times \cdots \times G \to \complexx \,.
  \ee
$\kappa$ is called normalized iff its value is $1\iN\complexx$
as soon as at least one of the arguments is the unit element $e\iN G$. 
The group of $k$-cochains is denoted by $C^k(G,\complexx)$. 
The coboundary operator
  \be  \rmd=\rmd_k:\quad   C^k(G,\complexx) \to\, C^{k+1}(G,\complexx)  
  \ee 
is defined by
  \be  \begin{array}{r}
  \rmd \kappa(g_1,g_2,...\,, g_{k+1}) 
  = \kappa(g_2,g_3,...\,, g_{k+1})\, \kappa(g_1g_2,g_3,...\,g_{k+1})^{-1}\,
  \kappa(g_1,g_2g_3,...\,, g_{k+1})
  \\{}\\[-.8em]
  \cdots\, \kappa(g_1,g_2,...\,,g_k g_{k+1})^{(-1)^k} 
  \kappa(g_1,g_2,...\,,g_k)^{(-1)^{k+1}} .
  \end{array} \ee
The elements in the kernel of ${\rm d}_k$ form a subgroup $Z^k(G,\complexx)$ of
$C^k(G,\complexx)$; they are called $k$-cocycles. The elements of the image 
of ${\rm d}_k$ form a subgroup $B^{k+1}(G,\complexx)$ of $C^{k+1}(G,\complexx)$; 
they are called $(k{+}1)$-coboundaries. The operator $\rmd$ squares to one, 
$\rmd_{k+1}{\circ}\,\rmd_k\eq 1 \iN\complex$. 
Thus $B^k(G,\complexx)\,{\le}\, Z^k(G,\complexx)$;
the cohomology $H^k$ is defined as the quotient group:
  \be  H^k(G,\complexx) := Z^k(G,\complexx)\, / B^{k}(G,\complexx) \,.
  \ee

For a deeper understanding of a cohomology theory, it is useful to have 
mathematical objects at one's command that are classified by the cohomology 
groups. The following is well known.
\nxt $H^1(G,\complexx)$ parametrizes group homomorphisms from $G$ to
     $\complexx$, and thus one-dimensional irreducible $G$-representations.
\nxt $H^2(G,\complexx)$ (modulo the action of outer automorphisms of $G$)
     parametrizes isomorphism classes of twisted group algebras.
     The group algebra $\complex G$ is the unital associative complex 
     algebra with a basis $\{b_g\}_{g\in G}$ labeled by $G$ and multiplication
     $ b_g \,{\star}\, b_{g'} \eq b_{g g'} $.
     Given a two-cocycle $\omega\iN Z^2(G,\complexx)$, one can twist the
     multipliciation by $\omega$ according to
  \be  b_g \star_\omega^{} b_{g'} := \omega(g,g')\, b_{g g'} \ee
to obtain the {\em twisted group algebra\/} $\complex_\omega G$. The fact
that $\omega$ is a cocycle is equivalent to the associativity of the product
$\star_\omega$. Cohomologous two-cocycles give rise to isomorphic algebras.
If $\omega$ is normalised, then $b_e$ is the unit element in $\complex_\omega G$.
Since every two-cocycle is cohomologous to a normalised two-cocycle,
$\complex_\omega G$ is an algebra with unit.
\nxt As we have seen in the main text, $H^3(G,\complexx)$ parametrizes marked
     categorifications of the group ring $\zet G$. The fact that 
     $\psi \iN Z^3(G,\complexx)$ is a 
     cocycle ensures that the tensor product on the category is associative. 
     Cohomologous three-co\-cycles give rise to equivalent categories.

\subsection{Abelian group cohomology}\label{Abeliangc}

{}From now on the finite group $G$ is assumed to be abelian. 
So far we have only used the associativity of the product of $G$, e.g.\
when checking that $\rmd$ is nilpotent. When $G$ is abelian, it is
natural to study also notions that in addition use the commutativity of
the product.
Indeed we also need {\em abelian group cohomology\/}; more precisely, we 
are only interested in $H^3_{{\rm ab}}(G,\complexx)$. 

An abelian two-cochain is an ordinary two-cochain. An 
{\em abelian three-co\-chain\/} is a pair $(\psi,\Mho)$ consisting of an ordinary
three-co\-chain $\psi$ and a two-cochain $\Mho$. It is called normalized iff
both $\psi$ and $\Mho$ are normalized as cochains. The {\em abelian coboundary\/}
$\rmd_{{\rm ab}}\kappa$ of an abelian two-cochain $\kappa$ is an
abelian three-co\-chain defined as 
   \be  \rmd_{{\rm ab}} \kappa := (\rmd \kappa\,, \kappa_{\rm comm} )
   \,,  \ee
i.e.\ one takes the pair consisting of the ordinary coboundary of the two-cochain
$\kappa$ and the {\em commutator cocycle\/} $\kappa_{\rm comm}$ of $\kappa$, 
defined by
  \be  \kappa_{\rm comm}(x,y) := \kappa(x,y)\, \kappa(y,x)^{-1} \,.  \labl{coco}
An abelian three-co\-chain $(\psi,\Mho)$ is an {\em abelian three-co\-cycle\/} iff 
$\psi$ is an ordinary three-co\-cycle, $\rmd \psi\eq 1$, and 
  \be \begin{array}{l}
  \psi(y,z,x)^{-1}\, \Mho(x,yz)\, \psi(x,y,z)^{-1}
  = \Mho(x,z)\, \psi(y,x,z)^{-1}\, \Mho(x,y) \,,
  \\{}\\[-.7em]
  \psi(z,x,y)\, \Mho(xy,z)\, \psi(x,y,z) = \Mho(x,z)\, \psi(x,z,y)\, \Mho(y,z) \,.
  \end{array} \labl{app1neu}
(In the notation used in \cite{macl2}, $\psi(x,y,z)\eq f(x,y,z)$ and 
$\Mho(x,y)\eq d(x\,|\,y)$, compare formulas (17)\,--\,(19) of \cite{macl2}.)

This means that for closed abelian three-co\-chains the deviation of $\Mho$
from being a {\em bihomomorphism\/} (i.e.\ from being multiplicative in both
arguments) is controlled by the ordinary three-co\-chain $\psi$. 
Finally, the abelian cohomology group $H^3_{{\rm ab}}(G,\complexx)$ is defined as 
the quotient of the group of normalized three-co\-cycles by its subgroup consisting
of coboundaries of normalized two-cochains.
Forgetting $\Mho$ provides a group homomorphism 
  \be \begin{array}{rcl}
  H^3_{{\rm ab}}(G,\complex) & \!\to\! & H^3(G,\complex)
  \\{}\\[-.9em]
  {}[(\psi,\Mho)] & \!\mapsto\! & [\psi] \, . 
  \end{array}  \ee
The fibers of this map are bihomomorphisms modulo alternating bihomomorphisms.
(A bihomomorphism $\zeta$ on $G$ is called alternating iff $\zeta(g,g)\eq 1$
for all $g\iN G$, see definition \ref{altbi}.)

As shown in the main text, the objects parametrized by the abelian cohomology
$H^3_{{\rm ab}}(G,\complexx)$ are the marked theta-categories that are marked
categorifications of $\zet G$. Another important aspect of the abelian group 
cohomology $H^3_{{\rm ab}}(G,\complexx)$ is its relation to quadratic forms on 
the abelian group $G$.  A {\em quadratic form\/} on $G$ with values in the 
multiplicative group $\complexx$ is a map
  \be  q: \quad G \to \complexx  \labl{quadform}
such that $q(g)\eq q(g^{-1})$ and such that the function
  \be \begin{array}{lrcl}
  \beta_q :\ &  G \times G &\!\to\!& \complexx
  \\{}\\[-.9em]&
  (g_1,g_2) &\!\mapsto\!& q(g_1g_2)\, q(g_1)^{-1} q(g_2)^{-1} 
  \end{array} \labl{A11}
is a bihomomorphism, which is called the {\em associated 
bihomomorphism\/}. The product of two quadratic forms is again a quadratic 
form, so that quadratic forms form an abelian group $\QF(G,\complexx)$.

Note that while the quadratic form completely determines the bihomomorphism,
the converse need not be true. Indeed, two quadratic forms $q$, $\tilde q$ 
on $G$ possess the same associated bihomomorphism 
$\beta_q \eq \beta_{\tilde q}$ iff $\tilde q \eq q \chi$ for some
$\chi \iN H^1(G,\zet_2)$. That $\chi$ takes values in 
$\zet_2$ follows because any quadratic form obeys $q(g) \eq q(g^{-1})$.

The map
  \be \begin{array}{lrcl}
  \EL:\ & H^3_{{\rm ab}}(G,\complex^*) &\!\to\!& \QF(G,\complexx)
  \\{}\\[-.9em]&
  [(\psi,\Mho)] &\!\mapsto\!& q \quad{\rm with}\quad q(g) := \Mho(g,g) 
  \end{array}  \labl{EL}
is a homomorphism of abelian groups.
In particular, the quadratic form depends only on the cohomology class.
It has been shown by Eilenberg and MacLane \cite{eima1x,macl2}
that $\EL$ is even an {\em iso\/}morphism.
Moreover, they show that the associated bihomomorphisms obeys
  \be  \beta_{q(\psi,\Mho)}(g,h) = \Mho(g,h)\, \Mho(h,g) \,. \labl{app2}

\sect{KS matrices versus KSBs} \label{ksmatrix}

Here we present the alternative description of KSBs that is due to Kreuzer and
Schellekens \cite{krSc}. For easier comparison with the original literature
\cite{krSc} we use CFT terminology, in particular the conformal weights 
$\Delta_i$ and the monodromy charges \erf{Qdef}. The conformal weights are 
only needed modulo $\zet$, so that it is sufficient to know the balancing 
phases $\th_i\eq\exp(-2\pi\ii \Delta_i)$ of the modular tensor category $\calc$.

\dtl{Definition}{KSM}
Let $H$ be a subgroup in the effective center $\Pico(\calc)$. Write
$H$ in the form 
  \be  H \,\cong\, \zet_{n_1} \times \zet_{n_2} \times \cdots \times
  \zet_{n_k} \,,  \labl{form2}
with generators $g_1,g_2,...\,,g_k$ of the cyclic factors.
A {\em KS matrix\/} for $H$ is a $k\,{\times k}\,$-matrix with entries 
$X_{ab}$, defined modulo $\zet$, that satisfy the constraints
\\[.2em]
  \be  \begin{array}{rl}
  {\rm i})  \ \   & X_{ab} + X_{ba} = Q_{g_a}(g_b) \bmod\zet \,
  \mbox{~~for all } a,b  \qquad{\rm and}
  \\{}\\[-.7em]
  {\rm ii}) \ \   & N_a\, X_{ab} \iN\zet \mbox{~~for all } a,b \,,\,
  \mbox{ with $N_a\eq N_{g_a}$ the order of $g_a$\,.} 
  \end{array}  \labl{matcond}
If the order $N_a$ of $g_a$ is even, then in addition 
  \be
  {\rm iii}) \quad\ \ \Delta_{g_a} = (N_a{-}1)\, X_{aa}\, \bmod\zet
  \mbox{~~for all } a\,. \hsp{8.5} \labl{addreq} 

\smallskip

\noindent
(In \erf{form2} we do not impose any divisibility conditions
on the $n_a$, so the decomposition is not unique.)

\medskip

To prove the relation between KSBs and KS matrices we use
the following properties of quadratic forms.
\\[-2.3em]

\dtl{Lemma}{lem:QF}
For $G$ a finitely generated abelian group with generators
$g_1,g_2,...\,,g_r$, let $q, q' \iN \QF(G,\complexx)$.
\\[.2em] 
(i)~\,If $q(g_i)\eq q'(g_i)$ and $q(g_i g_j) \eq q'(g_i g_j)$
for all $i,j\iN \{1,...\,,r\}$ with $i\,{\neq}\, j$, then $q\eq q'$.
\\[.2em] 
(ii)~For every $g\iN G$ we have $q(g^n) \eq q(g)^{n^2}$.

\medskip\noindent
Proof:\\
(i)~\,Since $q$ is a quadratic form, by definition $q(g)\eq q(g^{-1})$,
and $\beta_q(g,h) \eq q(gh) / (q(g)q(h))$ is a bihomomorphism. Expressing 
$g,h\iN G$ in terms of the generators as $g\eq g_1^{m_1} \cdots g_r^{m_r}$ 
and $h\eq g_1^{n_1} \cdots g_r^{n_r}$ and using the bihomomorphism property 
of $\beta_q$, we have
  \be
  \beta_q(g,h) = 
  \beta_q(g_1^{m_1} \cdots g_r^{m_r}, g_1^{n_1} \cdots g_r^{n_r}) =
  \prod_{i,j=1}^r \beta_q(g_i, g_j)^{m_i n_j}_{} \,,
  \ee
which when expressed in terms of $q$ amounts to
  \be
  q(gh) = q(g)\, q(h) \prod_{i,j=1}^r 
  \Big( \frac{ q(g_i g_j) }{q(g_i)\,q(g_j)} {\Big)}^{m_i n_j}_{} \,.
  \ee
$q(g)$ can therefore be determined recursively for all $g\iN G$
by starting from $q(g_i)$ and $q(g_i g_j)$.
\\
Next note that since $\beta_q$ is a bihomomorphism, we have
$\beta_q(g,g^{-1})\eq 1/\beta_q(g,g)$. Written in terms of $q$ this implies 
$q(g^2) \eq q(g)^4$. Hence $q(g_i g_i)$ can be expressed through $q(g_i)$. 
\\
These arguments show that a quadratic form $q$ is
uniquely determined for all $g\iN G$ once we prescribe
its values on $g \eq g_i$ and $g \eq g_i g_j$ for $i\,{\neq}\, j$. 
In particular, two quadratic forms that coincide on these
group elements are equal.
\\[3pt]
(ii)~Since $\beta_q$ is a bihomomorphism, we have
$\beta_q(g^n,g) \eq \beta_q(g,g)^n$. In terms of $q(g)$ this
implies, using also $q(g^2) \eq q(g)^4$,
  \be
  q(g^{n+1}) = q(g^n)\, q(g)^{2n+1}\,~.  \ee
This recursion relation has the unique solution $q(g^n) \eq q(g)^{n^2}$.
\qed

\medskip

The relation between between KSBs and KS matrices is now provided by
\\[-2.3em]

\dtl{Lemma}{lem-ksb-ksm}
Let $H$ be a subgroup in the effective center $\Pico(\calc)$, written
in the form \erf{form2}, and $X$ a $k\,{\times}\, k$-matrix. For any 
two elements $g,h$ of $H$, written in the form
$g\eq\prod_a (g_a)^{m_a}$ and $h\eq\prod_a (g_a)^{n_a}$, set
  \be \XI(g,h) := \exp(2\pi\ii \sum_{a,b=1}^k m_a X_{ab} n_b)\,.
  \labl{form}
Then $\XI$ is a KSB if and only if $X$ is a KS matrix.

\medskip \noindent
Proof: \\
Suppose that $\XI(g,h)$ is a KSB, i.e.\ $\XI$ is a bihomomorphism
with the additional property $\XI(g,g)\eq\theta_g$. Then in particular it is 
a well-defined map $G\Times G \,{\to}\, \complexx$. Since 
$g_a^{N_a}\eq e$, where $N_a$ is the order of the generator $g_a$, we have
  \be
  1 = \XI(g_a^{N_a}, g_b) = \exp(2\pi\ii N_a X_{ab})  \ee
for all $a,b$, i.e.\ $N_a X_{ab} \in \zet$, establishing
that the matrix $X$ has property (\ref{matcond}\,ii).
Further, by lemma \ref{lem:xi-xi=beta}, together
with \erf{mono} and \erf{Qdef} we know that
  \be
  \exp(2\pi\ii (X_{ab}{+}X_{ba})) = \XI(g_a, g_b)\, \XI(g_b, g_a) 
  = \beta(g_a,g_b) = \exp(2\pi\ii Q_{g_a}(g_b)) \,.  \ee
The matrix $X$ thus also has property (\ref{matcond}\,i).
Finally, property \erf{addreq} of $X$ follows from the fact
that a KSB by definition obeys $\XI(g_a,g_a)=\theta_{g_a}$, together
with $N_a X_{ab} \in \zet$. Thus $X$ is indeed a KS matrix.
\\[.2em]
Suppose now that $X$ is a KS matrix. We first check that
$\XI$ as defined in \erf{form} is a well-defined map
$G \Times G \,{\to}\, \complexx$. Note that the $m_a$ and $n_a$ in 
$g\eq\prod_a (g_a)^{m_a}$ and $h\eq\prod_a (g_a)^{n_a}$ are defined only 
mod $N_a$. Shifting $m_a \,\To\, m_a + k N_a$ changes the \rhs\ of
\erf{form} by $\exp(2\pi\ii k \sum_b N_a X_{ab} n_b)$, which is equal to 
one by property (\ref{matcond}\,ii) of a KS matrix.
Similarly, shifting $n_b \,\To\, n_b + k N_b$ changes the \rhs\ by
  \be  \bearll
  \eE^{2\pi\ii k \sum_a m_a X_{ab} N_b}  \!\!\!&
  = \eE^{2\pi\ii k \sum_a m_a Q_{g_a}(g_b) N_b}\,
  \eE^{-2\pi\ii k \sum_a N_b X_{ba} m_a} 
  \\{}\\[-.7em]&
  = \prod_{a} \beta(g_a,g_b)^{m_a N_b}
  = \prod_{a} \beta(g_a,g_b^{N_b})^{m_a} = 1 \,,
  \eear\ee
where in the first step (\ref{matcond}\,i) is used, in
the second step (\ref{matcond}\,ii) and \erf{mono},
and in the third and fourth step that $\beta$ is a bihomomorphism
and that $g_b^{N_b}\eq e$. 
\\[.2em]
Thus $\XI$ is well defined. That $\XI$ is a bihomomorphism
is then obvious. It follows that $q(g)\eq\XI(g,g)$ is a quadratic
form. To establish that $\XI$ is a KSB we must show that
$q$ coincides with the quadratic form $\delta(g)\eq\theta_g$. 
By lemma \ref{lem:QF} it is enough to verify
$q(g_a)\eq\delta(g_a)$ and $q(g_a g_b)\eq\delta(g_a g_b)$ for 
$a\,{\neq}\, b$. First note that by (\ref{matcond}\,i) and \erf{mono},
\be
  \XI(g_a,g_b)\, \XI(g_b,g_a) = \eE^{ 2\pi \ii (X_{ab} + X_{ba}) }
  = \beta(g_a,g_b) \,.  \labl{eq:KS-KSB-1}
By definition \erf{beta}, we have 
$\beta(g_a,g_a)\eq\delta(g_a^2)/\delta(g_a)^2$. Now, $\delta$
is a quadratic form, and by lemma \ref{lem:QF}(ii) we know
$\delta(g_a^2)\eq\delta(g_a)^4$. 
Evaluating \erf{eq:KS-KSB-1} for $a\eq b$ thus yields
  \be
  q(g_a)^2 = \delta(g_a)^2 \,.  \labl{eq:KS-KSB-2}
Furthermore,
  \be
  1 = q(g_a^{N_a}) = q(g_a)^{N_a^2} \qquad {\rm and} \qquad
  1 = \delta(g_a^{N_a}) = \delta(g_a)^{N_a^2} \,,
  \labl{eq:KS-KSB-3}
where each time in the first step it is used that $g_a$ has order $N_a$, 
while in the second step lemma \ref{lem:QF}(ii) is employed. Now note that 
for $N_a$ odd, at most one of the two roots $x$ of $x^2\eq\delta(q_a)^2$ 
can also satisfy $x^{N_a^2}\eq 1$.
Since both $q(g_a)$ and $\delta(g_a)$ obey both of these equalities,
they must be equal, $q(g_a)\eq\delta(g_a)$. On the other hand,
for even $N_a$, the equality $q(g_a)\eq\delta(g_a)$ is an immediate 
consequence of the properties (\ref{matcond}\,ii) and \erf{addreq},
  \be
  q(g_a) = \XI(g_a, g_a) = \exp(2 \pi \ii X_{aa})
  = \exp(- 2\pi\ii \Delta_{g_a}) = \theta_{g_a} \,.  \ee
Finally, for $q(g_a g_b)$ we find
  \be  \bearll
  q(g_a g_b) = \XI(g_a g_b, g_a g_b) \!\!\!&
  = \XI(g_a, g_a)\, \XI(g_b, g_b)\, \XI(g_a, g_b)\, \XI(g_b, g_a)
  \\{}\\[-.7em] &
  = \theta_{g_a} \theta_{g_b}\, \beta(g_a,g_b) = \theta_{g_a g_b} \,,
  \eear\ee
where in the second step it is used that $\XI$ is a bihomomorphism,
in the third step \erf{eq:KS-KSB-1} is substituted, and the
fourth step uses the definition of $\beta$ in \erf{beta}.
This shows that $q(g_a)\eq\delta(g_a)$ and 
$q(g_a g_b)\eq\delta(g_a g_b)$ for all $a,b$. It follows
that $q\eq\delta$ and hence $\XI(g,h)$ is indeed a KSB.
\qed

\sect{Conventions}\label{appC}

\subsection{Basis choices}

\paragraph{Simple currents.} Bases for the three-point coupling spaces
are denoted by
  \be  \bL_{g_1}{g_2} \,\in\, \Hom(\U_{g_1}\oti\U_{g_2},\U_{g_1g_2}) \,.  \ee
For $g\eq e$, we take the identity morphism, $\bL_{e}{g} \eq \id_{\U_g} \eq 
\bL_{g}{e}$ (here we use that $\U_e\eq\one$ is a {\em strict\/} tensor unit).

\paragraph{Algebras.} An adapted basis choice for the morphism spaces involving 
a \berta\ $(A,m,\eta,\Delta,\eps)$ and its simple subobjects is:
  \be  \begin{array}{lll}
  e_g \,\in\, \Hom(\U_g,A) \,,\qquad & r_g \,\in\, \Hom(A,\U_g)\,,  \qquad
  & r_g \cir e_g = \id_{\U_g} \,, \\{}\\[-.7em]
  e_1 = \eta \,, & r_1 = \Frac1{{\rm dim}(A)}\, \eps \,. \end{array}  \labl{e1r1}

\paragraph{Fixed points.} For a simple object $U$ with $\U_g\oti U \,{\cong}\,U$
 -- a fixed point of $g$ -- bases of the relevant morphism spaces are denoted by
  \be  
  \tildeb_g(U) \,\in\,  \Hom(U\oti\U_g ,U)  \qquad{\rm and}\qquad
  \bb_g(U) \,\in\, \Hom(\U_g \oti U ,U) \,.  \ee
The corresponding dual basis vectors
  \be
  \overline{\tildeb_g(U})\,\in\,\Hom(U,U \oti \U_g) \qquad{\rm and}\qquad
  \overline{\bb_g(U})\,\in\,\Hom(U,\U_g \oti U) \; \ee
are then uniquely determined via
  \be
  \tildeb_g(U)\cir\overline{\tildeb_g(U})\eq \id_U
  \qquad{\rm and}\qquad
  \bb_g(U)\cir\overline{\bb_g(U})\eq \id_U ~.
  \ee

\subsection{Pentagon, hexagons, and abelian three-co\-cycles}\label{app:PH-ab3}

In this appendix we compare the pentagon and the two hexagon
identities for the fusion and braiding matrices $\FF$, $\RR$
to the corresponding conditions for $(\psi,\Omega)$ to be
an abelian three-co\-cycle.

We first spell out the pentagon and the two hexagons 
for a ribbon category in which $\dim\Hom(U_i \oti U_j, U_k) \iN \{0,1\}$
for all triples of simple objects $U_i$, $U_j$ and $U_k$.
This holds in particular for theta-categories, which is the case we are 
interested in. By the definition of the 6j-symbols \FF\ (see (I:2.36)) we have
(compare also appendix II:A.1),
  \begin{eqnarray}
  \begin{picture}(400,70)(15,0)
  \put(0,0)      {\includeourbeautifulsmallpicture{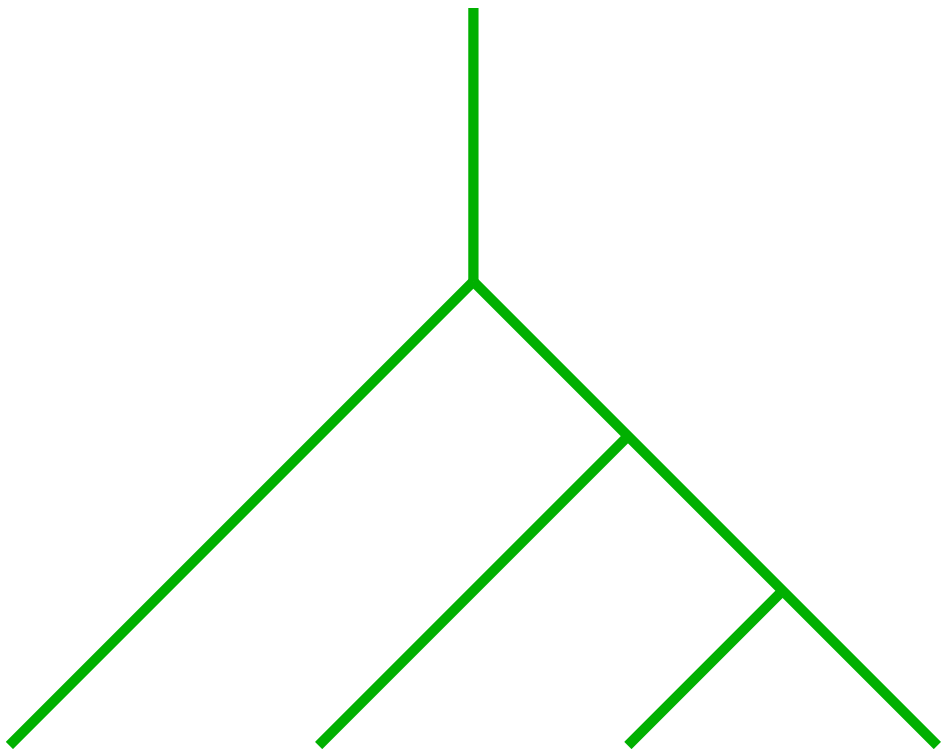}} 
  \put(0,0)      {\fusiYaYbYcd ijkl m pq} 
  \put(93,29)    {$ =\ \displaystyle\sum_r \Fs jkl p qr $}
  \put(163,0)    {\includeourbeautifulsmallpicture{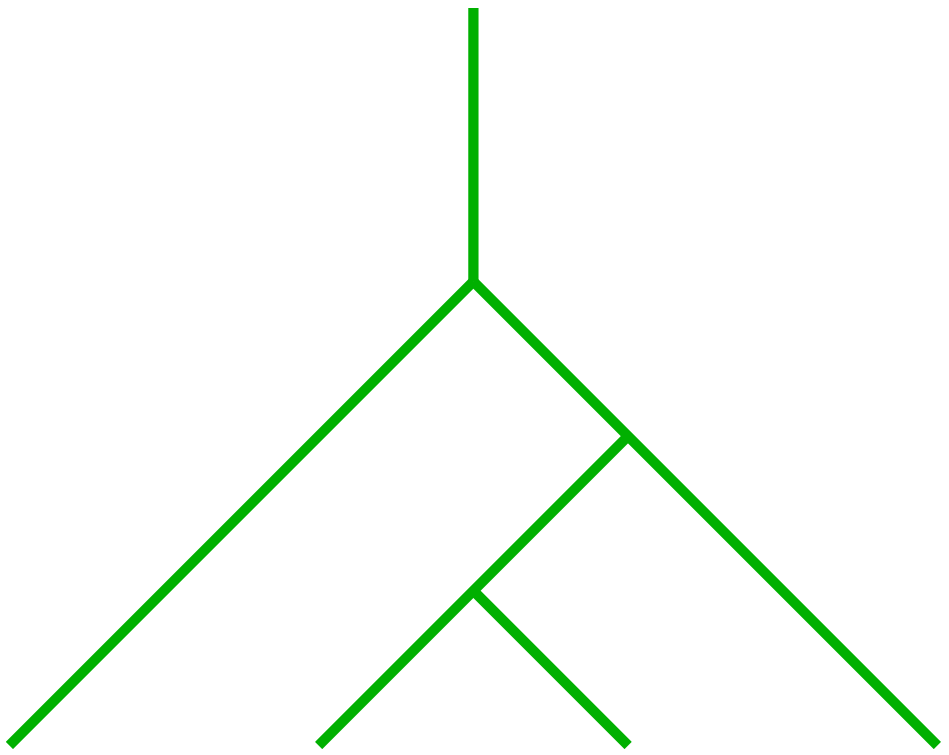}} 
  \put(163,0)    {\fusiYaYbcd ijkl m pr} 
  \put(249,29)   {$ =\ \displaystyle\sum_{r,s} \Fs jkl p qr\; \Fs irl m ps $}
  \put(368,0)    {\includeourbeautifulsmallpicture{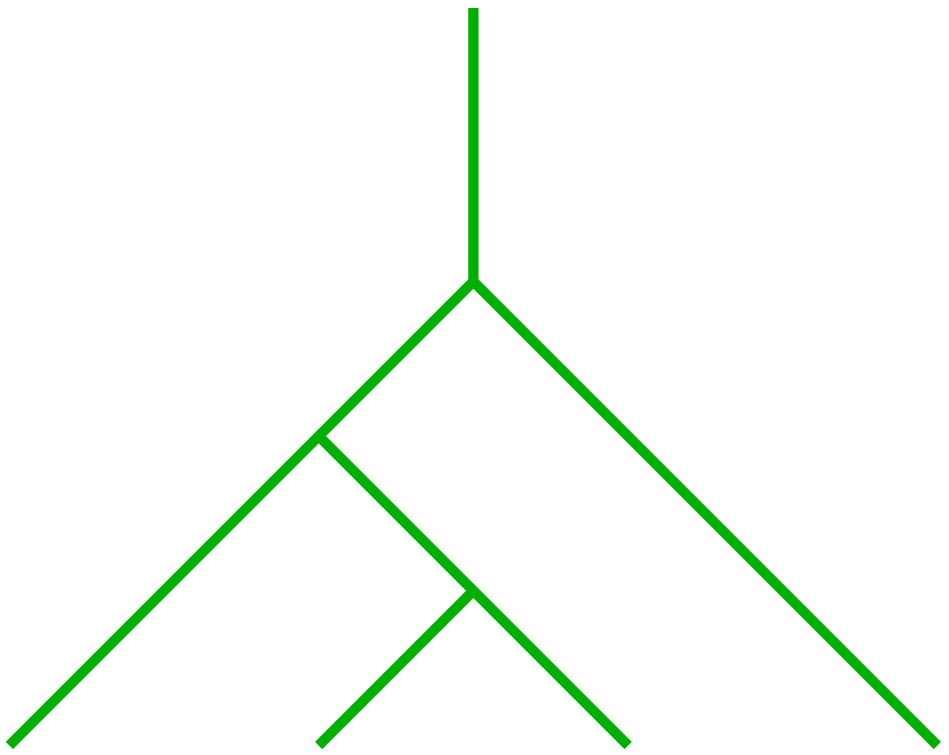}} 
  \put(368,0)    {\fusiYabcYd ijkl m sr}
  \end{picture}  \nonumber\\
  \begin{picture}(242,61)(0,29) 
  \put(0,29)     {$ =\ \displaystyle\sum_{r,s,t} \Fs jkl p qr\;
                                  \Fs irl m ps\; \Fs ijk s rt $}
  \put(155,0)    {\includeourbeautifulsmallpicture{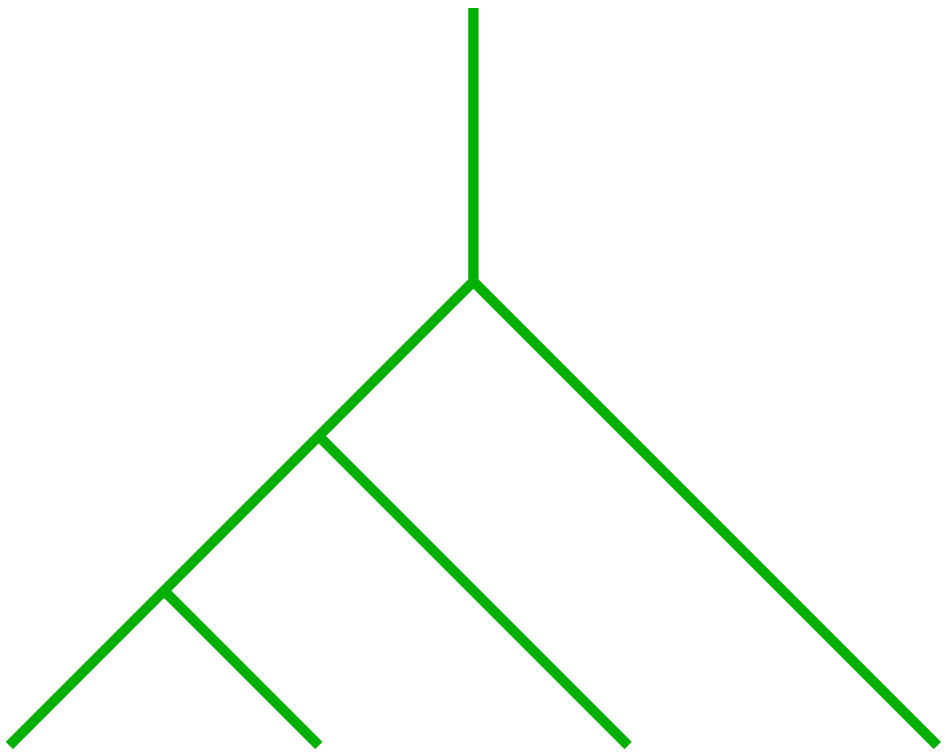}} 
  \put(155,0)    {\fusiYabYcYd ijkl m ts} 
  \end{picture}  
  \label{pentagon3} \\[.5em]{}\nonumber \end{eqnarray}
as well as
  \begin{eqnarray}
  \begin{picture}(400,70)(15,0)
  \put(0,0)      {\includeourbeautifulsmallpicture{FigFus_1_2_34}}
  \put(0,0)      {\fusiYaYbYcd ijkl m pq}
  \put(93,29)    {$ =\ \displaystyle\sum_t \Fs ijq m pt $}
  \put(163,0)    {\includeourbeautifulsmallpicture{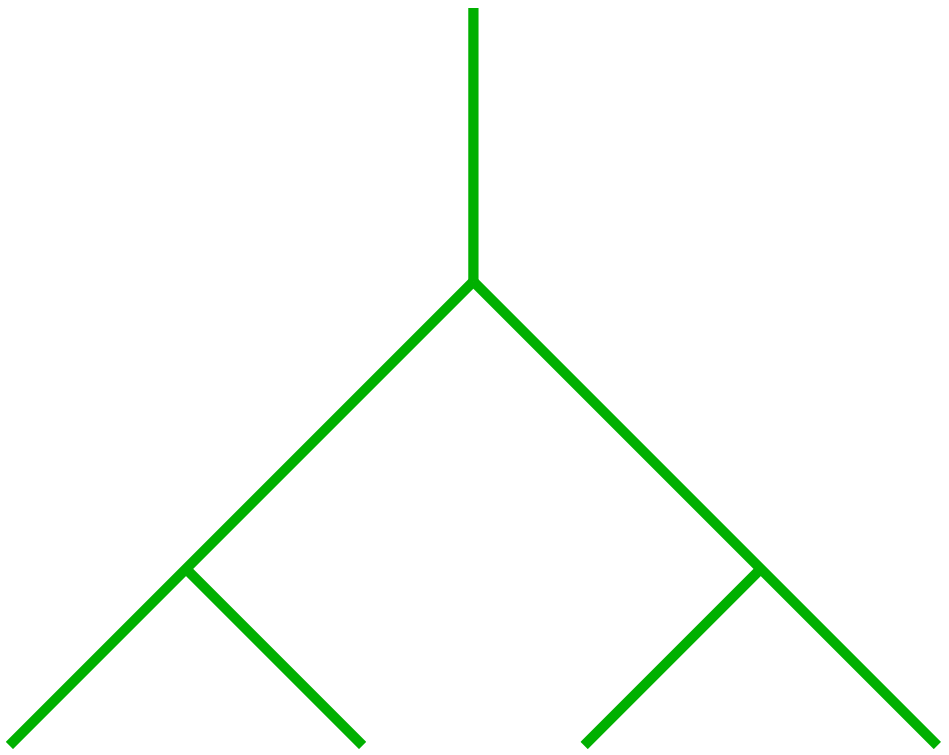}}
  \put(163,0)    {\fusiYabYcd ijkl m tq}
  \put(249,29)   {$ =\ \displaystyle\sum_{r,t} \Fs ijq m pt\; \Fs tkl m qs $}
  \put(368,0)    {\includeourbeautifulsmallpicture{FigFus_12_3_4}}
  \put(368,0)    {\fusiYabYcYd ijkl m ts}
  \end{picture}  \nonumber\\[1.1em]{}  
  \label{pentagon2}
  \\[-.97em]{}\nonumber \end{eqnarray}
Comparison yields the pentagon identity
  \be
  \sum_{r} \Fs jkl p qr\, \Fs irl m ps\, \Fs ijk s rt
  = \Fs ijq m pt\, \Fs tkl m qs \,.  \labl{pentagon}

Similarly, using also the definition of the braiding matrices \RR\ 
(see (I:2.41)) we have,
  \begin{eqnarray}
  \begin{picture}(400,76)(15,0)
  \put(0,0)      {\includeourbeautifulsmallpicture{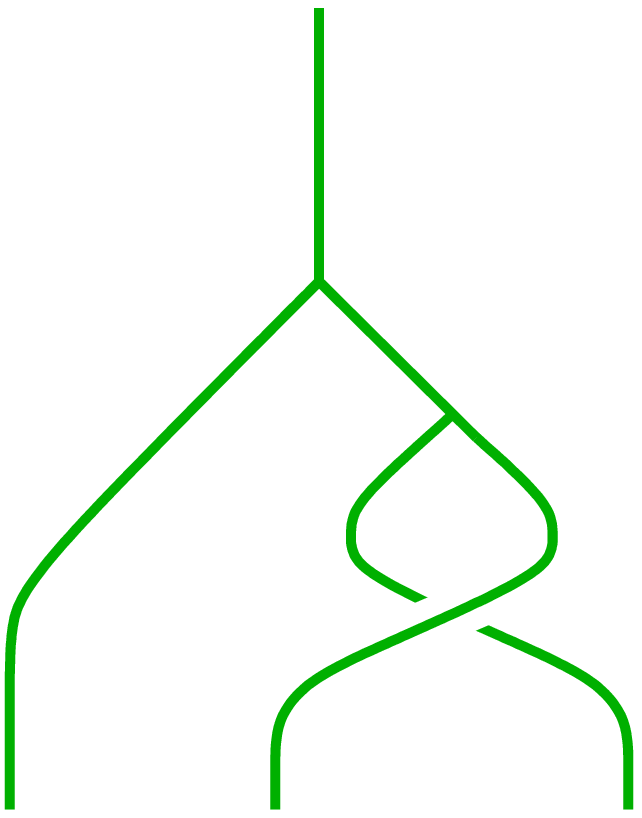}}
  \put(0,0)      {\brdiYaYbc ijk l p}
  \put(67,29)    {$ =\ \Rss jk p $}
  \put(120,0)    {\includeourbeautifulsmallpicture{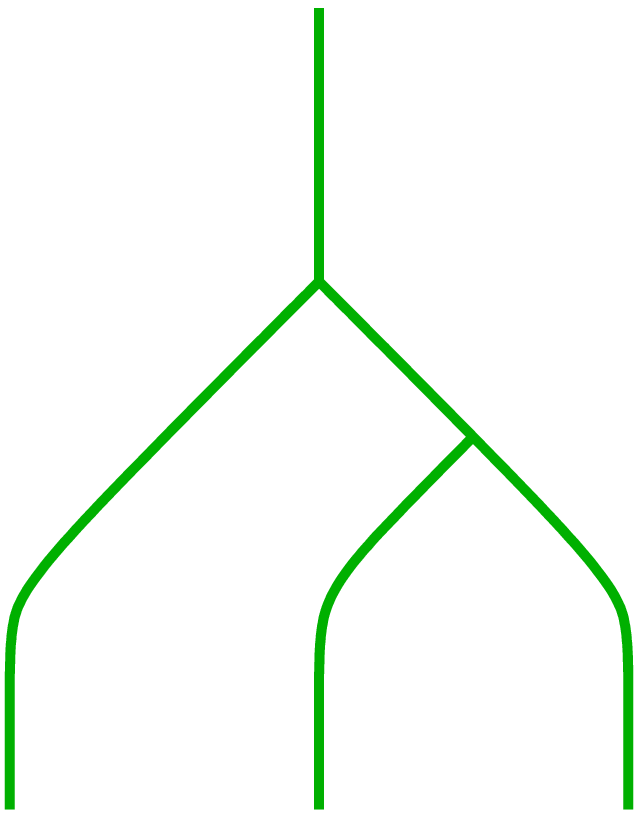}}
  \put(120,0)    {\fusiYaYbc ijk l p}
  \put(188,29)   {$ =\ \displaystyle\sum_q \Rss jk p\; \Fs ijk l pq $}
  \put(300,0)    {\includeourbeautifulsmallpicture{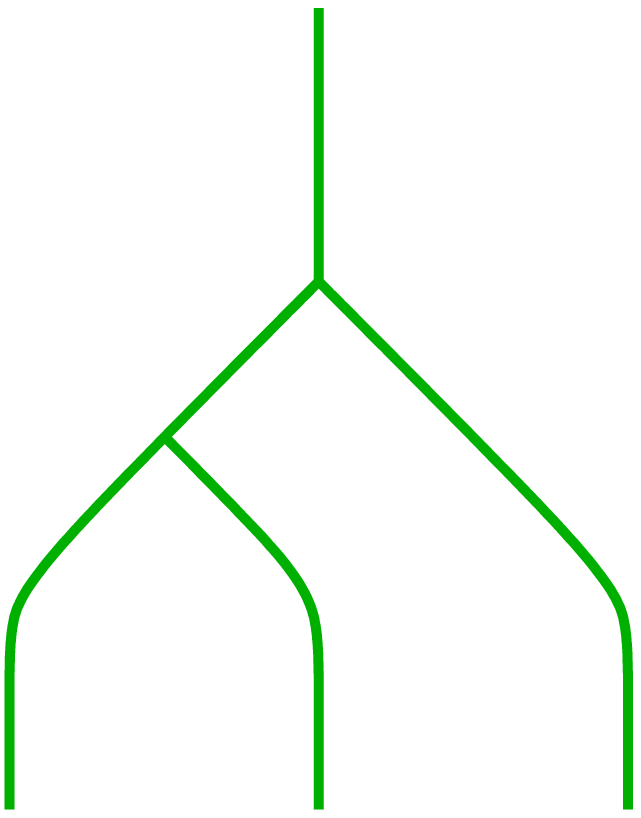}}
  \put(300,0)    {\fusiYabYc ijk l q}
  \end{picture}  \nonumber\\
  \begin{picture}(242,54)(0,29) 
  \put(0,29)     {$ =\ \displaystyle\sum_q \Rss jk p\;
                          \Fs ijk l pq\; \llb \Rsm ij q {\lrb}^{\!-1}_{} $}
  \put(172,0)    {\includeourbeautifulsmallpicture{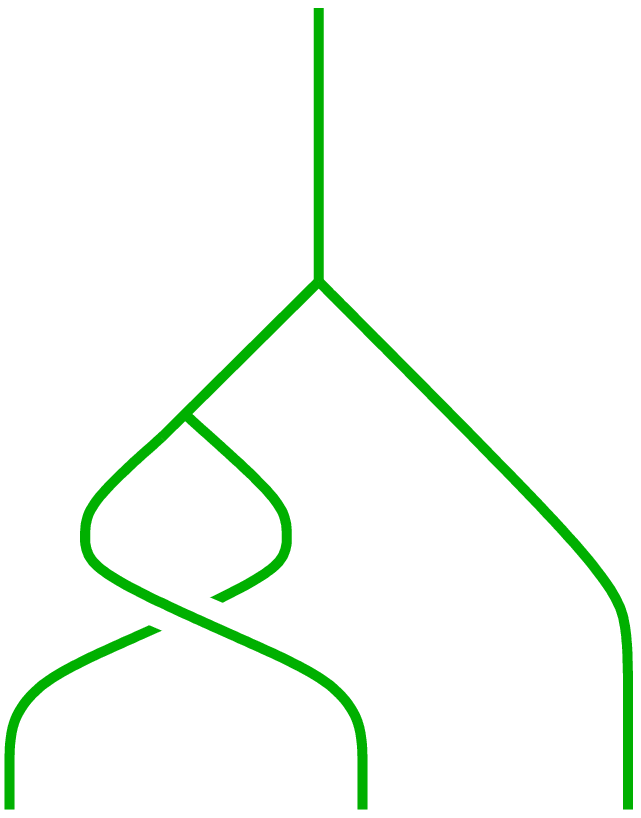}}
  \put(172,0)    {\brdiYabYc ijk l q}
  \end{picture}  
  \label{hexagonA} \\[.8em]{}\nonumber \end{eqnarray}
and
  \begin{eqnarray}
  \begin{picture}(400,77)(15,0)
  \put(0,0)      {\includeourbeautifulsmallpicture{FigBrd_1_23}}
  \put(0,0)      {\brdiYaYbc ijk l p}
  \put(67,29)    {$ =\ \displaystyle\sum_r \Fs ikj l pr $}
  \put(150,0)    {\includeourbeautifulsmallpicture{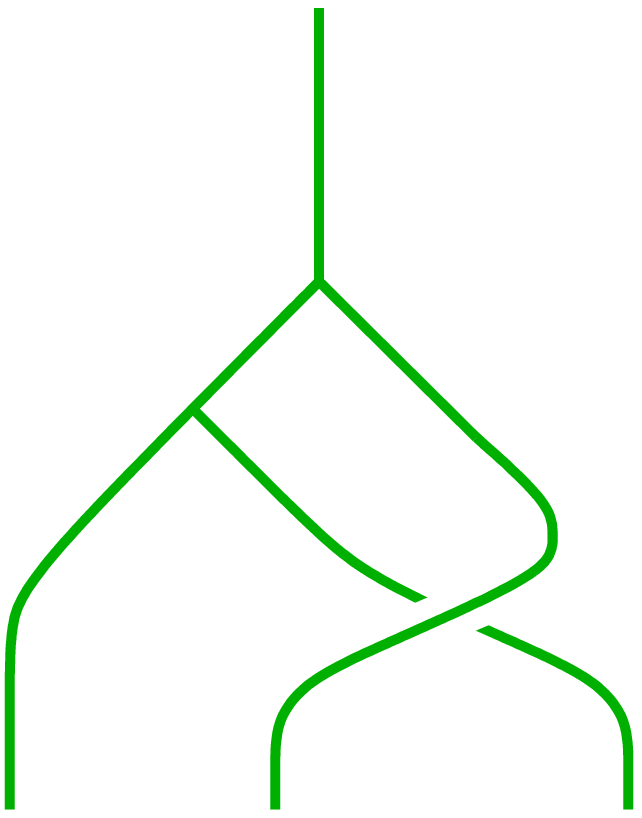}}
  \put(150,0)    {\brdiYacYb ijk l r}
  \put(217,29)   {$ =\ \displaystyle\sum_r \Fs ikj l pr\;
                                         \llb \Rsm rj l {\lrb}^{\!-1}_{} $}
  \put(355,0)    {\includeourbeautifulsmallpicture{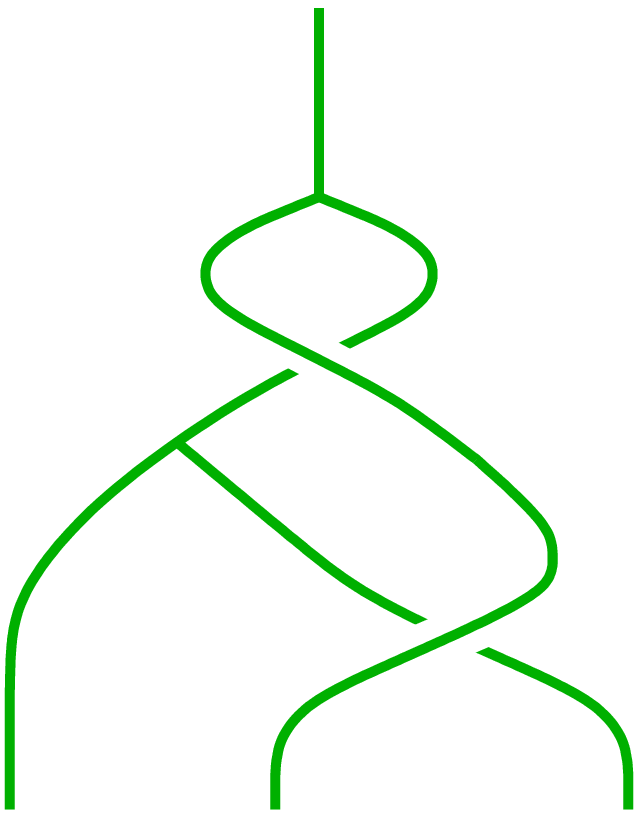}}
  \put(355,0)    {\brdiYbYac ijk l r}
  \end{picture}  \nonumber\\
  \begin{picture}(247,57)(0,30)
  \put(0,29)     {$ =\ \displaystyle\sum_{q,r} \Fs ikj l pr 
                    \llb \Rsm rj l {\lrb}^{\!-1}_{}\, \Fs jik l rq $}
  \put(177,0)    {\includeourbeautifulsmallpicture{FigBrd_12_3}}
  \put(177,0)    {\brdiYabYc ijk l q}
  \end{picture}  
  \label{hexagonB} \\[1.7em]{}\nonumber \end{eqnarray}
Using also that $\Rss mn t\, \Rsm nm t \eq 1$,
comparison yields the first of the two hexagon identities
  \be
  \Rss jk p\, \Fs ijk l pq\, \Rss ji q
  = \sum_r \Fs ikj l pr\, \Rss jr l\, \Fs jik l rq \,. 
  \labl{hexagon1}
The second hexagon identity is obtained by reversing the braiding in all
pictures. Explicitly,
  \be
  \llb \Rss kj p{\lrb}^{\!-1}_{}\, \Fs ijk l pq\, \llb \Rss ij q{\lrb}^{\!-1}
  = \sum_r \Fs ikj l pr\, \llb \Rss rj l{\lrb}^{\!-1}_{}\, \Fs jik l rq \,. 
  \labl{hexagon2}

\smallskip

Consider now the special case of a theta-category $\ircalc$ and denote 
$\Pic(\ircalc)$ by $G$. Then all simple objects are invertible, so that
the $r$-summations in \erf{pentagon}, \erf{hexagon1} and \erf{hexagon2} 
reduce to a single term. Two possible ways to define $(\psi,\Omega)$ in terms 
of $\FF$, $\RR$ are, first,
  \be 
  \Fs g{\,h}{\,k}{\,g{\cdot}h{\cdot}k}{h{\cdot}k}{\,\,g{\cdot}h} = \psi(g,h,k)
  \qquad {\rm and} \qquad \Rs{}g{\,h}{\,g{\cdot}h} = \Omega(h,g) \,,
  \ee 
and second,
  \be 
  \Fs g{\,h}{\,k}{\,g{\cdot}h{\cdot}k}{h{\cdot}k}{\,\,g{\cdot}h}
  = \psi(g,h,k)^{-1}_{}
  \qquad {\rm and} \qquad \Rs{}g{\,h}{\,g{\cdot}h} = \Omega(h,g)^{-1}_{}
  \,.  \labl{eq:FR-PsiOm-2}
Here $g,h,k \iN G$, and for better readability we spelled out the product 
`$\cdot$' of $G$ explicitly. One can now verify that for both choices 
the conditions $\rmd \psi\eq 1$ and \erf{app1neu} for
$(\psi,\Omega)$ to be an abelian 3-cocycle are equivalent
to the pentagon \erf{pentagon} and the two hexagon identities
\erf{hexagon1}, \erf{hexagon2}. 

In the definition of the theta-category $\calc(G,\psi,\Omega)$ in lemma \ref{xl12}
we have selected the second possible identification, i.e.\ \erf{eq:FR-PsiOm-2}.

\vskip 6em
\noindent
{\bf Acknowledgments.}\\[.2em] 
We are grateful to Terry Gannon for correspondence.
I.R.\ is supported by the DFG project KL1070/2-1.
C.S.\ is grateful to LPTHE, Universit\'e Paris 6 for hospitality while
part of this paper was written.  \\[.2em]
We dedicate this paper to Bert Schellekens, pioneer of simple currents, on 
the occasion of his 50th birthday.

 
\newcommand\wb{\,\linebreak[0]} \def\wB {$\,$\wb}
 \newcommand\Bi[1]    {\bibitem{#1}}
 \newcommand\J[5]     {{\sl #5\/}, {#1} {#2} ({#3}) {#4} }
 \newcommand\K[6]     {{\sl #6\/}, {#1} {#2} ({#3}) {#4}}
 \newcommand\Prep[2]  {{\sl #2\/}, pre\-print {#1}}
 \newcommand\BOOK[4]  {{\sl #1\/} ({#2}, {#3} {#4})}
 \newcommand\inBO[7]  {{\sl #7\/}, in:\ {\sl #1}, {#2}\ ({#3}, {#4} {#5}), p.\ {#6}}
 \newcommand\vyp[3]   {{#1} ({#2}) {#3} }
 \def\A     {Algebra }
 \def\dim   {dimension}
 \def\jf    {J.\ Fuchs}
 \def\adma  {Adv.\wb Math.}
 \def\anop  {Ann.\wb Phys.}
 \def\cocm  {Com\-mun.\wb Con\-temp.\wb Math.}
 \def\coia  {Com\-mun.\wB in\wB Algebra}
 \def\coma  {Con\-temp.\wb Math.}
 \def\comp  {Com\-mun.\wb Math.\wb Phys.}
 \def\fiic  {Fields\wB Institute\wB Commun.}
 \def\gatm  {Geom.\wB and\wB Topol.\wb Monogr.} 
 \def\hhaa  {Homol.\wb Homot.\wb Appl.}   
 \def\ijmb  {Int.\wb J.\wb Mod.\wb Phys.\ B}
 \def\ijmp  {Int.\wb J.\wb Mod.\wb Phys.\ A}
 \def\jams  {J.\wb Amer.\wb Math.\wb Soc.}
 \def\jlms  {J.\wB London\wB Math.\wb Soc.}
 \def\joal  {J.\wB Al\-ge\-bra}
 \def\josp  {J.\wb Stat.\wb Phys.}
 \def\jpaa  {J.\wB Pure\wB Appl.\wb Alg.}
 \def\maan  {Math.\wb Annal.}
 \newcommand\nqma[2] {\inBO{Non-perturbative QFT Methods and Their Applications}
              {Z.\ Horv\'ath and L.\ Palla, eds.} \WS\Si{2001} {{#1}}{{#2}} }
 \newcommand\nspq[2] {\inBO{ 
            New Symmetry Principles in Quantum Field Theory}
            {J.\ Fr\"ohlich et al., eds.} \PL\NY{1992} {{#1}}{{#2}}}
 \def\nupb  {Nucl.\wb Phys.\ B} 
 \def\pajm  {Pa\-cific\wB J.\wb Math.}
 \def\pams  {Proc.\wb Amer.\wb Math.\wb Soc.}
 \def\phlb  {Phys.\wb Lett.\ B}
 \def\phrd  {Phys.\wb Rev.\ D}
 \def\phrl  {Phys.\wb Rev.\wb Lett.}
 \def\pnas  {Proc.\wb Natl.\wb Acad.\wb Sci.\wb USA}
 \def\q     {quantum }
 \def\Q     {Quantum }
 \def\rims  {Publ.\wB RIMS}
 \def\rvmp  {Rev.\wb Math.\wb Phys.}
 \def\slnm  {Sprin\-ger\wB Lecture\wB Notes\wB in\wB Mathematics}
 \newcommand\Slnm[1] {{\rm[\slnm\ #1]}}
 \def\taac  {Theory\wB and\wB \wB Appl.\wb Cat.}
 \def\trgr  {Trans\-form.\wB Groups}
 \def\AMS    {{American Mathematical Society}}
 \def\IPC    {{International Press Company}}
 \def\NH     {{North Holland Publishing Company}}
 \def\PL     {{Plenum Press}}
 \def\PUP    {{Princeton University Press}}
 \def\SV     {{Sprin\-ger Ver\-lag}}
 \def\WS     {{World Scientific}}
 \def\Ad     {{Amsterdam}}
 \def\Be     {{Berlin}}
 \def\PR     {{Providence}}
 \def\pR     {{Princeton}}
 \def\Si     {{Singapore}}
 \def\NY     {{New York}}

\newpage

\end{document}